\documentclass[
amsmath,amssymb,
prd,nofootinbib,floatfix,12pt,
]{revtex4}

\usepackage{amssymb,amsmath,amsbsy}
\usepackage[dvips]{color}
\usepackage{xcolor}
\usepackage{dsfont}
\usepackage{mathrsfs}
\usepackage{float}

\def\signofmetric{0}

\definecolor{Red}{cmyk}{0,1,1,0}
\definecolor{BrickRed}{cmyk}{0,0.89,0.94,0.28}
\definecolor{Blue}{cmyk}{1,1,0,0}
\definecolor{Green}{cmyk}{1,0,1,0 }

\if0\signofmetric
\def\BDpos{}

\def\BDminus{-}
\fi

\if1\signofmetric
\def\BDpos{-}

\def\BDminus{+}
\fi

\if2\signofmetric
\def\BDpos{{\color{Red}\oplus}}

\def\BDminus{{\color{Blue}\ominus}}
\fi

\if3\signofmetric
\def\BDpos{{\color{Red}\ominus}}

\def\BDminus{{\color{Blue}\oplus}}
\fi

\newcommand\metric{\eta}
\def\newcdot{\kern.06em{\cdot}\kern.06em}
\newcommand\beq{\begin{eqnarray}}
\newcommand\eeq{\end{eqnarray}}

\def\lsim{\mathrel{\rlap{\lower4pt\hbox{$\sim$}}
    \raise1pt\hbox{$<$}}}                
\def\gsim{\mathrel{\rlap{\lower4pt\hbox{$\sim$}}
    \raise1pt\hbox{$>$}}}            

\allowdisplaybreaks
\usepackage{graphicx}
\usepackage{graphics}
\usepackage{setspace}

\newcommand{\nocontentsline}[3]{}
\newcommand{\tocless}[2]{\bgroup\let\addcontentsline=\nocontentsline#1{#2}\egroup}

\begin{document}
\renewcommand{\theequation}{\arabic{section}.\arabic{equation}}
\renewcommand{\thefigure}{\arabic{section}.\arabic{figure}}
\renewcommand{\thetable}{\arabic{section}.\arabic{table}}

\title{\large \baselineskip=20pt 
Signal-background
interference for digluon resonances\\ at the Large Hadron Collider}

\author{Prudhvi N.~Bhattiprolu and Stephen P.~Martin}
\affiliation{\it
Department of Physics, Northern Illinois University, DeKalb IL 60115}

\begin{abstract}\normalsize \baselineskip=15.5pt
We study the interference between the amplitudes for $gg \rightarrow X \rightarrow gg$, 
where $X$ is a new heavy digluon resonance, and the QCD background $gg \rightarrow gg$, 
at the Large Hadron Collider. The interference produces a large low-mass tail and 
a deficit of events above the resonance mass, compared to the naive pure resonance peak. 
For a variety of different resonance quantum numbers and 
masses, we evaluate the signal-background interference contribution 
at leading order, including showering, hadronization, and detector effects. The resulting new physics dijet mass 
distribution may have a shape that appears, after QCD background fitting and subtraction, to resemble 
an enhanced peak, a shelf, a peak/dip, or even a pure dip. We argue 
that the true limits on new digluon resonances are 
likely to differ significantly from the limits obtained when interference is neglected, 
especially if the branching ratio to $gg$ is less than 1. \end{abstract}

\maketitle
\vspace{-1.1cm}

\makeatletter
\def\l@subsubsection#1#2{}
\makeatother

\baselineskip=18pt

\tableofcontents

\baselineskip=15.2pt

\setcounter{footnote}{1}
\setcounter{figure}{0}
\setcounter{table}{0}
\newpage

\section{Introduction\label{sec:introduction}}
\setcounter{equation}{0}
\setcounter{figure}{0}
\setcounter{table}{0}
\setcounter{footnote}{1}

Understanding the dijet invariant mass spectrum provides an essential way to discover or set limits on 
certain types of new physics beyond the Standard Model (SM), 
as new particles could decay primarily to $gg$, $q\bar{q}$, $qq$, or $qg$. 
The most recent Large Hadron Collider (LHC) searches in the dijet channel 
can be found in \cite{CMSdijet2018,CMSdijet2019} (CMS) and \cite{ATLASdijet2017,ATLASdijet2019} (ATLAS),
based in part on strategies developed in 
earlier searches in \cite{Chatrchyan:2013qha}-\cite{Sirunyan:2016iap}.
Some models of phenomenological interest that yield dijet signals include the following.
Chiral color \cite{Pati:1975ze,Preskill:1980mz,Hall:1985wz,Frampton:1987dn,Frampton:1987ut,Bagger:1987fz,Frampton:alternatechiralcolor}, flavor-universal coloron \cite{Chivukula:1996yr,Simmons:1996fz,Bai:2010dj,Chivukula:2013xka,Chivukula:2013xla,Chivukula:2014rka,
Chen:2014haa,Chivukula:2015kua,Bai:2017zhj,Agrawal:2017ksf,Bai:2018jsr,Bai:2018wnt}, and certain supersymmetric models \cite{Martin:2019su3su3} predict the existence of massive color-octet gauge bosons, because of the embedding of QCD within a broken symmetry of $SU(3) \times SU(3)$ gauge group beyond the TeV scale. These massive gauge bosons are known as axigluons (axial-vectors) or colorons (vectors), which typically decay to a $q \bar{q}$ pair, or topgluons \cite{Hill:1991at,Martin:1992aq,Hill:1993hs,
Dobrescu:1997nm,Chivukula:1998wd}, which can preferentially couple to $t$ quarks and 
appear in topcolor and similar models of dynamical electroweak symmetry breaking.
Some $E_6$ grand unified theories predict diquarks \cite{Hewett:1989e6gut} which decay to $qq$.
Models with new electroweak gauge bosons $W^{\prime}$ and $Z^{\prime}$ \cite{Eichten:1984scp,Dobrescu:2013leptophobiczprime}, on top of the Standard Model, could also decay to a $q \bar{q}$ pair. 
Excited or composite quarks \cite{Baur:1987eqp,Baur:1987eqlp} could decay to $qg$ or $\bar{q}g$.
String Regge resonances \cite{Anchordoqui:2008strings, Cullen:2000strings} of the quark and the gluon could decay to a quark and a gluon.
Scalar color-octets, which occur in models of technicolor \cite{Hill:2003sdesb}, and universal extra dimensions \cite{Dobrescu:2007mcob,Dobrescu:2007lplhc}, can be considered as digluon resonances.
The Randall-Sundrum (RS) model \cite{Randall:1999rsmodel}, which provides a possible solution to the Planck scale hierarchy problem by adding extra dimensions, predicts RS gravitons which decay to gluon pairs or quark pairs.
The Kaluza-Klein (KK) states interact with the Standard Model fields, thus the KK gravitons could contribute to the dijet spectrum \cite{Han:1998}.
Also, models with dark matter mediators \cite{Chala:2015csd,Abercrombie:2019dmbm,Abdallah:2015smdm,Busoni:2016rplhc} which also couple to quarks, predict dijet signatures.
A general classification and study of dijet resonances for the LHC has been given in 
\cite{Han:2010rf}.

One of the biggest challenges in limiting or discovering a new dijet resonance is 
dealing with the huge QCD background, which is imperfectly known but should be 
a smooth function of the invariant mass in the range appropriate for new physics searches.
In order to effectively deal with the background, for lower masses 
CMS uses the data scouting \cite{Mukherjee:2017datascouting} technique to reconstruct 
or save only the crucial information to do analyses, thus allowing them to record more events. 
In the CMS  and ATLAS
searches that set limits on dijet resonances, 
the interference between the resonant amplitude and the QCD background was not considered, which could have a significant impact on the experimental limits.
As we will see below, the interference effect means that dijet resonances need not necessarily produce a peak in the dijet mass distributions in the vicinity of their mass, especially once the QCD background
fitting and subtraction are implemented.
The effects of interference are likely to be most pronounced in the digluon channel, 
where the QCD background
amplitudes are large compared to the new-physics amplitudes. A preliminary study of this kind of 
interference effect for digluon resonances, done only at parton level with smearing and only
for the case of a spin-0 color-singlet with mass near 750 GeV (motivated in large part by an infamous possible diphoton signal that turned out to be a fluctuation), was performed in ref.~\cite{Martin:2016bgw}.

A similar, but much smaller, effect on the diphoton lineshape for the Higgs boson due to interference with
the quark-loop induced Standard Model amplitudes $gg \rightarrow \gamma\gamma$ has been studied in
refs.~\cite{Dicus:1987fk}-\cite{Cieri:2017kpq}. As noted in \cite{Martin:2012xc}, there is a shift in the diphoton mass peak which can eventually be observable, and can be used \cite{Dixon:2013haa} to bound the Higgs boson width. Another important case of 
interference involving Standard Model Higgs boson-mediated amplitudes
and the continuum contributions occur for the processes $gg \rightarrow ZZ$ and
$gg \rightarrow W^+ W^-$, which have been studied in refs.~\cite{Glover:1988fe,
Glover:1988rg,
Seymour:1995np,
Binoth:2006mf,
Accomando:2007xc,
Campbell:2011cu,
Kauer:2012ma,
Passarino:2012ri,
Kauer:2012hd,
Campanario:2012bh,
Bonvini:2013jha,
Caola:2013yja,
Campbell:2013una,
Chen:2013waa,
Kauer:2013qba,
Campbell:2013wga,
Englert:2014aca,
Campbell:2014gua,
Englert:2014ffa,
Logan:2014ppa,
Kauer:2015pma,
Li:2015jva}. In particular, as noted in \cite{Kauer:2012hd}, 
enhanced contributions occur for invariant masses far off the Higgs mass-shell,
despite its narrow width, and are reduced by the interference.
This effect can be, as shown in  \cite{Caola:2013yja,Campbell:2013una} (see refs.~\cite{Chen:2013waa,
Kauer:2013qba,
Campbell:2013wga,
Englert:2014aca,
Campbell:2014gua,
Englert:2014ffa,
Logan:2014ppa,
Kauer:2015pma,
Li:2015jva} for further important developments), 
and has been \cite{Khachatryan:2014iha,Aad:2015xua,Khachatryan:2015mma,Khachatryan:2016ctc,Aaboud:2018puo,Sirunyan:2019twz},
used to bound the Higgs width from studying $VV$ events in the invariant mass region far above the Higgs mass at the LHC.
Other aspects of resonance-continuum interferences as a probe of new physics at the LHC, with approaches similar or complementary to the present paper, have been given in
\cite{Wprimeinterference,Zprimeinterference,ttinterferences,Jung:2015etr,Jung:2015gta,Craig:2016iea,Kauer:2019qei}.
Also, reference~\cite{Choudhury:2011cg} discussed the importance of non-interference off-shell effects in spin-1 digluon resonances at hadron colliders.

In this paper, we consider digluon resonances of various spin and color quantum numbers, whose existence need not necessarily be justified by any particular model. We study the importance of the interference between the digluon resonant signal $g g \rightarrow X \rightarrow g g$ and the QCD background $g g \rightarrow gg$ when 
setting limits on the digluon resonances, where $X$ couples to gluon pairs by non-renormalizable operators, subject to QCD gauge invariance. Here and from now on, $X$ refers to any such digluon resonance. 

At leading order (LO) the interference terms change the naive Breit-Wigner resonance peak, at dijet invariant mass close to the resonance mass ($m_{jj} \approx M_X$), to a peak just below and a dip just above the resonance mass.
(As a caution, we note that a next-to-leading order (NLO) calculation with virtual 1-loop and real emission of an extra jet 
would provide a more realistic estimate; NLO effects can be quite significant for the interference in the analogous case of diphoton Higgs signal/background interference 
\cite{deFlorian:2013psa}-\cite{Coradeschi:2015tna}, especially
when there is an additional jet with high $p_T$.) 
The magnitude of these interference effects are dependent on the spin and the color structures of the digluon resonances.
The interference effects are studied for scalar and pseudo-scalar resonances in both singlet and octet color representations, spin-1 color-octets, and color-singlet massive gravitons.
Although the Landau-Yang theorem forbids the decay of a massive spin-1 particle into two on-shell photons, it does not forbid the decay of an odd parity massive color-octet spin-1 particle into a pair of on-shell gluons in a non-Abelian $SU(N_c)$ Yang-Mills theory \cite{Ma:2014oha,Chivukula:2013xla,Beenakker:2015mra,Cacciari:2015ela,Bai:2014fkl}, as in our 
case.\footnote{The Landau-Yang theorem also rules out the possibility of colored or colorless massive pseudo-vectors (i.e. even-parity spin-1 particles) decaying to a pair of on-shell massless gauge bosons.
In general, this selection rule does not forbid the decay of a massive spin-1 particle to two massless gauge bosons if one of the three bosons is off-shell \cite{Cacciari:2015ela,Bai:2014fkl}.}

The previous study \cite{Martin:2016bgw} mentioned above for a 750 GeV color-singlet 
spin-0 digluon resonance
used parton-level differential cross-sections which were then 
smeared by convolution with an assumed approximate detector 
response function.
Here, we repeat that type of analysis, but also obtain the detector-level dijet invariant mass distributions for both the signal process alone, and its interference with the QCD background using Monte Carlo event generators including showering and hadronization and detector simulation.
The latter method serves as a validation of the qualitative results obtained by the simpler method used in ref.~\cite{Martin:2016bgw}.

The rest of this paper is structured as follows. In Section~\ref{sec:models}, we introduce the considered models of various spin and color quantum numbers, along with their effective interaction Lagrangians.
To elucidate the importance of the interference effects, we then consider a few benchmark examples for various resonance masses $M_X=(750,1000,1500,2000,2500,3000)$ GeV, such that their resonant production cross sections are close to the present claimed exclusions of the CMS experiment as given in the most recent reported searches~\cite{CMSdijet2019,CMSdijet2018}.
In Section~\ref{sec:methods}, we discuss the methods and techniques used to obtain smeared parton-level 
and full event simulated results.
We then present the results for the considered benchmark models in section \ref{sec:results}.
We start by assuming that $X$ almost always decays to a pair of gluons. We then show 
in section \ref{sec:resultsforotherdecays} that the interference
effects are even more dramatic if the digluon resonance has other non-detectable decays 
contributing to its width. (These could include invisible or multi-jet final states from each $X$ decay.)
Finally, in Section~\ref{sec:outlook}, we conclude the paper by summarizing 
the significance of the signal/background interference for digluon resonances.

\section{Digluon resonances and benchmark models\label{sec:models}}
\setcounter{equation}{0}
\setcounter{figure}{0}
\setcounter{table}{0}
\setcounter{footnote}{1}
The models considered in this paper are described in the following.
In all the models, $X$ is assumed to couple to gluons with non-renormalizable operators invariant under QCD gauge transformations.
The couplings $c_i$ are dimensionless, possibly complex, form factors, and $\Lambda$ is a mass scale associated with new physics.
The effective form-factor couplings will be suppressed by masses of heavier particles, if the interaction is loop-induced.

\begin{figure}
\centering
\includegraphics[width=12cm]{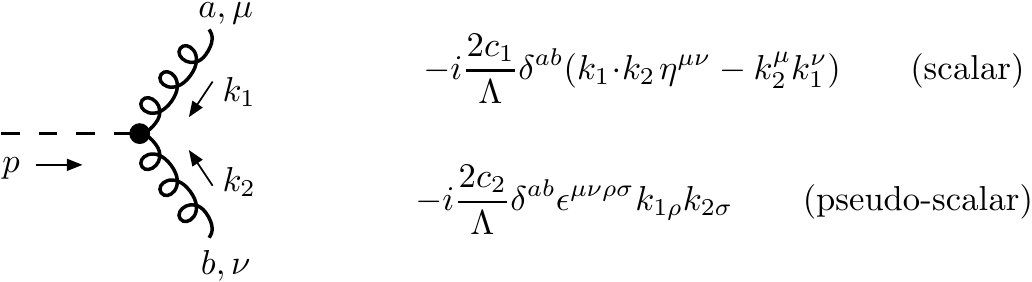}
\caption{Feynman rule for the effective coupling of parity-even (top) and parity-odd (bottom) spin-0 color-singlet resonance with a gluon pair, with $p^\mu + k_1^\mu + k_2^\mu = 0$. Here, $c_1$ and $c_2$ are dimensionless form factors, and $\Lambda$ is the mass scale associated with the new physics.}\label{fig:s0c1_feynrule}
\end{figure}

\paragraph*{Spin 0, color singlet:}
The effective Lagrangian for an even parity (scalar) resonance $X$ can be written as:
\beq
{\cal L} &=& \frac{c_{1}}{2 \Lambda} X F_{\mu\nu}^a F^{a\mu\nu}
,
\eeq
and for an odd parity (pseudo-scalar) resonance as:
\beq
{\cal L} &=& -\frac{c_{2}}{4 \Lambda} X \epsilon^{\mu\nu\rho\sigma}
F_{\mu\nu}^a F^{a}_{\rho\sigma}
, 
\eeq
where $F_{\mu\nu}^a = \partial_{\mu} A_{\nu}^a - \partial_{\nu} A_{\mu}^a - g_s f^{abc} A_{\mu}^b A_{\nu}^c$ is the QCD field strength tensor for $a,b,c = 1,2,\ldots, 8$, and $f^{abc}$ are the anti-symmetric structure constants of $SU(3)_c$, and $g_s$ is the strong coupling constant.
The corresponding Feynman rules for $X$-$g$-$g$ couplings, for both color-singlet scalar and pseudo-scalar, are shown in Figure~\ref{fig:s0c1_feynrule}.

\begin{figure}
\centering
\includegraphics[width=12cm]{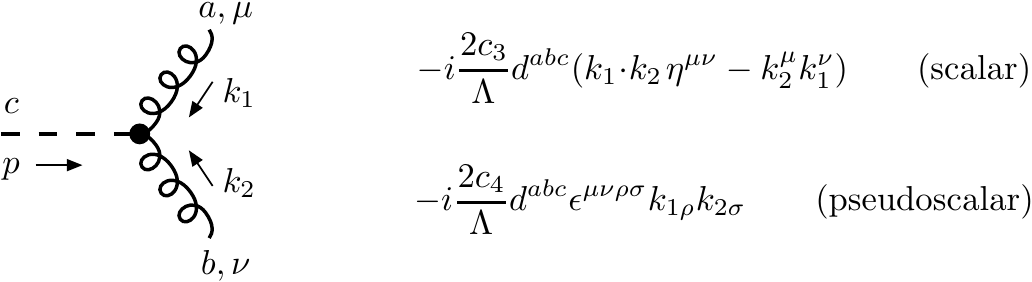}
\caption{Feynman rule for the effective coupling of parity-even (top) and parity-odd (bottom) spin-0 color-octet resonance with a gluon pair, with $p^\mu + k_1^\mu + k_2^\mu = 0$. Here, $c_3$ and $c_4$ are dimensionless form factors, and $\Lambda$ is the mass scale associated with the new physics.}\label{fig:s0c8_feynrule}
\end{figure}

\paragraph*{Spin 0, color octet:}
The effective Lagrangian for an even parity (scalar) resonance $X$ is:
\beq
{\cal L} &=& \frac{c_{3}}{2 \Lambda} d^{abc} X^c F_{\mu\nu}^a F^{b\mu\nu}
,
\eeq
and for an odd parity (pseudo-scalar) resonance it is:
\beq
{\cal L} &=& -\frac{c_{4}}{4\Lambda} d^{abc} X^c \epsilon^{\mu\nu\rho\sigma}
F_{\mu\nu}^a F^{b}_{\rho\sigma}
,
\eeq
where the symmetric anomaly coefficients of $SU(3)_c$ are defined as:
\beq
d^{abc} = 2 {\rm Tr} [\{T^a, T^b\} T^c],
\eeq
with the usual normalization for the fundamental representation matrices
\beq
{\rm Tr}[T^a T^b] = \frac{1}{2} \delta^{ab},
\eeq
so that
\beq
d^{abc} d^{abe} = \frac{N_c^2-4}{N_c} \,\delta^{ce}  = \frac{5}{3} \delta^{ce}
. 
\eeq
The corresponding Feynman rules for the effective couplings of color-octet scalar and pseudo-scalar with two gluons are shown in Figure~\ref{fig:s0c8_feynrule}.

\begin{figure}
\centering
\includegraphics[width=12cm]{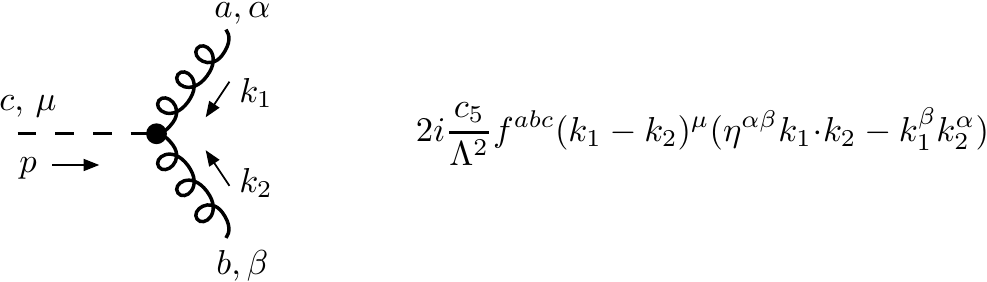}
\caption{Feynman rule for the effective coupling of a spin-1 color-octet resonance with a gluon pair, with $p^\mu + k_1^\mu + k_2^\mu = 0$. Here, $c_5$ is a dimensionless form factor, and $\Lambda$ is the mass scale associated with the new physics.}\label{fig:s1c8_feynrule}
\end{figure}

\paragraph*{Spin 1, color octet:}
The Landau-Yang theorem does not forbid the decay of a color-octet massive vector to an on-shell gluon pair \cite{Ma:2014oha,Chivukula:2013xla,Beenakker:2015mra,Cacciari:2015ela,Bai:2014fkl}. 
The effective Lagrangian describing the non-trivial coupling
of two gluons to a massive odd-parity spin-1, color-octet 
resonance $X^{a}_{\mu}$ is \cite{Cacciari:2015ela}:
\beq
{\cal L} &=& \frac{c_{5}}{\Lambda^2} f^{abc} (D_\mu X_\nu^a - D_\nu X_\mu^a)
F^{b\nu\rho} F^{c\mu}_\rho
\label{eq:Leffspin1octet}
\eeq
where $D_\mu$ is the gauge-covariant derivative.
(For the $X$-gluon-gluon 
interaction, only the ordinary partial derivative part of this is 
pertinent, so one can replace
$D_\mu$ by $\partial_\mu$.)
Note that dimensional analysis says that if we want $c_5$ to be dimensionless,
we now need $\Lambda^2$ in the denominator, where $\Lambda$ is the new physics scale. 
The Feynman rule for a massive color-octet vector coupling to a gluon pair is shown in Figure~\ref{fig:s1c8_feynrule}.
In the Feynman rule, the contributions proportional to $k_1^\alpha$ and $k_2^\beta$
have been dropped, as they never contribute to amplitudes because of
$\epsilon_1 \cdot k_1 = \epsilon_2 \cdot k_2 = 0$.
It should be noted that our treatment of the spin-1 color-octet is not the same as that of the
usual ``axigluon/coloron" models as in
\cite{Frampton:1987dn,Simmons:1996fz,Chivukula:2013xla}.
In those models, the spin-1 color-octet appears primarily as a $q\overline q$ resonance; 
although the $gg$ production channel is non-zero, 
in specific models it is 
small compared to the $q\overline q$ production channel. In this paper, we instead focus on the case
that the production of the resonance is mostly through a large coupling to $gg$.

\paragraph*{Spin 2, color singlet:}

The free Lagrangian for a massive spin-2 resonance $X^{\mu \nu}$, first derived by Markus Fierz and Wolfgang Pauli, can be written as \cite{schwartz:qft}:
\beq
{\cal L}_{f} = \frac{1}{2}X^{\mu \nu} \partial_{\alpha} \partial^{\alpha} X_{\mu \nu} - X^{\mu \nu} \partial_{\mu} \partial^{\alpha} X_{\nu \alpha} &+& X \partial_{\mu} \partial_{\nu} X^{\mu \nu} 
-\frac{1}{2} X \partial_{\mu} \partial^{\mu} X 
\notag
\\
&+& \frac{1}{2} M_X^2 \left [ X^{\mu \nu} X_{\mu \nu} - X^2 \right ],
\eeq
where $X = X^{\alpha}_{ \ \alpha}$.

\begin{figure}
\centering
\includegraphics[width=9cm]{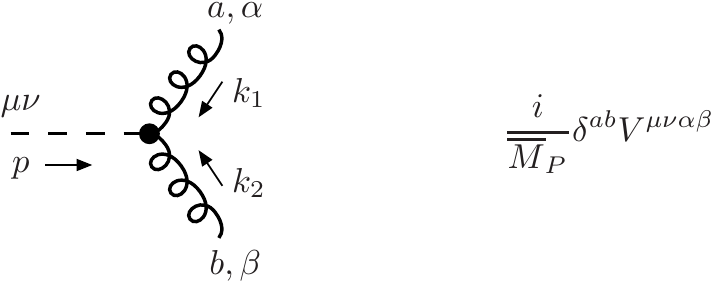}
\caption{Feynman rule for the effective coupling of spin-2 color-singlet resonance with a gluon pair, with $p^\mu + k_1^\mu + k_2^\mu = 0$. Here, $\overline{M}_P$ is a new mass scale, and $V^{\mu \nu \alpha \beta}$ is defined in eq.~(\ref{eq:s2c1_Vtensor}).}\label{fig:s2c1_feynrule}
\end{figure}

The effective interaction Lagrangian for an even parity color-singlet spin-2 resonance $X^{\mu\nu}$
(for example, a Kaluza-Klein (KK) or Randall-Sundrum (RS) graviton) can be written as \cite{Hagiwara:2008,Han:1998,Giudice:1998ck}:
\beq
{\cal L} &=& 
\BDpos \frac{1}{\overline M_P}  X^{\mu\nu} \left [ F_{\mu\rho}^a F^{a\rho}_\nu
- \frac{1}{4} \metric_{\mu\nu} F^a_{\rho\sigma} F^{a\rho\sigma} \right ]
,
\label{eq:Leffspin2}
\eeq
where $\overline{M}_P$ is a new mass scale.
The Feynman rule for the effective coupling of a massive color-singlet spin-2 particle with two gluons is shown in Figure~\ref{fig:s2c1_feynrule}, where
\beq
V^{\mu\nu\alpha\beta} &=& 
k_1 \hspace{-2pt}\cdot\hspace{-2pt} k_2
(\metric^{\mu\nu} \metric^{\alpha\beta} - \metric^{\mu\alpha} \metric^{\nu\beta} - 
\metric^{\nu\alpha} \metric^{\mu\beta}) - \metric^{\mu\nu} k_1^\beta k_2^\alpha
-(k_1^\mu k_2^\nu + k_2^\mu k_1^\nu) \metric^{\alpha\beta} 
\nonumber \\ && 
+ k_1^\mu k_2^\alpha \metric^{\nu\beta}
+ k_1^\nu k_2^\alpha \metric^{\mu\beta}
+ k_1^\beta k_2^\nu \metric^{\mu\alpha}
+ k_1^\beta k_2^\mu \metric^{\nu\alpha} .
\label{eq:s2c1_Vtensor}
\eeq
Note this satisfies tracelessness and QCD gauge invariance conditions:
\beq
\metric_{\mu\nu} V^{\mu\nu\alpha\beta} &=& 0,
\\
k_{1\alpha} V^{\mu\nu\alpha\beta} &=& 0,
\\
k_{2\beta} V^{\mu\nu\alpha\beta} &=& 0.
\eeq
One could also consider a theory where the second term (with $\metric_{\mu\nu}$) is
omitted from the Lagrangian eq.~(\ref{eq:Leffspin2}). In that case, the terms
containing $\metric_{\mu\nu}$ are removed from $V^{\mu\nu\alpha\beta}$,
and the tracelessness condition is not satisfied. One could also consider a general
linear combination of these terms. There are a few other terms
that could be written down, involving higher derivatives, but they are omitted from study here.

\begin{figure}
\centering
\includegraphics[width=8cm]{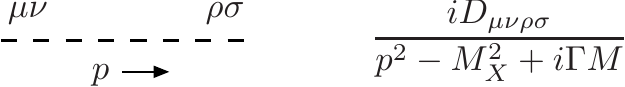}
\caption{Propagator for a massive spin-2 resonance. $D_{\mu \nu \rho \sigma}$ is defined in eq.~(\ref{eq:s2c1_Dtensor}).}\label{fig:s2c1_prop}
\end{figure}

The propagator for the massive spin-2 resonance is shown in Figure~\ref{fig:s2c1_prop},
where
\beq
D_{\mu\nu\rho\sigma} &=& \frac{1}{2} G_{\mu\rho} G_{\nu\sigma} +
\frac{1}{2} G_{\mu\sigma} G_{\nu\rho} - \frac{1}{3} G_{\mu\nu} G_{\rho\sigma} ,
\label{eq:s2c1_Dtensor}
\\
G_{\mu\nu} &=& \metric_{\mu\nu} \BDminus p_\mu p_\nu/M_X^2.
\label{eq:defG}
\eeq
The tensor in the numerator of the propagator  can be related to a basis for 
the five symmetric, traceless, orthonormal external state polarization tensors 
for the spin-2 particle, which can be written as:
\beq
&&\epsilon^{(1)}_{\mu\nu} = \frac{1}{\sqrt{2}} 
\begin{pmatrix}
0 & 0 & 0 & 0 \\[-6pt]
0 & 0 & 1 & 0 \\[-6pt]
0 & 1 & 0 & 0 \\[-6pt]
0 & 0 & 0 & 0 
\end{pmatrix}
,
\qquad
\epsilon^{(2)}_{\mu\nu} = \frac{1}{\sqrt{2}} 
\begin{pmatrix}
0 & 0 & 0 & 0 \\[-6pt]
0 & 0 & 0 & 1 \\[-6pt]
0 & 0 & 0 & 0 \\[-6pt]
0 & 1 & 0 & 0 
\end{pmatrix}
,
\qquad
\epsilon^{(3)}_{\mu\nu} = \frac{1}{\sqrt{2}} 
\begin{pmatrix}
0 & 0 & 0 & 0 \\[-6pt]
0 & 0 & 0 & 0 \\[-6pt]
0 & 0 & 0 & 1 \\[-6pt]
0 & 0 & 1 & 0 
\end{pmatrix}
,
\nonumber
\\
&&\epsilon^{(4)}_{\mu\nu} = \frac{1}{\sqrt{2}} 
\begin{pmatrix}
0 & 0 & 0 & 0 \\[-6pt]
0 & 1 & 0 & 0 \\[-6pt]
0 & 0 & -1 & 0 \\[-6pt]
0 & 0 & 0 & 0 
\end{pmatrix}
,
\qquad
\epsilon^{(5)}_{\mu\nu} = \frac{1}{\sqrt{6}} 
\begin{pmatrix}
0 & 0 & 0 & 0 \\[-6pt]
0 & 1 & 0 & 0 \\[-6pt]
0 & 0 & 1 & 0 \\[-6pt]
0 & 0 & 0 & -2 
\end{pmatrix} .
\eeq
These satisfy:
\beq
\epsilon_{\mu\nu}^{(i)}
\epsilon^{(j)\mu\nu}
&=& \delta^{ij}
\qquad\quad (i,j = 1,\ldots,5)
,
\eeq
and
\beq
\sum_{i=1}^5 \epsilon_{\mu\nu}^{(i)} \epsilon_{\rho\sigma}^{(i)} &=&
D_{\mu\nu\rho\sigma}
.
\eeq
which imply:
\beq
\epsilon_{\mu\nu}^{(i)}  D^{\mu\nu\rho\sigma}
&=& \epsilon^{(i){\rho\sigma}}
\qquad\quad (i = 1,\ldots,5) .
\eeq
For simplicity, we consider here only the case that the spin-2 resonance coupling to two partons is 
only (or mainly) to 
two gluons. 
A KK or RS graviton would also couple to $q \overline q$, which would 
result in a more complicated analysis. The interference effect would be smaller 
in those cases relative to the resonant dijet process, because the $q\overline q$
initial and final states for $X$ production and decay of course do not 
interfere with the large $gg \rightarrow gg$ QCD amplitude.

\begin{figure}[!tb]
  \begin{minipage}[]{0.8\linewidth}
    \includegraphics[width=13.0cm,angle=0]{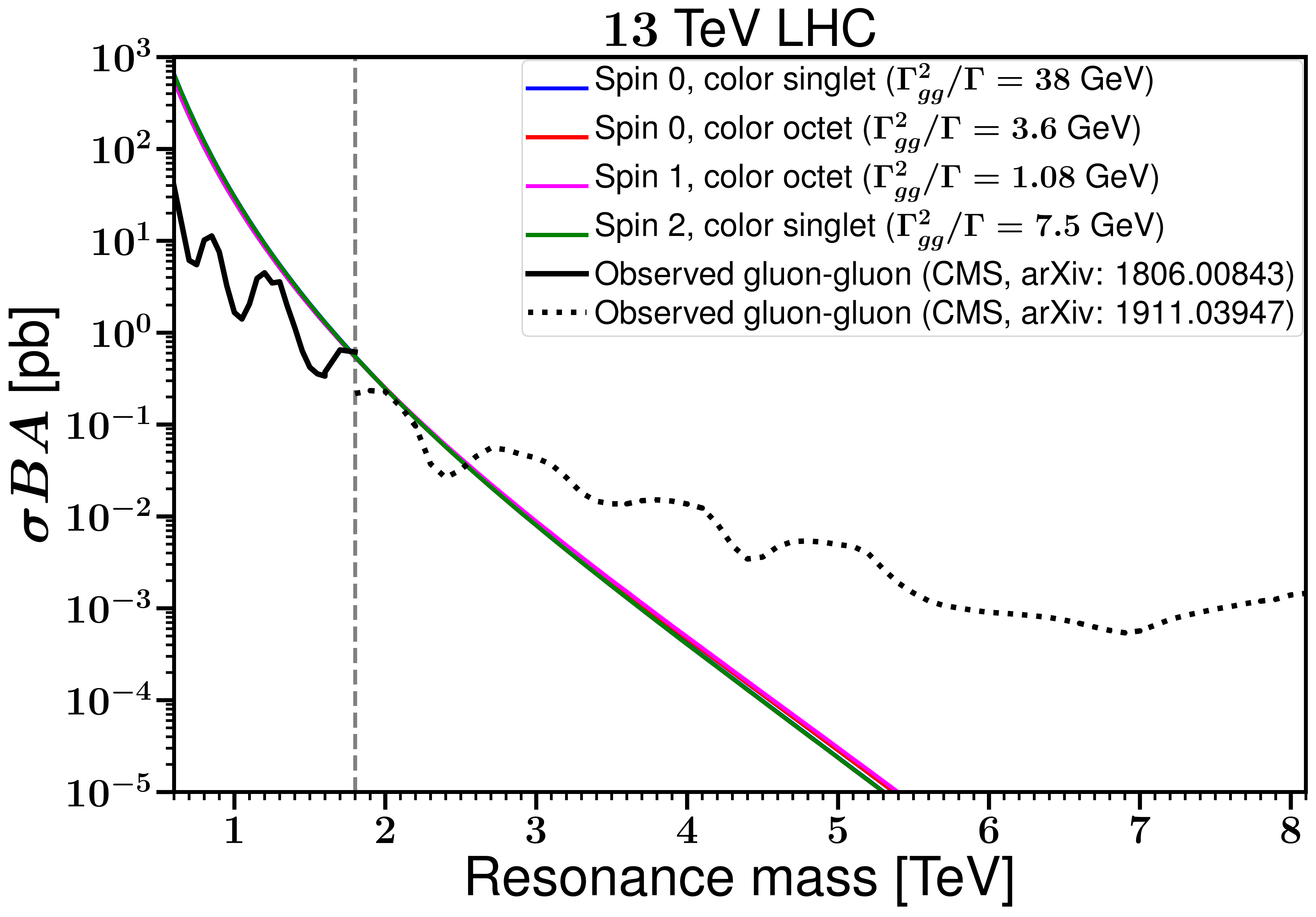}
  \end{minipage}
\begin{center}\begin{minipage}[]{0.95\linewidth}
\caption{\label{fig:benchmarks} 
The observed 95\% CL upper limits on the product of the cross section ($\sigma$), branching fraction ($B$), and acceptance ($A$) for dijet resonances decaying to a gluon pair from ref.~\cite{CMSdijet2019} (solid black line, left of vertical dashed gray line) and ref.~\cite{CMSdijet2018} (dotted black line, right of vertical dashed gray line). The solid colored lines show the leading order $s$-channel resonant cross sections, in the narrow-width approximation, for the benchmarks for a resonance mass of 2000 GeV 
in each of the considered models of digluon resonances.}
\end{minipage}\end{center}
\end{figure}

We chose benchmark models as specified in Table~\ref{tab:benchmarks} such that the $s$-channel 
resonant-only cross sections are close to the claimed exclusions by 
CMS in refs.~\cite{CMSdijet2019,CMSdijet2018}. 
Specifically, for the resonance masses $M_X=(750, 1000, 1500, 2000, 2500, 3000)$ GeV, the observed 95\% CL upper limits by CMS in refs.~\cite{CMSdijet2019,CMSdijet2018} are (5.5, 1.66, 0.42, 0.22837, 0.032155, 0.043386)  pb. In Table~\ref{tab:benchmarks}, $\Gamma_{gg}^2/\Gamma M_X$ increases as we move to higher masses, except from  $M_X = 2000$ GeV to 2500 GeV; this is because the observed limit at $M_X = 2500$ GeV is evidently a downward fluctuation compared to the limits obtained at $M_X = 2000$ and 3000 GeV.
For the $s$-channel resonant cross sections, we chose a $K$ factor of 1.5 and 
an acceptance of 0.5 for the purposes of choosing the benchmarks.
The claimed limits as a function of mass, and the predicted resonance cross-sections 
for an example $M_X = 2000$ GeV, are shown in Figure~\ref{fig:benchmarks}. 
For our first set of studies below, 
we chose the digluon partial width of $X$ (denoted in this paper as $\Gamma_{gg}$) 
to be the same as the total width ($\Gamma$). More generally, 
if $X$ can decay to final states that are more difficult to detect for some
reason or are just not part of the dijet search
(for example, $X \rightarrow jjj$ or $X \rightarrow jjjj$ or $X \rightarrow$ invisibles), then the total width 
$\Gamma$ might exceed
$\Gamma_{gg}$, and
${\rm BR}(X \rightarrow gg) = \Gamma_{gg}/\Gamma$. 

The  resonance partonic total cross-section
(after angular integration, and with no cuts) in the narrow-width approximation is equal to
\beq
\hat\sigma(gg \rightarrow X \rightarrow gg) &=& (2j+1) k \frac{\Gamma_{gg}^2}{\Gamma M_X} \frac{\pi^2}{4} \delta(\hat s - M_X^2)
,
\label{eq:sigmanarrowwidth}
\eeq
where $j = 0,1,2$ and $k=1,8$ are the spin and color of $X$, and $\sqrt{\hat s}$ is the partonic invariant mass.
This can be checked as a limit in each of the special case results of the next section, and 
reflects the fact that there are $(2j+1)k$ times more 
$spin \otimes color$ states for a spin $j$, color $k$ resonance than for a spin-0, color-singlet.
Therefore, for our models with $j\not= 0$ and/or $k\not=1$, 
we chose benchmark points that have $\Gamma_{gg}^2/\Gamma$ 
approximately $(2j+1)k$ times smaller than for a spin-0, color-singlet of the same mass, 
in order to maintain (roughly) the same resonant total cross section.

\begin{table}
\caption{Our choice of benchmark masses and widths for the digluon resonances in four ($J$=spin, color) representations considered in this paper. Here $M_X$, $\Gamma$, and $\Gamma_{gg}$ are the mass, total width, and 
 digluon partial width of $X$, respectively. 
\label{tab:benchmarks}}
\begin{center}
\begin{tabular}{|c|c|c|c|c|}
\hline
 
 ~~Resonance Mass~~   & \multicolumn{4}{c|}{$\Gamma_{gg}^2 / \Gamma M_X$}\\
\cline{2-5} 
~~$M_X$ (GeV)~~  &  ~~$J=0$, singlet~~ & ~~$J=0$, octet~~ & ~~ $J=1$, octet ~~& ~~$J=2$, singlet~~ \\
\hline
750  &  0.0015 &    0.00016 &   0.00005  &  0.0003  \\
\hline
1000 &  0.002  &    0.0002  &   0.000065 &  0.00041 \\
\hline
1500 &  0.005  &    0.0005  &   0.00015  &  0.001   \\
\hline
2000 &  0.019  &    0.0018  &   0.00054  &  0.00375 \\
\hline
2500 &  0.0108 &    0.001   &   0.0003   &  0.0022  \\
\hline
3000 &  0.07   &    0.006   &   0.00183  &  0.014   \\
\hline
\end{tabular}
\end{center}
\end{table}

\newpage
\section{Signal-background interference: parton-level and Monte Carlo methods \label{sec:methods}}
\setcounter{equation}{0}
\setcounter{figure}{0}
\setcounter{table}{0}
\setcounter{footnote}{1}

\subsection{Parton-level approximation} \label{subsec:partonlevel}
For both the resonant signal $g g \rightarrow X \rightarrow g g$ and the continuum QCD background $g g \rightarrow g g$ processes, we can write an amplitude, for each choice of external gluon polarizations, in terms of a redundant basis of color combinations:
\beq
{\cal A}^{abcd} &=& 
  a_1 f^{abe} f^{cde} 
+ a_2 f^{ace} f^{bde}
+ a_3 f^{ade} f^{bce}
\nonumber \\ && 
+ a_4 \delta^{ab} \delta^{cd}
+ a_5 \delta^{ac} \delta^{bd}
+ a_6 \delta^{ad} \delta^{bc}
\nonumber \\ && 
+ a_7 d^{abe} d^{cde} 
+ a_8 d^{ace} d^{bde}
+ a_9 d^{ade} d^{bce} .
\label{eq:colorbasis}
\eeq
which then leads to the color sum:
\beq
\sum_{a,b,c,d} |{\cal A}^{abcd}|^2 &=&
72 \left (|a_1|^2 + |a_2|^2 + |a_3|^2 \right ) + 72\, {\rm Re}[a_1 a_2^* - a_1 a_3^* + a_2 a_3^*] 
\nonumber \\ && 
+ 64 \left (|a_4|^2 + |a_5|^2 + |a_6|^2 \right ) 
+ 16 \, {\rm Re}[a_4 a_5^* + a_4 a_6^* + a_5 a_6^*] 
\nonumber \\ && 
+ \frac{200}{9} \left (|a_7|^2 + |a_8|^2 + |a_9|^2 \right ) 
- \frac{40}{3} \, {\rm Re}[a_7 a_8^* + a_7 a_9^* + a_8 a_9^*] 
\nonumber \\ &&
+ 48 \, {\rm Re}[a_1 a_5^* - a_1 a_6^* + a_2 a_4^* - a_2 a_6^* + a_3 a_4^* - a_3 a_5^* ]
\nonumber \\ &&
+ 40 \, {\rm Re}[a_1 a_8^* - a_1 a_9^* + a_2 a_7^* - a_2 a_9^* + a_3 a_7^* - a_3 a_8^* ]
\nonumber \\ &&
+ \frac{80}{3} \, {\rm Re}[a_4 a_8^* + a_4 a_9^* + a_5 a_7^* + a_5 a_9^* + a_6 a_7^* + a_6 a_8^* ]
.
\label{eq:colorsum}
\eeq
Of the nine coefficients $a_i$ in eq.~(\ref{eq:colorbasis}), only five are independent.
This redundancy is reflected in the following QCD 
identities \cite{Haber:sunidentities,MacFarlane:1968qcdidentities}:
\beq
f^{abe} f^{cde} + f^{ace} f^{dbe} + f^{ade} f^{bce} &=& 0
,
\label{eq:jacobiidentity}
\\
d^{abe} d^{cde} - \frac{1}{3} \left(f^{ace} f^{bde} + f^{ade} f^{bce}\right) 
&=&  \frac{1}{3} \left( \delta^{ac} \delta^{bd} + \delta^{ad} \delta^{bc} - \delta^{ab} \delta^{cd} \right)
,
\label{eq:qcdidentity_su3}
\\
f^{abe} f^{cde} - d^{ace} d^{bde} + d^{bce} d^{ade} &=& \frac{2}{N_c} \left( \delta^{ac} \delta^{bd} - \delta^{ad} \delta^{bc} \right)
,
\label{eq:qcdidentity_sun}
\\
d^{abe} d^{cde} + d^{ace} d^{bde} + d^{ade} d^{bce} &=& \frac{1}{3} \left( \delta^{ab} \delta^{cd} + \delta^{ac} \delta^{bd} + \delta^{ad} \delta^{bc} \right)
.
\label{eq:dabcidentity}
\eeq
where eq.~(\ref{eq:jacobiidentity}) is the Jacobi identity, which holds true for the structure constants of any $SU(N_c)$ group.
Eq.~(\ref{eq:qcdidentity_sun}) also holds true for any $SU(N_c)$.
On the other hand, the identities in eqs.~(\ref{eq:qcdidentity_su3}) and (\ref{eq:dabcidentity}) are special to $SU(3)$.
It should be noted that the above four identities are not independent of each other. Namely, eq.~(\ref{eq:qcdidentity_su3}) is obtained by using eq.~(\ref{eq:qcdidentity_sun}) in conjunction with eq.~(\ref{eq:dabcidentity}).
Also, eq.~(\ref{eq:qcdidentity_su3}) along with eq.~(\ref{eq:qcdidentity_sun}) gives eqs.~(\ref{eq:jacobiidentity}) and (\ref{eq:dabcidentity}).
Using eqs.~(\ref{eq:jacobiidentity}) and (\ref{eq:qcdidentity_su3}), we can eliminate the coefficients $a_7, a_8, a_9$, and one of $a_1, a_2$, or $a_3$ in eq.~(\ref{eq:colorbasis}). Or we can use eqs.~(\ref{eq:qcdidentity_sun}) and (\ref{eq:dabcidentity}) to eliminate the coefficients $a_1, a_2, a_3$, and one of $a_7, a_8$, or $a_9$. In any case, we can eliminate four out of the nine coefficients $a_i$ in eq.~(\ref{eq:colorbasis}).

Although the QCD identities in eqs.~(\ref{eq:jacobiidentity})-(\ref{eq:dabcidentity}) can be used to rearrange the coefficients, there are natural choices that follow from the Feynman rules. 
After including the contributions from $t$- and $u$-channel exchanges of $X$, 
and the interferences with QCD amplitudes,
for color-singlet digluon resonances one naturally has $a_7 = a_8 = a_9 = 0$. 
For color-octet digluon resonances with spin 0, one has $a_4 = a_5 = a_6 =0$. 
For color-octet digluon resonances with spin 1, one has $a_4 = a_5 = a_6 = a_7 = a_8 = a_9 = 0$ in eq.~(\ref{eq:colorbasis}),

Eq.~(\ref{eq:colorsum}), which is understood to also include the interference terms, can then be summed over all the final states, and averaged over all the initial states for both color and spin to finally obtain the complete LO partonic differential cross section of all $X$ exchange diagrams (signal), pure QCD background, and the interference between the signal and the background processes. We then define:
\beq
\frac{d \hat \sigma}{dz} = \frac{d \hat \sigma_s}{dz} & + & \frac{d \hat \sigma_t}{dz} + \frac{d \hat \sigma_u}{dz} + \frac{d \hat \sigma_{s,t}}{dz} + 
\frac{d \hat \sigma_{t,u}}{dz} + \frac{d \hat \sigma_{s,u}}{dz} \notag
 \\
& + & \frac{d \hat \sigma_{s,QCD}}{dz} + \frac{d \hat \sigma_{t,QCD}}{dz} + \frac{d \hat \sigma_{u,QCD}}{dz},
\label{eq:dsigdz}
\eeq
excluding the pure QCD contribution. In the above definition, $d \hat \sigma_{s,QCD}/dz$, for example, stands for the interference of the $s$-channel $X$ exchange diagram with the pure QCD amplitude. Also, $z$ is the cosine of the gluon scattering angle in the partonic center-of-momentum frame. 
Analytic formulas for all the components of $d \hat \sigma/dz$ as defined in eq.~(\ref{eq:dsigdz}), 
for each of the digluon resonances considered, are listed next.

\paragraph*{Spin 0, color singlet:}
The resonant and interference partonic cross sections at leading order for both parity-even and parity-odd spin 0, color singlets (i.e. $i = 1,2$) are:
\beq
\frac{d\hat \sigma_{\rm s}}{dz} &=& \frac{|c_i|^4 \hat s^3}{32 \pi \Lambda^4 D(\hat s)} ,
\label{eq:resS0C1}
\\
\frac{d\hat \sigma_{\rm t}}{dz} + \frac{d\hat \sigma_{\rm u}}{dz} &=& \frac{|c_i|^4}{32 \pi \Lambda^4 \hat s} \left [ \frac{\hat t^4}{D(\hat t)}+\frac{\hat u^4}{D(\hat u)} \right ] ,
\\
\frac{d\hat \sigma_{\rm s,t}}{dz} &=& \frac{|c_i|^4 \hat s \hat t^2}{256 \pi \Lambda^4 D(\hat s) D(\hat t)} \left [(\hat s - M_X^2) (\hat t - M_X^2) + \Gamma^2 M_X^2 \right ],
\\
\frac{d\hat \sigma_{\rm t,u}}{dz} &=& \frac{|c_i|^4 \hat t^2 \hat u^2}{256 \pi \Lambda^4 \hat s D(\hat t) D(\hat u)} \left [(\hat t - M_X^2) (\hat u - M_X^2) + \Gamma^2 M_X^2 \right ],
\\
\frac{d\hat \sigma_{\rm s,u}}{dz} &=& \frac{|c_i|^4 \hat s \hat u^2}{256 \pi \Lambda^4 D(\hat s) D(\hat u)} \left [(\hat s - M_X^2) (\hat u - M_X^2) + \Gamma^2 M_X^2 \right ],
\\
\frac{d\hat \sigma_{\rm s,QCD}}{dz} &=& -
\frac{3\alpha_S \hat s}{8\Lambda^2 D(\hat s) (1 - z^2) } \left \{
{\rm Re}[c_i^2] (\hat s - M_X^2) + {\rm Im}[c_i^2] \Gamma M_X \right \},
\label{eq:intS0C1}
\\
\frac{d\hat \sigma_{\rm t,QCD}}{dz} &=& \frac{3\alpha_S \hat s (1-z)^4 }{256\Lambda^2 D(\hat t) (1+z) } \left \{
{\rm Re}[c_i^2] (\hat t - M_X^2) + {\rm Im}[c_i^2] \Gamma M_X \right \},
\\
\frac{d\hat \sigma_{\rm u,QCD}}{dz} &=& \frac{3\alpha_S \hat s (1+z)^4}{256\Lambda^2 D(\hat u) (1-z) } \left \{
{\rm Re}[c_i^2] (\hat u - M_X^2) + {\rm Im}[c_i^2] \Gamma M_X \right \},
\label{eq:uqcdintS0C1}
\eeq
where $\hat t = \hat s (z-1)/2$ and $\hat u = -\hat s (z+1)/2$, and
\beq
D(\hat s) &\equiv& (\hat s - M_X^2)^2 + \Gamma^2 M_X^2 .
\eeq
Also, the LO partial width into the digluon final state is
\beq
\Gamma_{gg} &=& \frac{|c_i|^2 M_X^3}{2\pi \Lambda^2}
.
\eeq
\paragraph*{Spin 0, color octet:}
The resonant and interference partonic cross sections at leading order for both scalar and pseudo-scalar color octets (i.e. $i = 3,4$) are:
\beq
\frac{d\hat \sigma_{\rm s}}{dz} &=& \frac{25 |c_i|^4 \hat s^3}{2304 \pi \Lambda^4 D(\hat s)} ,
\label{eq:resS0C8}
\\
\frac{d\hat \sigma_{\rm t}}{dz} + \frac{d\hat \sigma_{\rm u}}{dz} &=& \frac{25 |c_i|^4}{2304 \pi \Lambda^4 \hat s} \left [ \frac{\hat t^4}{D(\hat t)}+\frac{\hat u^4}{D(\hat u)} \right ] ,
\\
\frac{d\hat \sigma_{\rm s,t}}{dz} &=& - \frac{5 |c_i|^4 \hat s \hat t^2}{1536 \pi \Lambda^4 D(\hat s) D(\hat t)} \left [(\hat s - M_X^2) (\hat t - M_X^2) + \Gamma^2 M_X^2 \right ],
\\
\frac{d\hat \sigma_{\rm t,u}}{dz} &=& - \frac{5 |c_i|^4 \hat t^2 \hat u^2}{1536 \pi \Lambda^4 \hat s D(\hat t) D(\hat u)} \left [(\hat t - M_X^2) (\hat u - M_X^2) + \Gamma^2 M_X^2  \right ],
\\
\frac{d\hat \sigma_{\rm s,u}}{dz} &=& - \frac{5 |c_i|^4 \hat s \hat u^2}{1536 \pi \Lambda^4 D(\hat s) D(\hat u)} \left [(\hat s - M_X^2) (\hat u - M_X^2) + \Gamma^2 M_X^2 \right ],
\\
\frac{d\hat \sigma_{\rm s,QCD}}{dz} &=& -\frac{5\alpha_S \hat s}{16\Lambda^2 D(\hat s) (1 - z^2) } \left \{
{\rm Re}[c_i^2] (\hat s - M_X^2) + {\rm Im}[c_i^2] \Gamma M_X \right \},
\label{eq:intS0C8}
\\
\frac{d\hat \sigma_{\rm t,QCD}}{dz} &=& \frac{5\alpha_S \hat s (1 - z)^4}{512 \Lambda^2 D(\hat t) (1 + z) } \left \{
{\rm Re}[c_i^2] (\hat t - M_X^2) + {\rm Im}[c_i^2] \Gamma M_X \right \},
\\
\frac{d\hat \sigma_{\rm u,QCD}}{dz} &=& \frac{5\alpha_S \hat s (1 + z)^4}{512 \Lambda^2 D(\hat u) (1 - z) } \left \{
{\rm Re}[c_i^2] (\hat u - M_X^2) + {\rm Im}[c_i^2] \Gamma M_X \right \},
\label{eq:uqcdintS0C8}
\eeq
and the partial width into the digluon final state is
\beq
\Gamma_{gg} &=& \frac{5 |c_i|^2 M_X^3}{48 \pi \Lambda^2}
.
\eeq
In the case that the decay width is entirely due to digluons, we can compare the $s$-channel resonant
production cross-section to that of the spin-0, color-singlet case, using
\beq
\frac{1}{D(\hat s)} &=& \frac{\pi}{\Gamma M_X} \delta(\hat s - M_X^2).
\label{eq:narrowwidth}
\eeq
This means that the $s$-channel resonant cross-sections can be written as
\beq
\frac{d\hat \sigma_{s}}{dz} &=& 
n_i^{s} \frac{\pi |c_i|^2 M_X^2}{\Lambda^2} \delta(\hat s - M_X^2)
,
\eeq
where for spin 0, color singlets,
\beq
n_1^{s} = n_2^{s} = 1/16,
\eeq
and for spin 0, color octets,
\beq
\qquad\quad n_3^{s} = n_4^{s} = 5/48.
\label{eq:fiveoverfortyeight}
\eeq
To have the same resonant cross-section, 
we could take $|c_3|^2 = (3/5) |c_1|^2$. 
In that case, comparing
eqs.~(\ref{eq:intS0C1})-(\ref{eq:uqcdintS0C1}) and (\ref{eq:intS0C8})-(\ref{eq:uqcdintS0C8}), 
we see that the pre-factor for the 
interference cross-section 
would be half as big for the color-octet case as for the color-singlet case.
So, all other things being equal, the importance of the interference terms relative to the resonance terms is half as big
in the color-octet spin-0 case as in the color-singlet spin-0 case.

\paragraph*{Spin 1, color octet:}
The resonant and interference partonic cross sections at leading order for a spin-1, color-octet resonance are:
\beq
\frac{d\hat \sigma_{\rm s}}{dz} &=& \frac{9 |c_5|^4 \hat s^5 z^2}{256 \pi \Lambda^8 D(\hat s)} ,
\label{eq:resS1C8}
\\
\frac{d\hat \sigma_{\rm t}}{dz} + \frac{d\hat \sigma_{\rm u}}{dz} &=& \frac{9 |c_5|^4 \hat s}{1024 \pi \Lambda^8} \left [ \frac{\hat t^4 (3+z)^2}{D(\hat t)}+\frac{\hat u^4 (3-z)^2}{D(\hat u)} \right ] ,
\\
\frac{d\hat \sigma_{\rm s,t}}{dz} &=& \frac{9 |c_5|^4 \hat s^3 \hat t^2 z (3+z)}{1024 \pi \Lambda^8 D(\hat s) D(\hat t)} \left [(\hat s - M_X^2) (\hat t - M_X^2) + \Gamma^2 M_X^2 \right ],
\\
\frac{d\hat \sigma_{\rm t,u}}{dz} &=& \frac{9 |c_5|^4 \hat s \hat t^2 \hat u^2 (9-z^2)}{2048 \pi \Lambda^8 D(\hat t) D(\hat u)} \left [(\hat t - M_X^2) (\hat u - M_X^2) + \Gamma^2 M_X^2 \right ],
\\
\frac{d\hat \sigma_{\rm s,u}}{dz} &=& - \frac{9 |c_5|^4 \hat s^3 \hat u^2 z (3-z)}{1024 \pi \Lambda^8 D(\hat s) D(\hat u)} \left [(\hat s - M_X^2) (\hat u - M_X^2) + \Gamma^2 M_X^2 \right ],
\\
\frac{d\hat \sigma_{\rm s,QCD}}{dz} &=& -\frac{9 \alpha_S \hat s^2 z^2}{16 \Lambda^4 D(\hat s) (1-z^2)} \left \{
{\rm Re}[c_5^2] (\hat s - M_X^2) + {\rm Im}[c_5^2] \Gamma M_X \right \},
\\
\frac{d\hat \sigma_{\rm t,QCD}}{dz} &=& -\frac{9 \alpha_S \hat s^2 (3+z)^2 (1-z)^3 }{1024 \Lambda^4 D(\hat t) (1+z)} \left \{
{\rm Re}[c_5^2] (\hat t - M_X^2) + {\rm Im}[c_5^2] \Gamma M_X \right \},
\\
\frac{d\hat \sigma_{\rm u,QCD}}{dz} &=& -\frac{9 \alpha_S \hat s^2 (3-z)^2 (1+z)^3 }{1024 \Lambda^4 D(\hat u) (1-z)} \left \{
{\rm Re}[c_5^2] (\hat u - M_X^2) + {\rm Im}[c_5^2] \Gamma M_X \right \},
\label{eq:intS1C8}
\eeq
and the partial width into the digluon final state is
\beq
\Gamma_{gg} &=& \frac{|c_5|^2 M_X^5}{16 \pi \Lambda^4}
.
\eeq
\paragraph*{Spin 2, color singlet:}
The resonant and interference partonic cross sections at leading order for a spin-2, color-singlet resonance are:
\beq
\frac{d\hat \sigma_{\rm s}}{dz} &=& 
\frac{\hat s^3 (1 + 6 z^2 + z^4)}{512 \pi \overline M_P^4D(\hat s)} ,
\label{eq:resS2C1}
\\
\frac{d\hat \sigma_{\rm t}}{dz}  &=& \frac{ \hat s^3 (17+4z+6z^2+4z^3+z^4)}{1024 \pi \overline M_P^4 D(\hat t)} ,
\\
\frac{d\hat \sigma_{\rm u}}{dz} &=& \frac{ \hat s^3 (17-4z+6z^2-4z^3+z^4)}{1024 \pi \overline M_P^4 D(\hat u)} ,
\\
\frac{d\hat \sigma_{\rm s,t}}{dz} &=& \frac{ \hat s^3 (1+z)^4}{4096 \pi \overline M_P^4 D(\hat s) D(\hat t)} \left [(\hat s - M_X^2) (\hat t - M_X^2) + \Gamma^2 M_X^2 \right ],
\\
\frac{d\hat \sigma_{\rm t,u}}{dz} &=& \frac{ \hat s^3}{256 \pi \overline M_P^4 D(\hat t) D(\hat u)} \left [(\hat t - M_X^2) (\hat u - M_X^2) + \Gamma^2 M_X^2 \right ],
\\
\frac{d\hat \sigma_{\rm s,u}}{dz} &=& \frac{ \hat s^3 (1-z)^4}{4096 \pi \overline M_P^4 D(\hat s) D(\hat u)} \left [(\hat s - M_X^2) (\hat u - M_X^2) + \Gamma^2 M_X^2 \right ],
\\
\frac{d\hat \sigma_{\rm s,QCD}}{dz} &=& -\frac{3 \alpha_S \hat s (\hat s - M_X^2) (1 + 6 z^2 + z^4)}{64 \overline M_P^2 D(\hat s) (1-z^2)},
\\
\frac{d\hat \sigma_{\rm t,QCD}}{dz} &=& \frac{3 \alpha_S \hat s (\hat t - M_X^2) (17 + 4z + 6z^2 + 4z^3 + z^4) }{256 \overline M_P^2 D(\hat t) (1-z^2)} ,
\\
\frac{d\hat \sigma_{\rm u,QCD}}{dz} &=& \frac{3 \alpha_S \hat s (\hat u - M_X^2) (17 - 4z + 6z^2 - 4z^3 + z^4)}{256 \overline M_P^2 D(\hat u) (1-z^2)} .
\label{eq:intS2C1}
\eeq
The pure $X$-exchange parts and their corresponding interference-with-QCD terms have common 
angular factors of $(1+6z^2 +z^4), (17 + 4z + 6z^2 + 4z^3 + z^4),$ and 
$(17 - 4z + 6z^2 - 4z^3 + z^4)$ for $s$-, $t$-, and $u$-channels respectively.

The digluon partial width of the resonance is
\beq
\Gamma_{gg} &=& \frac{M_X^3}{10 \pi \overline M_P^2} .
\eeq
In the narrow width approximation of eq.~(\ref{eq:narrowwidth}), assuming that $X$ has no
other decays, we therefore have
\beq
\frac{d\hat \sigma_{s}}{dz} &=& 
\frac{5 \pi M_X^2}{256 \overline M_P^2} (1 + 6 z^2 + z^4) \delta(\hat s - M_X^2)
.
\eeq
Having obtained the analytic formulas for all the components of $d \hat \sigma/dz$, for all four models considered, we can then compute the LO digluon production cross section at the LHC,
\beq
\frac{d \sigma_{p p \rightarrow g g }}{d (\sqrt{\hat s})} &=& \sqrt{\hat s} \int_{x_{-}}^{x_{+}} \frac{dx}{xs} g(x) g\left (\frac{\hat s}{x s} \right ) \int_{-z_{\rm cut}}^{z_{\rm cut}} dz \frac{d \hat \sigma}{dz}
.
\label{eq:dsigdm}
\eeq
Here, $x$ is the longitudinal momentum fraction for the parton, $g(x)$ is the gluon parton distribution function (PDF) obtained from the NNPDF2.3 LO PDF set \cite{nnpdf23lo} with factorization scale $\mu_F = M_X$, $\sqrt{s}$ is the total energy of the $pp$ collisions at the LHC, and $\sqrt{\hat s}$ is the invariant mass of the gluon pairs in both initial and final states.
In eq.~(\ref{eq:dsigdm}), we have imposed 
cuts on the transverse momentum and the pseudo-rapidity of the gluons:
\beq 
p_{T_j} > {p^{\rm cut}_{T_j}} &=& \mbox{100 GeV},
\label{eq:pTjcut}
\\
|\eta_j| < \eta^{\rm cut}_{j} &=& 2.5,
\label{eq:etajcut}
\eeq
respectively. 
Also, in order to increase the significance of resonance signal,
CMS has defined signal regions that cut on the difference between the
pseudo-rapidities of the two jets:
\beq
\Delta \eta &=& |\eta_{j_1} - \eta_{j_2}| < (\Delta \eta)^{\rm cut}
.
\label{eq:Deltaetajcut}
\eeq
We will follow CMS by choosing
$(\Delta \eta)^{\rm cut} = 1.1$ for $M_X > 1800$ GeV as in \cite{CMSdijet2019},
and
$(\Delta \eta)^{\rm cut} = 1.3$ for $M_X < 1800$ GeV as in \cite{CMSdijet2018}.
For a parton level analysis, these cuts can be imposed by using 
\beq
x_{\pm} &=& e^{\pm \eta^{\rm cut}_{j}} \sqrt{\hat s/s} 
,
\\
z_{\rm cut} &=& {\rm Min} \left [ \sqrt{1 - {4 {{p^{\rm cut}_{T_j}}}}^2 / \hat s}
,
\>\, 
\tanh \Bigl( \eta^{\rm cut}_{j} - 
\frac{1}{2} \left | \ln \left(x^2 s/\hat s \right) \right | \Bigr ) 
,
\>\,
\tanh \left( (\Delta\eta)^{\rm cut}/2 \right )
\right ]
\eeq
in eq.~(\ref{eq:dsigdm}).

\begin{figure}[!t]
\begin{minipage}[]{0.6\linewidth}
\begin{flushright}
\includegraphics[width=\linewidth,angle=0]{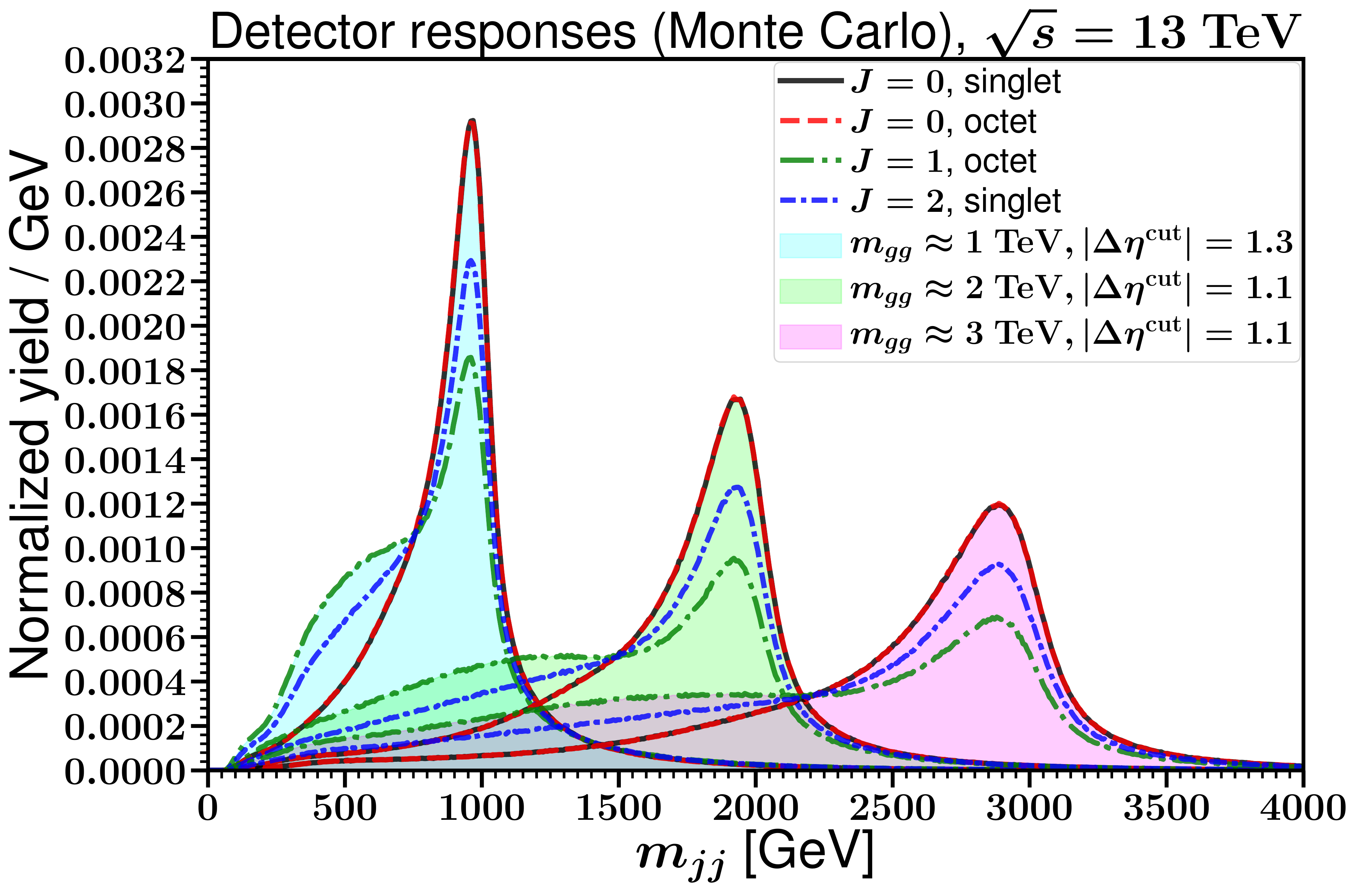}
\end{flushright}
\end{minipage}
\begin{minipage}[]{0.39\linewidth}
\begin{minipage}[]{0.04\linewidth}
\end{minipage}
\begin{minipage}[]{0.90\linewidth}
\caption{\label{fig:yield}
The normalized detector response dijet mass distributions for two wide jets, for digluon masses 
$m_{gg} = 1000$, $2000$, and $3000$ GeV and various spin and color quantum numbers, 
generated and analyzed using {\sc Pythia 8} and {\sc Delphes}.}
\end{minipage}\end{minipage}
\end{figure}

Then to obtain more realistic distributions roughly approximating the experimental ones, 
the parton-level distributions of $d \sigma_{p p \rightarrow g g }/d (\sqrt{\hat s})$ in eq.~(\ref{eq:dsigdm}) can be smeared by convolution with an approximate detector response dijet mass distribution, shown for the case of a digluon invariant mass $m_{gg} = (1000$, $2000$, $3000$) GeV in Figure~\ref{fig:yield}.
To obtain the approximate detector responses for the dijet invariant mass distributions, we used 
a sample of detector-level events, generated with {\sc Pythia 8.2} 
\cite{Sjostrand:2006za,Sjostrand:2015za} and {\sc Delphes 3.4} \cite{deFavereau:2013fsa}, 
obtained from (at least $1.5 \times 10^7$) parton-level events for the resonant process
$g g \rightarrow X \rightarrow g g$ with the required digluon invariant mass, here $\sqrt{\hat s}=m_{gg} = 1000$, $2000$, and $3000$ GeV, using the same cuts and procedure for combining jets into wide jets
as described in the next subsection. This was done for each of the four (spin, color) quantum number combinations described above.

The results obtained by the above simple procedure will be referred to below as the parton-level approximation (with smearing). This method can be viewed as a quick approximation and qualitative 
independent cross-check of the more involved method we describe next, which is much more demanding of computer resources.

\subsection{Showering, hadronization, and detector-level simulation\label{subsec:detectorlevel}}

For a much more realistic approximation, referred to below simply as the detector-level simulation, we used {\sc MadGraph5\_aMC@NLO} v2.6.6 \cite{Alwall:2011uj} for all our LO event simulations, which uses a {\it Universal FeynRules Output (UFO)} file generated by {\sc FeynRules} v2.3 \cite{Alloul:2013bka}, a {\sc Mathematica} package to get the Feynman rules from an input Lagrangian.
We used {\sc Pythia 8.2} for showering and hadronization, and {\sc Delphes 3.4} for detector simulation. 

Using {\sc Madgraph}, our goal was not only to generate detector-level events for the signal, including all $X$ exchange diagrams, but also for the interference between the signal and the  QCD background $g g \rightarrow g g$.
One of the the challenges to generate detector-level events for the 
interference terms is that some of the generated events have negative cross-sections.
(Simply generating a full QCD+$X$ sample and then subtracting the pure QCD part is not practical, because of the very poor statistics. The actual LHC experiment has much better statistics than our simulations can provide.)
In order to keep track of the events with positive and negative cross-sections, the parton-level events for the interference terms are divided into two sets, one with positive cross-sections, and the other with negative cross-sections.
The detector-level events for both sets are independently obtained.
Then, to obtain the dijet invariant mass distributions for the interference terms, the distributions with negative cross-section events are subtracted bin-by-bin from the ones with positive cross-section events.

\begin{figure}
  \begin{minipage}[]{0.495\linewidth}
    \includegraphics[width=7.5cm,angle=0]{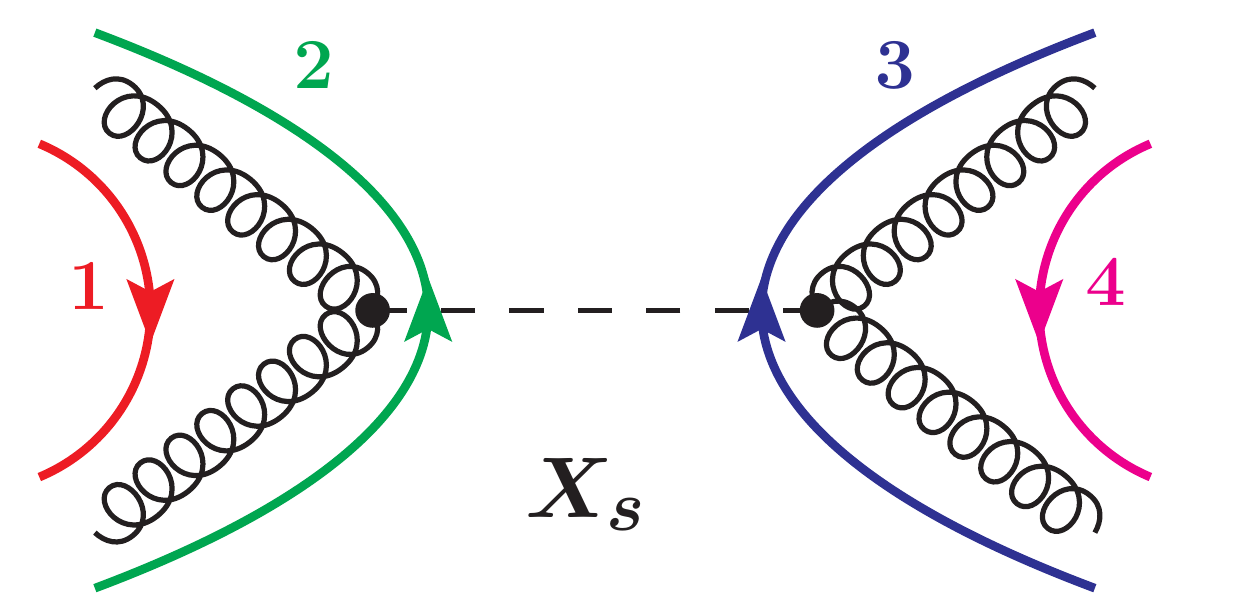}
  \end{minipage}
  \begin{minipage}[]{0.495\linewidth}
    \includegraphics[width=7.5cm,angle=0]{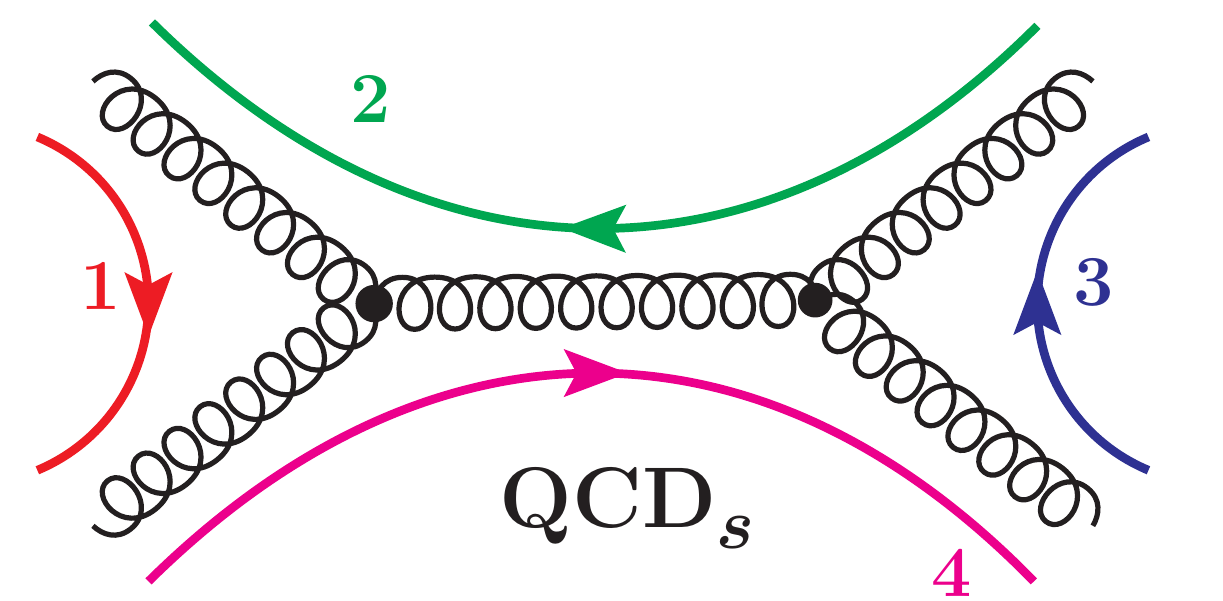}
  \end{minipage}
  
\vspace{0.2cm}
    \begin{minipage}[]{0.495\linewidth}
    \includegraphics[width=4.7cm,angle=0]{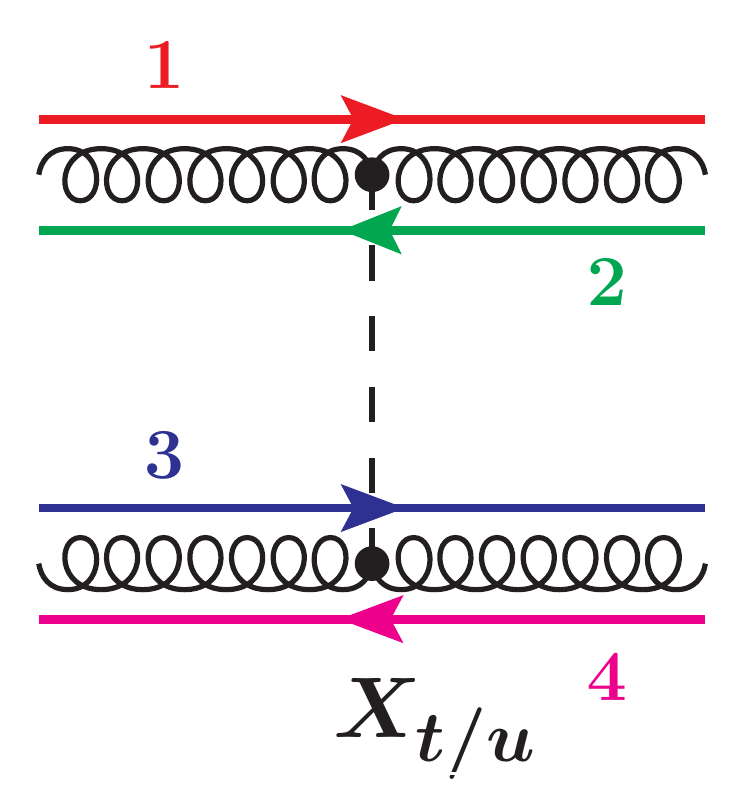}
  \end{minipage}
    \begin{minipage}[]{0.495\linewidth}
    \includegraphics[width=5.0cm,angle=0]{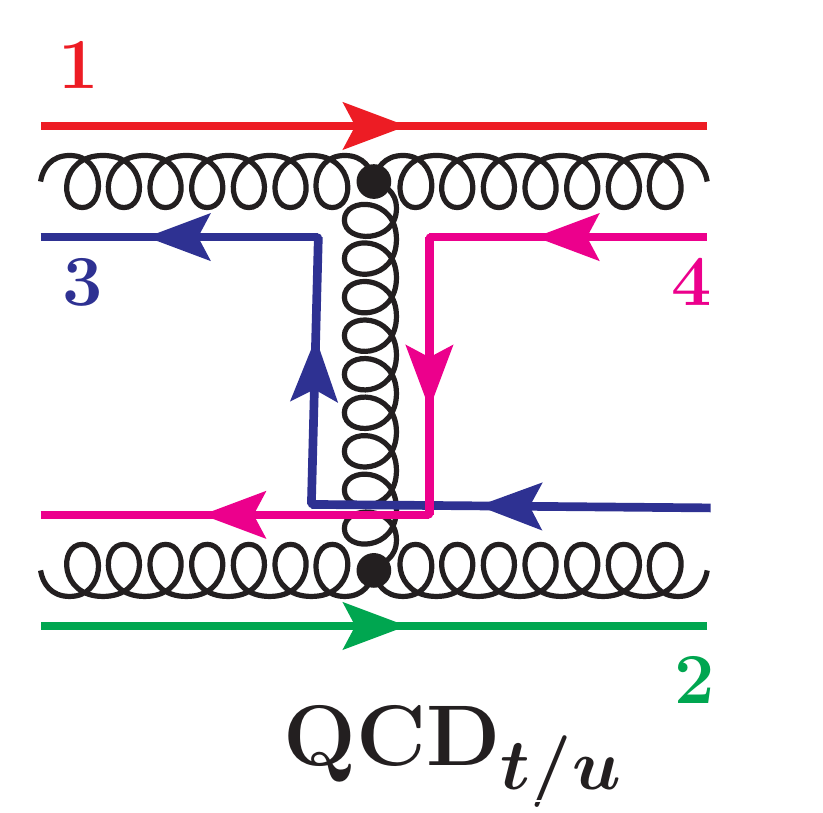}
  \end{minipage}
\begin{center}\begin{minipage}[]{0.95\linewidth}
\caption{\label{fig:colorflow} 
The four distinct color flows for the parton-level events for the interference between the color-singlet digluon resonant process $g g \rightarrow X \rightarrow g g$ and the QCD process $g g \rightarrow g g$. 
These color flows are referred to below in the text as $X_s$, $X_{t/u}$, QCD$_s$, and QCD$_{t/u}$, as labeled. The integer tags (e.g.~1-4) for the color flow lines are required by the showering and hadronization event generators to begin the parton shower. For example, in {\sc Pythia}, the integer tags are typically assigned as 501-504.} 
\end{minipage}\end{center}
\end{figure}

Another problem is that {\sc Madgraph} cannot assign a unique color flow \cite{colorflowmadgraph,colorflowqcdamplitudes,colorflowpeskin} for each parton-level event for the interference between the QCD background $g g \rightarrow g g$ and color-singlet resonant signal $g g \rightarrow X \rightarrow g g$, thus disabling {\sc Pythia} from showering and hadronization, thereby precluding the generation of any detector-level events.
To get around this, we independently generated detector-level events for the interference between the  
QCD background and the color-singlet resonant signal by simply 
assigning a color flow by fiat to each event.
There are four physically distinct color-flow possibilities, as shown in Fig.~\ref{fig:colorflow}.
We therefore repeated the analysis four times, 
each time with the same color flow assigned to every event.
Thus, for color-singlet resonances, 
we obtain a spread of the possible differential cross sections by taking the maximum and minimum of the 
four possibilities in each invariant mass bin, as will be shown below in Figs.~\ref{fig:gggg_s0c1}, \ref{fig:gggg_s2c1}, \ref{fig:gggg_s0c1_otherdecays}, and \ref{fig:gggg_s2c1_otherdecays}. On the other hand, this problem does not occur for color-octet resonances, as they 
are uniquely determined to have the same color charge as that of a gluon.

We generated at least $5 \times 10^6$ events for both the digluon resonant signal and the interference between the signal and the QCD background, for $pp$ collisions at $\sqrt{s}=13$ TeV, for each of the considered digluon resonant benchmark mass models.
While generating events with {\sc Madgraph}, we used  {\sc MadGraph}'s default parton distribution functions (PDFs) based on NNPDF2.3 LO set \cite{nnpdf23lo} with factorization and renormalization scales $\mu_F = \mu_R = M_X$. 

In our analysis, all the jets are required to satisfy a cut on pseudo-rapidity $|\eta_j| < 2.5$
as noted above in eq.~(\ref{eq:etajcut}).
To obtain the dijet mass ($m_{jj}$), we follow the procedure used by CMS in refs.~\cite{CMSdijet2019,CMSdijet2018} to reduce the sensitivity to radiation of additional gluons from the final-state gluons in the hard-scattering event. Specifically,
we start with the two leading $p_T$ jets, and the four-vectors of all other jets within $\Delta R = \sqrt{(\Delta \eta)^2 + (\Delta \phi)^2} < 1.2$ of the two leading $p_T$ jets are added to the nearest leading jet to obtain two wide jets. 
These wide jets are then required to satisfy the cut
$p_{T_j} > p^{\text{wide-jets}}_{T_j} = 100 \text{ GeV}$,
as noted above in eq.~(\ref{eq:pTjcut}). Finally, as also already mentioned in the previous section, we 
follow CMS by defining the signal region to have a cut on the difference in
pseudo-rapidity for the two wide jets, $|\Delta \eta_{jj}| < 1.3$ (for resonance masses smaller than
1.8 TeV as in \cite{CMSdijet2018}), and $|\Delta \eta_{jj}| < 1.1$ (for resonance masses larger than
1.8 TeV as in \cite{CMSdijet2019}). This preferentially eliminates the pure background events, and
tends to improve the efficiency for pure resonance events compared to interference. However, as we will see it certainly does not reduce the effects of interference to a negligible level.

\section{Results for $\Gamma = \Gamma_{gg}$\label{sec:results}}
\setcounter{equation}{0}
\setcounter{figure}{0}
\setcounter{table}{0}
\setcounter{footnote}{1}

In this section, we show results obtained for our benchmark model cases. In each case, we start with results obtained using the simple method of parton-level generation with smearing as outlined in subsection \ref{subsec:partonlevel}. These are followed by results for the full detector simulation of subsection \ref{subsec:detectorlevel}. In this section,
we consider the case $\Gamma_{gg} = \Gamma$, for which the relative effects of the interference are
minimized. 

\subsection{Spin 0, color singlet}

\subsubsection{Parton-level with smearing}

\begin{figure}[!tb]
  \begin{minipage}[]{0.495\linewidth}
    \includegraphics[width=8.0cm,angle=0]{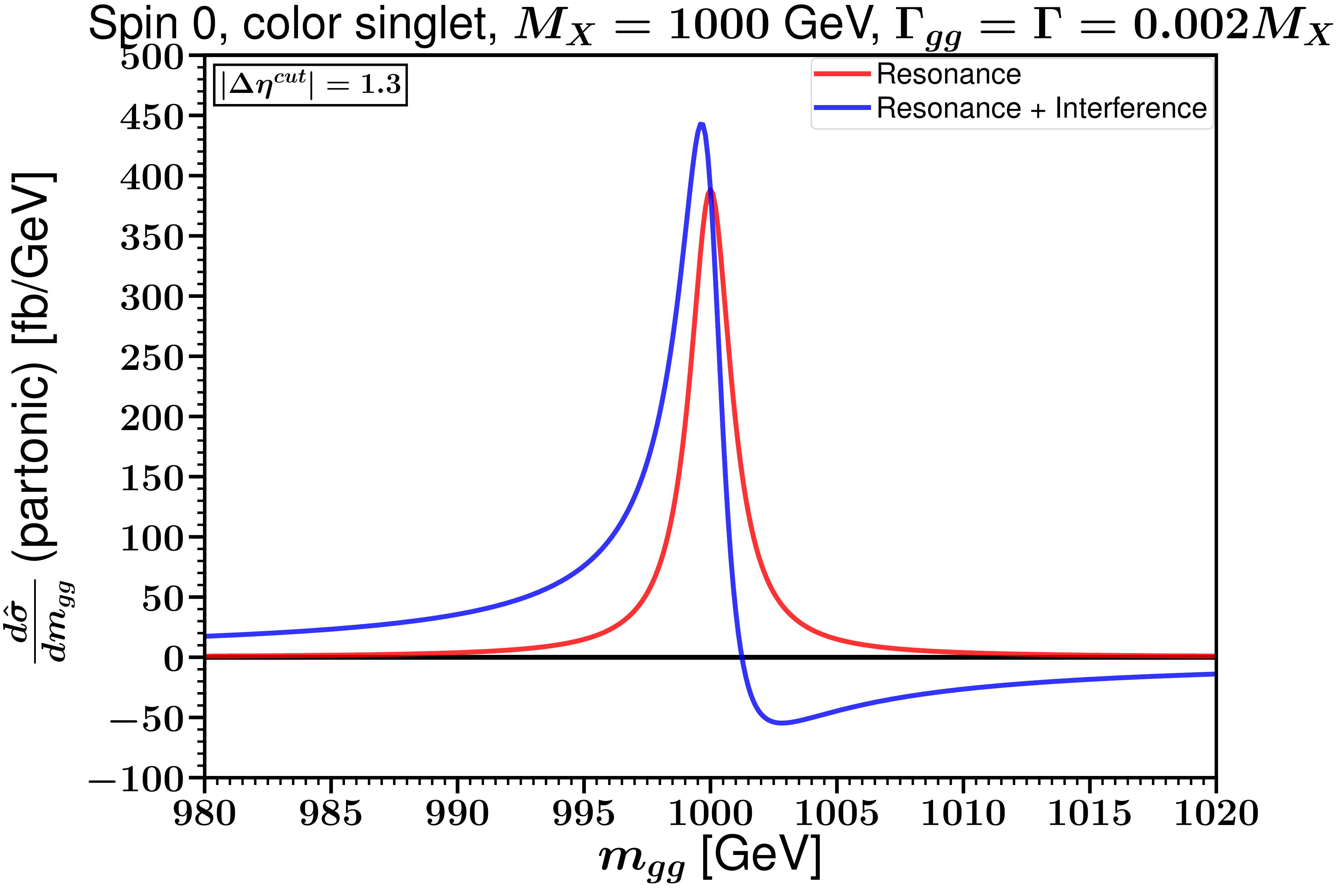}
  \end{minipage}
    \begin{minipage}[]{0.495\linewidth}
    \includegraphics[width=8.0cm,angle=0]{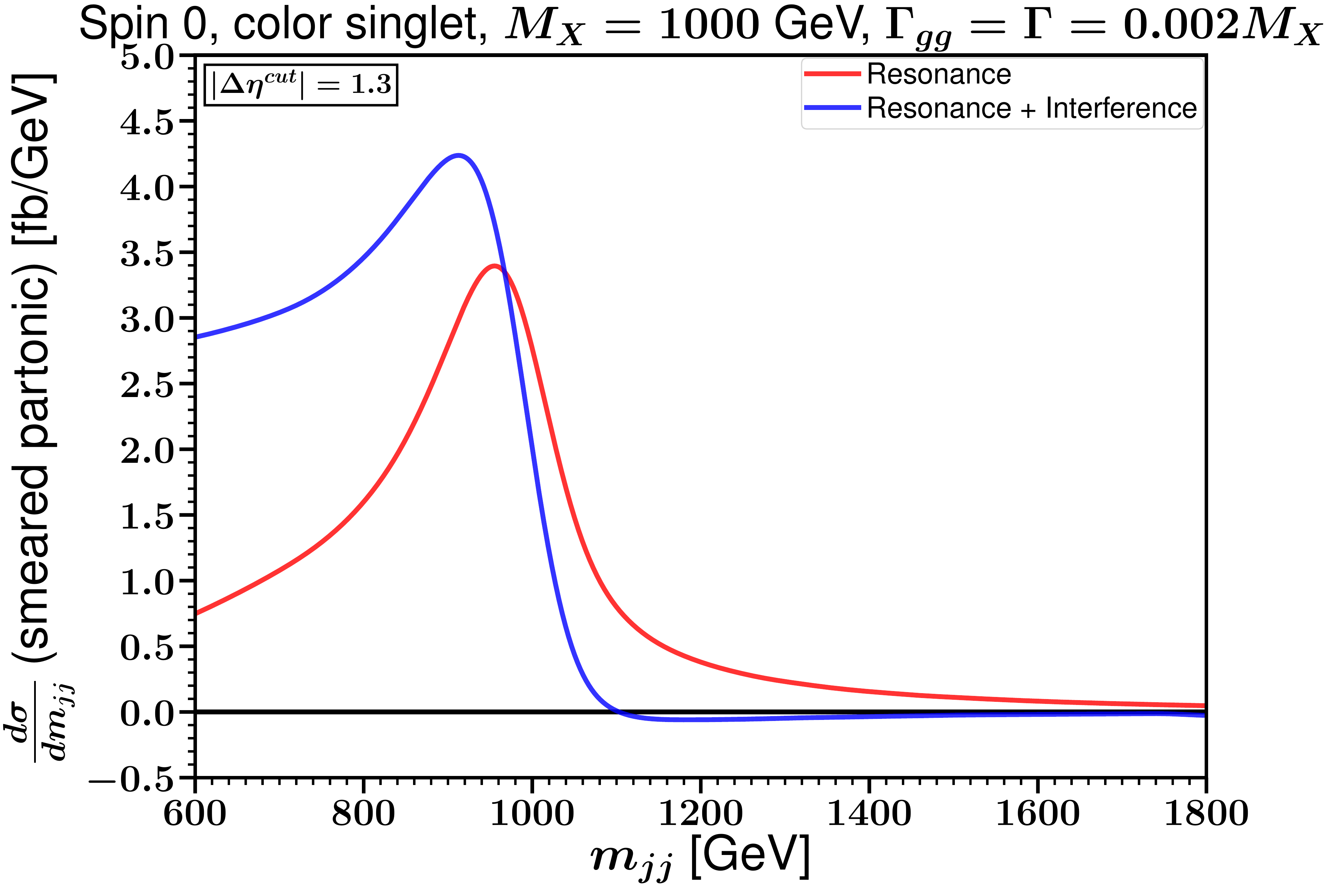}
  \end{minipage}

  \vspace{0.15cm}
  
  \begin{minipage}[]{0.495\linewidth}
    \includegraphics[width=8.0cm,angle=0]{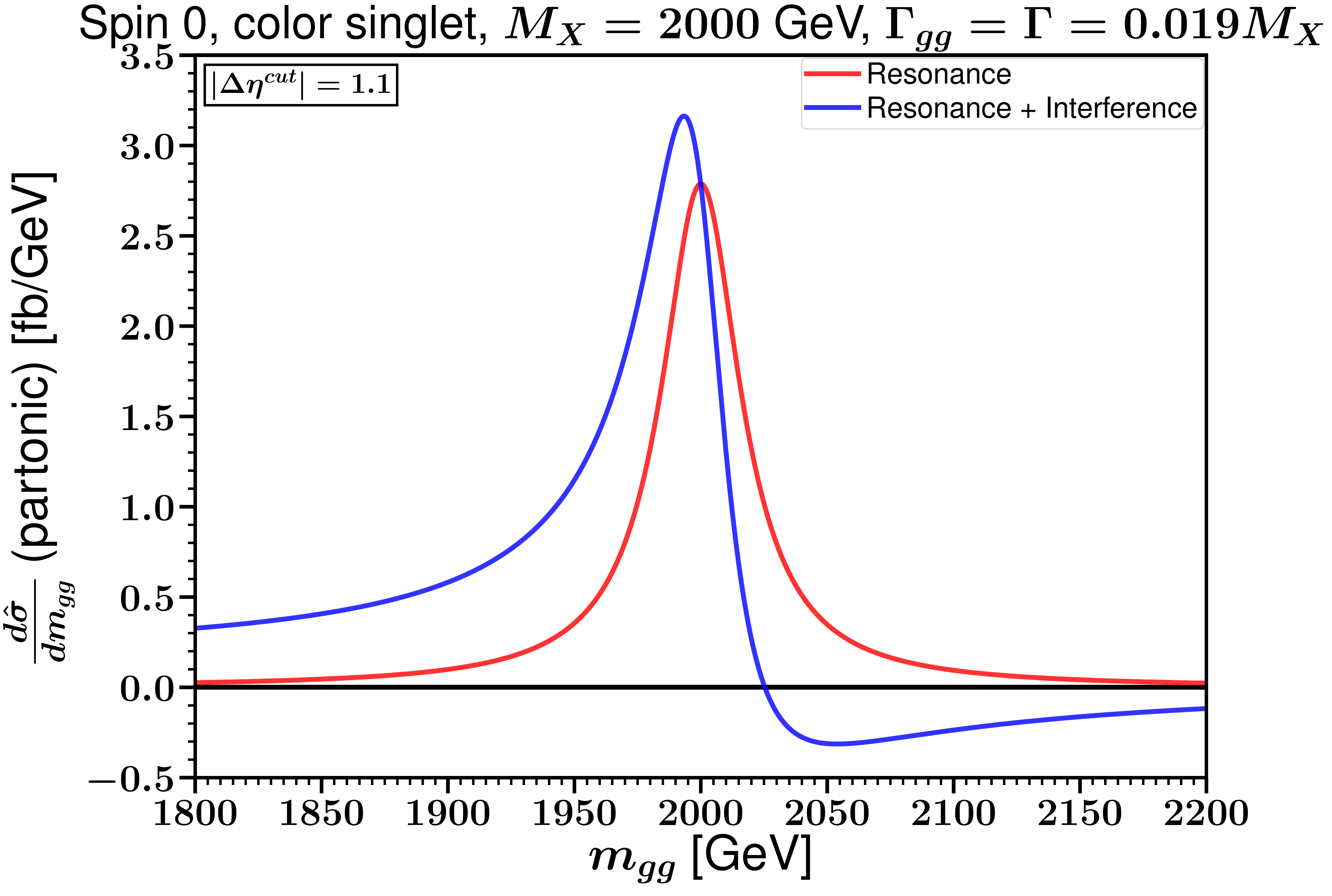}
  \end{minipage}
    \begin{minipage}[]{0.495\linewidth}
    \includegraphics[width=8.0cm,angle=0]{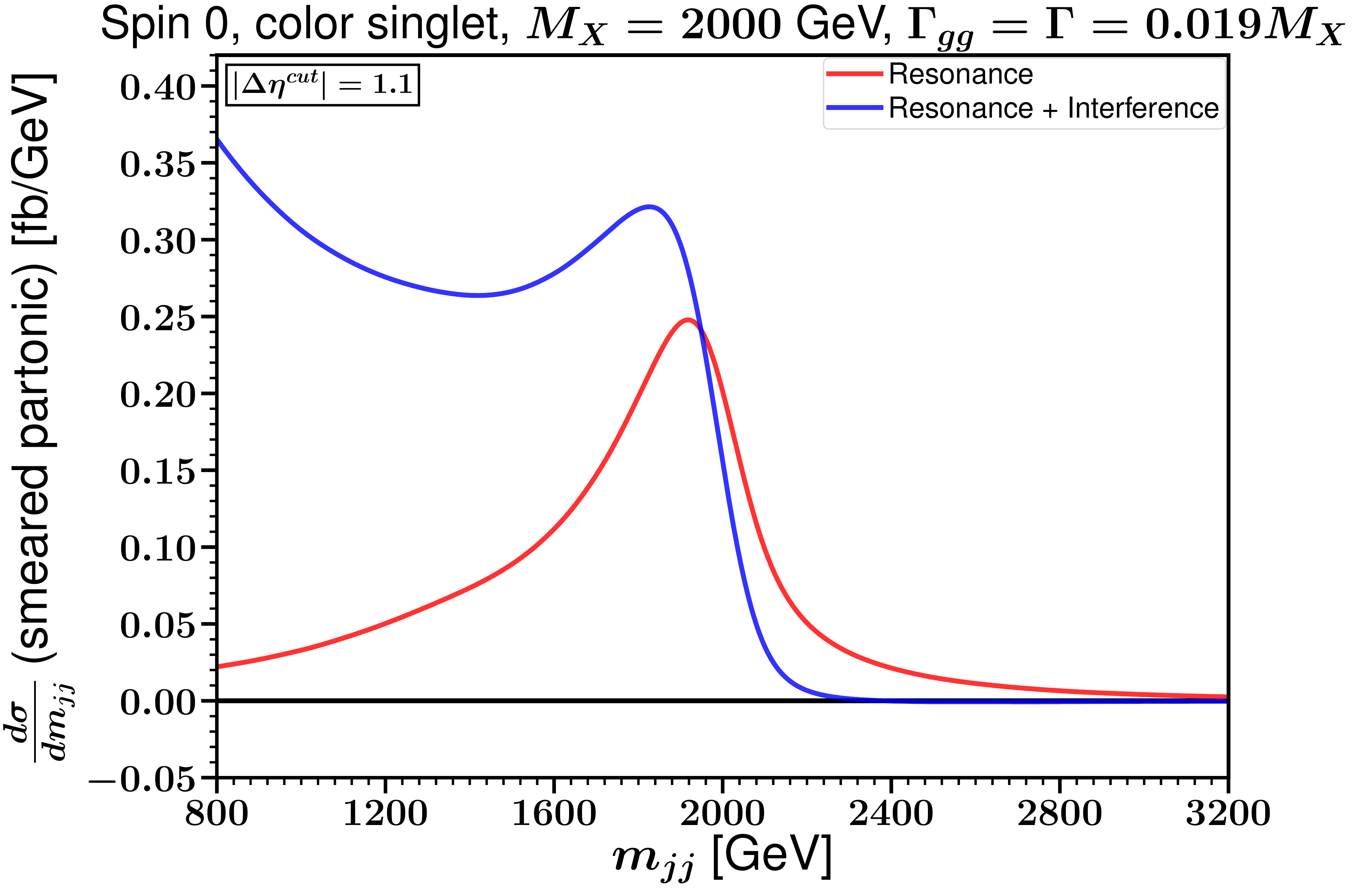}
  \end{minipage}

  \vspace{0.15cm}
  
  \begin{minipage}[]{0.495\linewidth}
    \includegraphics[width=8.0cm,angle=0]{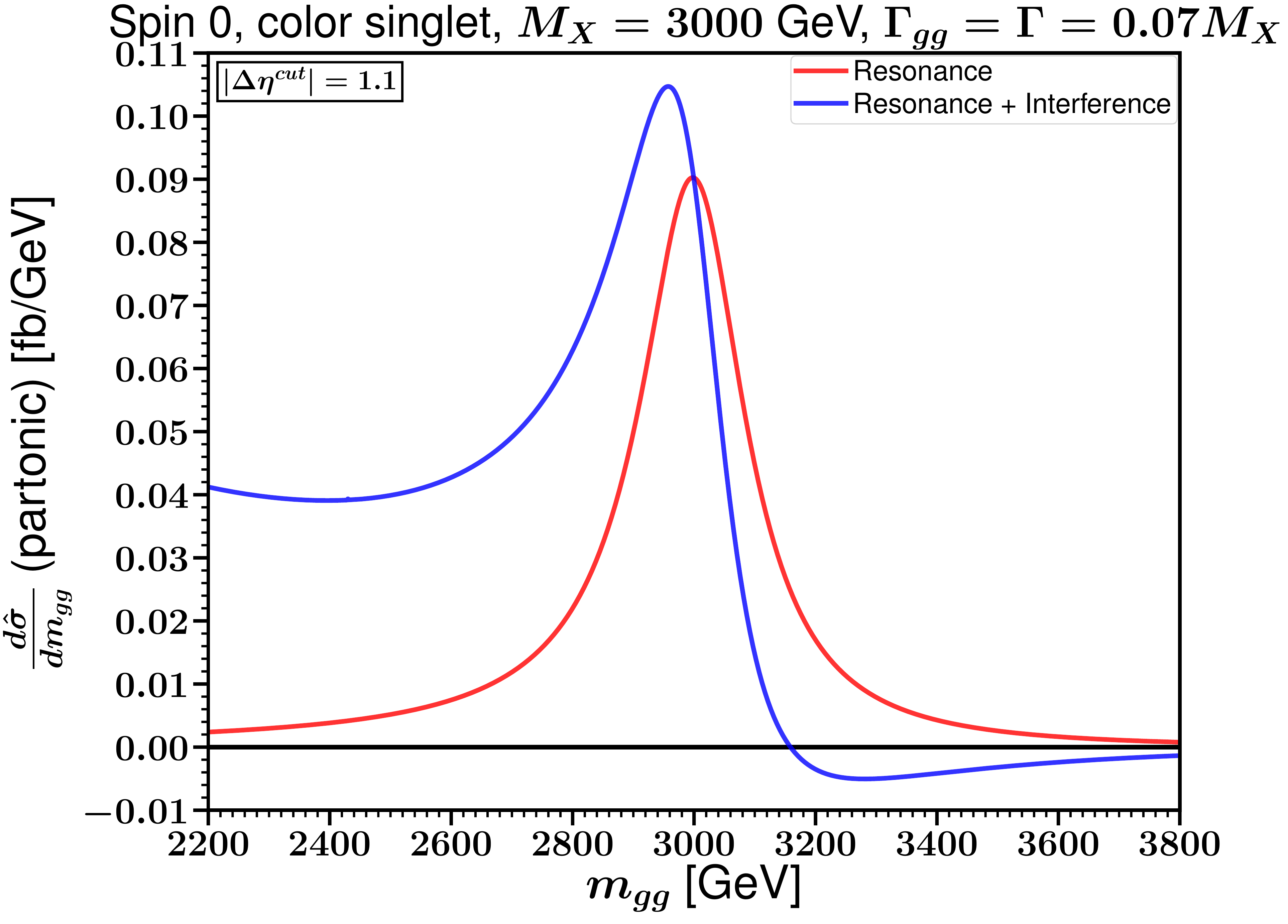}
  \end{minipage}
    \begin{minipage}[]{0.495\linewidth}
    \includegraphics[width=8.0cm,angle=0]{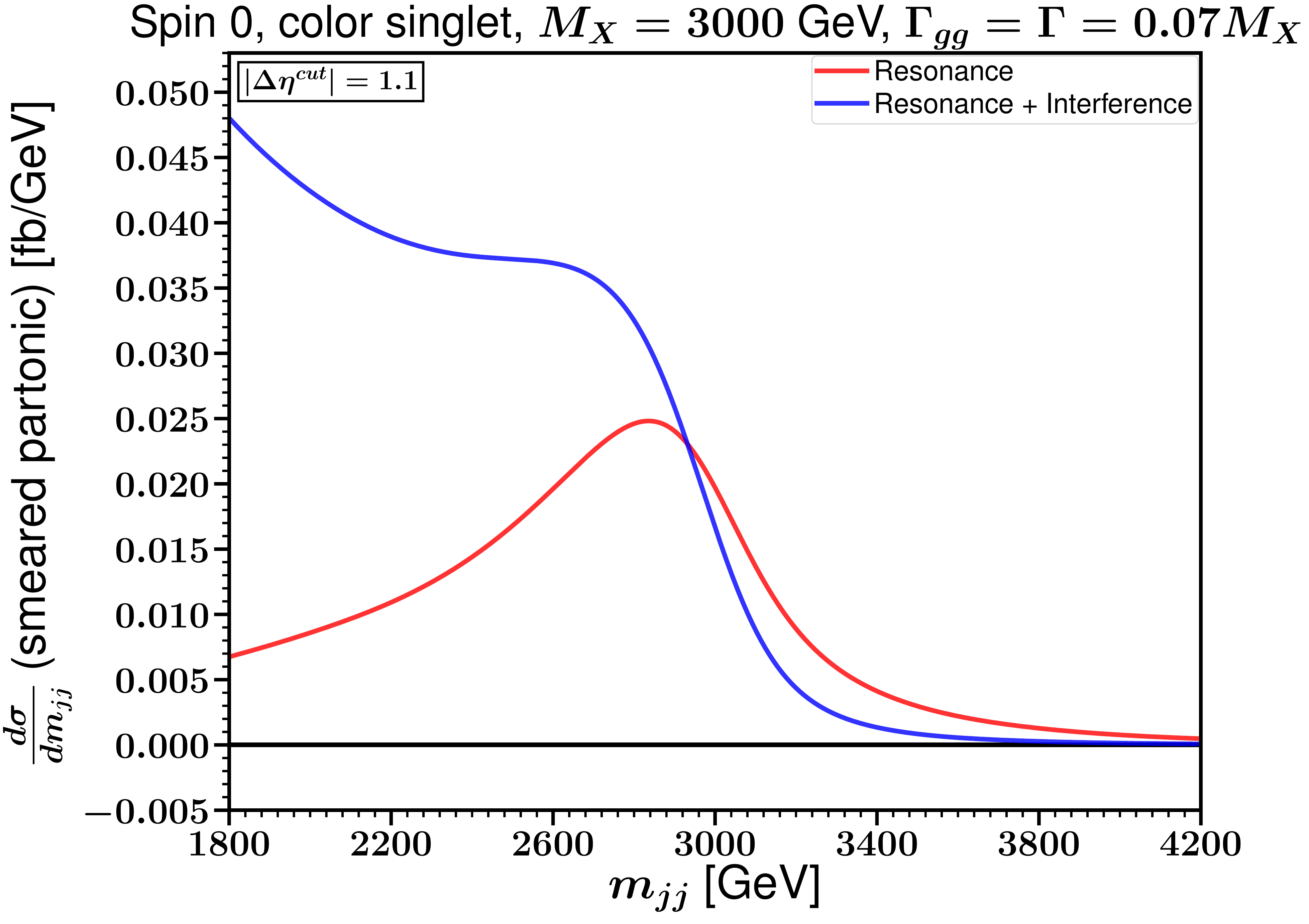}
  \end{minipage}
\begin{center}\begin{minipage}[]{0.95\linewidth}
\caption{\label{fig:gggg_s0c1_analytic}
Digluon invariant mass distributions, at the 13 TeV LHC, for spin-0, color-singlet resonances with
$(M_X, \Gamma/M_X) =$ 
(1000 GeV, 0.002) [top row] and 
(2000 GeV, 0.019) [middle row] and 
(3000 GeV, 0.07) [bottom row].
The parton-level distributions are shown in the left column panels. These are are then smeared by convolution with the estimated detector responses shown in Figure~\ref{fig:yield} to obtain the dijet invariant mass distributions in the panels of the right columns. In all six panels, the red lines show the naive results for the resonant signal $g g \rightarrow X \rightarrow g g$, while the blue lines show the full results including the interferences with the  QCD background $g g \rightarrow g g$.}
\end{minipage}\end{center}
\end{figure}

The left panels of Figure~\ref{fig:gggg_s0c1_analytic} show the leading order digluon invariant mass distributions, obtained by parton-level leading order calculation, for $pp$ collisions at $\sqrt{s}=13$ TeV, for a spin-0, color-singlet resonance, for benchmark examples from Table~\ref{tab:benchmarks} with resonance masses of 1000, 2000, and 3000 GeV in the top, middle and bottom rows, respectively. 
We remind that the chosen benchmark values of $\Gamma_{gg} = \Gamma$ predict a resonant-only
cross-section about equal to the current CMS limit from ref.~\cite{CMSdijet2019}.
The parton-level distributions are in the left column, and the right column shows the results after  smearing by convolution with approximate detector responses (as shown in Figure~\ref{fig:yield}), to obtain a rough estimate of dijet mass distributions.
In all six panels, the red lines show the naive results for the resonant signal $g g \rightarrow X \rightarrow g g$ with all $X$ exchange diagrams, while the blue lines show the full results including the interferences with the QCD background $g g \rightarrow g g$. (The pure QCD background contribution is much larger and is not shown, here or in the following.)

From Figure~\ref{fig:gggg_s0c1_analytic}, we see that when interferences with  QCD amplitudes are included, the dijet mass distributions (blue lines) are both qualitatively
and quantitatively different from that of distributions with only the $X$ exchange diagrams (red lines). 
In the parton-level results in the left column panels, with interference included, we see a peak below and a dip above the resonance mass rather than just a resonance peak. The origin of the peak/dip signature can be traced back to the term 
$\hat s - M_{X}^2$ in $d\hat \sigma_{\rm s,QCD}/dz$, as given in eq.~(\ref{eq:intS0C1}). 
The magnitude of the interference is, in general, enhanced for $\sqrt{\hat s} $ below $M_X$ because of the steeply falling gluon PDFs.
After smearing (right column panels), the dip at higher invariant masses manifests as a deficit of events compared to the naive resonance results.
For the  results with interference included, we can also note that the peak below the resonance mass is significantly larger
than the naive resonance peak that we get from not including the interference terms. 
However, the larger peak is counteracted by the fact that it is connected to a large low-energy tail. If the low-energy tail is absorbed into a 
QCD background fit (which we do not attempt in this paper, and can probably be done in a variety of distinct ways), an effective dip for $m_{jj} > M_X$ would probably result.
We have checked that the interference effect would have been larger without
the imposition of the $\Delta\eta$ cut. 
Correctly deriving a limit on these distribution shapes as new physics sources might require a more flexible analysis strategy than just modeling a
resonance peak, even if the pure resonance contribution is very narrow compared to the experimental resolution. 

As the experimental data sets are increased in the future, 
so that one is probing models with even smaller 
cross-sections, we have checked that the relative importance of the interference 
and the pure resonance at a 
given $M_X$ stays nearly constant. We have also checked that this feature holds true for the other digluon resonances considered in this paper. On the other hand, as the mass is increased, Figure~\ref{fig:gggg_s0c1_analytic} shows that the relative importance
of the interference tends to increase substantially, particularly in the low-mass tail.

\subsubsection{Monte Carlo results with showering, hadronization, and detector simulation}

\begin{figure}[!tb]
  \begin{minipage}[]{0.495\linewidth}
    \includegraphics[width=8.0cm,angle=0]{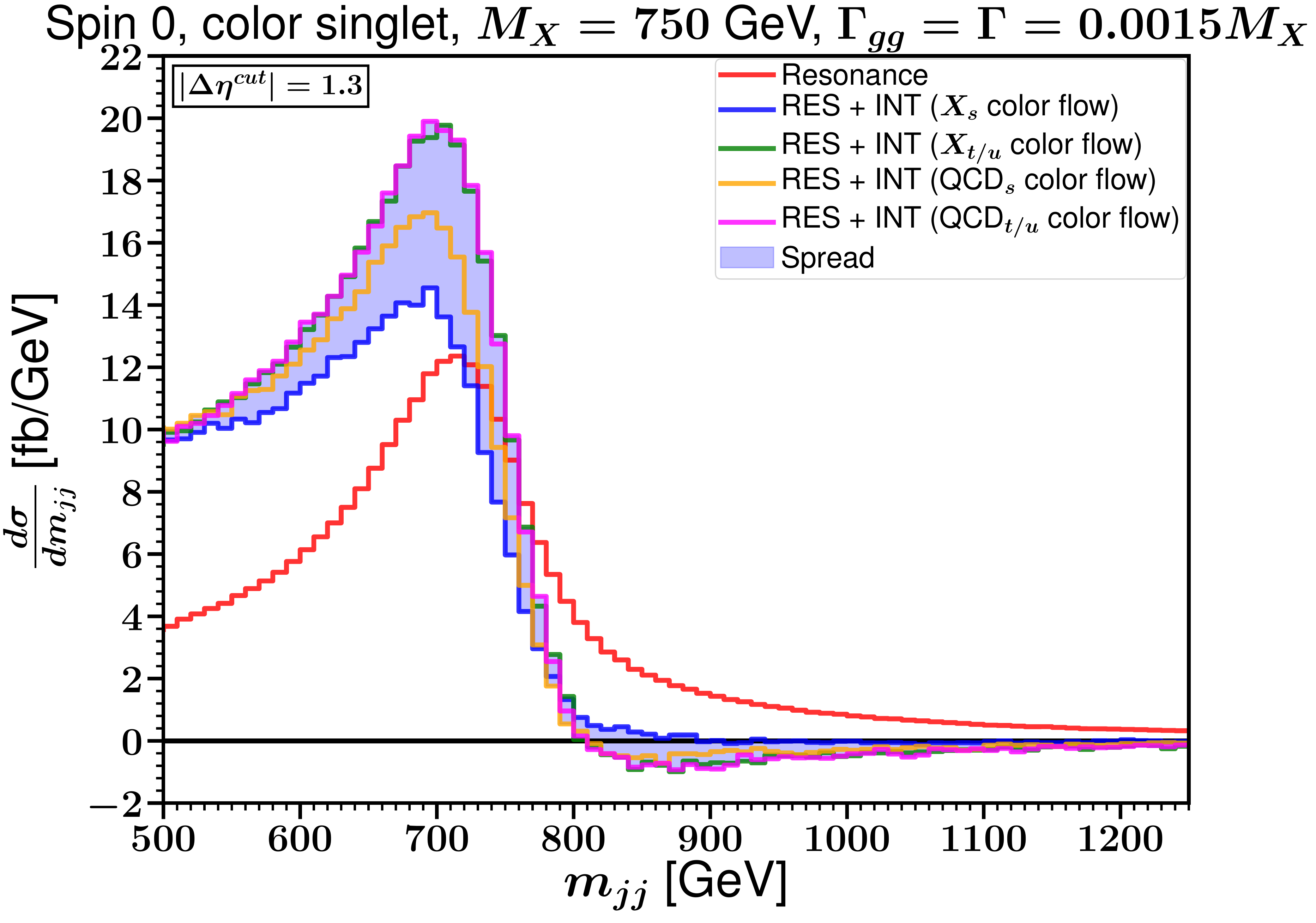}
  \end{minipage}
  \begin{minipage}[]{0.495\linewidth}
    \includegraphics[width=8.0cm,angle=0]{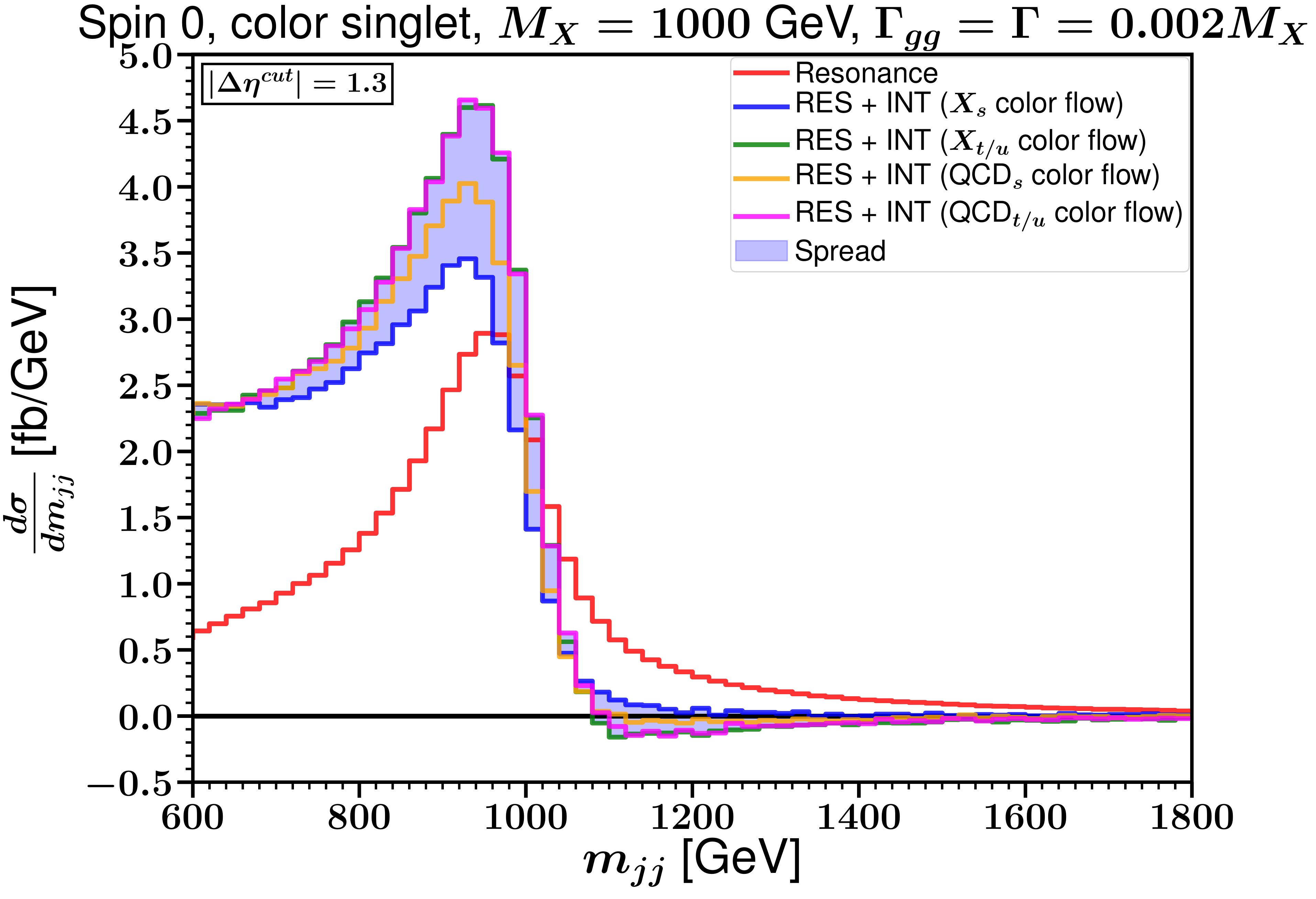}
  \end{minipage}

  \vspace{0.15cm}
    
  \begin{minipage}[]{0.495\linewidth}
    \includegraphics[width=8.0cm,angle=0]{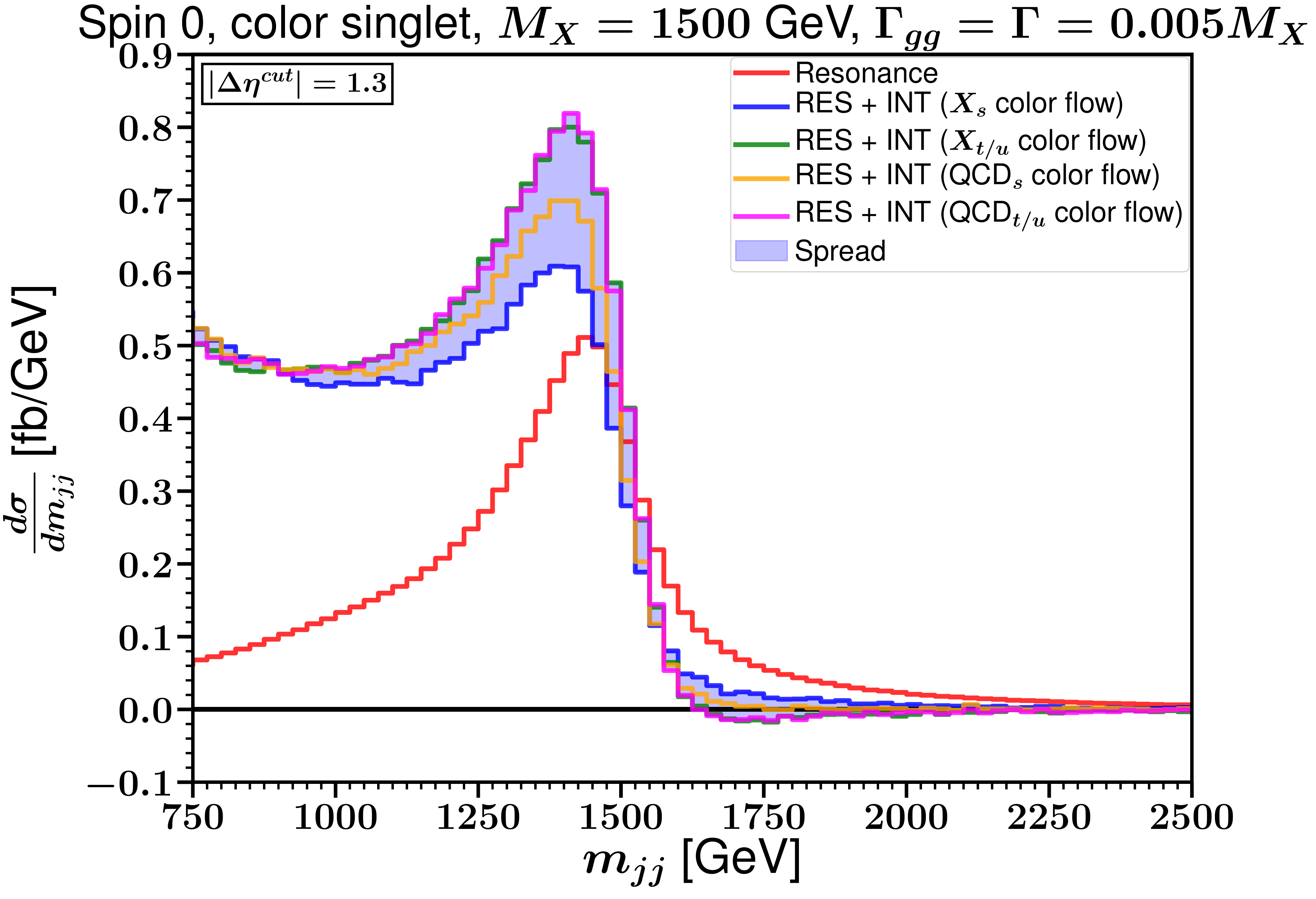}
  \end{minipage}
    \begin{minipage}[]{0.495\linewidth}
    \includegraphics[width=8.0cm,angle=0]{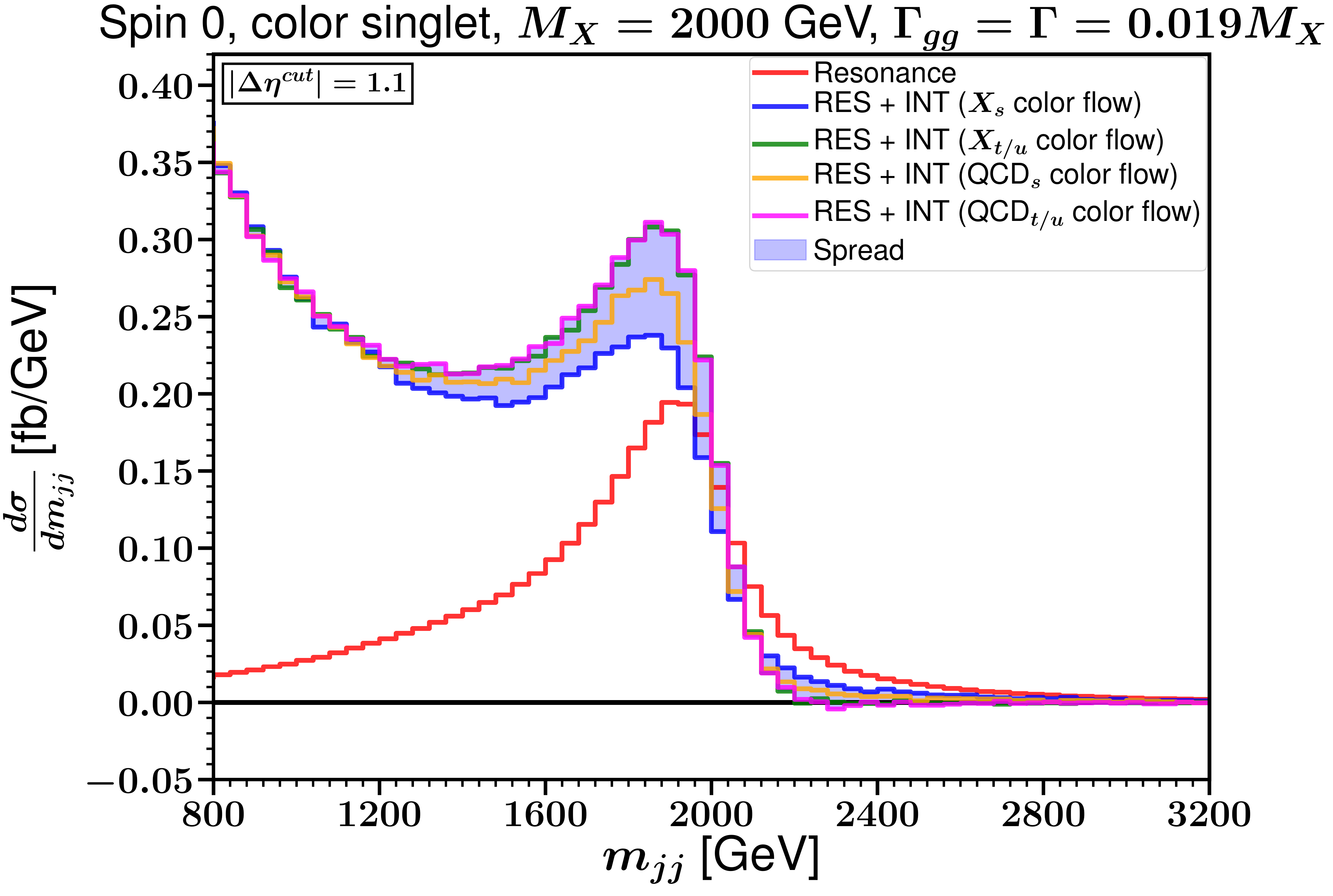}
  \end{minipage}

  \vspace{0.15cm}
    
  \begin{minipage}[]{0.495\linewidth}
    \includegraphics[width=8.0cm,angle=0]{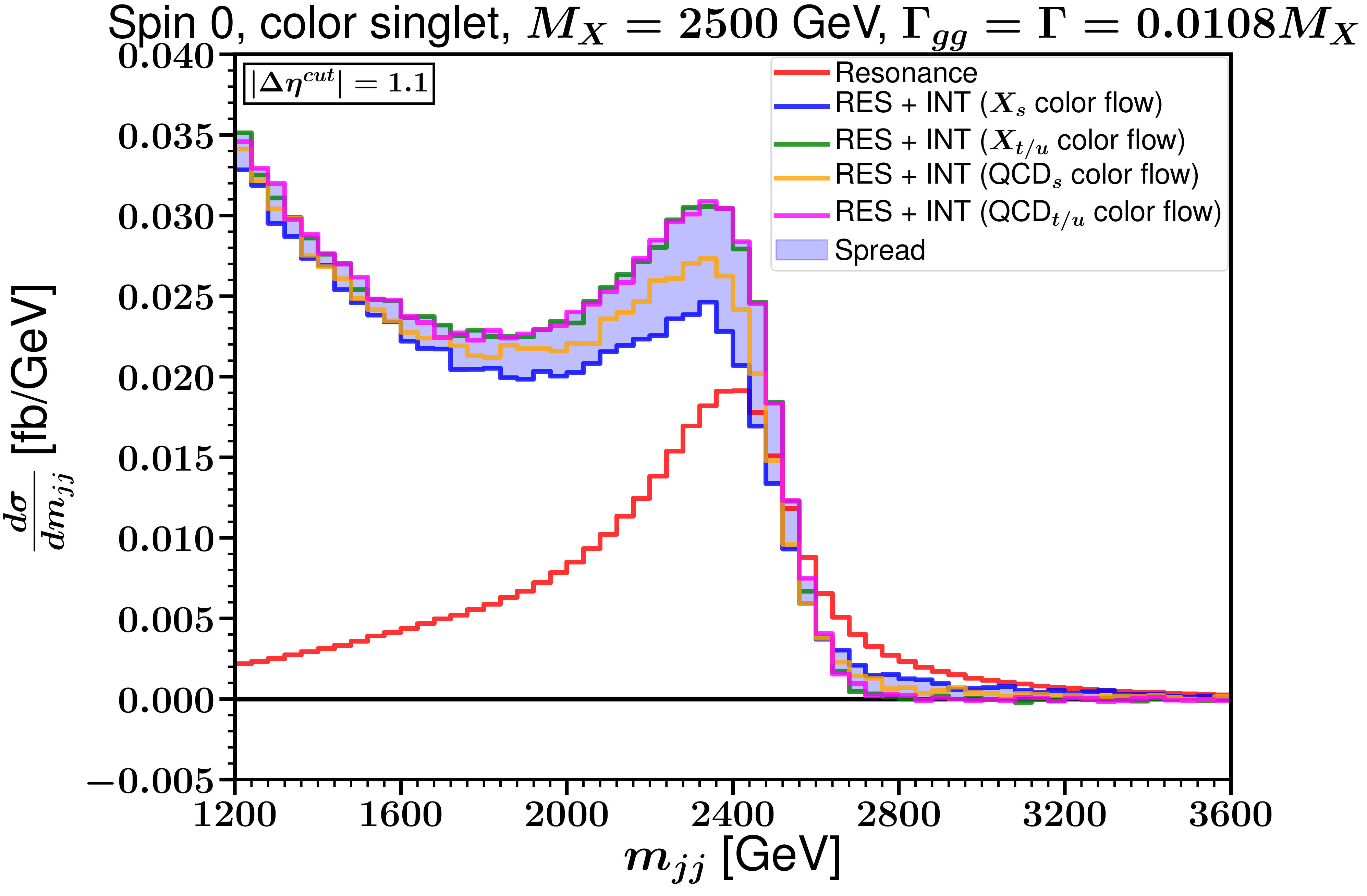}
  \end{minipage}
    \begin{minipage}[]{0.495\linewidth}
    \includegraphics[width=8.0cm,angle=0]{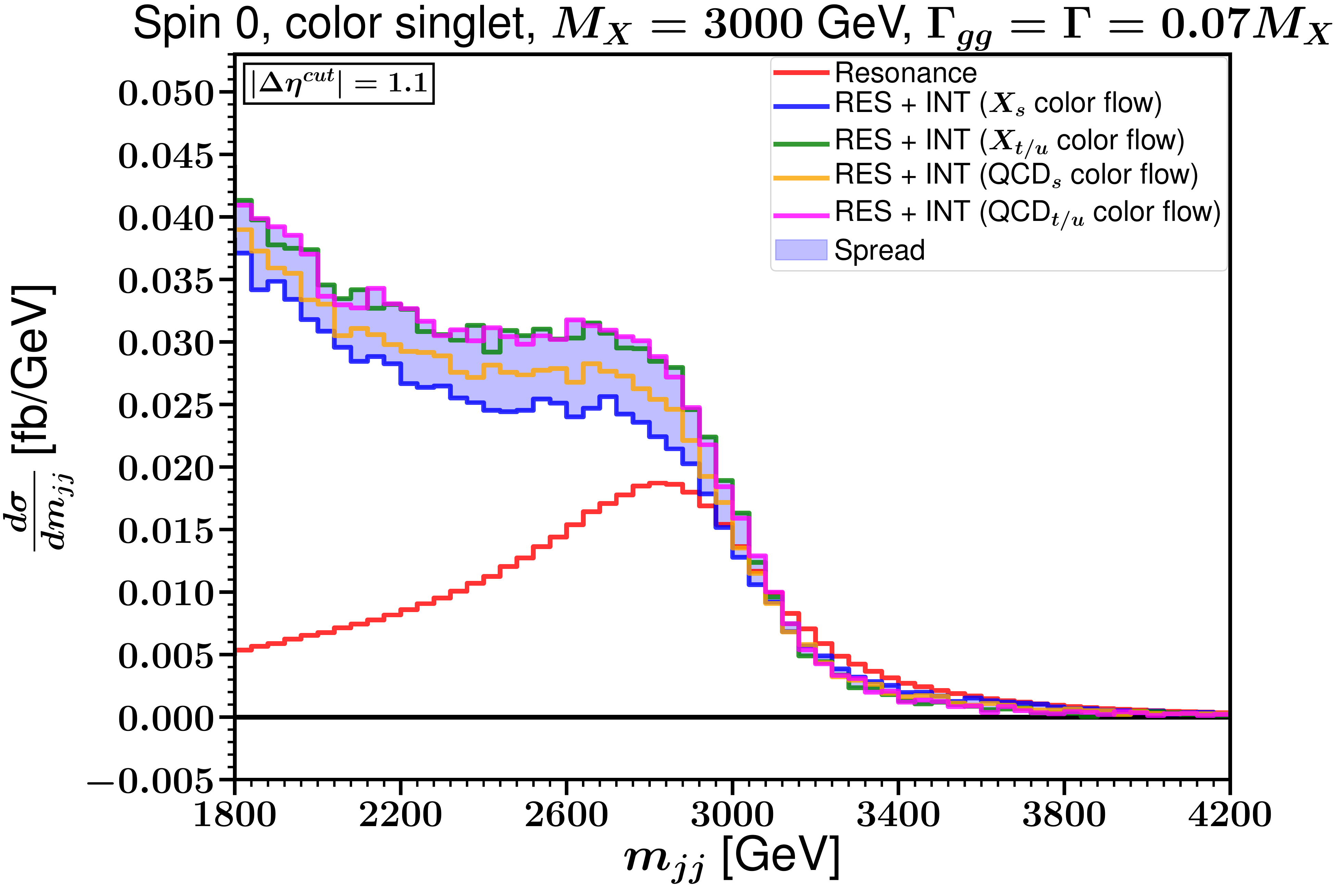}
  \end{minipage}
\begin{center}\begin{minipage}[]{0.95\linewidth}
\caption{\label{fig:gggg_s0c1} 
Dijet invariant mass distributions for the spin-0, color-singlet benchmarks of Table~\ref{tab:benchmarks} with $\Gamma=\Gamma_{gg}$, at the 13 TeV LHC, obtained with showering, hadronization and detector simulation. The red lines show the naive results with only the resonance diagrams of $g g \rightarrow X \rightarrow g g$ process (RES), which include the $s$-, $t$-, and $u$-channel exchanges of $X$, while the other four colored lines show the full results including interferences with the continuum QCD $g g \rightarrow g g$ amplitudes (INT) for all four color flows shown in Figure \ref{fig:colorflow}, as labeled. The shaded region shows the spread in the full result in each invariant mass bin for different color flow choices. The cases shown in the right column can be compared directly
to those in the right column of the previous Figure \ref{fig:gggg_s0c1_analytic} based on the more simplistic method of parton level with smearing.}
\end{minipage}\end{center}
\end{figure}

In Figure~\ref{fig:gggg_s0c1}, we show the dijet invariant mass distributions for the considered benchmarks of Table~\ref{tab:benchmarks} with $\Gamma=\Gamma_{gg}$, at 13 TeV LHC, for spin-0, color-singlet resonances, this time obtained using Monte Carlo generation of events followed by simulations with showering, hadronization and detector simulation.\footnote{As a check of our implementation of the $X$ interactions using {\sc FeynRules}, we verified that the parton-level distributions obtained using {\sc Madgraph} (not shown here) closely match the parton-level distributions obtained by our calculations as described in Section~\ref{subsec:partonlevel} and shown in Figure \ref{fig:gggg_s0c1_analytic}.
A similar check was done for each of the other (spin, color) combinations for $X$.}
The results are shown for a digluon resonant process $g g \rightarrow X \rightarrow g g$, which include 
only the $s$-, $t$-, and $u$-channel exchanges of $X$ (red line), and the full results. Here, 
the shaded blue region is the envelope of the results for the four possible color flows, as discussed
above in reference to Figure \ref{fig:colorflow}.

From Figure~\ref{fig:gggg_s0c1}, we note that the impact of QCD interference is least if we assume
the color flow is always the one that we have labeled $X_s$ (corresponding to the $s$-channel resonant process), and is greatest if we instead assume either the $X_{t/u}$ and QCD$_{t/u}$ color flows, which produce almost identical results. The color flow labeled QCD$_s$  produces intermediate results.
In any case, the QCD interference again is seen to change 
the naively expected resonance peak shape in a way consistent with the previous discussion.
Again, we note that the relative importance of interference seems to increase with the resonance mass.
The right-hand panels of Figure~\ref{fig:gggg_s0c1} can be compared to the right-hand panels of Figure~\ref{fig:gggg_s0c1_analytic}, as both have the same masses $M_X = 1000, 2000, 3000$ GeV. 
The match is of course not an exact one, because the smearing method using the 
detector responses in Figure~\ref{fig:yield}, is only a very rough approximation to the full-fledged event generation with showering, hadronization and detector simulation, and furthermore the color-flow uncertainty 
is evidently a non-trivial one. The more complete treatment tends to give lower 
yields in this case. However, the qualitative similarity between the results of the two 
methods is a useful check. 
Even the more complete detector simulation is of course
different from the true CMS and ATLAS detector responses.

\clearpage

\subsection{Spin 0, color octet}

\subsubsection{Parton-level with smearing}

\begin{figure}[!tb]
  \begin{minipage}[]{0.495\linewidth}
    \includegraphics[width=8.0cm,angle=0]{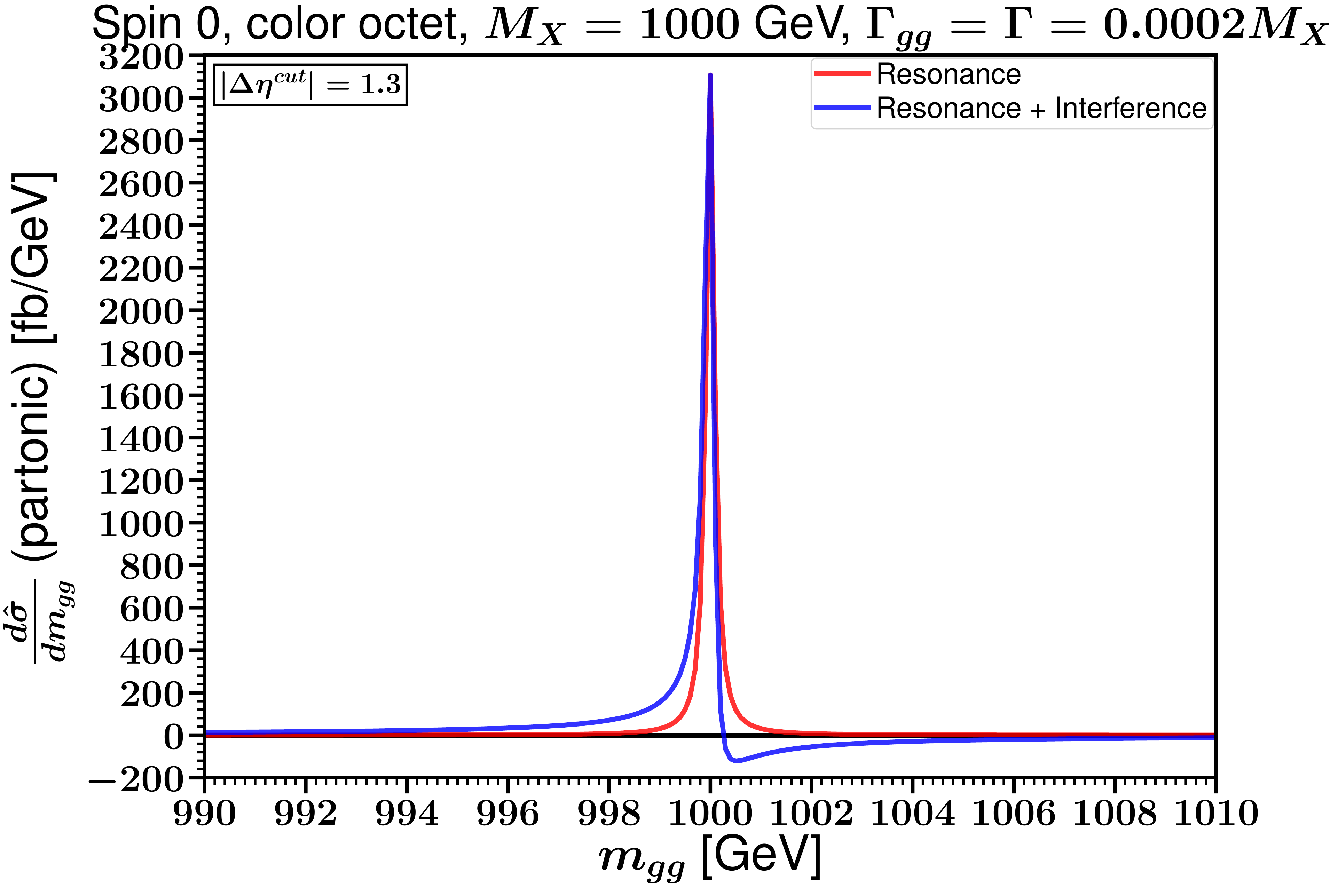}
  \end{minipage}
  \begin{minipage}[]{0.495\linewidth}
    \includegraphics[width=8.0cm,angle=0]{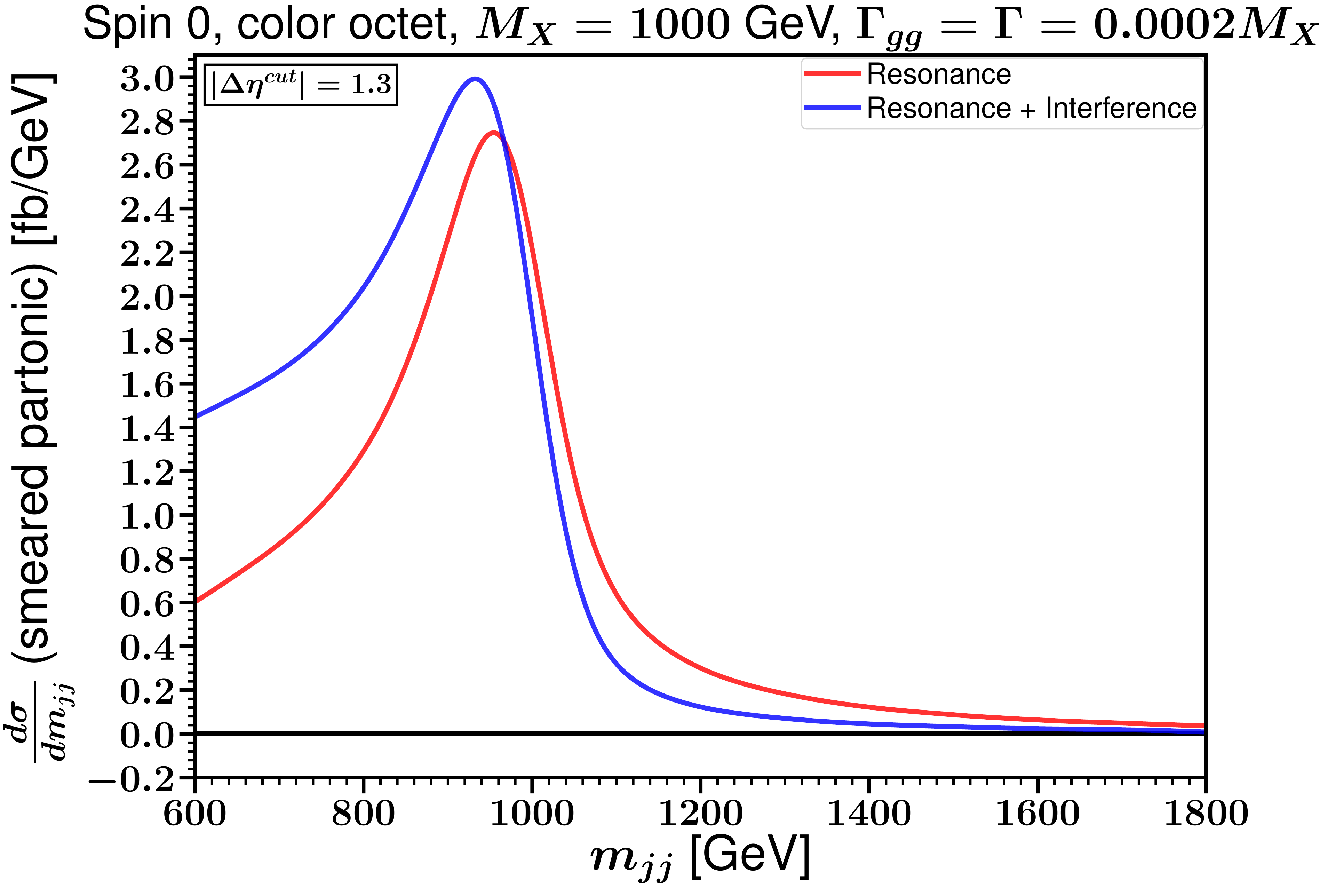}
  \end{minipage}

  \vspace{0.15cm}
  
  \begin{minipage}[]{0.495\linewidth}
    \includegraphics[width=8.0cm,angle=0]{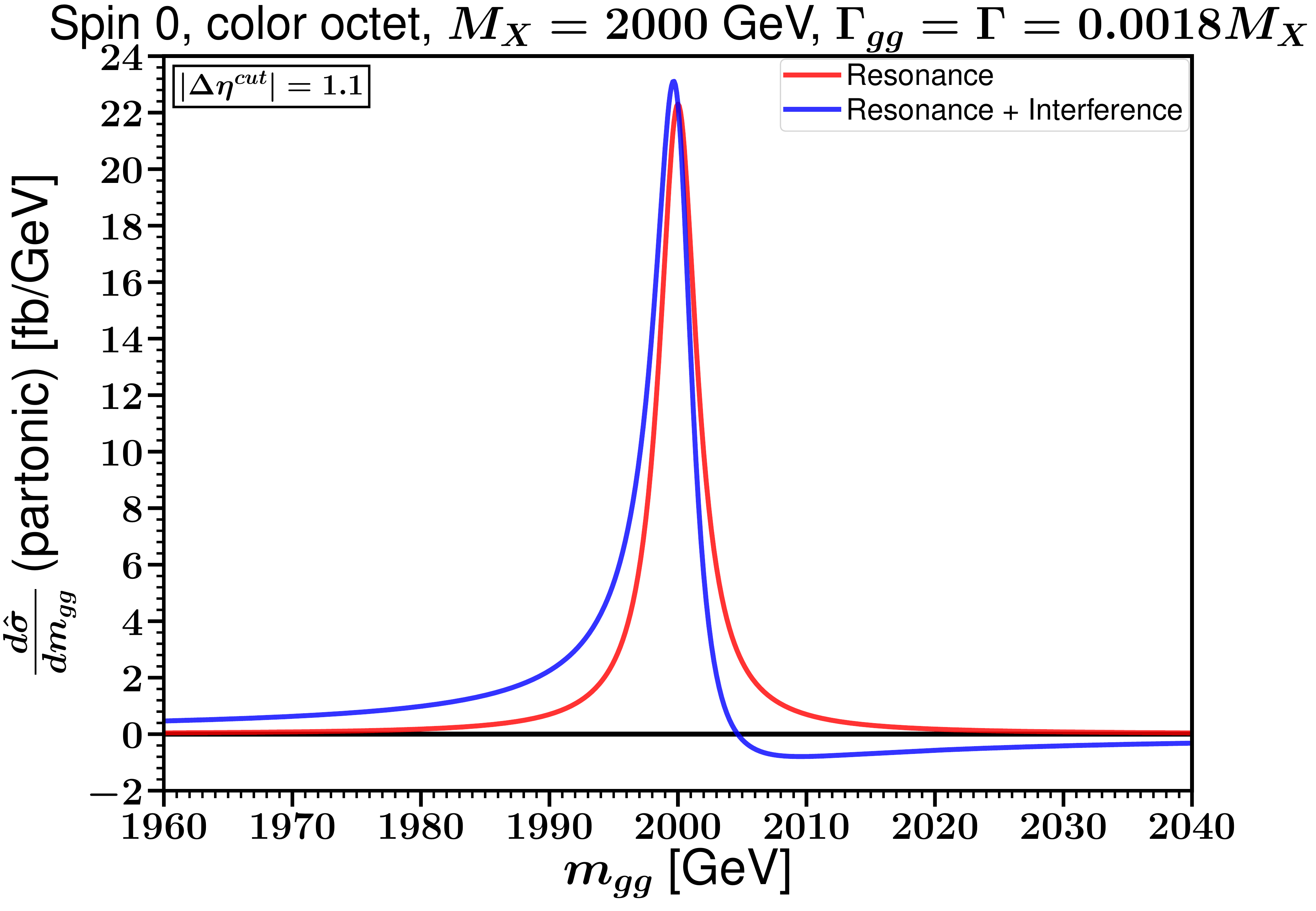}
  \end{minipage}
    \begin{minipage}[]{0.495\linewidth}
    \includegraphics[width=8.0cm,angle=0]{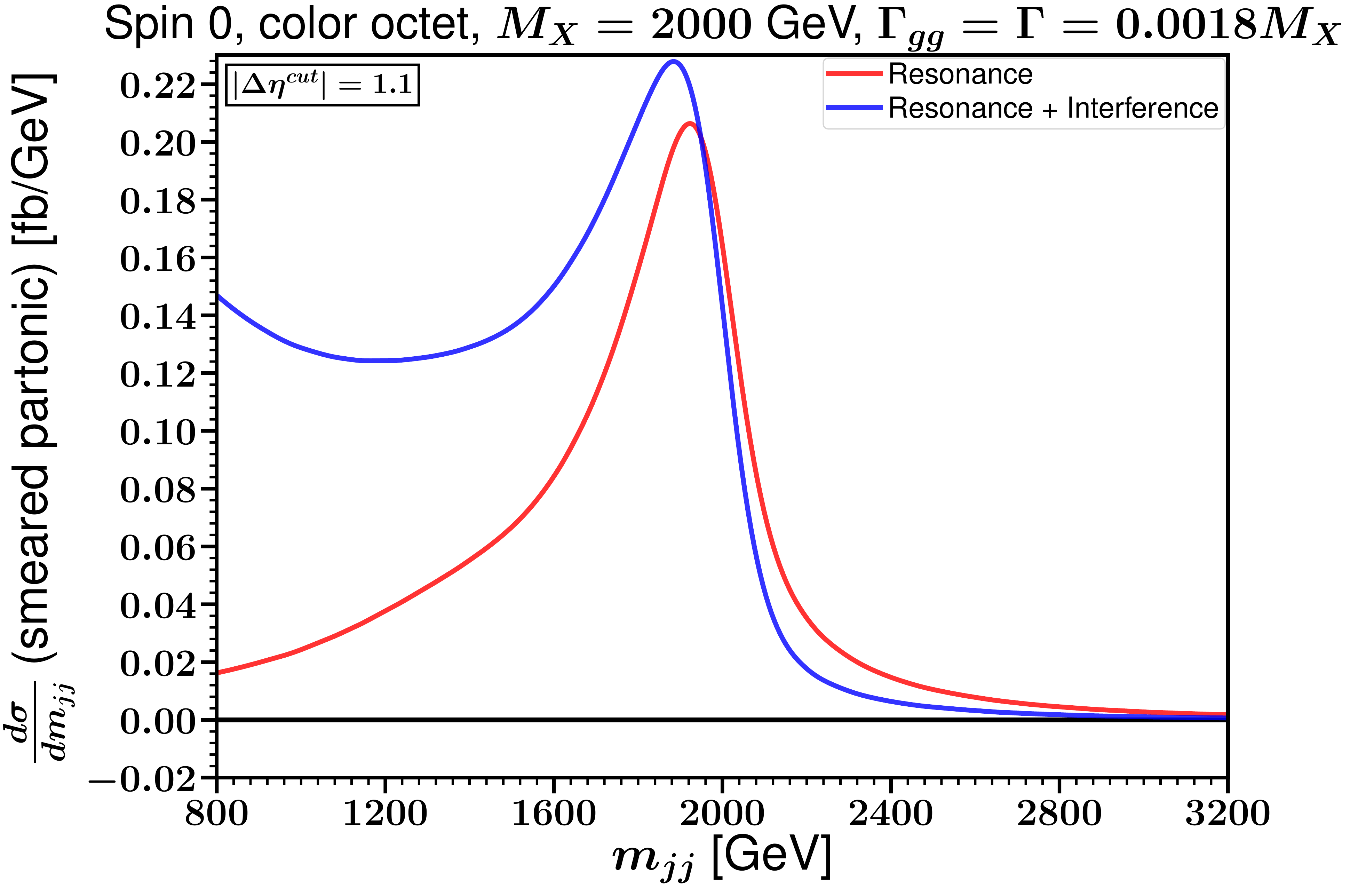}
  \end{minipage}

  \vspace{0.15cm}
  
  \begin{minipage}[]{0.495\linewidth}
    \includegraphics[width=8.0cm,angle=0]{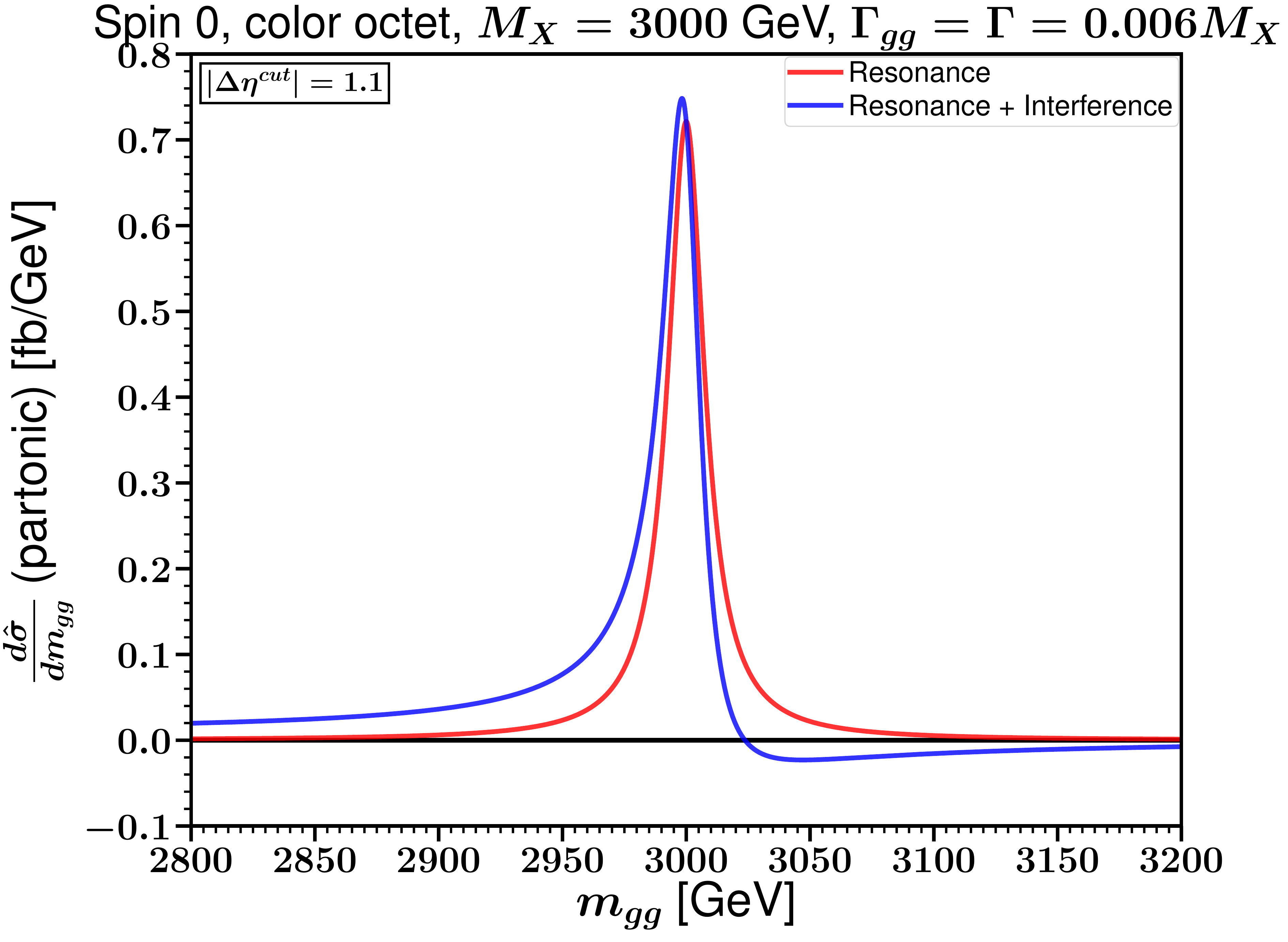}
  \end{minipage}
    \begin{minipage}[]{0.495\linewidth}
    \includegraphics[width=8.0cm,angle=0]{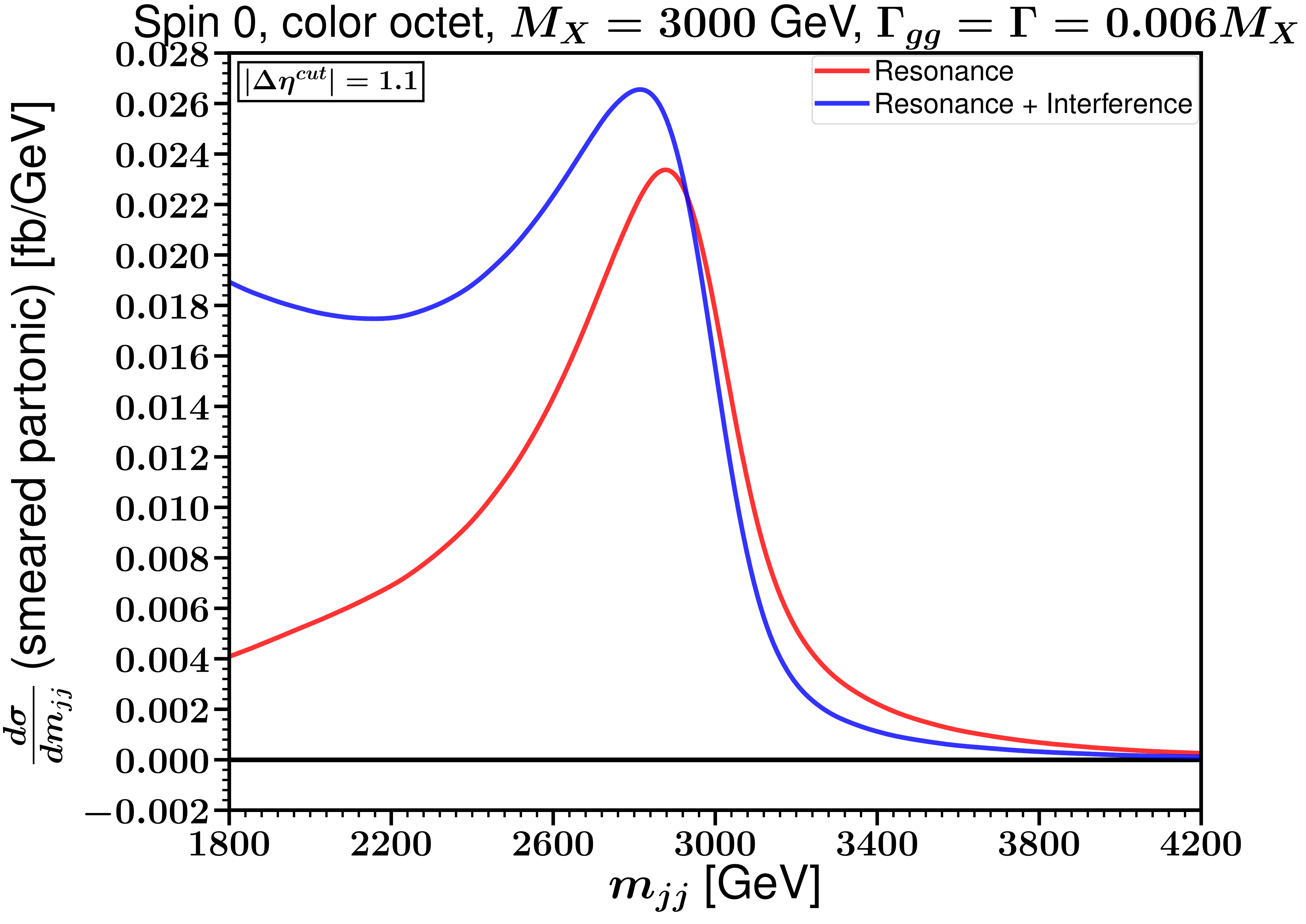}
  \end{minipage}
\begin{center}\begin{minipage}[]{0.95\linewidth}
\caption{\label{fig:gggg_s0c8_analytic} 
Digluon invariant mass distributions, at the 13 TeV LHC, for spin-0, color-octet resonances 
with $(M_X, \Gamma/M_X) =$ (1000 GeV, 0.0002) [top row], 
(2000 GeV, 0.0018) [middle row], and (3000 GeV, 0.006) [bottom row].
The parton-level distributions are shown in the left column panels. These are then smeared by convolution with the detector responses shown in Figure~\ref{fig:yield}, to obtain the dijet invariant mass distributions in the right panels. In all six panels, the red lines show the naive results for the resonant signal $g g \rightarrow X \rightarrow g g$, while the blue lines show the full results including the interferences with the  QCD background $g g \rightarrow g g$.}
\end{minipage}\end{center}
\end{figure}

In Figure~\ref{fig:gggg_s0c8_analytic}, we show the parton-level digluon invariant mass distributions 
for $pp$ collisions at $\sqrt{s}=13$ TeV, for massive spin-0, color-octet resonances, for the benchmark examples of Table~\ref{tab:benchmarks} with masses 1000, 2000, and 3000 GeV in the top, middle and bottom
rows, respectively.
As in the previous subsection, 
the parton-level distributions before smearing are shown in the left panels, and the right panels show
the distributions after convolution with approximate detector responses shown in Figure~\ref{fig:yield}, as described in Section~\ref{subsec:partonlevel}.
In all six panels, the red lines show the results for the resonant signal $g g \rightarrow X \rightarrow g g$ with all $X$ exchange diagrams, while the blue lines show the full results including the interferences with the  QCD amplitudes $g g \rightarrow g g$.

Just as was the case for spin-0, color-singlet resonances, the parton-level results before smearing show 
a peak below $m_{jj} < M_X$ and a dip above $m_{jj} > M_X$ for the full result (blue line), as opposed to a pure peak (red line) that we get without including the interference terms.  After smearing, the deficit above the input resonance mass, compared to the naive result, does not appear as considerable as in the color-singlet case. The main feature is again the presence of the low-mass
positive tail. Depending on how this would be absorbed into the QCD background fit, this 
could again lead to both a peak slightly below $M_X$ and an apparent dip 
in the differential distribution above $M_X$. However, the interference effects for spin-0, color-octet resonances are not as big as for spin-0, color-singlets, as we anticipated above in the discussion immediately following eq.~(\ref{eq:fiveoverfortyeight}).

\subsubsection{Monte Carlo results with showering, hadronization, and detector simulation}
\begin{figure}[!tb]
  \begin{minipage}[]{0.495\linewidth}
    \includegraphics[width=8.0cm,angle=0]{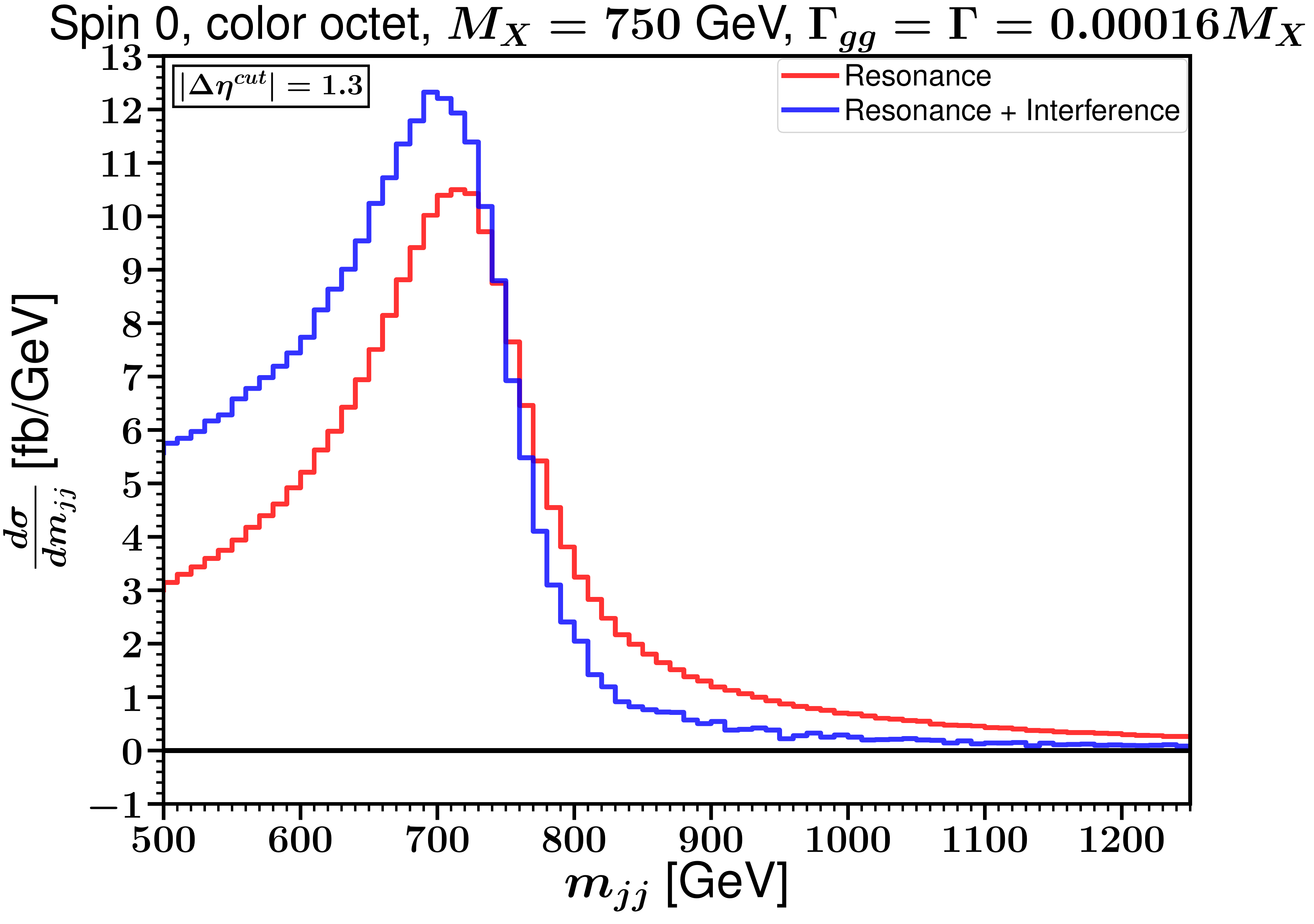}
  \end{minipage}
    \begin{minipage}[]{0.495\linewidth}
    \includegraphics[width=8.0cm,angle=0]{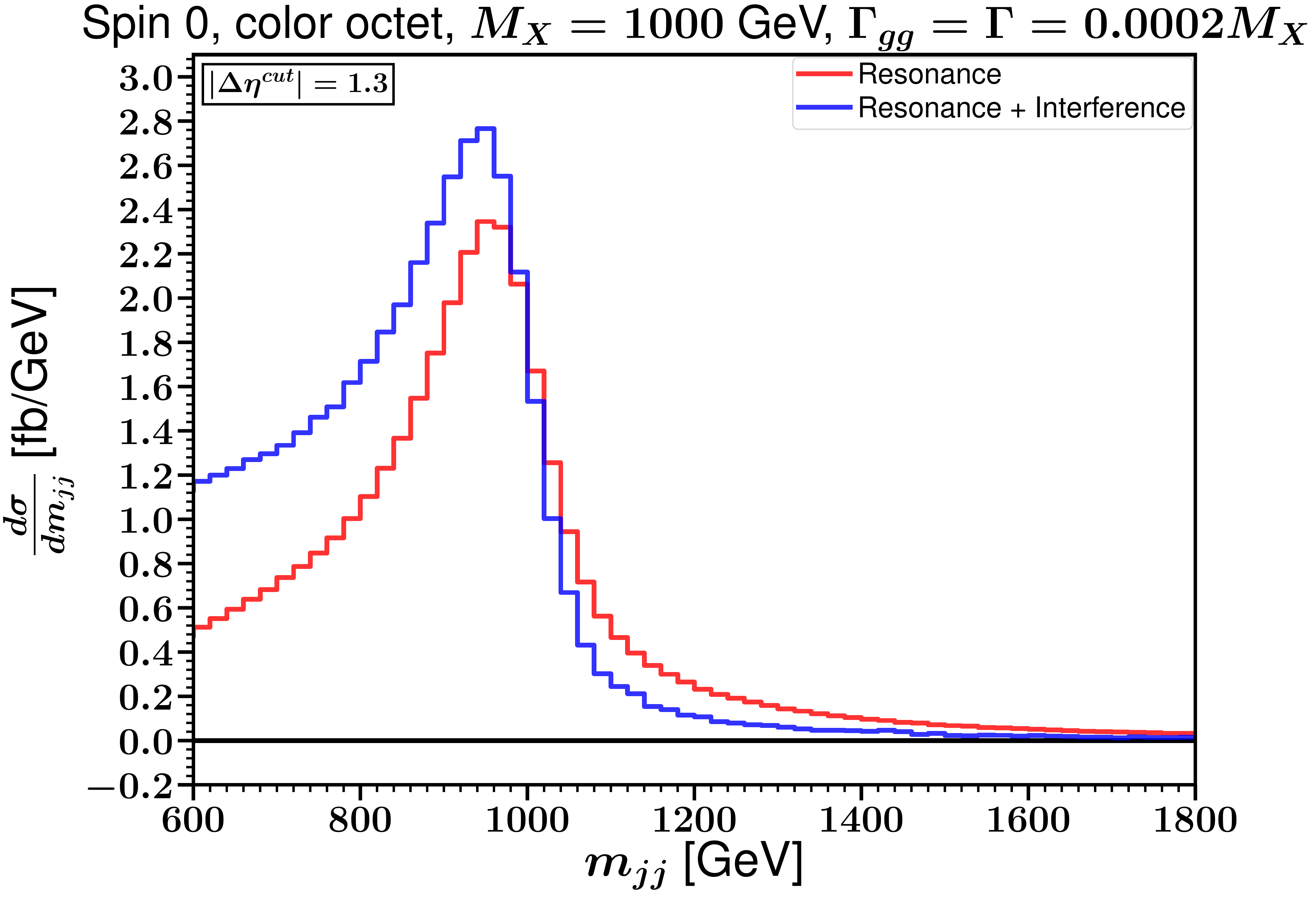}
  \end{minipage}

  \vspace{0.15cm}
  
  \begin{minipage}[]{0.495\linewidth}
    \includegraphics[width=8.0cm,angle=0]{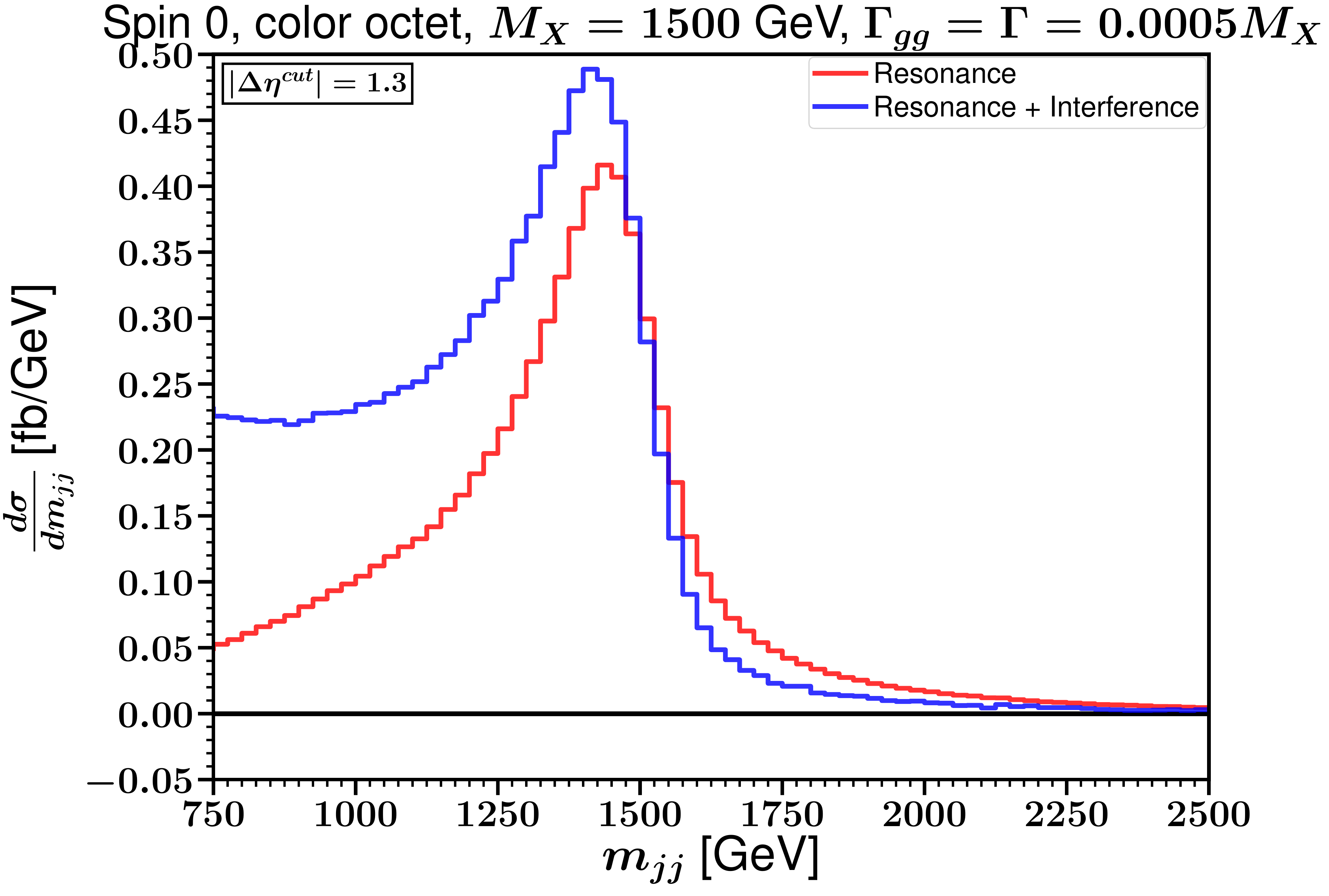}
  \end{minipage}
    \begin{minipage}[]{0.495\linewidth}
    \includegraphics[width=8.0cm,angle=0]{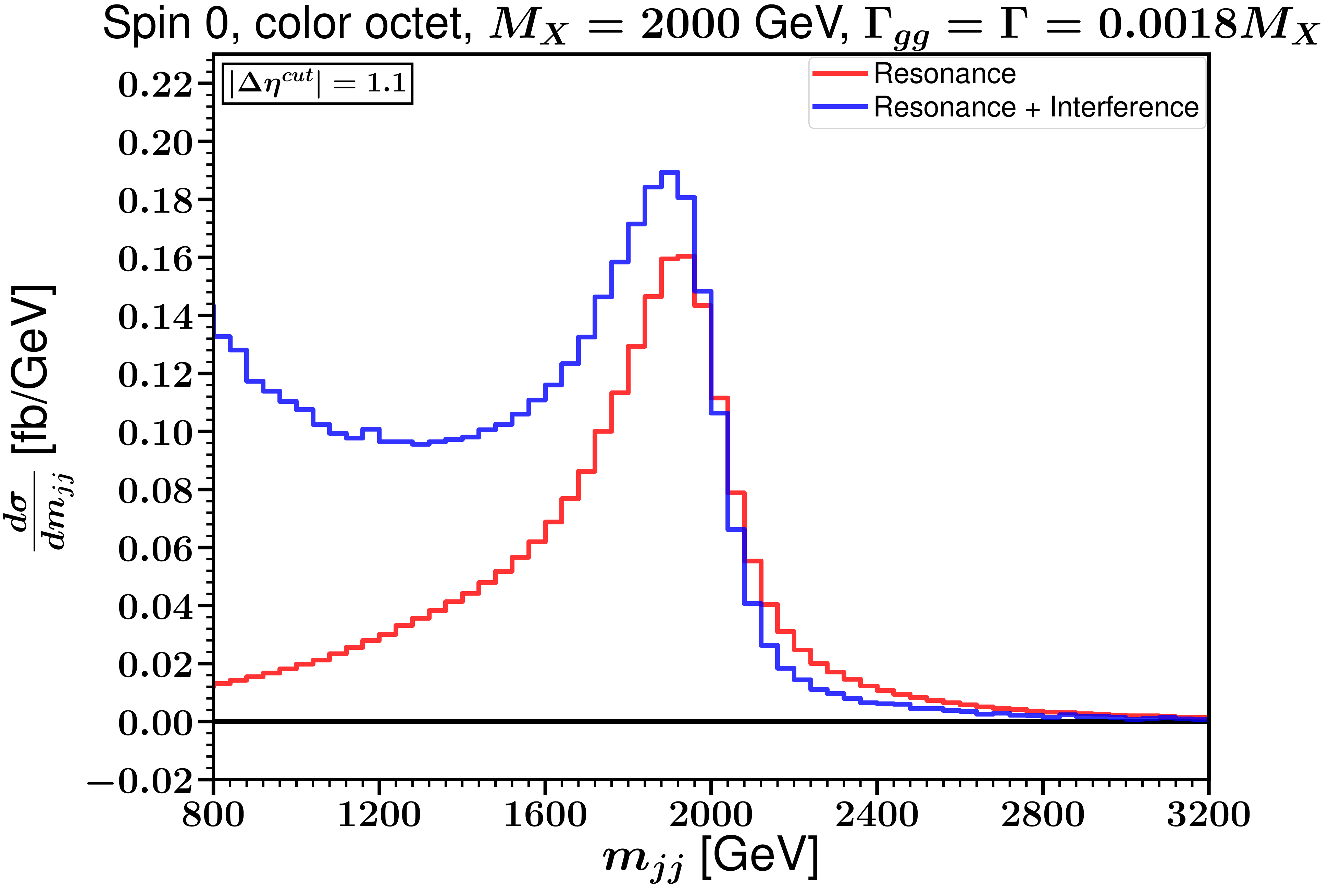}
  \end{minipage}

  \vspace{0.15cm}
  
  \begin{minipage}[]{0.495\linewidth}
    \includegraphics[width=8.0cm,angle=0]{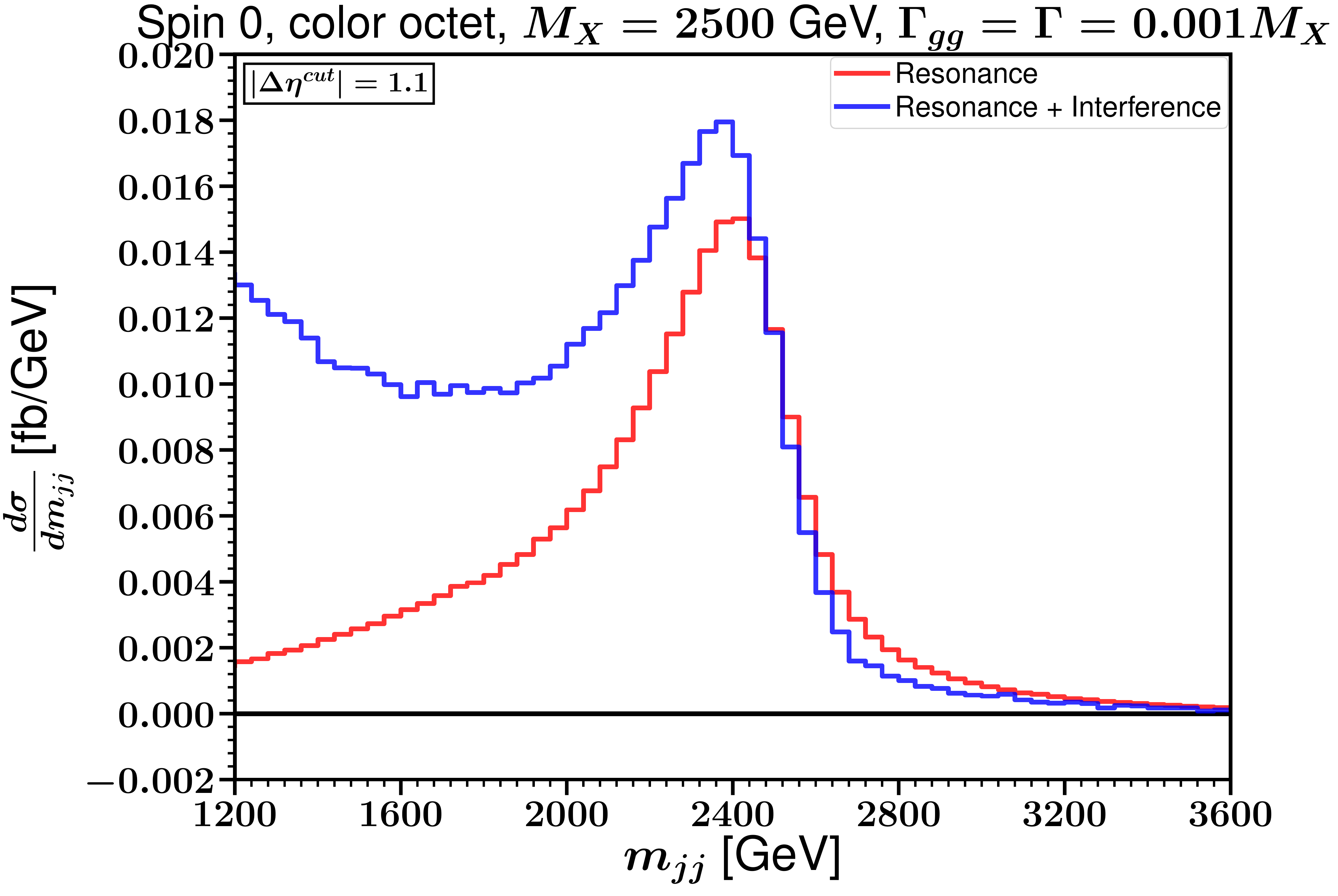}
  \end{minipage}
    \begin{minipage}[]{0.495\linewidth}
    \includegraphics[width=8.0cm,angle=0]{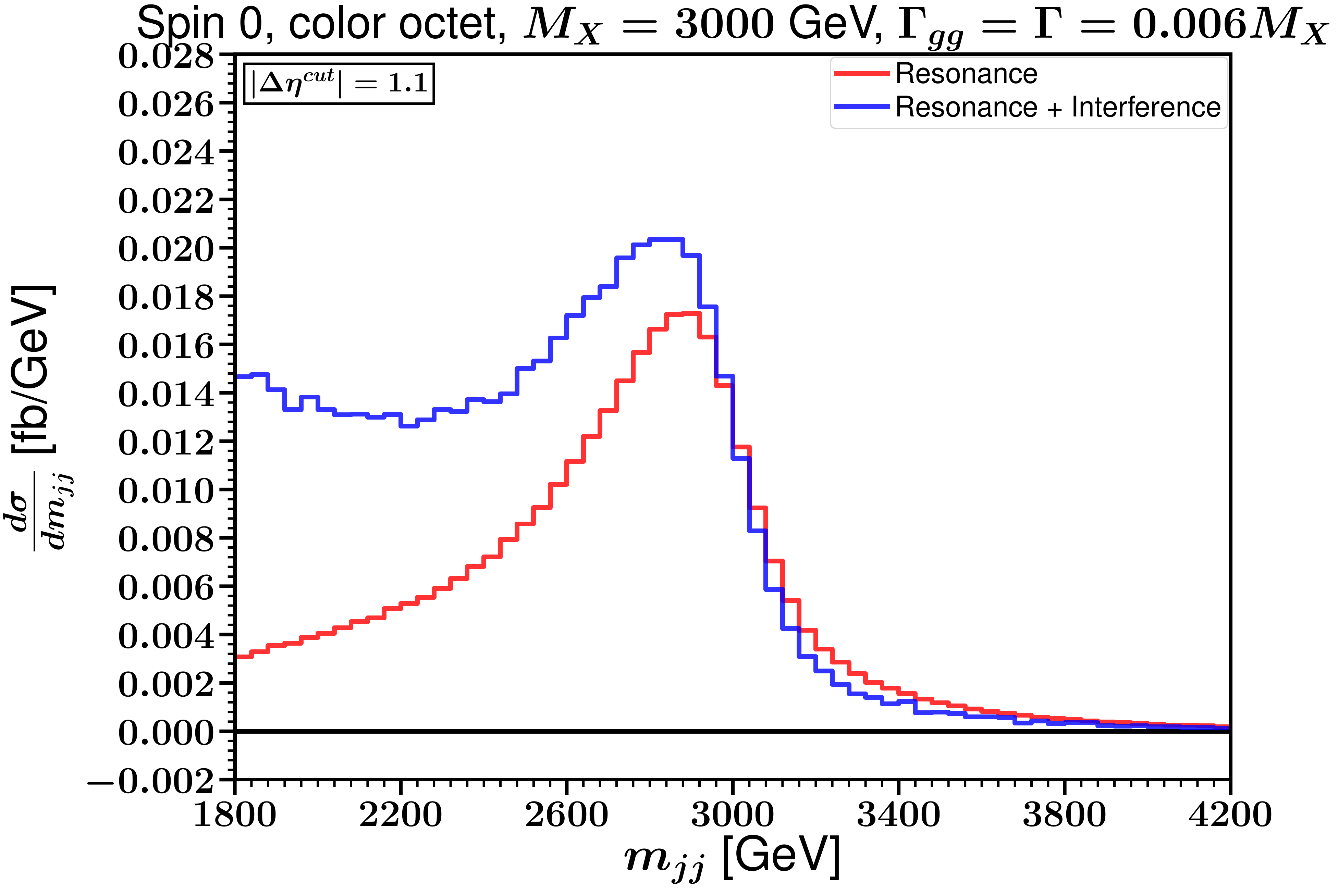}
  \end{minipage}
\begin{center}\begin{minipage}[]{0.95\linewidth}
\caption{\label{fig:gggg_s0c8} 
Dijet invariant mass distributions for the spin-0, color-octet benchmarks of Table~\ref{tab:benchmarks} with $\Gamma=\Gamma_{gg}$, at the 13 TeV LHC, obtained with showering, hadronization and detector simulation. The red lines show the naive results with the resonance diagrams of $g g \rightarrow X \rightarrow g g$ process (RES), which include the $s$-, $t$-, and $u$-channel exchanges of $X$, while the blue lines show the full results including interferences with the continuum QCD $g g \rightarrow g g$ amplitudes (INT).
The cases shown in the right column can be compared directly
to those in the right column of the previous Figure \ref{fig:gggg_s0c8_analytic} based on the more simplistic method of parton level with smearing.}
\end{minipage}\end{center}
\end{figure}

Figure~\ref{fig:gggg_s0c8} shows the dijet invariant mass distributions for the considered benchmarks of Table~\ref{tab:benchmarks} with $X$ assumed to always decay to a pair of gluons, at 13 TeV LHC, for spin-0, color-octet resonances, obtained using Monte Carlo simulations.
The results are shown for the naive resonant process $g g \rightarrow X \rightarrow g g$, and for the full results, which also include the interferences with the continuum QCD $g g \rightarrow g g$ amplitudes.

From Figure~\ref{fig:gggg_s0c8}, the QCD interference with a spin-0, color-octet resonance has the aforementioned feature of having less dramatic positive tails in the region $m_{jj} < M_X$ with almost no negative tails in the region $m_{jj} > M_X$.
Unlike the case of a color-singlet scalar, here we have exactly one result for the QCD interferences because the color flow is uniquely determined. 
The relative importance of the interference again increases as one moves to higher resonance masses $M_X$.
Again, comparing the right column panels of Figure~\ref{fig:gggg_s0c8} to the corresponding right column panels of Figure~\ref{fig:gggg_s0c8_analytic}, which have the same masses 1000, 2000, and 3000 GeV, the fully simulated results don't exactly match with the smeared parton-level results, with the more complete simulation producing a lower yield, but the shapes are reassuringly qualitatively similar.

\clearpage

\subsection{Spin 1, color octet}

\subsubsection{Parton-level with smearing}

\begin{figure}[!tb]
  \begin{minipage}[]{0.495\linewidth}
    \includegraphics[width=8.0cm,angle=0]{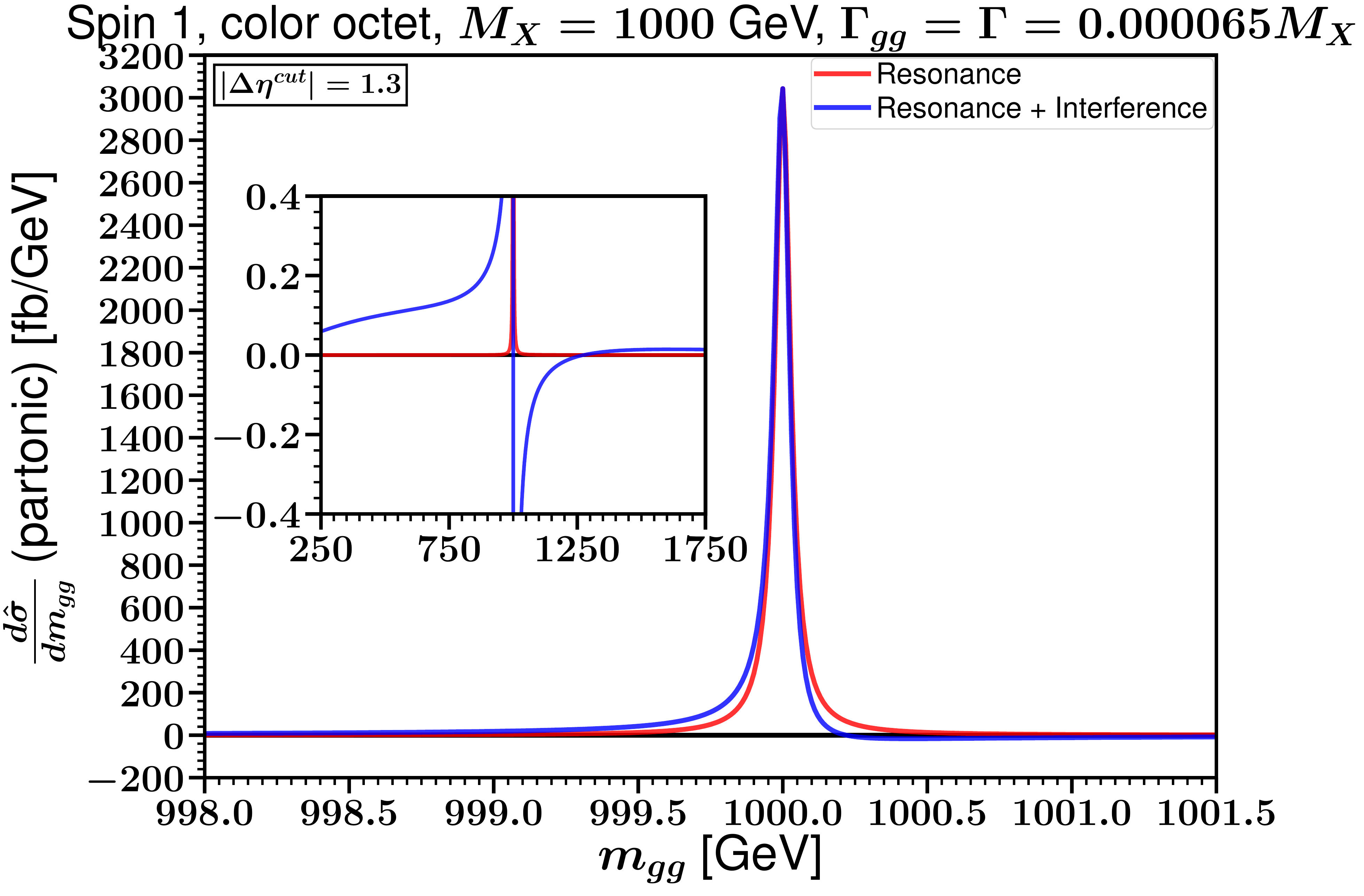}
  \end{minipage}
    \begin{minipage}[]{0.495\linewidth}
    \includegraphics[width=8.0cm,angle=0]{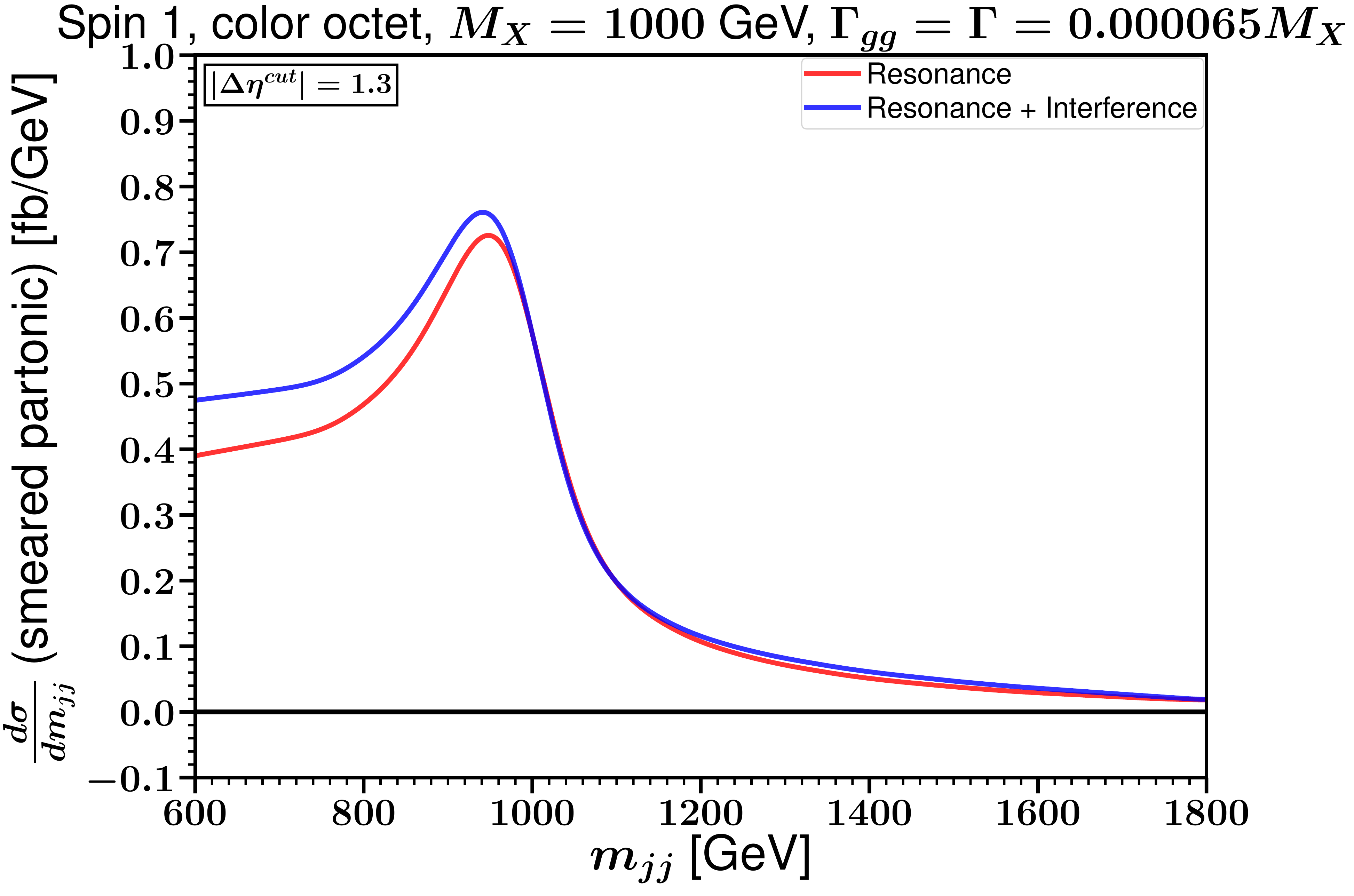}
  \end{minipage}

  \vspace{0.15cm}
  
  \begin{minipage}[]{0.495\linewidth}
    \includegraphics[width=8.0cm,angle=0]{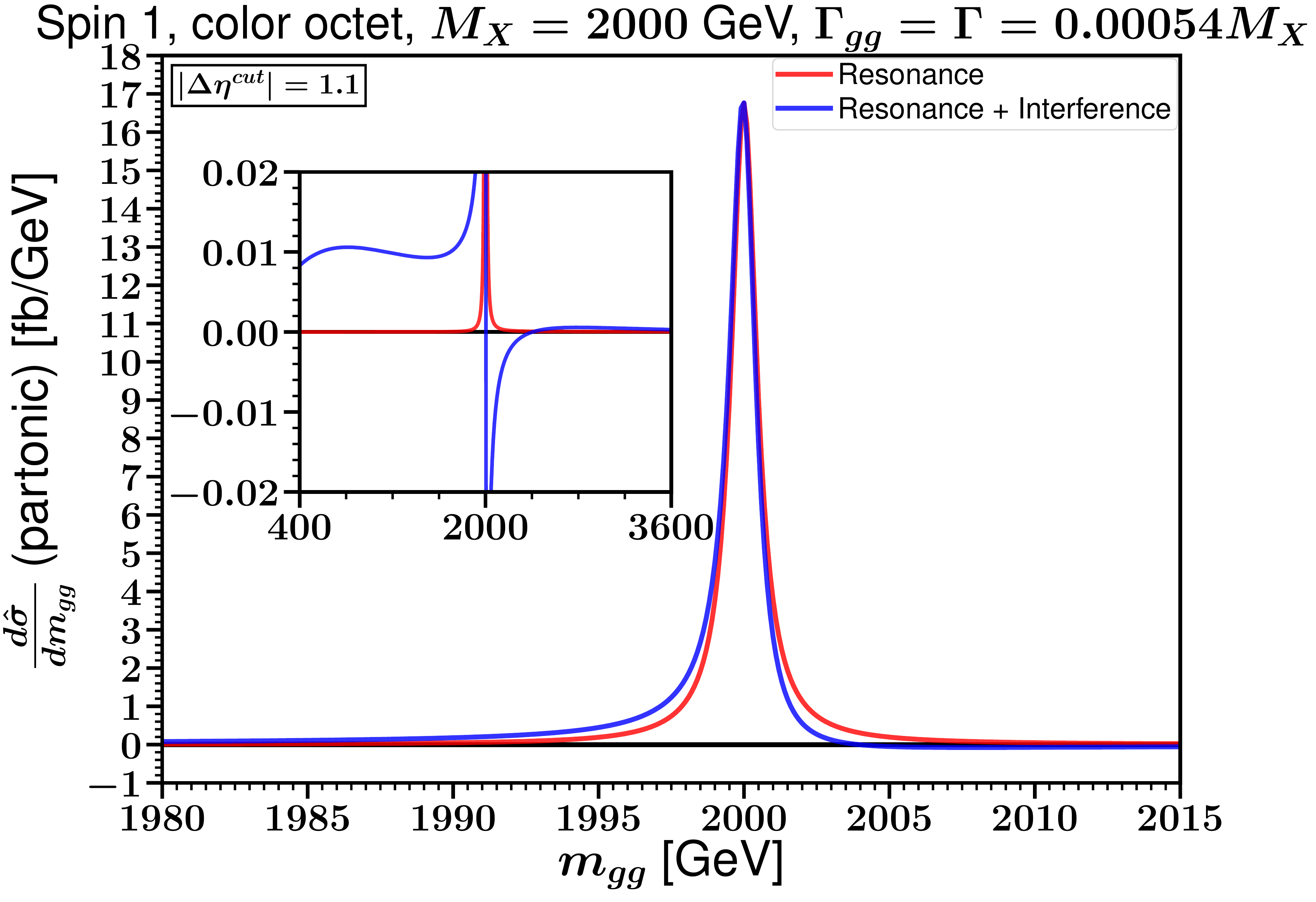}
  \end{minipage}
    \begin{minipage}[]{0.495\linewidth}
    \includegraphics[width=8.0cm,angle=0]{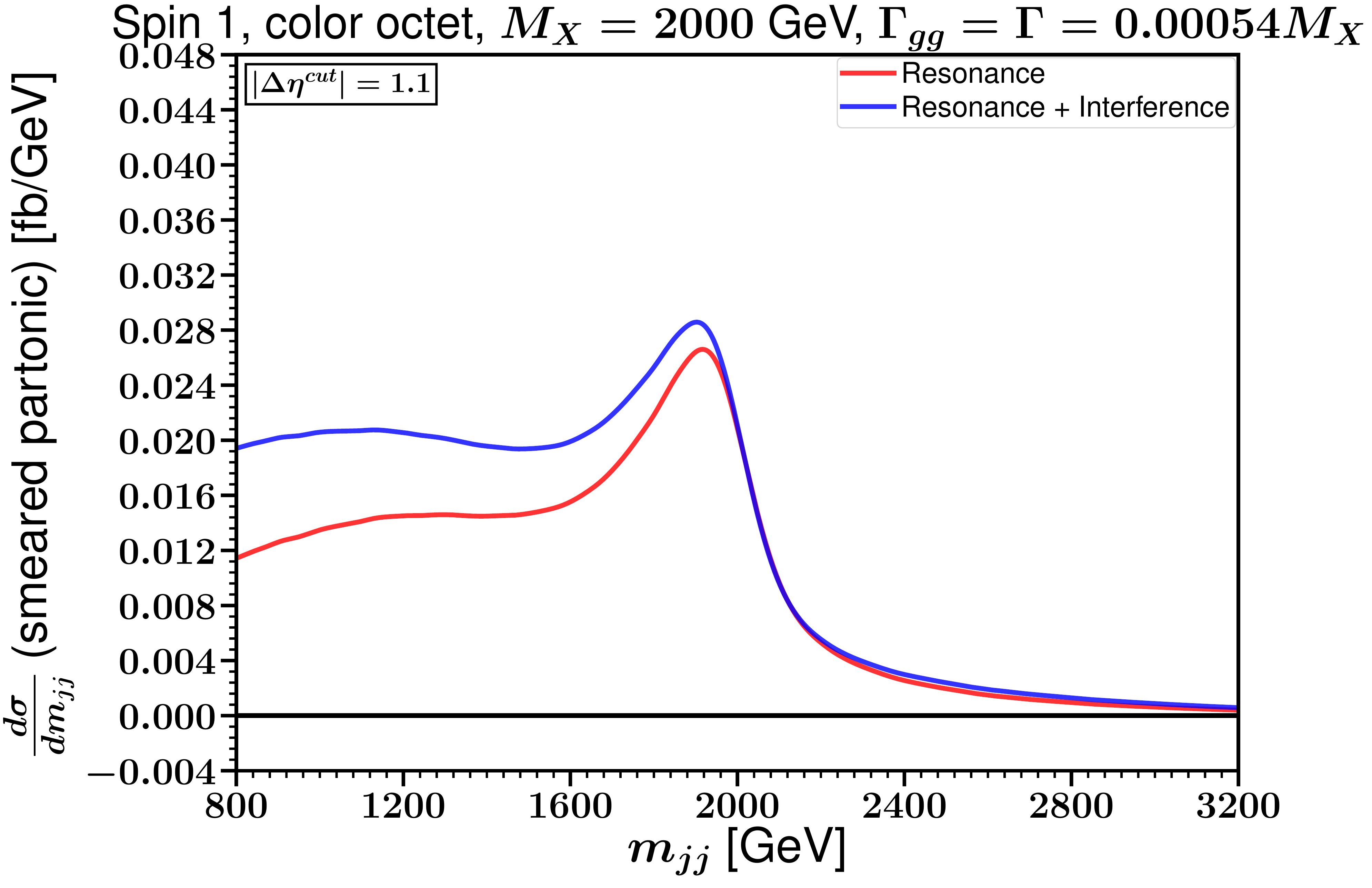}
  \end{minipage}

  \vspace{0.15cm}
  
  \begin{minipage}[]{0.495\linewidth}
    \includegraphics[width=8.0cm,angle=0]{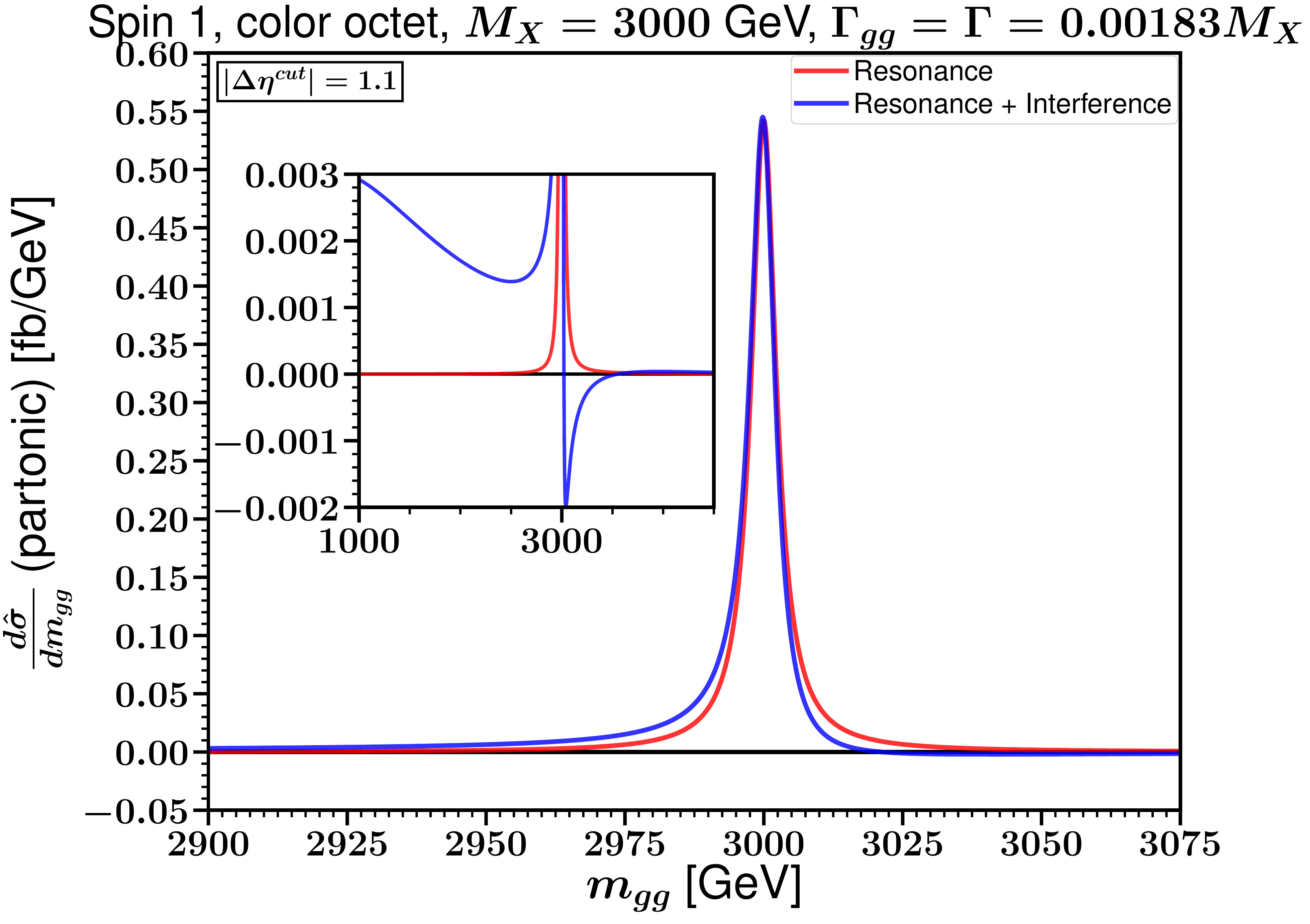}
  \end{minipage}
    \begin{minipage}[]{0.495\linewidth}
    \includegraphics[width=8.0cm,angle=0]{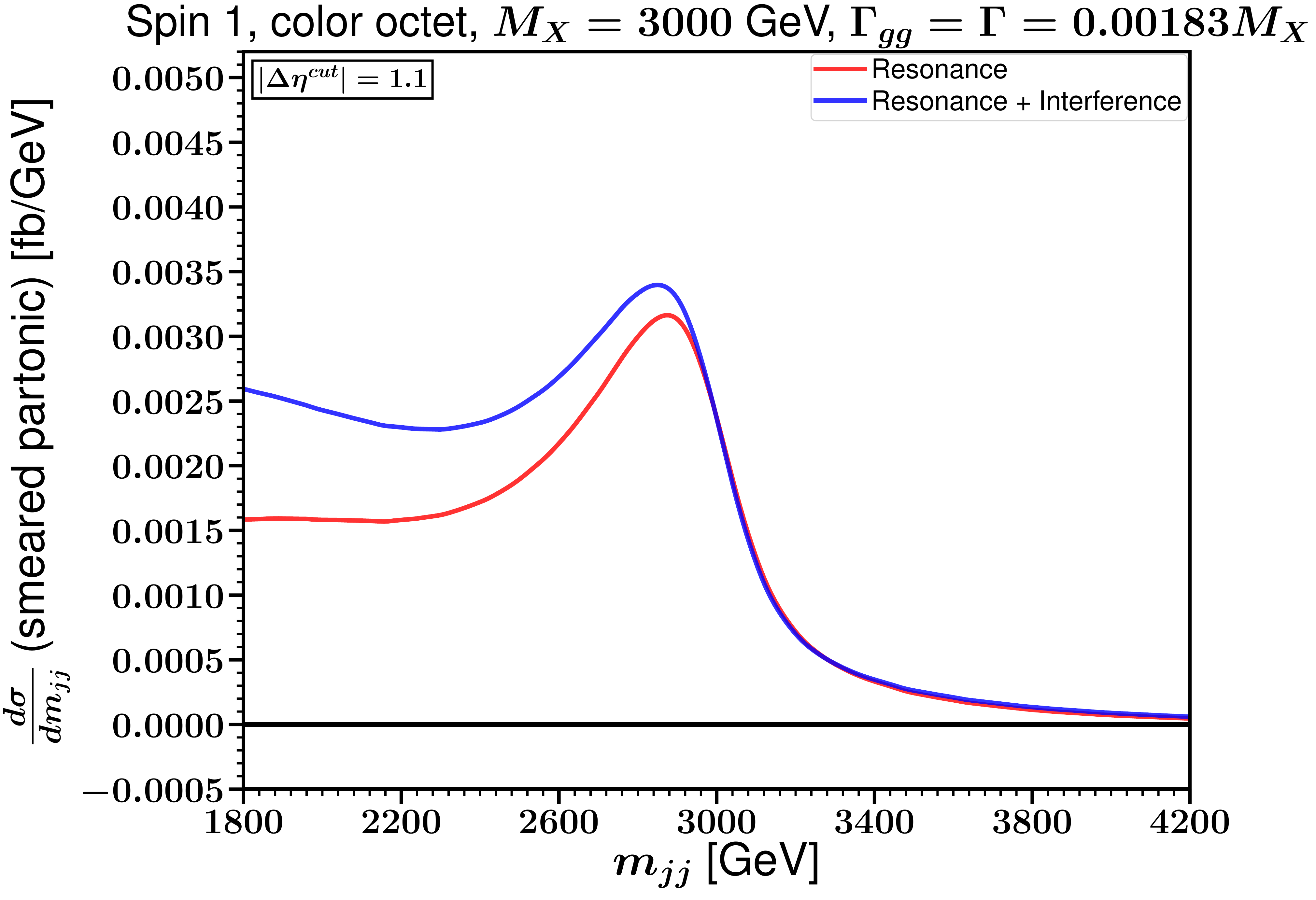}
  \end{minipage}
\begin{center}\begin{minipage}[]{0.95\linewidth}
\caption{\label{fig:gggg_s1c8_analytic} 
Digluon invariant mass distributions, at the 13 TeV LHC, 
for spin-1 color-octet resonances with 
$(M_X, \Gamma/M_X)=$ (1000 GeV, $6.5 \times 10^{-5}$) [top row], 
(2000 GeV, $5.4 \times 10^{-4}$) [middle row] and 
(3000 GeV, 0.00183) [bottom row].
The parton-level distributions (left panels) are smeared by convolution with the detector response, shown in Figure~\ref{fig:yield}, to obtain the dijet invariant mass distributions (right panels). 
In all six panels, the red lines show the naive results for the resonant signal $g g \rightarrow X \rightarrow g g$, while the blue lines show the full results including the interferences with the  QCD background $g g \rightarrow g g$.
The inset plots within the left panels and their enclosing plots show the same data but with different scales on the axes.}
\end{minipage}\end{center}
\end{figure}

Figure~\ref{fig:gggg_s1c8_analytic} shows the digluon invariant mass distributions, for $pp$ collisions at $\sqrt{s}=13$ TeV, for spin-1 color-octet resonances with benchmark examples of Table \ref{tab:benchmarks},
namely
$(M_X, \Gamma/M_X)=$ (1000 GeV, $6.5 \times 10^{-5}$) [top row], 
(2000 GeV, $5.4 \times 10^{-4}$) [middle row] and 
(3000 GeV, 0.00183) [bottom row].
As before, the parton-level results before smearing are shown in the left panels, and
the corresponding mass distributions after smearing are shown in the right panels.
In all panels, the red lines show the results for the resonant signal $g g \rightarrow X \rightarrow g g$ with all $s,t,u$-channel $X$ exchange diagrams, while the blue lines show the full results including the interferences with the  QCD background $g g \rightarrow g g$. In this case, the relative effects of the interference
are seen to be of a similar character, but smaller than, the spin-0 cases discussed above.

\subsubsection{Monte Carlo results with showering, hadronization, and detector simulation}

\begin{figure}[!tb]
  \begin{minipage}[]{0.495\linewidth}
    \includegraphics[width=8.0cm,angle=0]{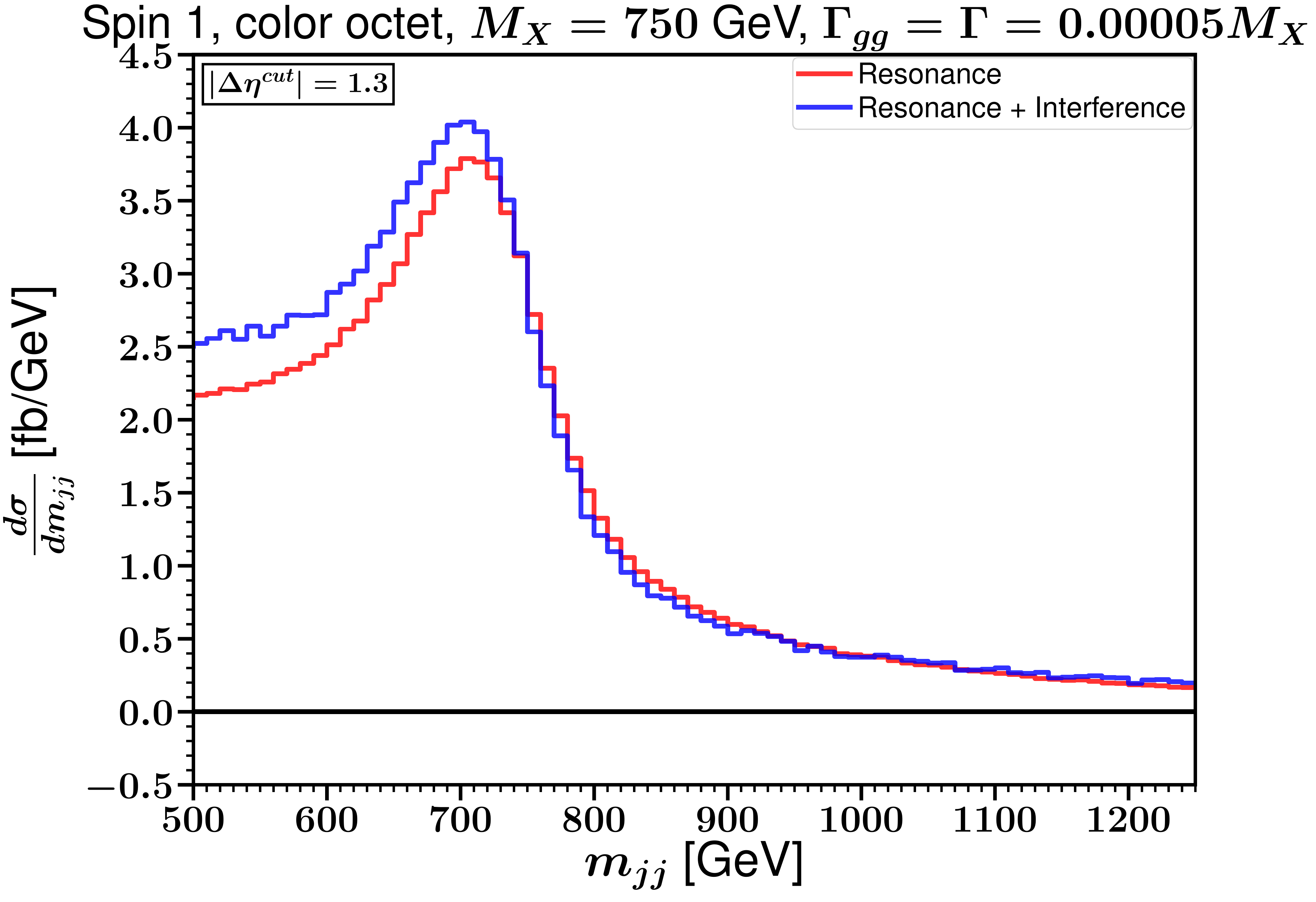}
  \end{minipage}
  \begin{minipage}[]{0.495\linewidth}
    \includegraphics[width=8.0cm,angle=0]{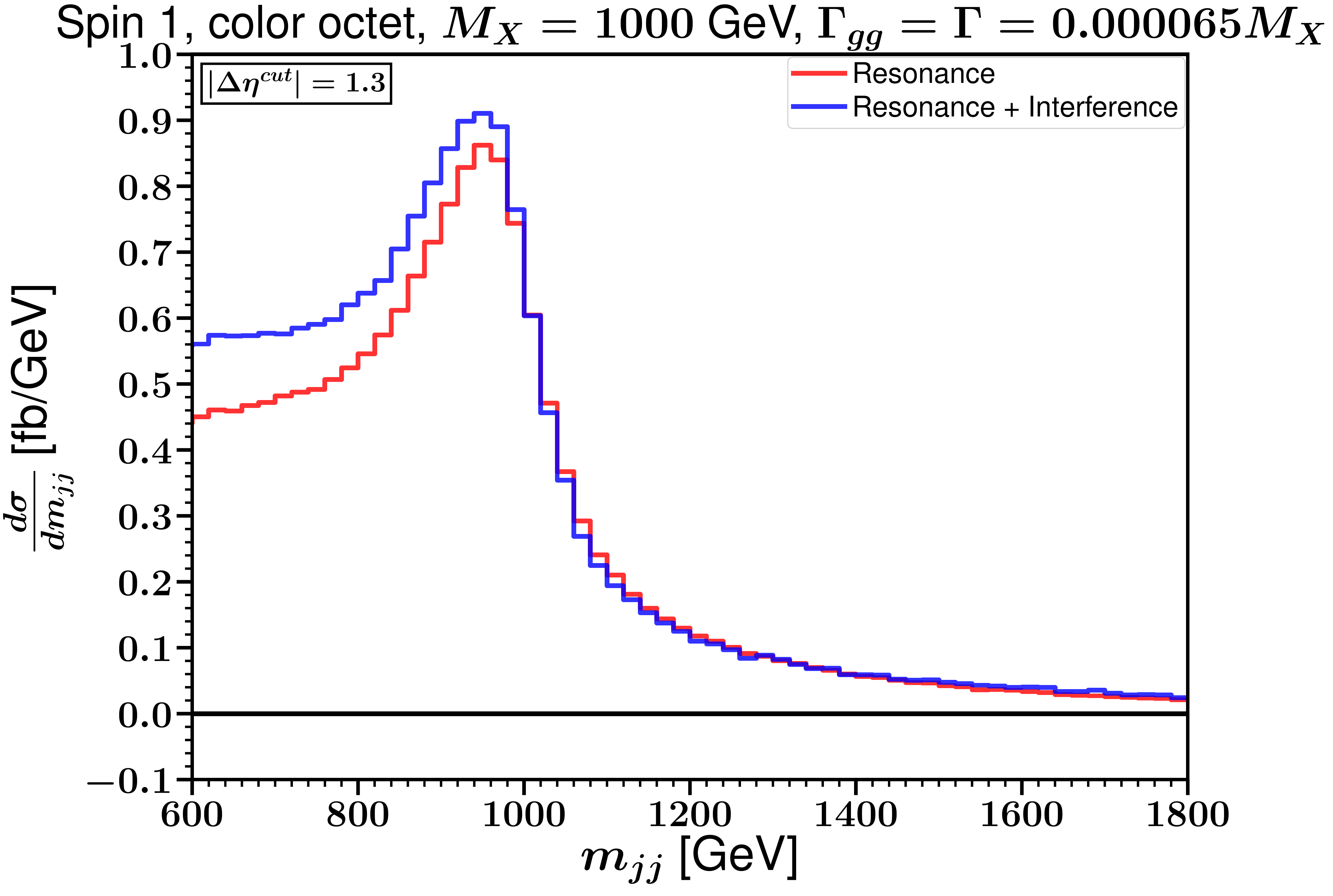}
  \end{minipage}

  \vspace{0.15cm}
  
  \begin{minipage}[]{0.495\linewidth}
    \includegraphics[width=8.0cm,angle=0]{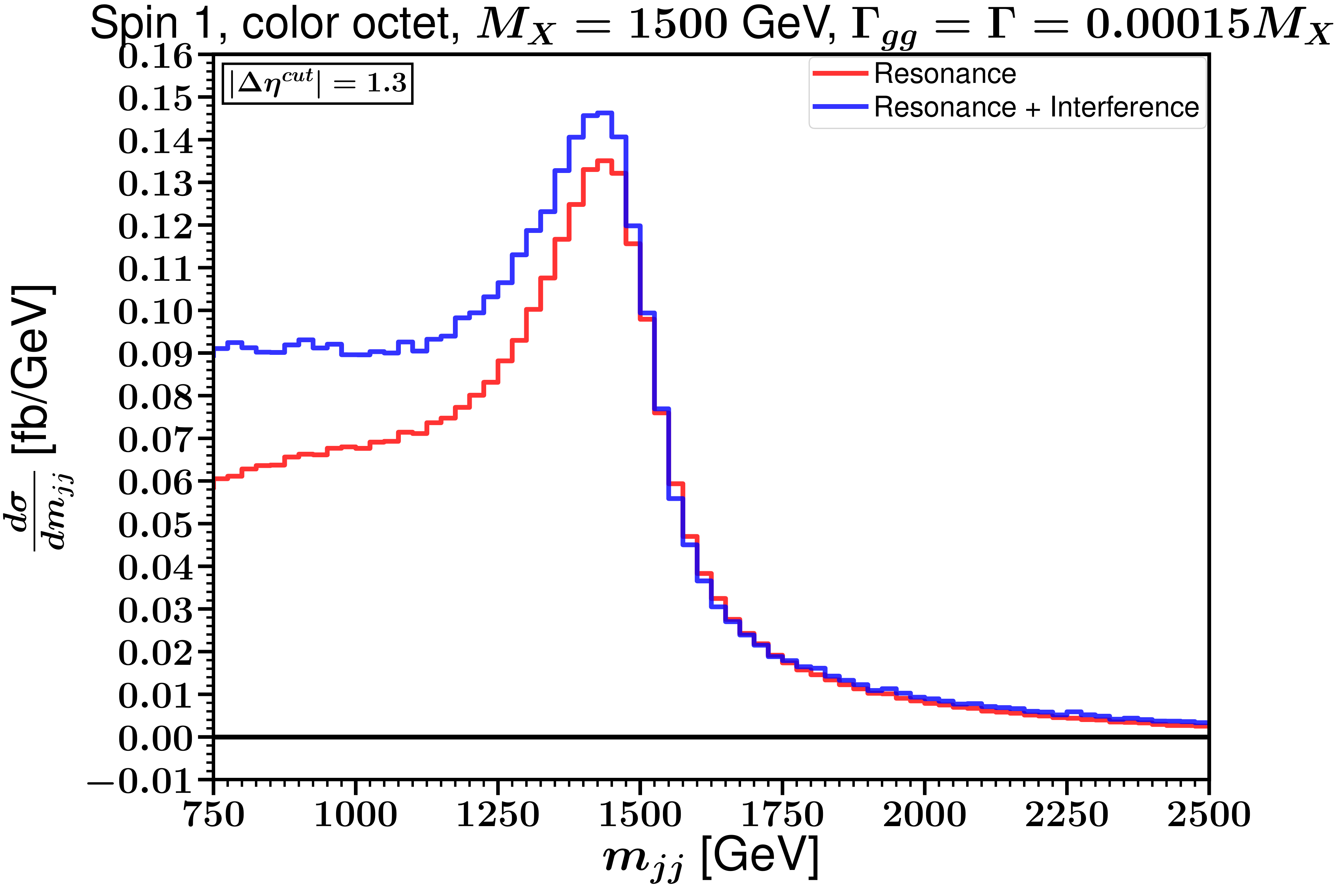}
  \end{minipage}
  \begin{minipage}[]{0.495\linewidth}
    \includegraphics[width=8.0cm,angle=0]{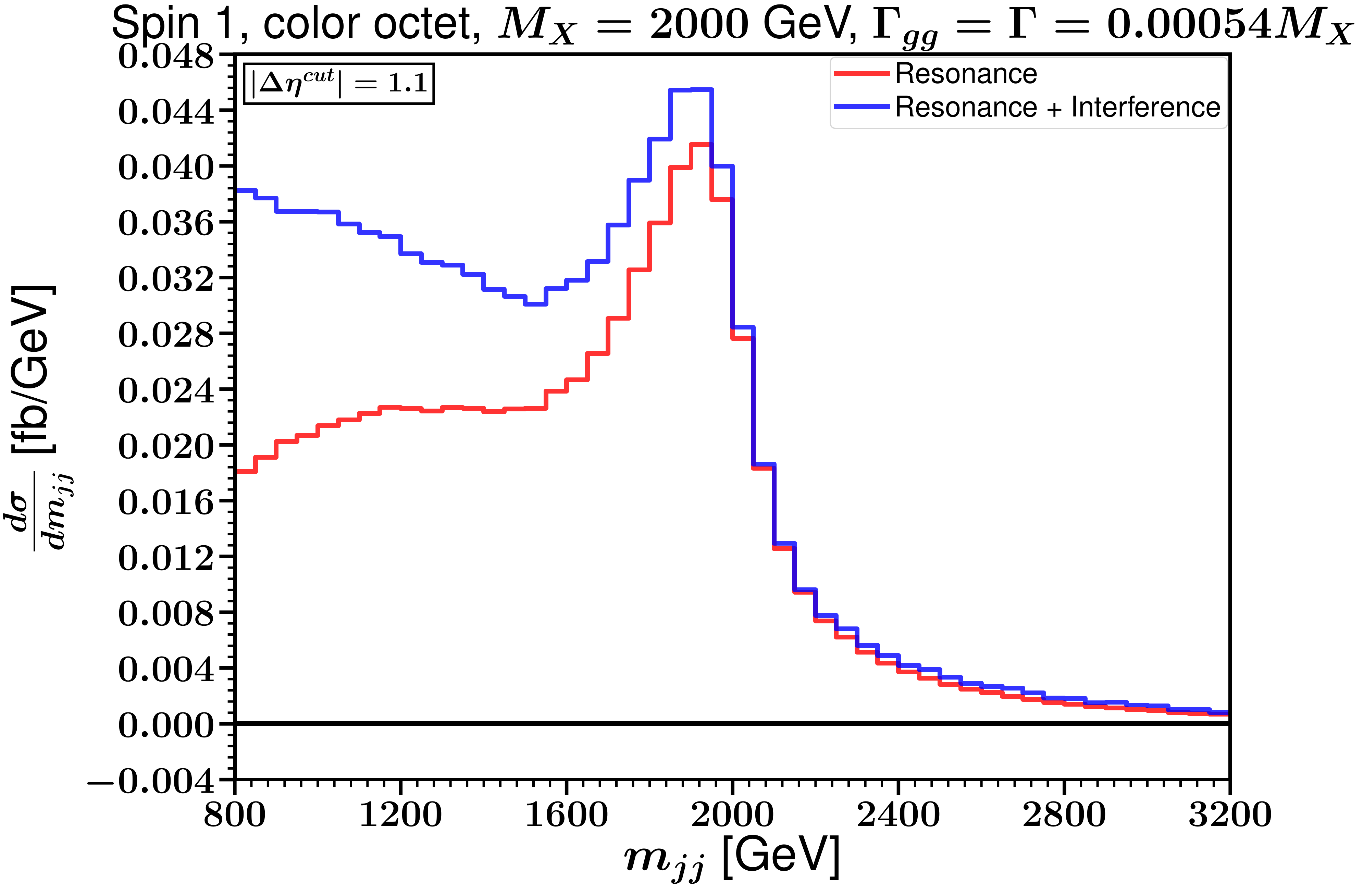}
  \end{minipage}

  \vspace{0.15cm}
  
  \begin{minipage}[]{0.495\linewidth}
    \includegraphics[width=8.0cm,angle=0]{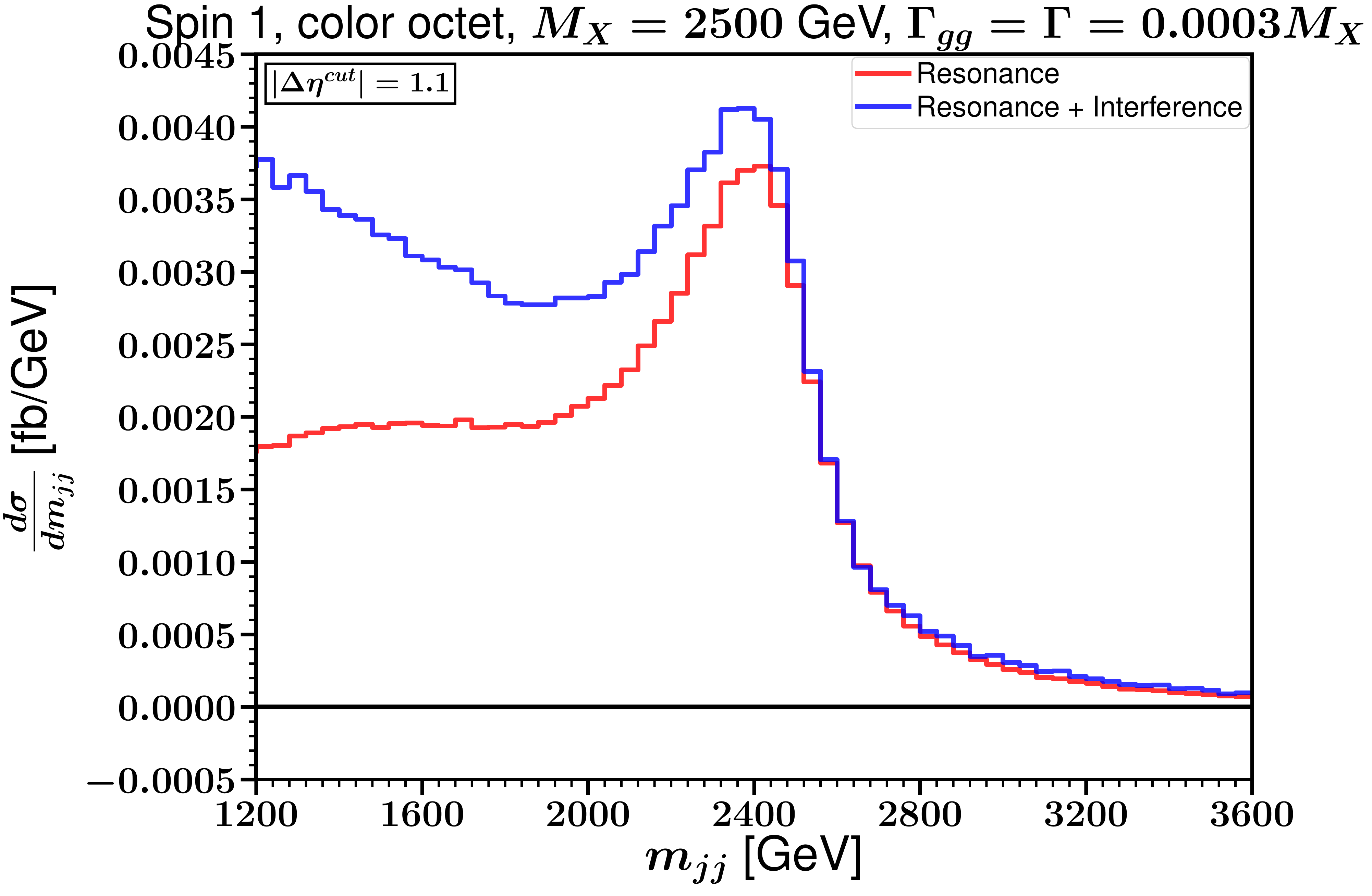}
  \end{minipage}
  \begin{minipage}[]{0.495\linewidth}
    \includegraphics[width=8.0cm,angle=0]{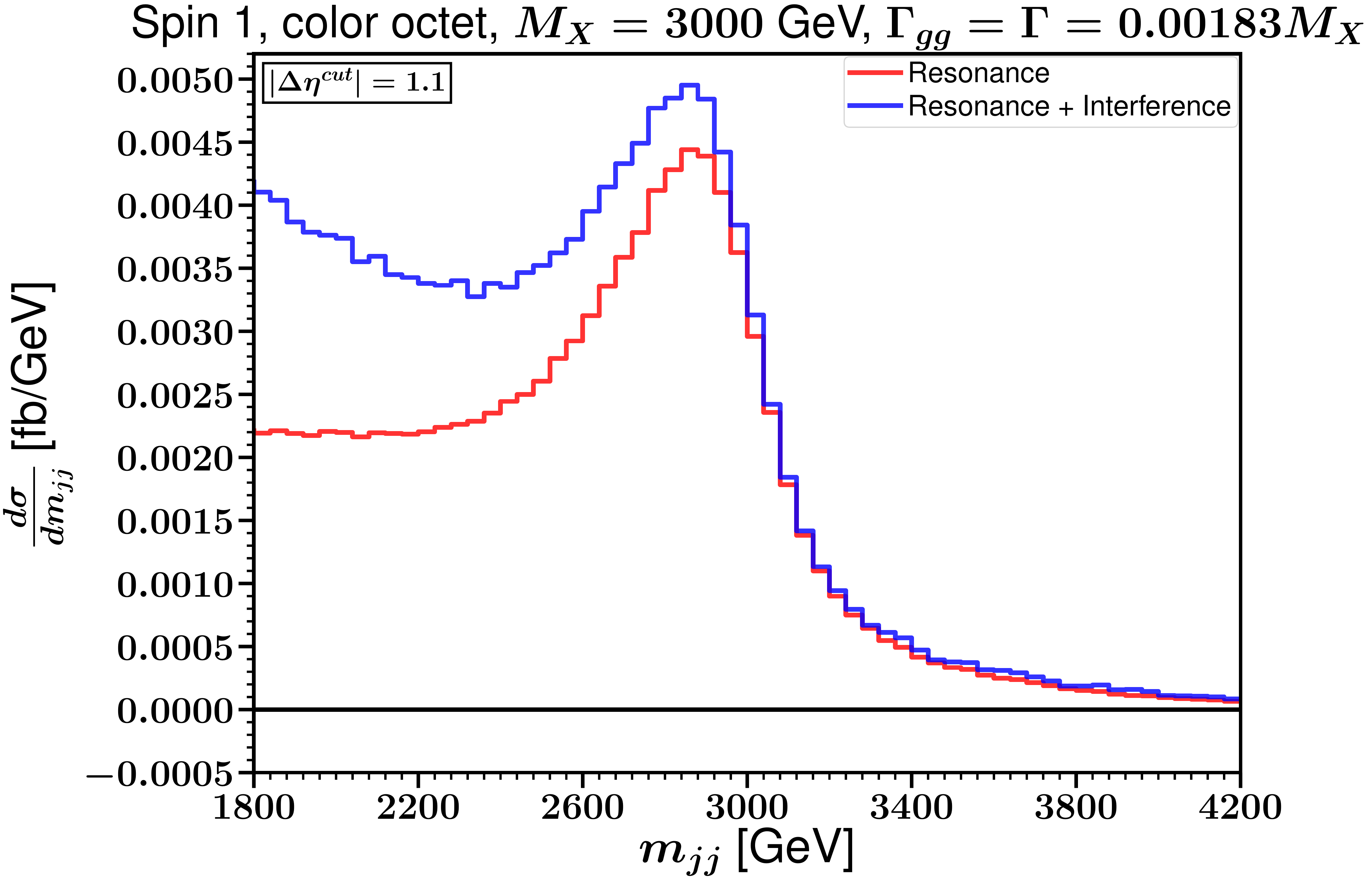}
  \end{minipage}
\begin{center}\begin{minipage}[]{0.95\linewidth}
\caption{\label{fig:gggg_s1c8} 
Dijet invariant mass distributions for the spin-1 color-octet benchmarks of Table~\ref{tab:benchmarks} with $\Gamma=\Gamma_{gg}$, at the 13 TeV LHC, obtained with showering, hadronization and detector simulation. The red lines show the naive results with the resonance diagrams of $g g \rightarrow X \rightarrow g g$ process (RES), which include the $s$-, $t$-, and $u$-channel exchanges of $X$, while the blue lines show the full results including interferences with the continuum QCD $g g \rightarrow g g$ amplitudes (INT).
The cases shown in the right column can be compared directly
to those in the right column of the previous Figure \ref{fig:gggg_s1c8_analytic} based on the more simplistic method of parton level with smearing.}
\end{minipage}\end{center}
\end{figure}

Figure~\ref{fig:gggg_s1c8} shows all of the the dijet invariant mass distributions for the considered benchmarks of Table~\ref{tab:benchmarks} with $\Gamma=\Gamma_{gg}$, at 13 TeV LHC, for spin-1, color-octet resonances, obtained using Monte Carlo simulations. The results are shown for a digluon resonant process 
$g g \rightarrow X \rightarrow g g$, and the full results, which also include the interferences with the continuum QCD $g g \rightarrow g g$ amplitudes.

From Figure~\ref{fig:gggg_s1c8} we see that the QCD interferences with spin-1, color-octet (blue lines) has a peak below the resonance mass, which is almost comparable to the pure peak, obtained by excluding the interference terms. The differential cross sections in the region $m_{jj} > M_X$ are almost unaffected
by including the interference.
However, as before, the presence of the large low-mass tail means that after fitting the QCD background, the residual distribution may have
an apparent deficit of events above $M_X$. Comparing the right columns of
Figures \ref{fig:gggg_s1c8_analytic} and \ref{fig:gggg_s1c8}, we note that in this case 
the difference between the shapes found with the parton-level smearing method  and the full event simulation
method is more significant than in the spin-0 case, this time with a larger yield and a more pronounced
low mass tail using the latter, presumably more accurate, method.
As was the case with the color-singlet resonances, the QCD interference seems to have relatively larger impact at higher resonance masses than at smaller $M_X$.

\clearpage

\subsection{Spin 2, color singlet}

\subsubsection{Parton-level with smearing}
\begin{figure}[!tb]
  \begin{minipage}[]{0.495\linewidth}
    \includegraphics[width=8.0cm,angle=0]{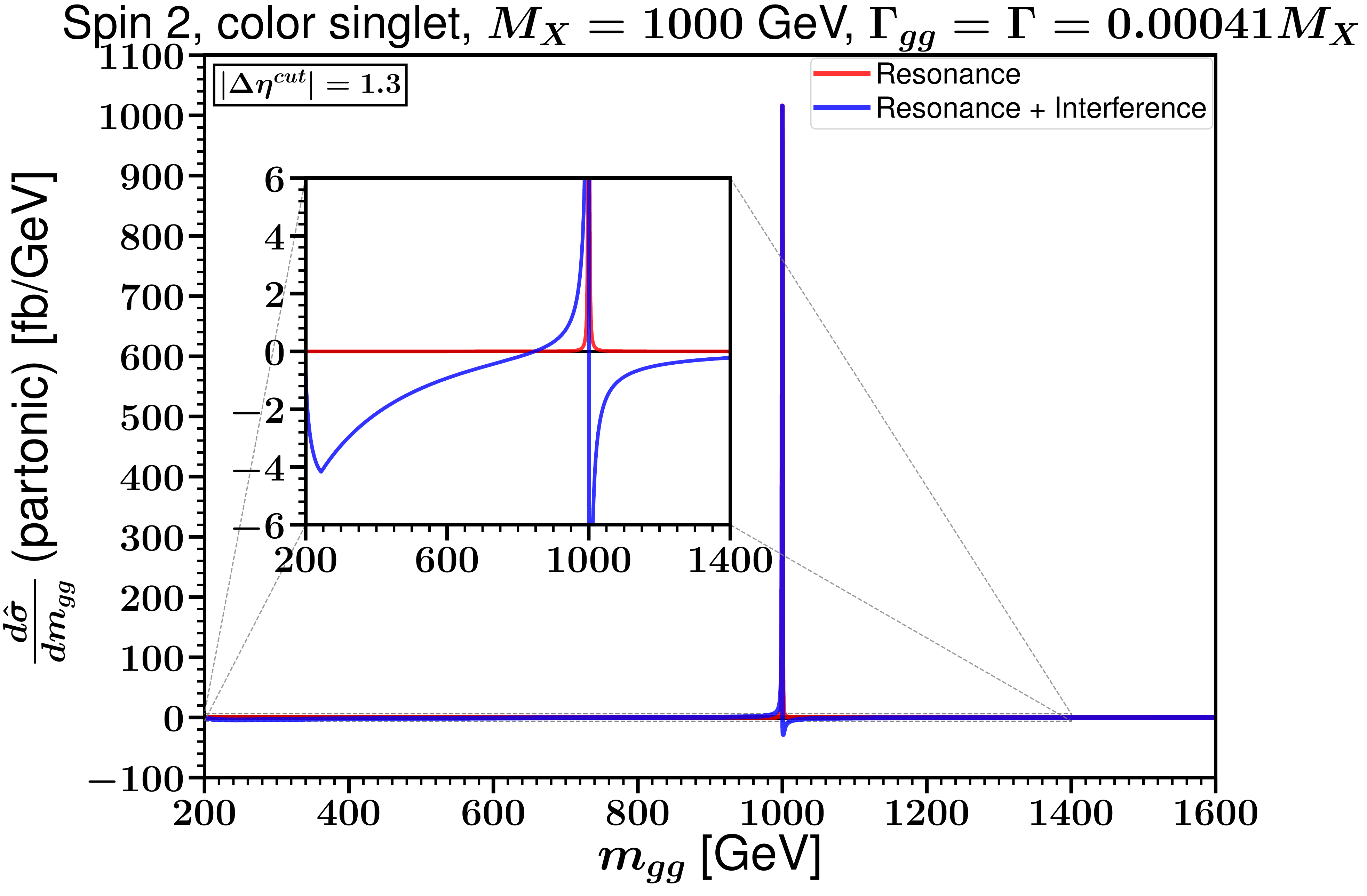}
  \end{minipage}
    \begin{minipage}[]{0.495\linewidth}
    \includegraphics[width=8.0cm,angle=0]{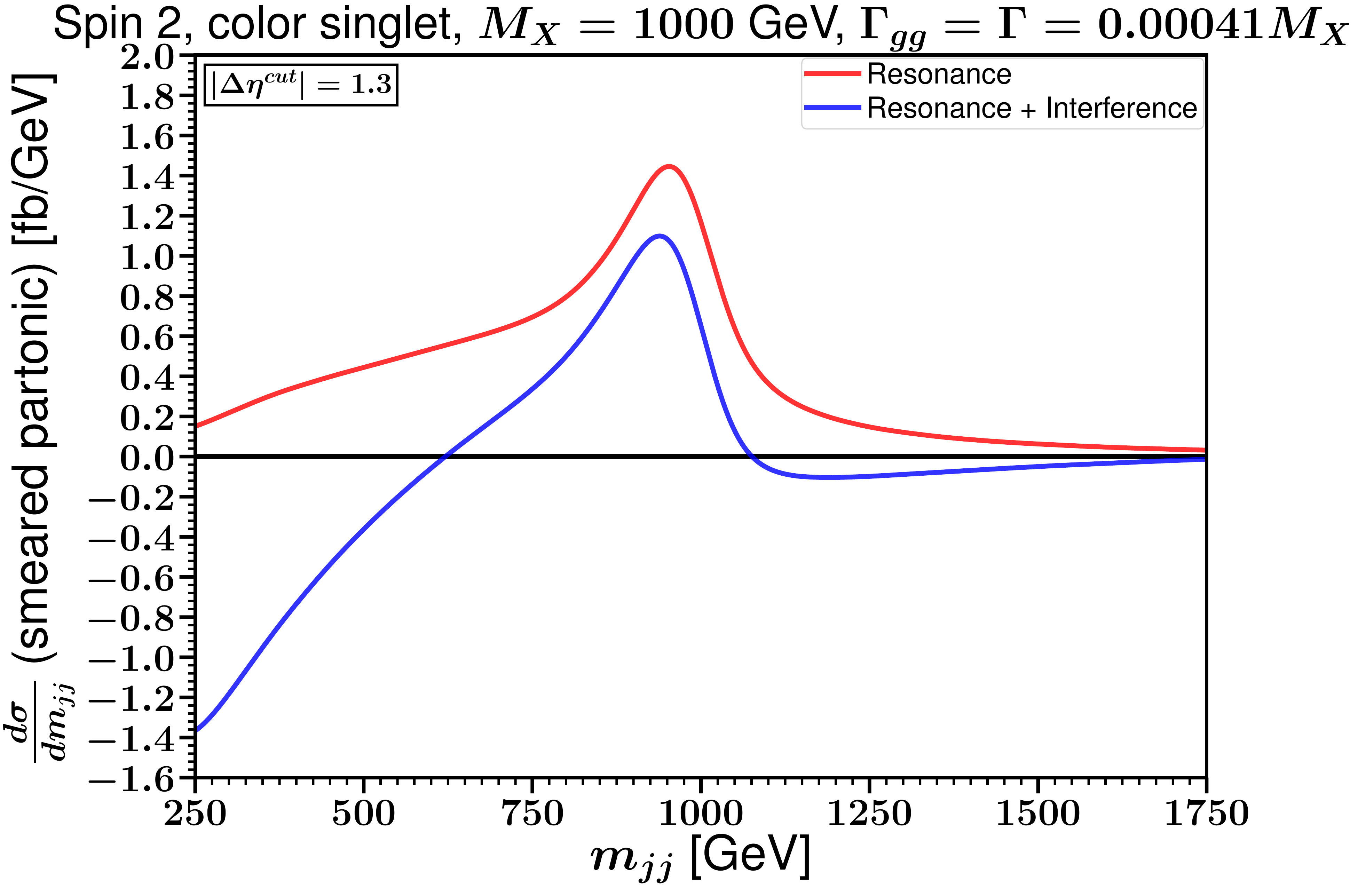}
  \end{minipage}

  \vspace{0.15cm}
  
  \begin{minipage}[]{0.495\linewidth}
    \includegraphics[width=8.0cm,angle=0]{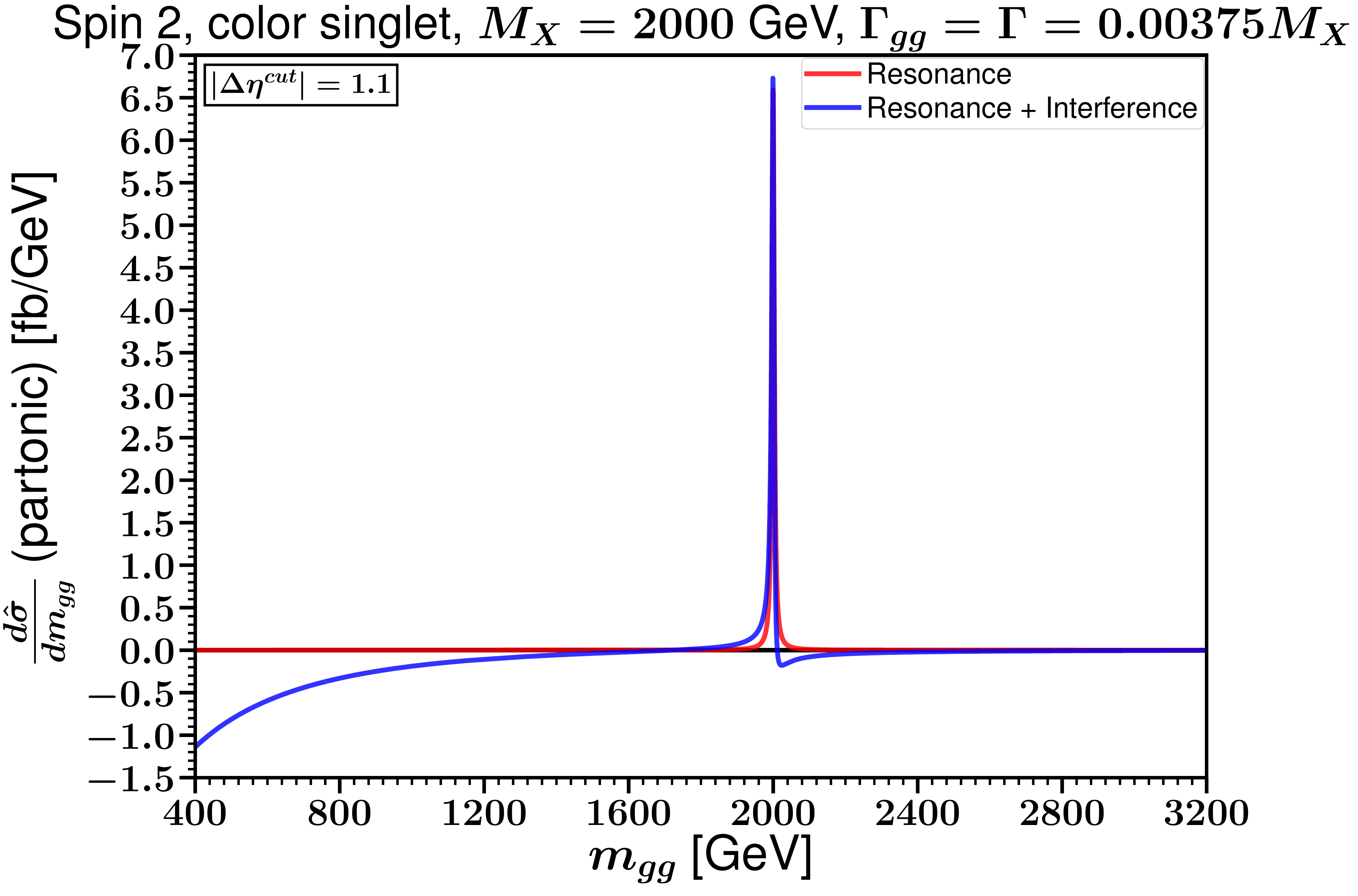}
  \end{minipage}
    \begin{minipage}[]{0.495\linewidth}
    \includegraphics[width=8.0cm,angle=0]{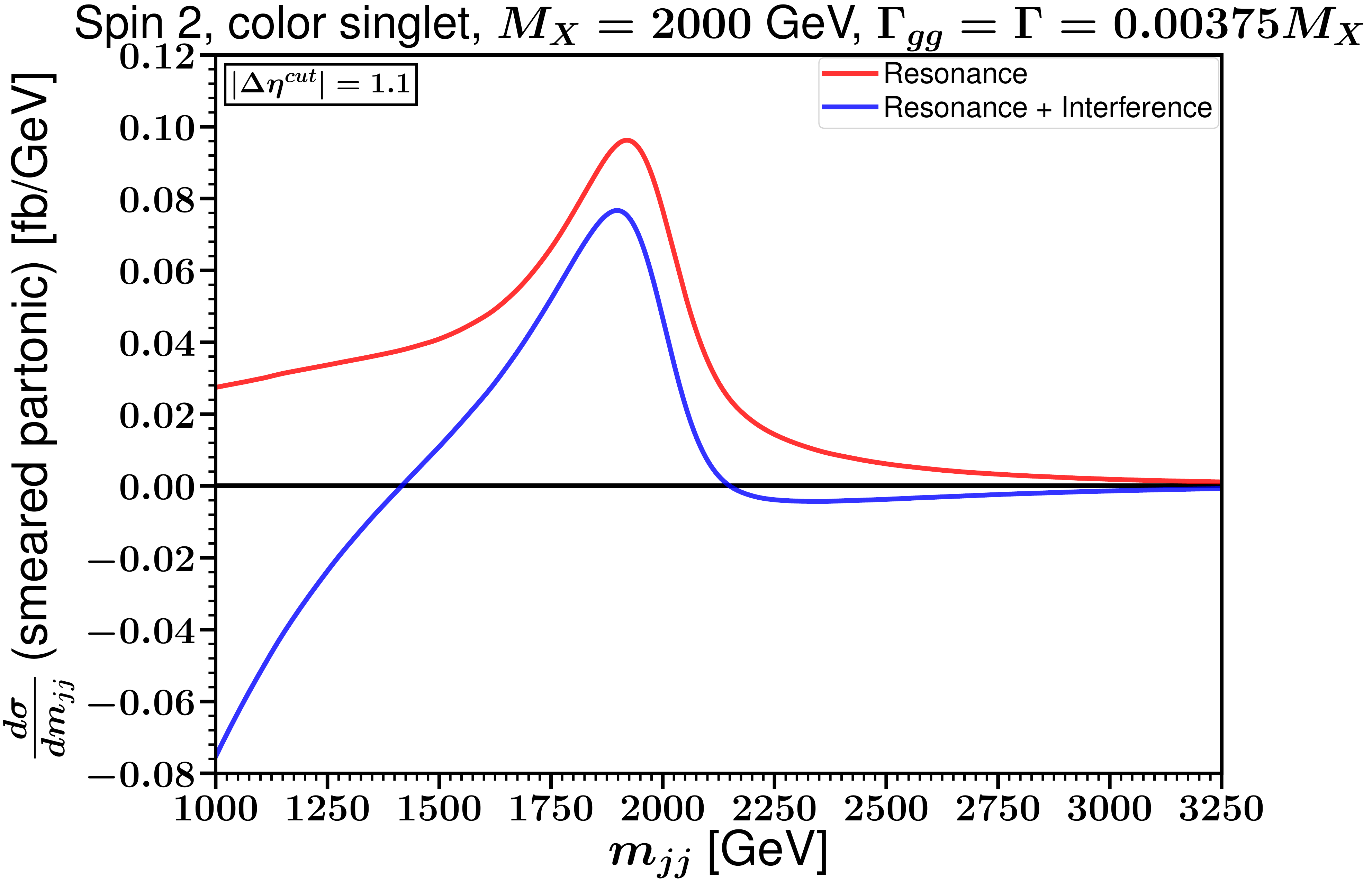}
  \end{minipage}

  \vspace{0.15cm}
  
  \begin{minipage}[]{0.495\linewidth}
    \includegraphics[width=8.0cm,angle=0]{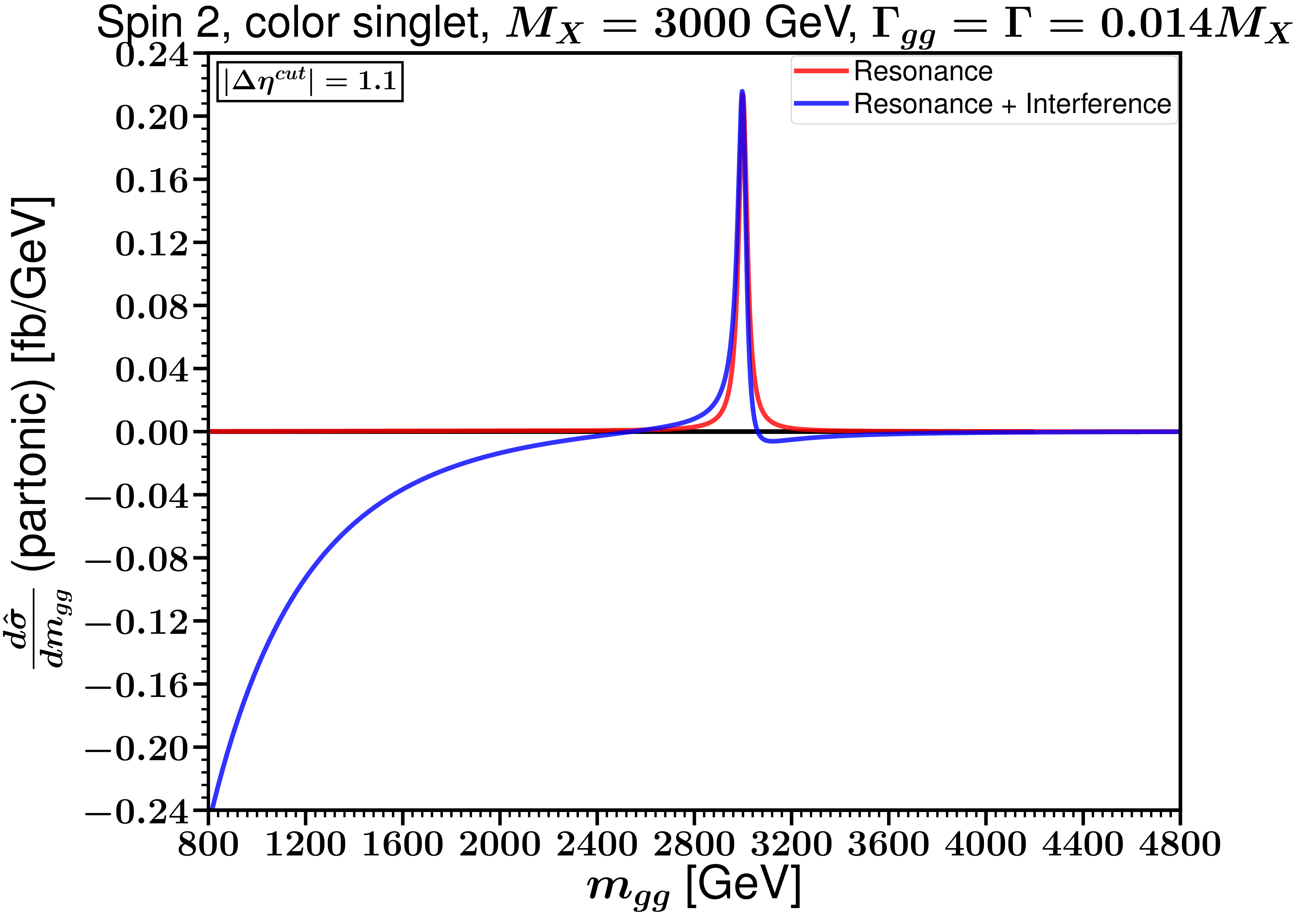}
  \end{minipage}
  \begin{minipage}[]{0.495\linewidth}
    \includegraphics[width=8.0cm,angle=0]{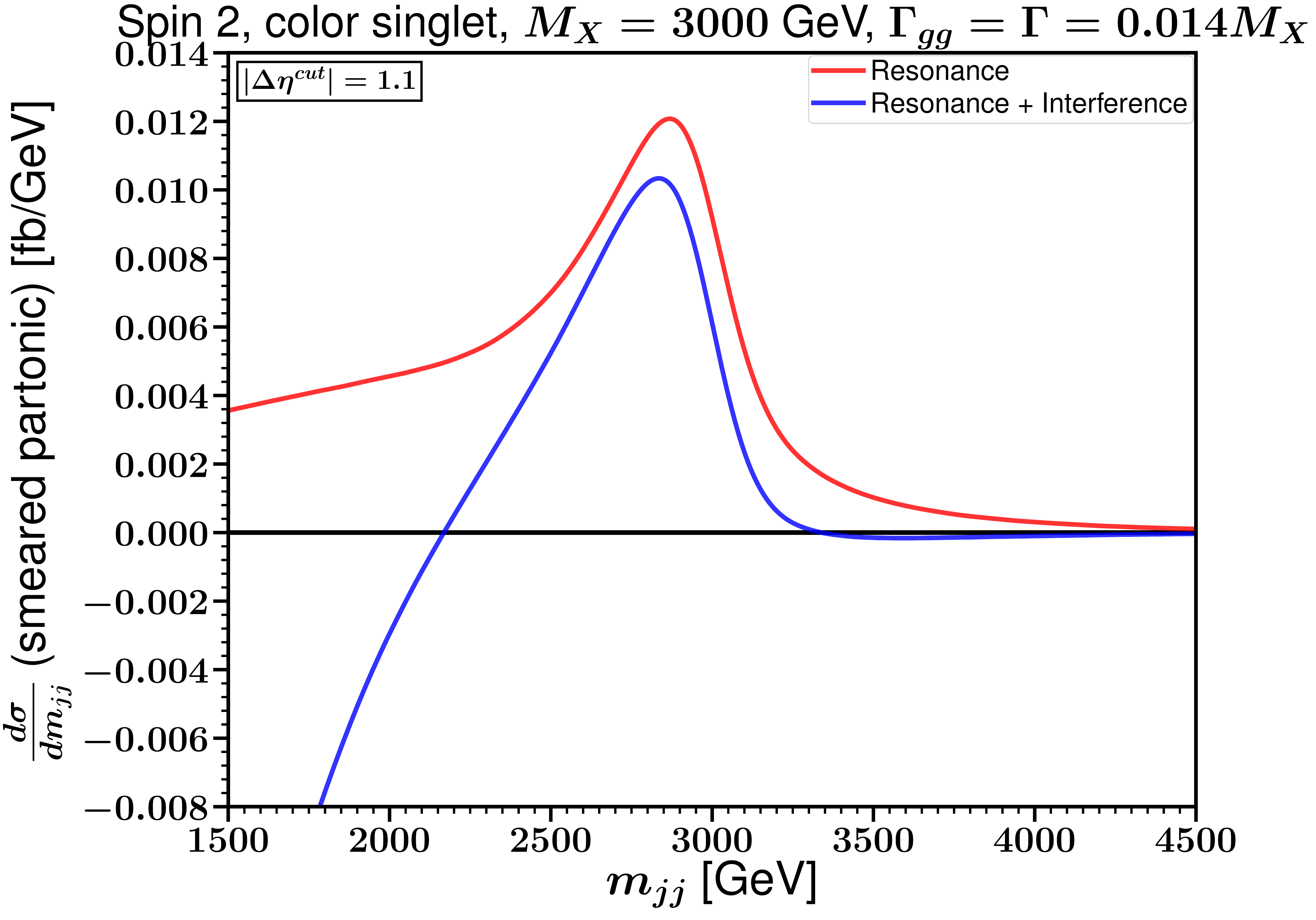}
  \end{minipage}
\begin{center}\begin{minipage}[]{0.95\linewidth}
\caption{\label{fig:gggg_s2c1_analytic} 
Digluon invariant mass distributions, at the 13 TeV LHC, for spin-2, color-singlet resonances with
$(M_X, \Gamma/M_X) = $
(1000 GeV, 0.00041) [top row],
(2000 GeV, 0.00375) [middle row], and
(3000 GeV, 0.014) [bottom row].
The parton-level distributions (left panels) are smeared by convolution with the detector response, shown in Figure~\ref{fig:yield}, to obtain the dijet invariant mass distributions (right panels). In all six panels, the red lines show the naive results for the resonant signal $g g \rightarrow X \rightarrow g g$, while the blue lines show the full results including the interferences with the  QCD background $g g \rightarrow g g$.
The inset plot within the top-left panel shows the same data as its enclosing plot but with different scales on the axes. The negative tails at small invariant mass come from the interference between the QCD amplitudes and the $t$- and $u$-channel $X$ exchange diagrams.}
\end{minipage}\end{center}
\end{figure}

We now turn to the case of a massive spin-2, color-singlet digluon resonance.
In Figure~\ref{fig:gggg_s2c1_analytic}, we show the parton-level digluon invariant mass distributions for $pp$ collisions at $\sqrt{s}=13$ TeV, for $(M_X, \Gamma/M_X) = $
(1000 GeV, 0.00041) [top row],
(2000 GeV, 0.00375) [middle row], and
(3000 GeV, 0.014) [bottom row]. As before,
the results before smearing are shown in the left panels, and 
mass distributions after smearing are shown in the right panels.
In all panels, the red lines show the results for the resonant signal $g g \rightarrow X \rightarrow g g$ with all $X$ exchange diagrams, while the blue lines show the full results including the interferences with the  QCD $g g \rightarrow g g$ amplitudes.

In the spin-2 case, we note that there is a unique feature not found in the previous cases: the effect of the interference is negative
for all dijet invariant masses well below $M_X$
after smearing (but with a magnitude that of course varies with the mass).
This can be traced in part
to a large negative interference effect in the parton-level results for $m_{gg}$ well below $M_X$, 
due to the contributions from the interference between $t$-, and $u$-channel $X$ exchange diagrams and the QCD diagrams.
Thus, at the parton level for spin-2 color-singlets, there is an interference dip in the regions $m_{gg} \ll M_X$ (although 
it is small compared to the large QCD background in that range of $m_{gg}$). 
This translates into a substantial negative low-mass tail compared to the naive pure resonance result.
There is then a steady rise until 
$m_{jj}$ is slightly less than $M_X$, followed by a dip in the regions $m_{jj} > M_X$.
It would be interesting to see whether this pattern is maintained after including higher order contributions. 

\subsubsection{Monte Carlo results with detector simulation}
\begin{figure}[!tb]
  \begin{minipage}[]{0.495\linewidth}
    \includegraphics[width=8.0cm,angle=0]{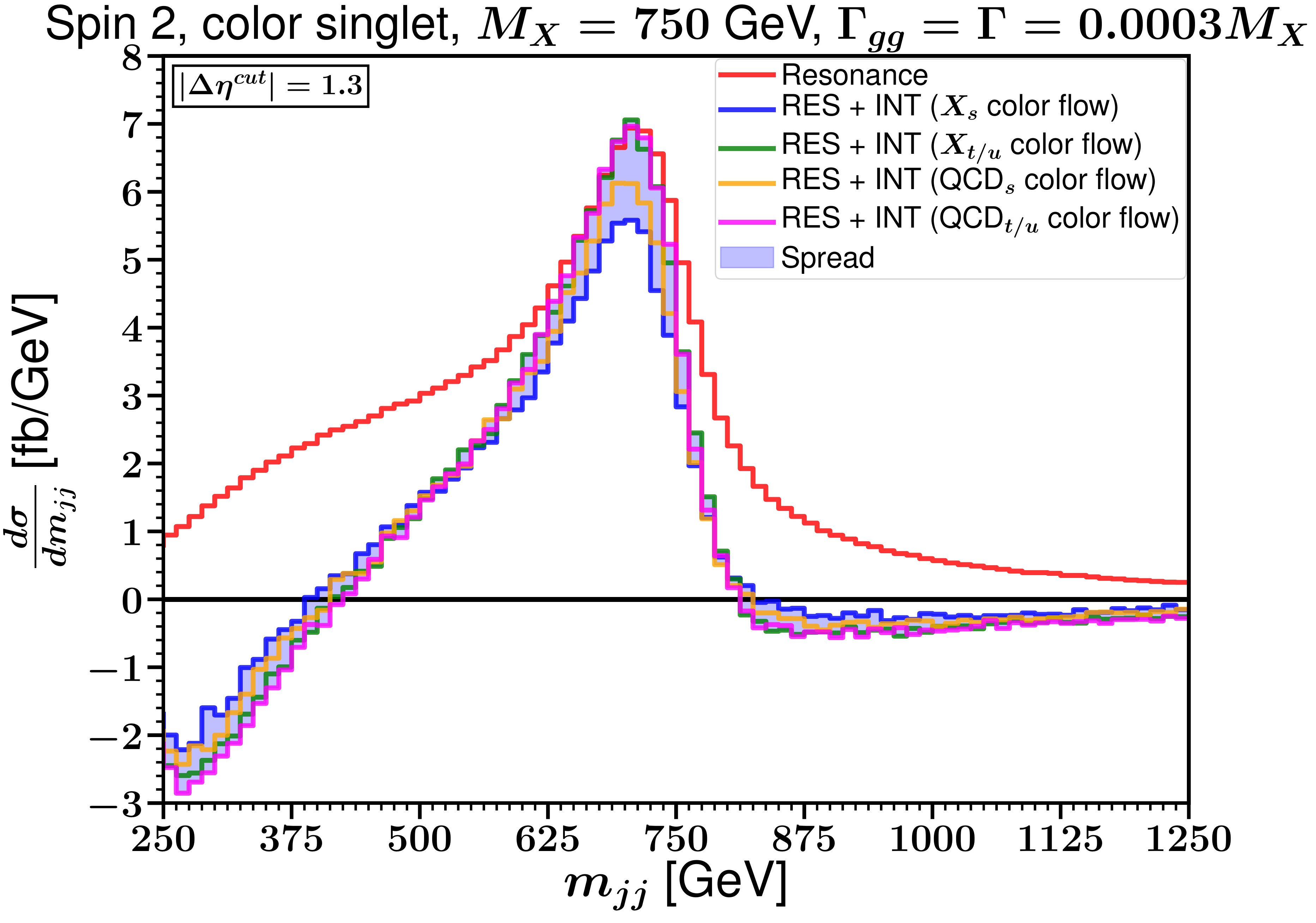}
  \end{minipage}
  \begin{minipage}[]{0.495\linewidth}
    \includegraphics[width=8.0cm,angle=0]{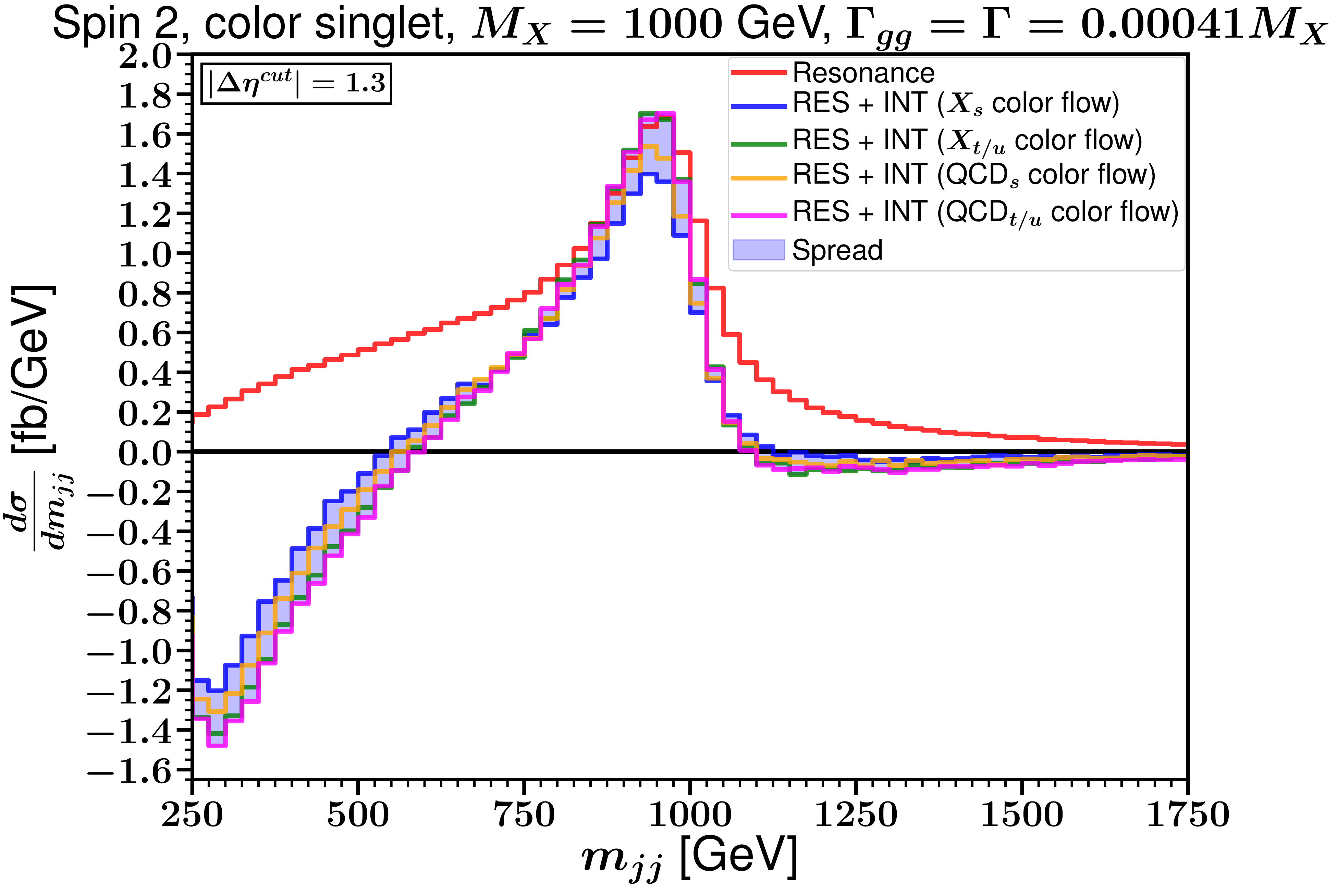}
  \end{minipage}

  \vspace{0.15cm}
  
  \begin{minipage}[]{0.495\linewidth}
    \includegraphics[width=8.0cm,angle=0]{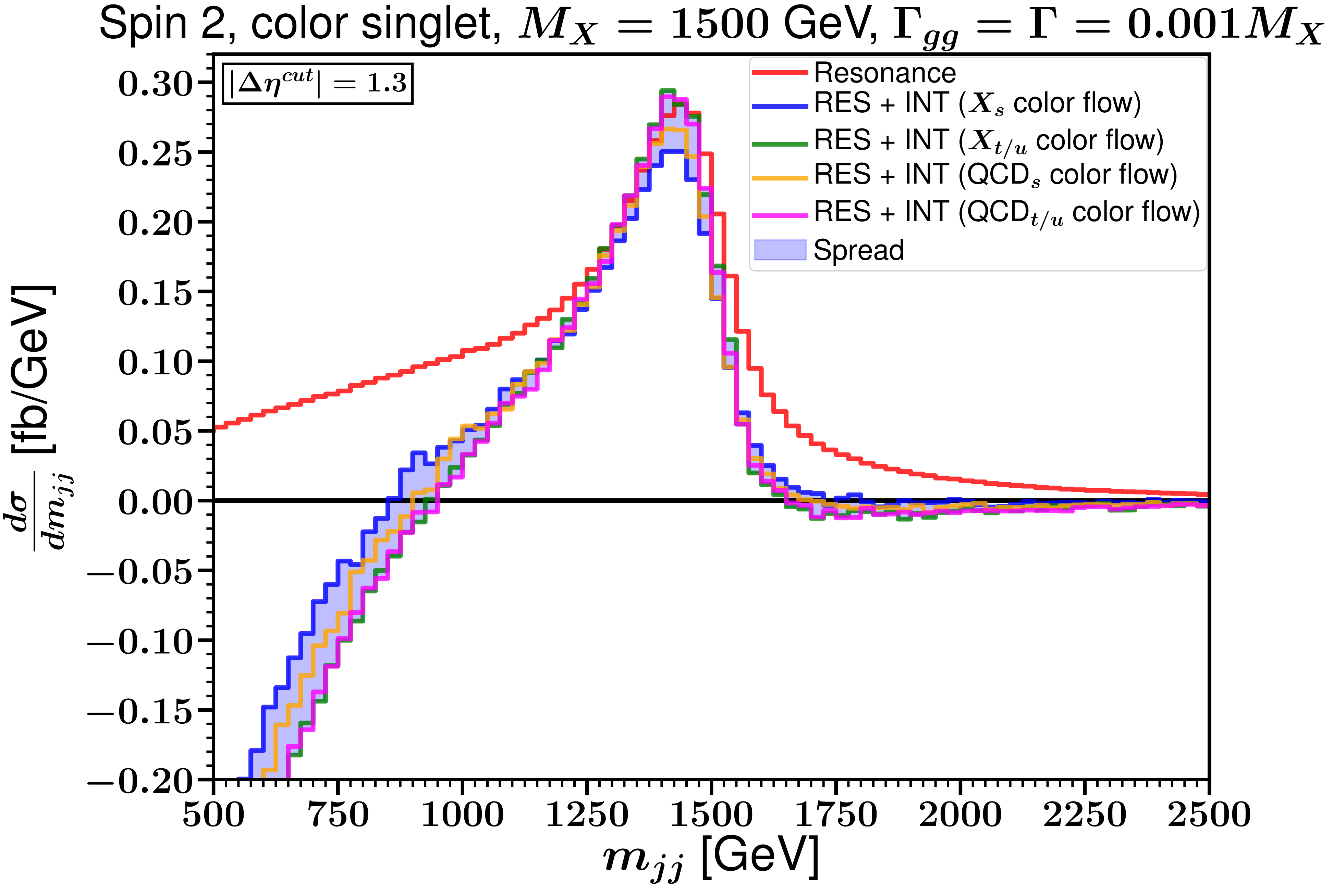}
  \end{minipage}
    \begin{minipage}[]{0.495\linewidth}
    \includegraphics[width=8.0cm,angle=0]{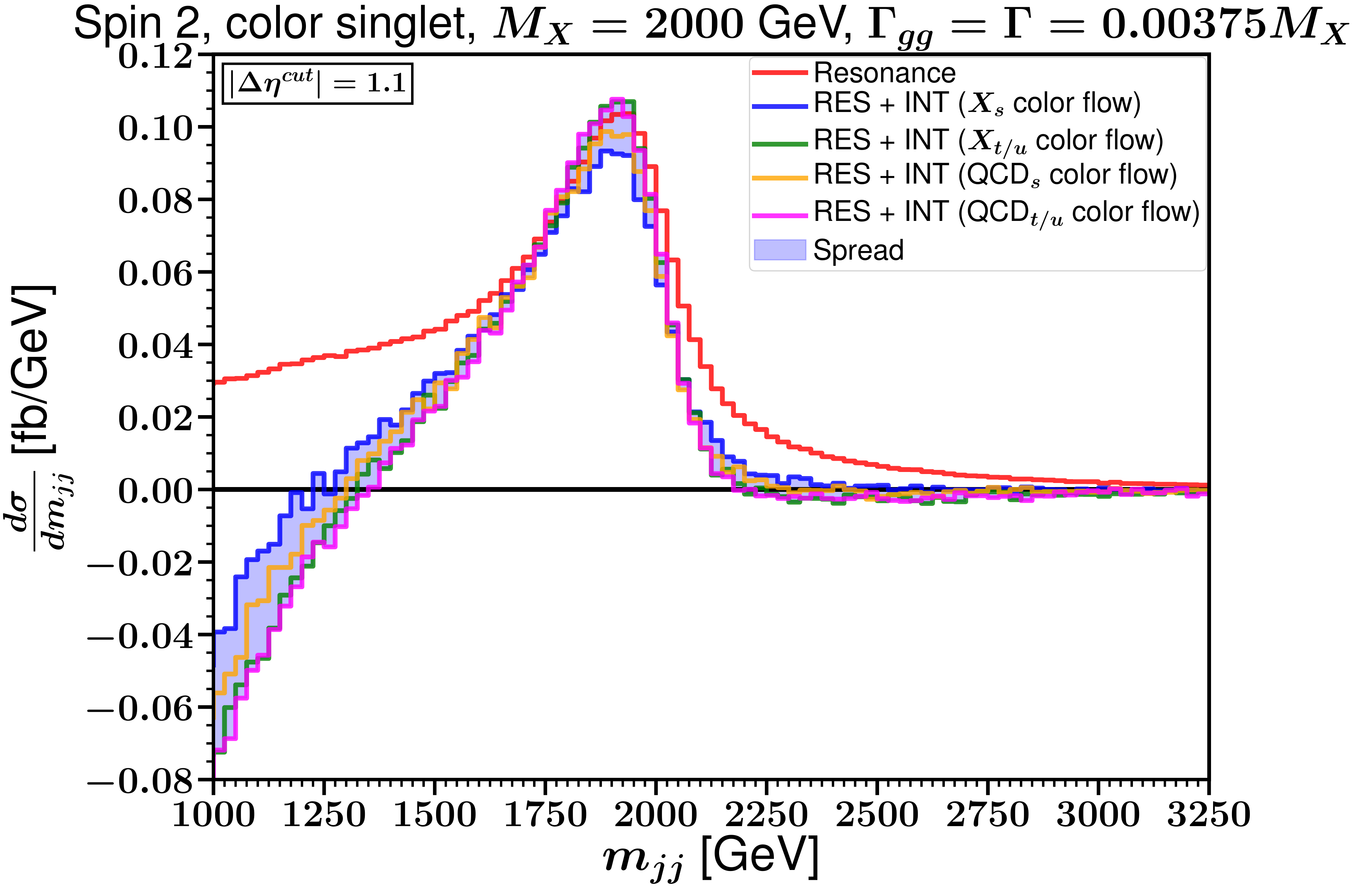}
  \end{minipage}

  \vspace{0.15cm}
  
  \begin{minipage}[]{0.495\linewidth}
    \includegraphics[width=8.0cm,angle=0]{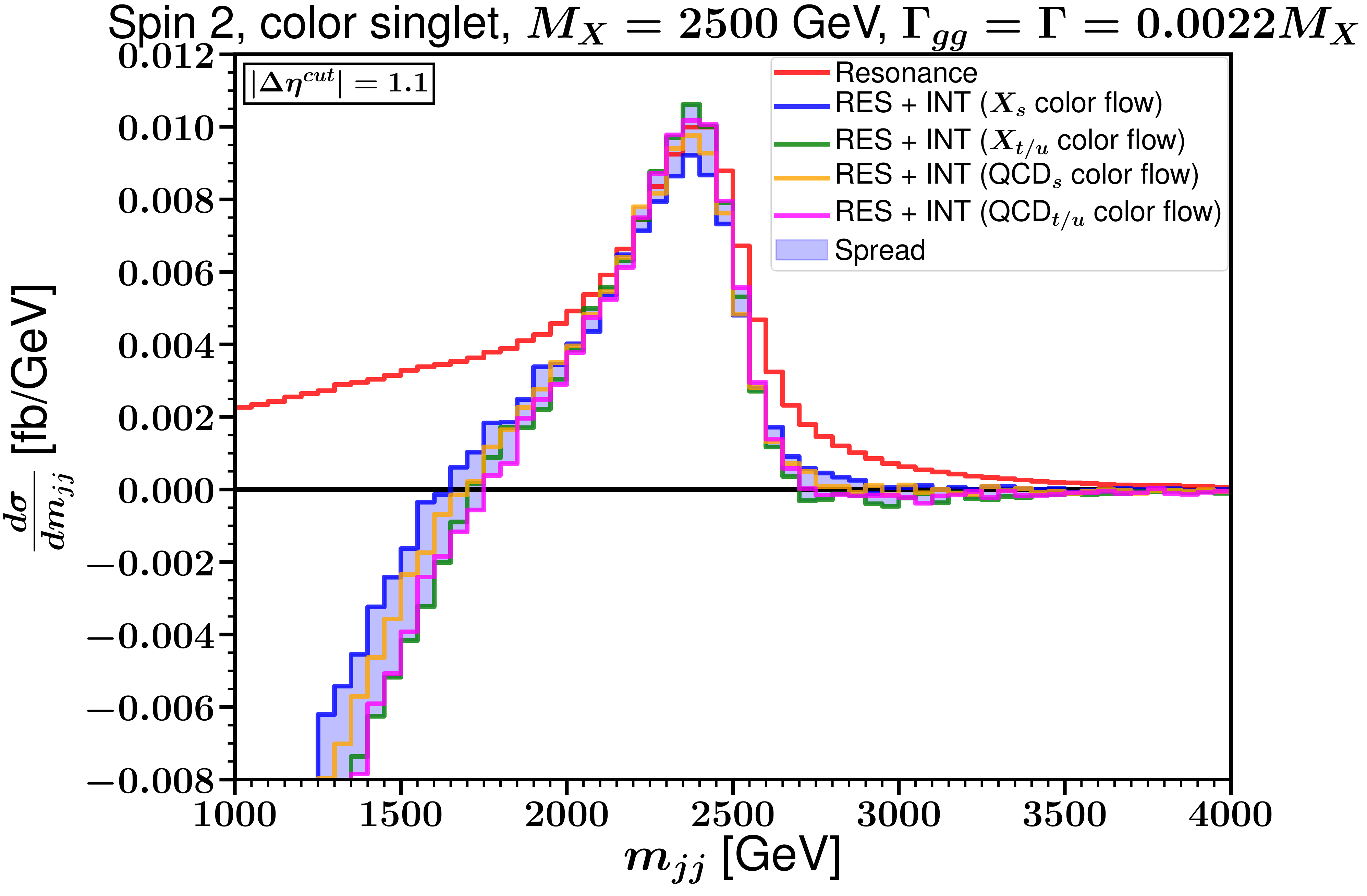}
  \end{minipage}
    \begin{minipage}[]{0.495\linewidth}
    \includegraphics[width=8.0cm,angle=0]{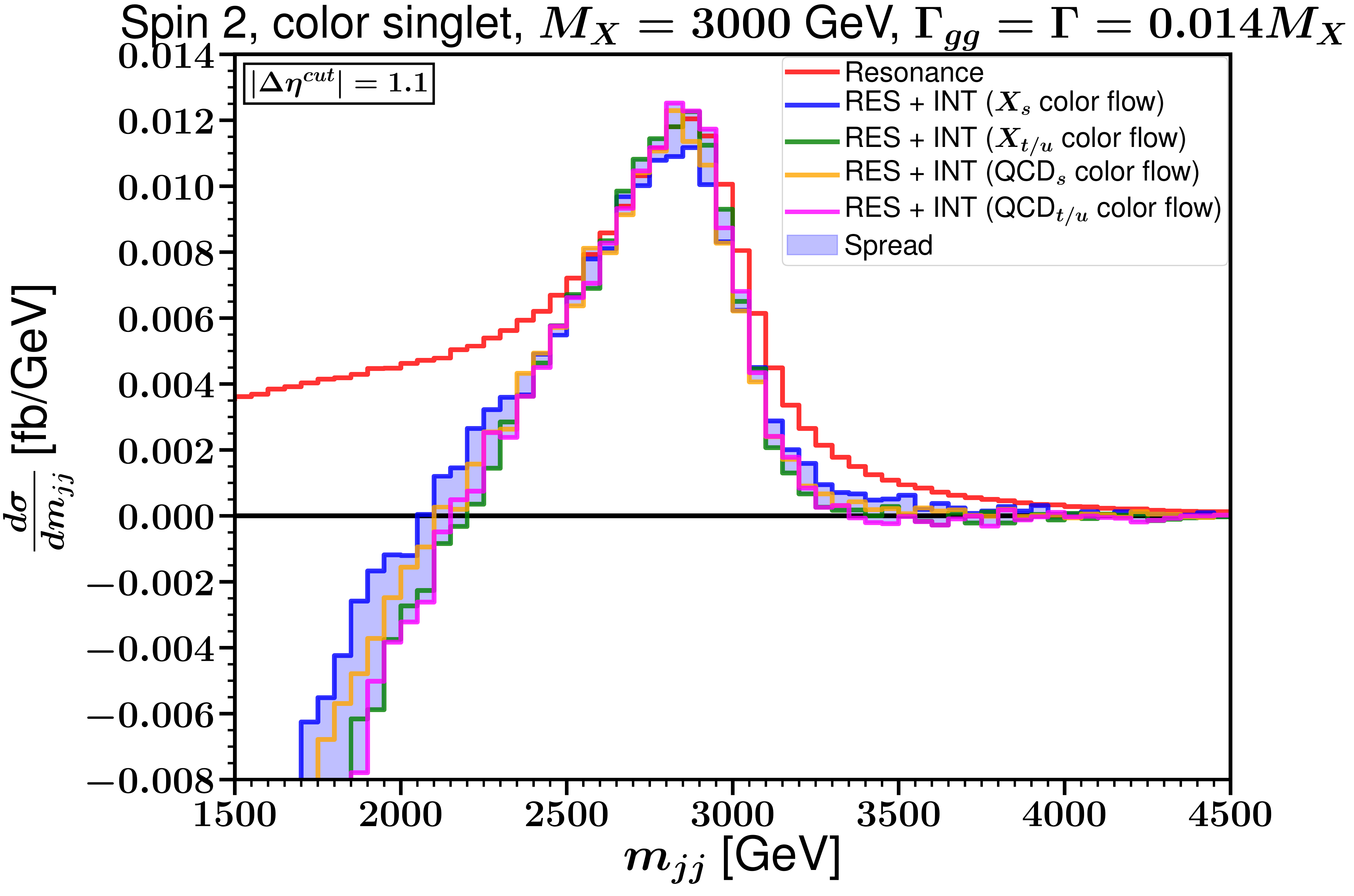}
  \end{minipage}
\begin{center}\begin{minipage}[]{0.95\linewidth}
\caption{\label{fig:gggg_s2c1} 
Dijet invariant mass distributions for the spin-2, color-singlet resonance benchmarks of Table~\ref{tab:benchmarks} with $\Gamma=\Gamma_{gg}$, at 13 TeV LHC, obtained with showering, hadronization and detector simulation. The red lines show the naive results with the resonance diagrams of $g g \rightarrow X \rightarrow g g$ process (RES), which include the $s$-, $t$-, and $u$-channel exchanges of $X$, while the other four colored lines show the full results including interferences with the continuum QCD $g g \rightarrow g g$ amplitudes (INT) for all four color flows shown in Figure \ref{fig:colorflow}, as labeled. The shaded region shows the spread in the full result in each invariant mass bin, for the four different color flow choices.
The negative tails at small invariant mass come from the interference between the QCD amplitudes and the $t$- and $u$-channel $X$ exchange diagrams.
The cases shown in the right column can be compared directly
to those in the right column of the previous Figure \ref{fig:gggg_s2c1_analytic} based on the more simplistic method of parton level with smearing.}
\end{minipage}\end{center}
\end{figure}

Figure~\ref{fig:gggg_s2c1} shows the dijet invariant mass distributions for the spin-2, color-singlet resonance benchmarks of Table~\ref{tab:benchmarks} with $\Gamma=\Gamma_{gg}$, at 13 TeV LHC, obtained using Monte Carlo simulations.
The results are shown for a digluon resonant process $g g \rightarrow X \rightarrow g g$, and the full results, which also include the interferences with the continuum QCD $g g \rightarrow g g$ amplitudes for all four color flows shown in Figure \ref{fig:colorflow}.

As was the case with the color-singlet scalars, from Figure~\ref{fig:gggg_s2c1}, the interference effects have the least impact for the color flow that we call $X_s$. On the other hand, interference effects are more pronounced for the $t/u$-channel color flow of both QCD and resonant processes. Also, we confirm that QCD interferences have negative cross section for smaller invariant masses $m_{jj} \ll M_X$. Then, similar to the case of spin-0 color-singlet resonances, there is a peak/dip pattern around the resonance mass $m_{jj} \approx M_X$, in agreement with the results at parton-level with smearing found 
in Fig.~\ref{fig:gggg_s2c1_analytic}. The net effect of negative interference both above and below
the $M_X$ means that, after fitting to the QCD background, the resonance peak could actually stand out more prominently than predicted by the naive pure-resonance prediction. 
Once again, the relative importance of the interference for $m_{jj}<M_X$ increases with $M_X$.

\clearpage

\section{Results for $\Gamma = 5 \Gamma_{gg}$\label{sec:resultsforotherdecays}}
\setcounter{equation}{0}
\setcounter{figure}{0}
\setcounter{table}{0}
\setcounter{footnote}{1}

As noted in ref.~\cite{Martin:2016bgw}, 
the interference effects are expected to be relatively enhanced for smaller branching ratios BR$(X \rightarrow gg) = \Gamma_{gg}/\Gamma$,
for a given fixed resonant production cross-section. The reason for this is that to reach the same 
cross-section,
both $\Gamma_{gg}$ and $\Gamma$ must be larger than if they were equal
[see eq.~(\ref{eq:sigmanarrowwidth})], leading to much larger Breit-Wigner
tails away from the resonance region, which then produce larger interference with the QCD amplitude.
Thus, the case with $\Gamma = \Gamma_{gg}$ studied above actually has the minimal impact on the interference with QCD, compared to the general case $\Gamma_{gg} < \Gamma$. 
In this section, we illustrate this by considering, somewhat arbitrarily, 
the case that $\Gamma_{gg} = \Gamma/5$.

\begin{figure}[!tb]
  \begin{minipage}[]{0.495\linewidth}
    \includegraphics[width=8.0cm,angle=0]{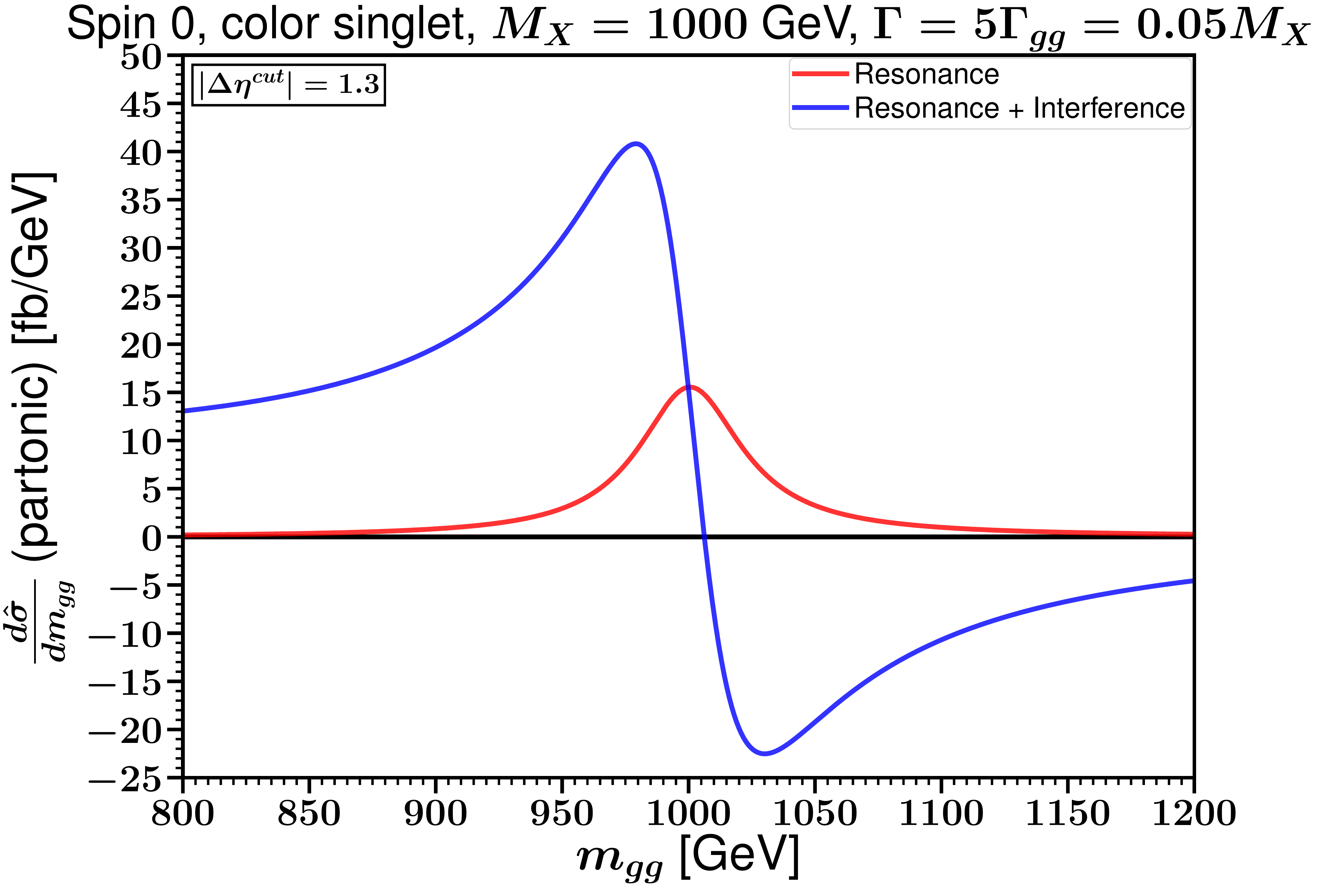}
  \end{minipage}
    \begin{minipage}[]{0.495\linewidth}
    \includegraphics[width=8.0cm,angle=0]{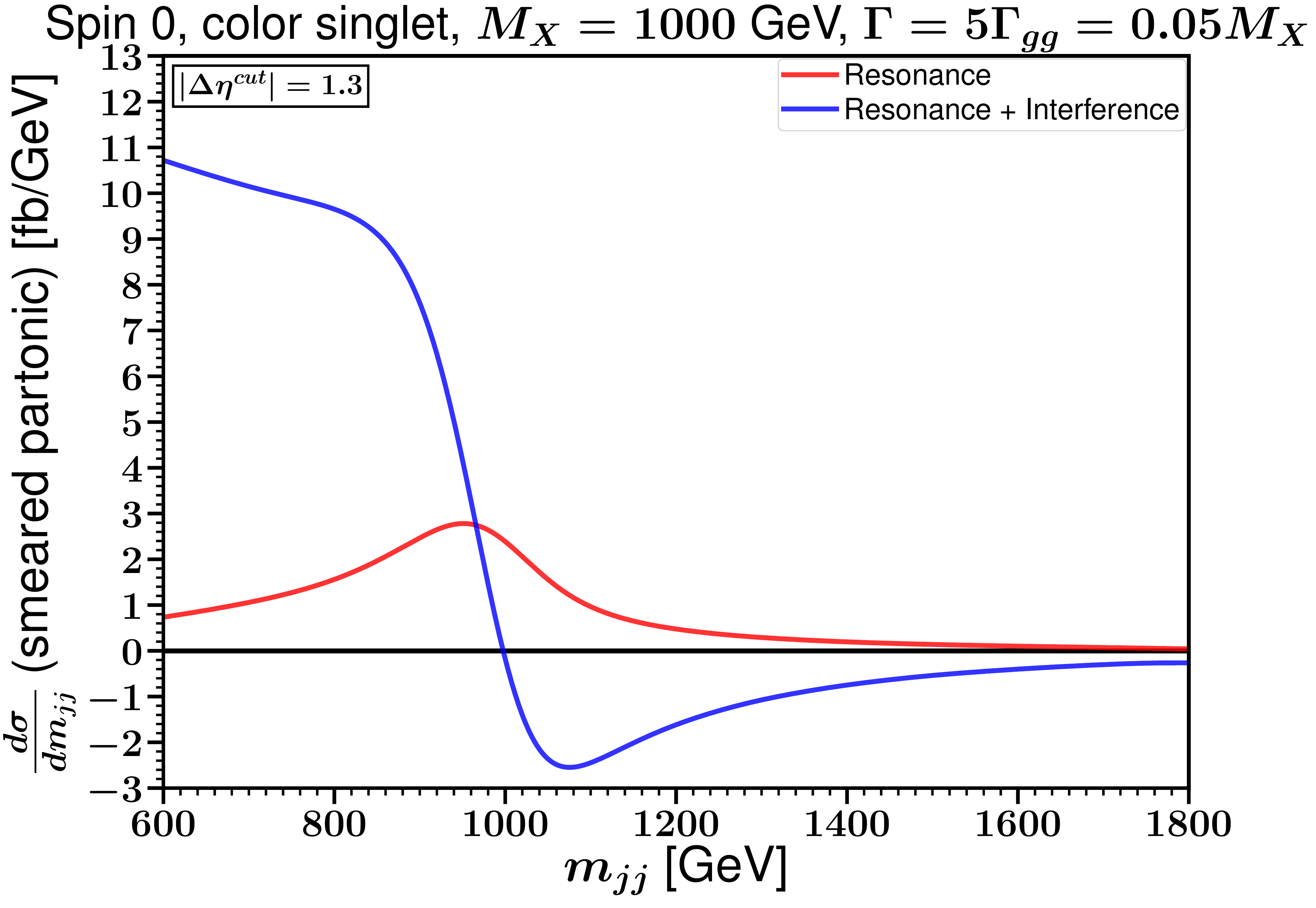}
  \end{minipage}

  \vspace{0.15cm}
  
  \begin{minipage}[]{0.495\linewidth}
    \includegraphics[width=8.0cm,angle=0]{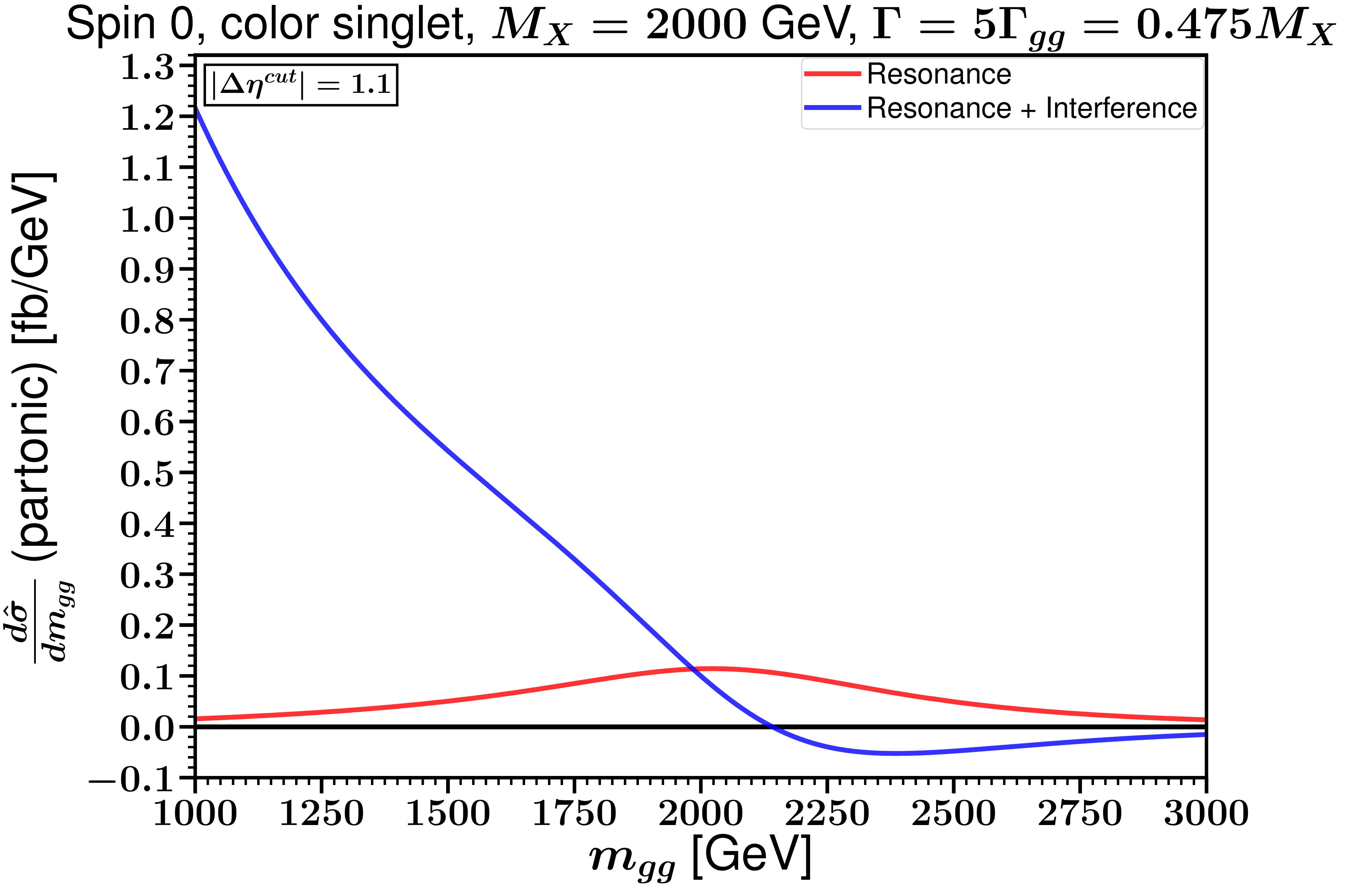}
  \end{minipage}
    \begin{minipage}[]{0.495\linewidth}
    \includegraphics[width=8.0cm,angle=0]{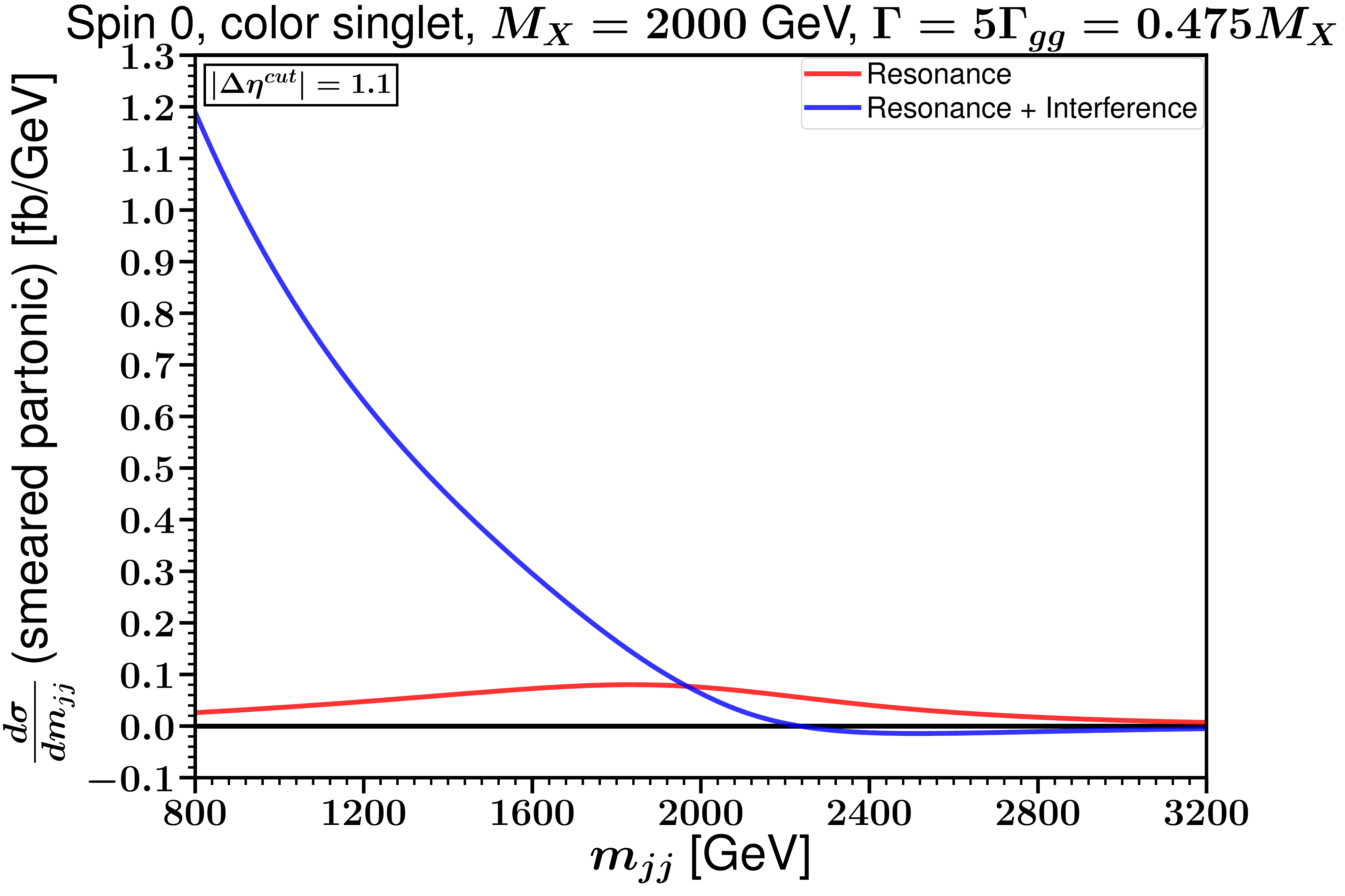}
  \end{minipage}

  \vspace{0.15cm}
  
\begin{center}\begin{minipage}[]{0.95\linewidth}
\caption{\label{fig:gggg_s0c1_analytic_otherdecays}
Digluon invariant mass distributions, at the 13 TeV LHC, for benchmark spin-0, color-singlet resonances 
from Table \ref{tab:benchmarks}, with $\Gamma_{gg} = \Gamma/5$, and 
$M_X = 1000$ GeV (top row), and 
2000 GeV (bottom row).
The parton-level distributions are shown in the left column panels. These are are then smeared by convolution with the estimated detector responses shown in Figure~\ref{fig:yield} to obtain the dijet invariant mass distributions in the right column panels. In all four panels, the red lines show the naive results for the resonant signal $g g \rightarrow X \rightarrow g g$, while the blue lines show the full results including the interferences with the  QCD background $g g \rightarrow g g$.
}
\end{minipage}\end{center}
\end{figure}

\begin{figure}[!tb]
  \begin{minipage}[]{0.495\linewidth}
    \includegraphics[width=8.0cm,angle=0]{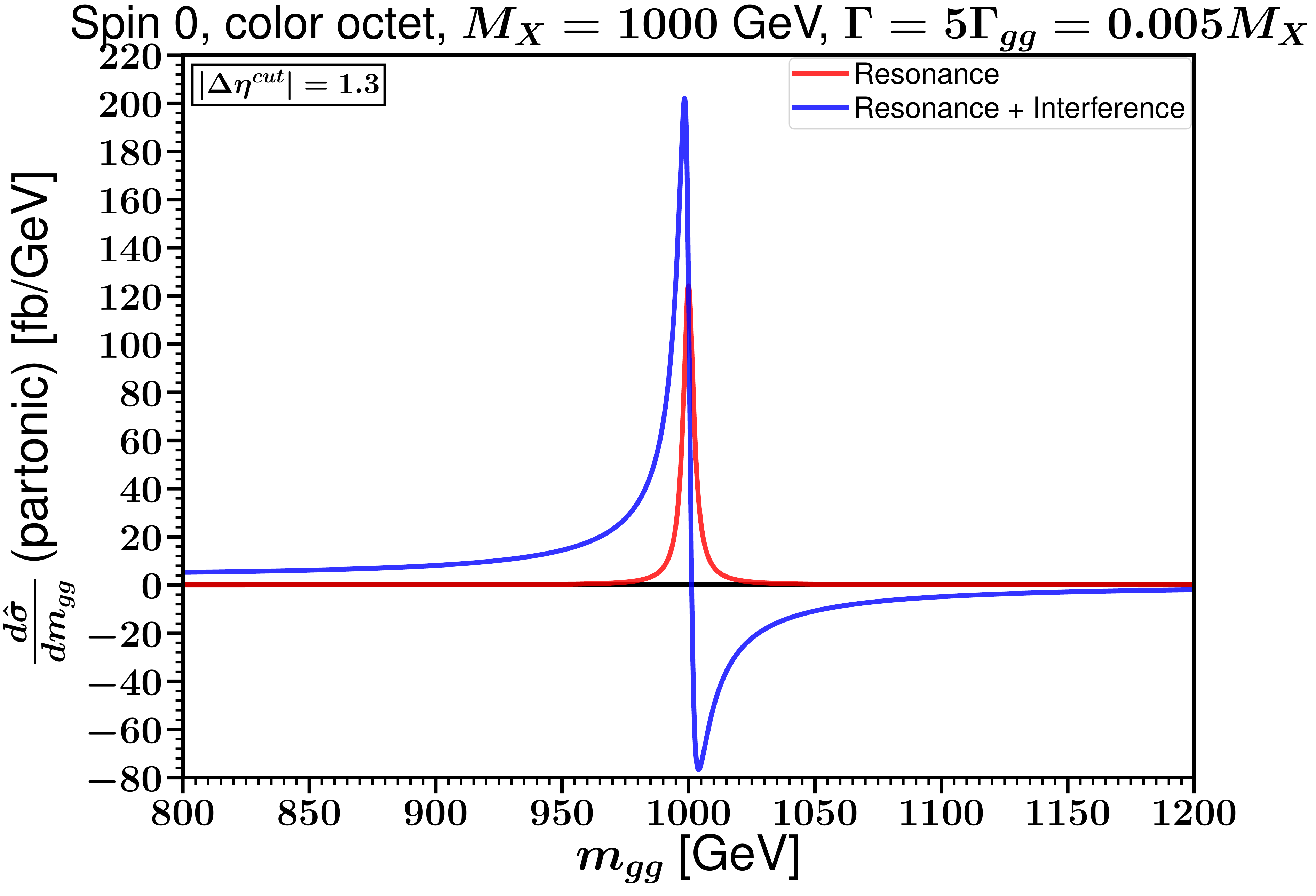}
  \end{minipage}
    \begin{minipage}[]{0.495\linewidth}
    \includegraphics[width=8.0cm,angle=0]{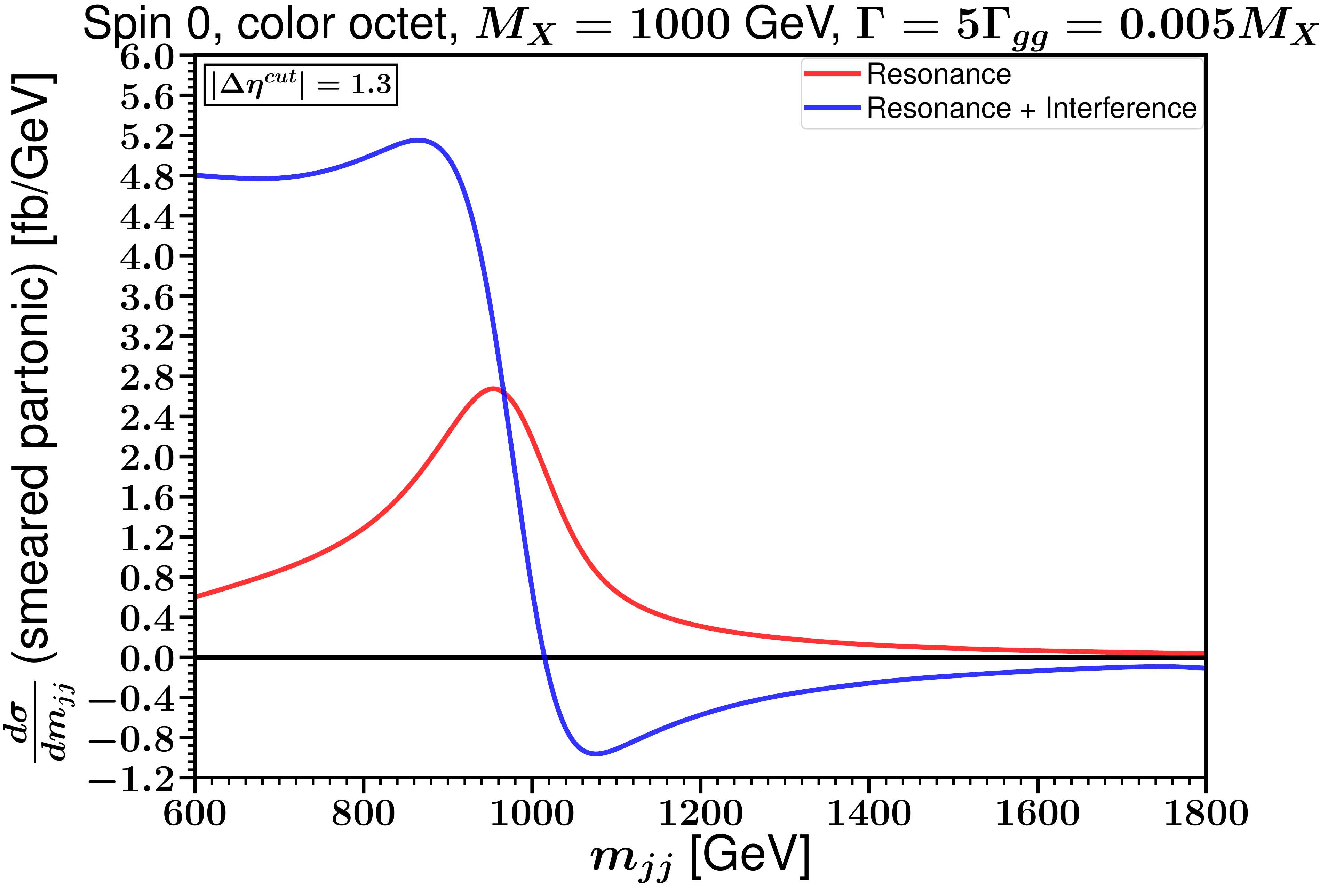}
  \end{minipage}

  \vspace{0.15cm}
  
  \begin{minipage}[]{0.495\linewidth}
    \includegraphics[width=8.0cm,angle=0]{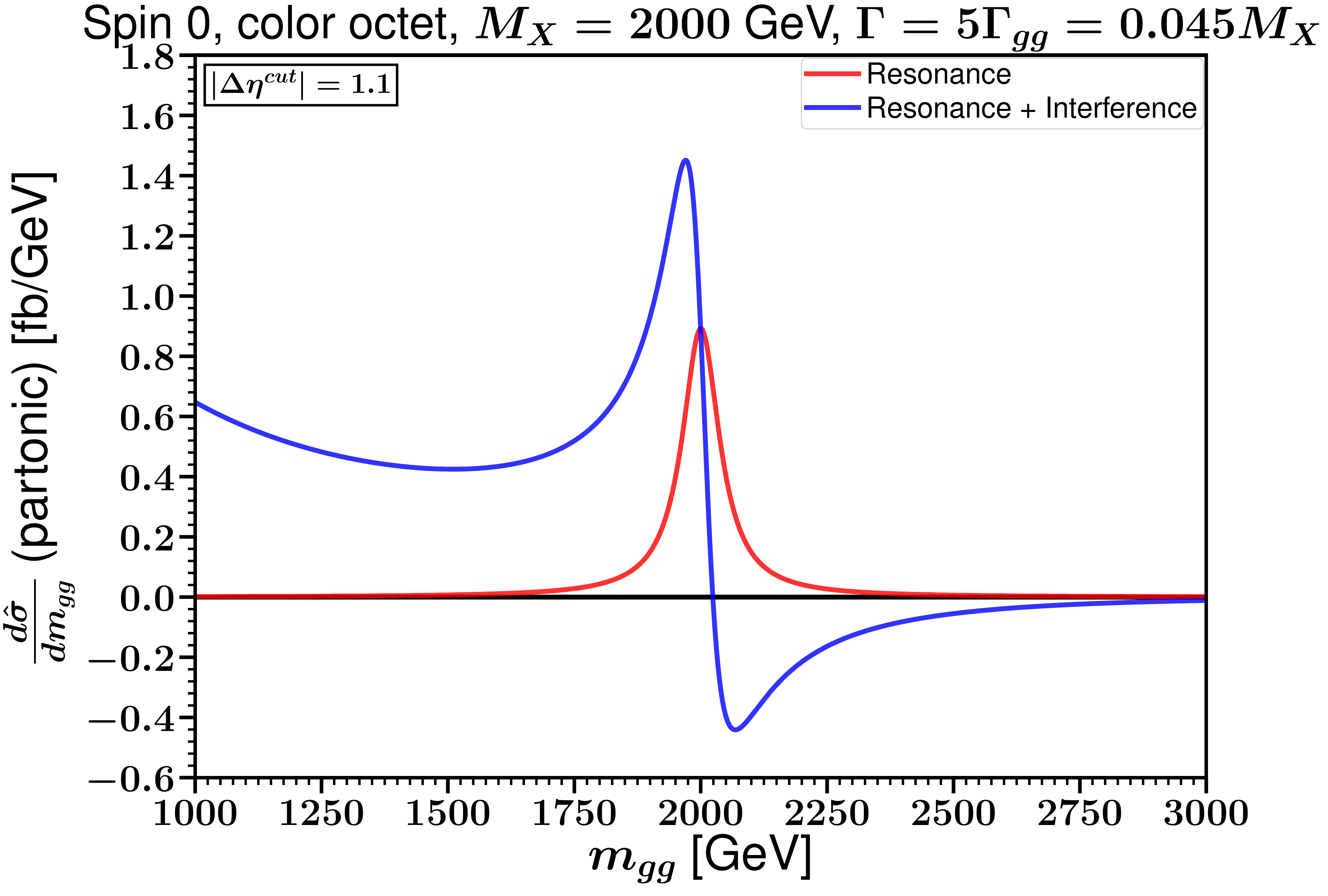}
  \end{minipage}
    \begin{minipage}[]{0.495\linewidth}
    \includegraphics[width=8.0cm,angle=0]{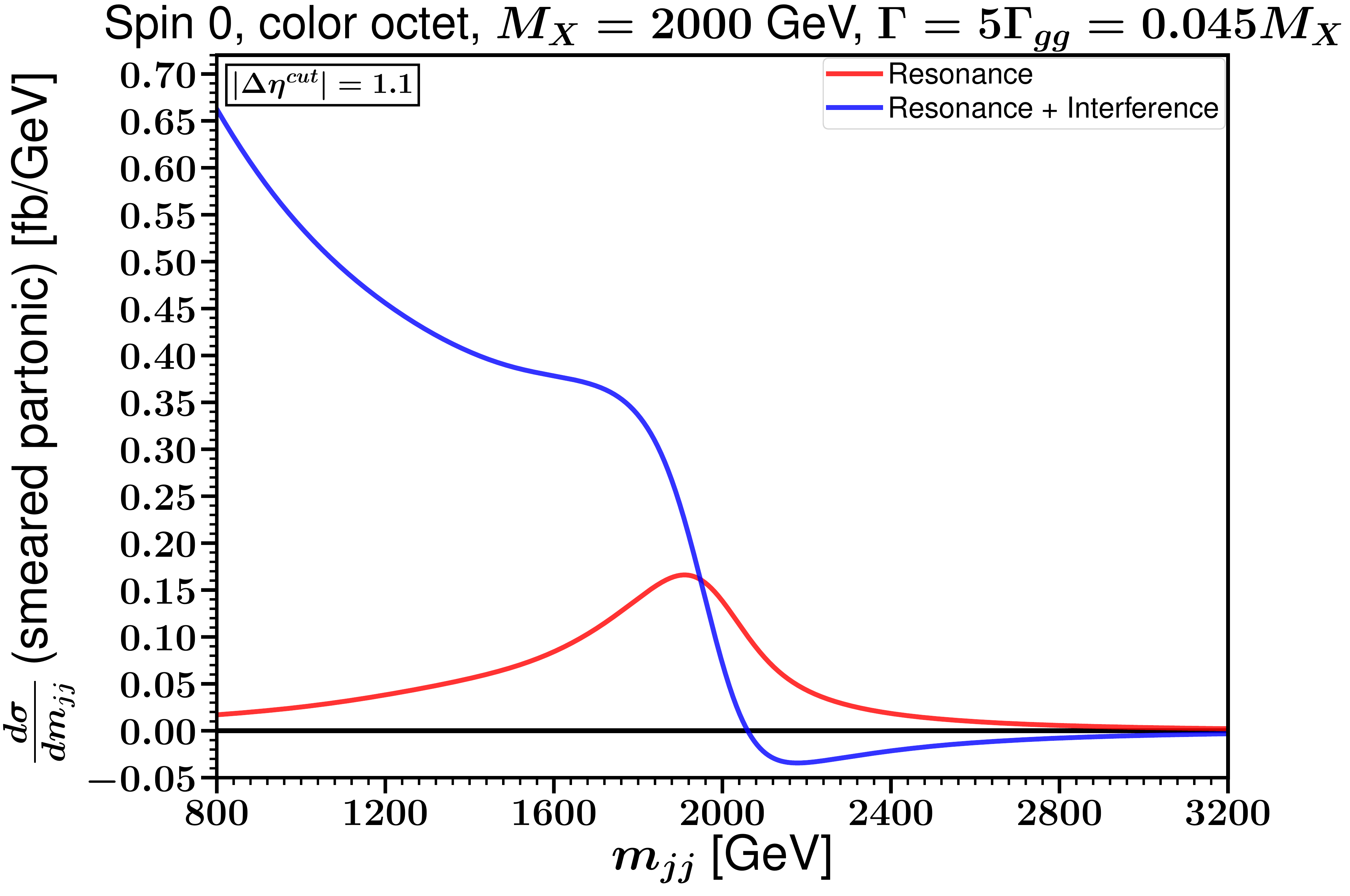}
  \end{minipage}

  \vspace{0.15cm}
  
  \begin{minipage}[]{0.495\linewidth}
    \includegraphics[width=8.0cm,angle=0]{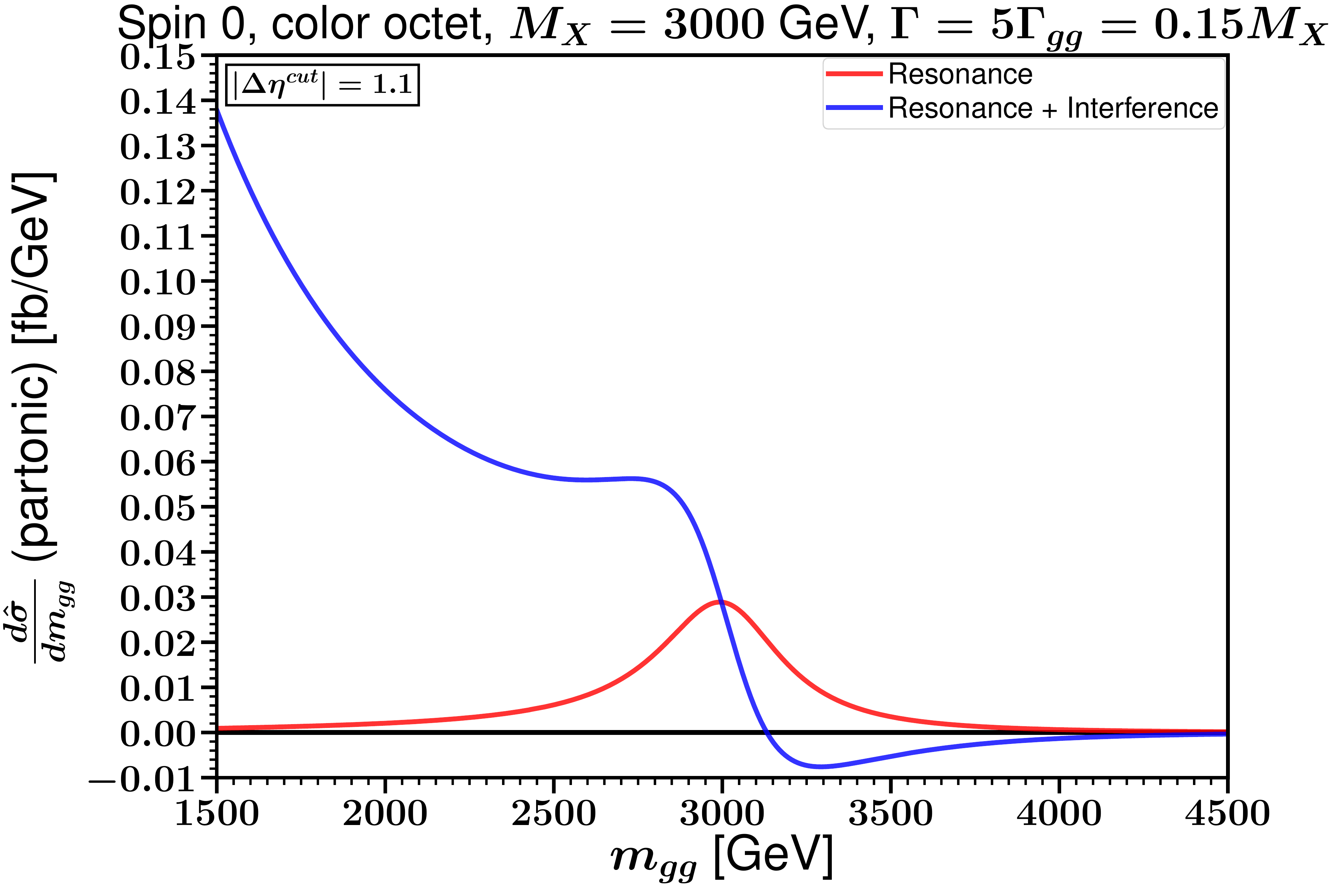}
  \end{minipage}
    \begin{minipage}[]{0.495\linewidth}
    \includegraphics[width=8.0cm,angle=0]{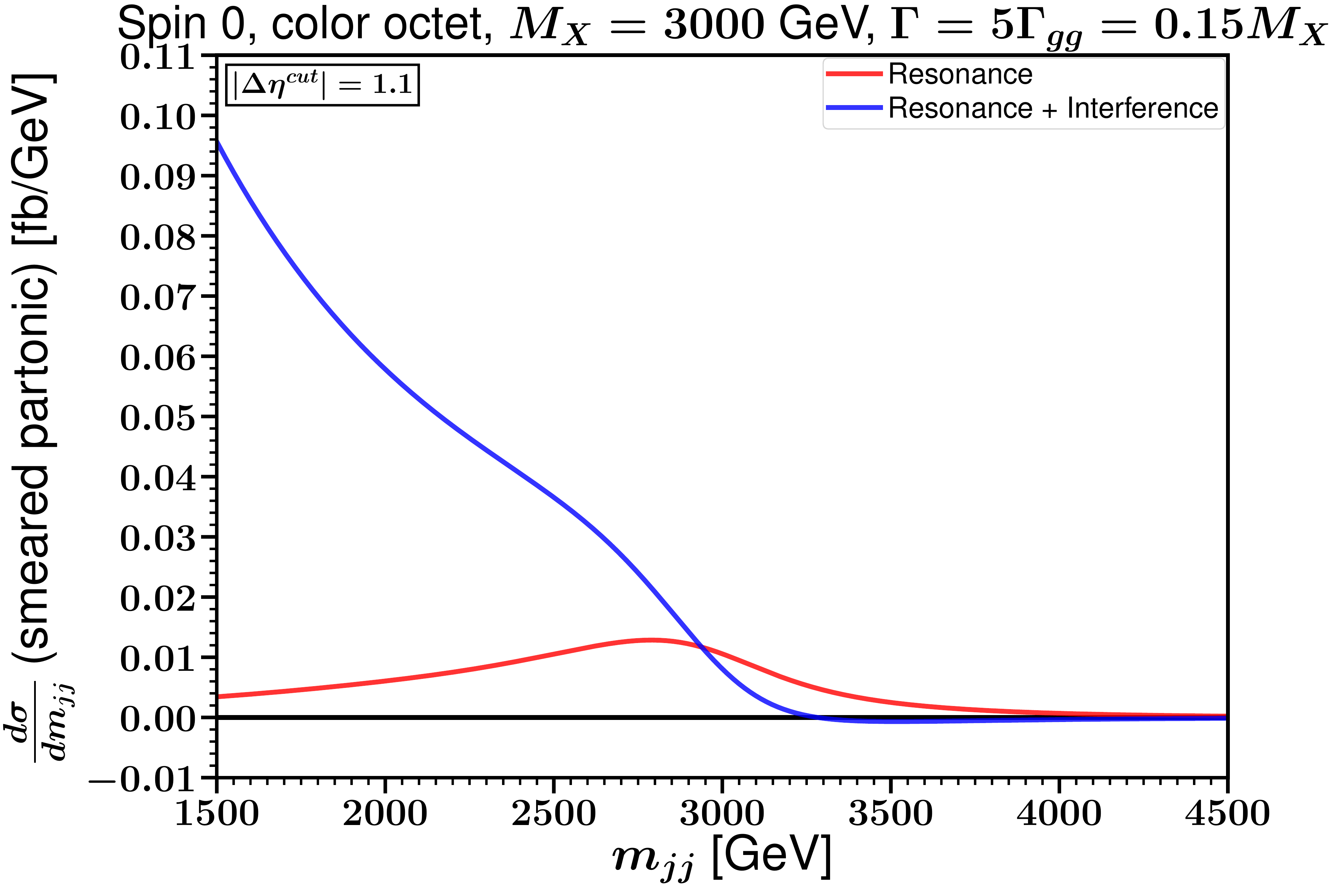}
  \end{minipage}
\begin{center}\begin{minipage}[]{0.95\linewidth}
\caption{\label{fig:gggg_s0c8_analytic_otherdecays}
Digluon invariant mass distributions, at the 13 TeV LHC, for benchmark spin-0, color-octet resonances 
from Table \ref{tab:benchmarks}, with $\Gamma_{gg} = \Gamma/5$, and 
$M_X = 1000$ GeV (top row), and 
2000 GeV (middle row) and 
3000 GeV (bottom row).
The parton-level distributions are shown in the left column panels. These are are then smeared by convolution with the estimated detector responses shown in Figure~\ref{fig:yield} to obtain the dijet invariant mass distributions in the right column panels. In all six panels, the red lines show the naive results for the resonant signal $g g \rightarrow X \rightarrow g g$, while the blue lines show the full results including the interferences with the  QCD background $g g \rightarrow g g$.}
\end{minipage}\end{center}
\end{figure}

\begin{figure}[!tb]
  \begin{minipage}[]{0.495\linewidth}
    \includegraphics[width=8.0cm,angle=0]{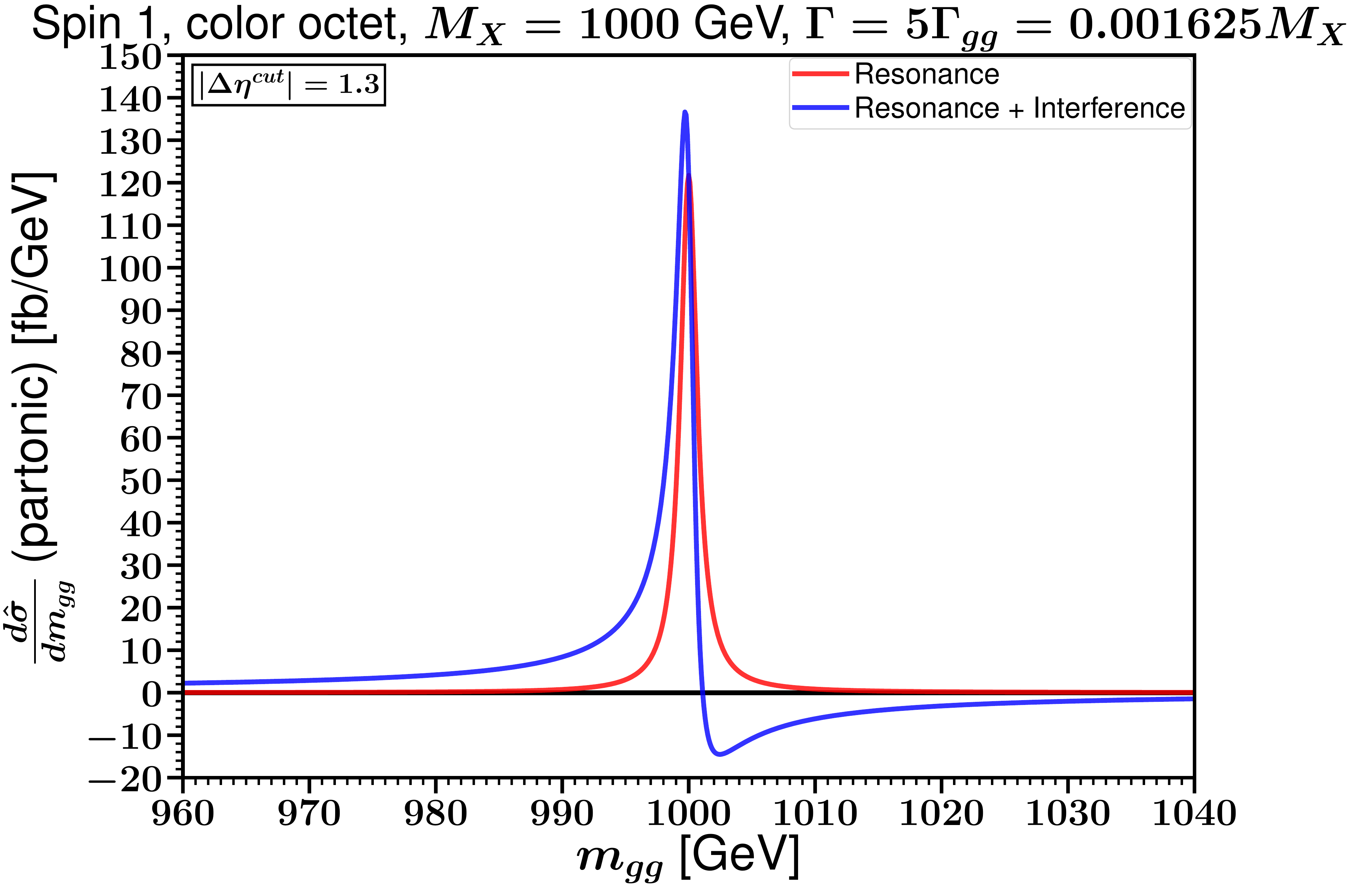}
  \end{minipage}
    \begin{minipage}[]{0.495\linewidth}
    \includegraphics[width=8.0cm,angle=0]{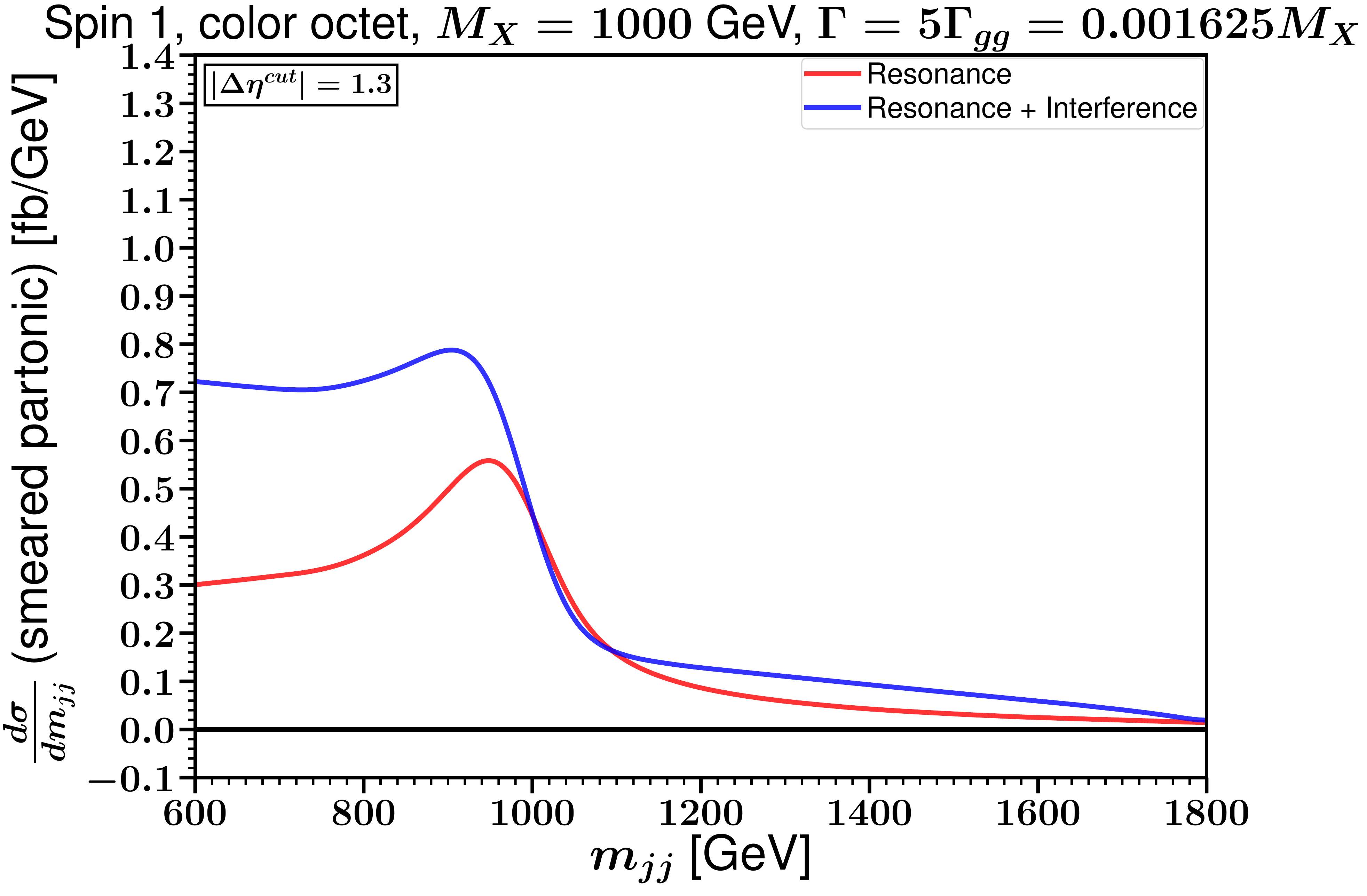}
  \end{minipage}

  \vspace{0.15cm}
  
  \begin{minipage}[]{0.495\linewidth}
    \includegraphics[width=8.0cm,angle=0]{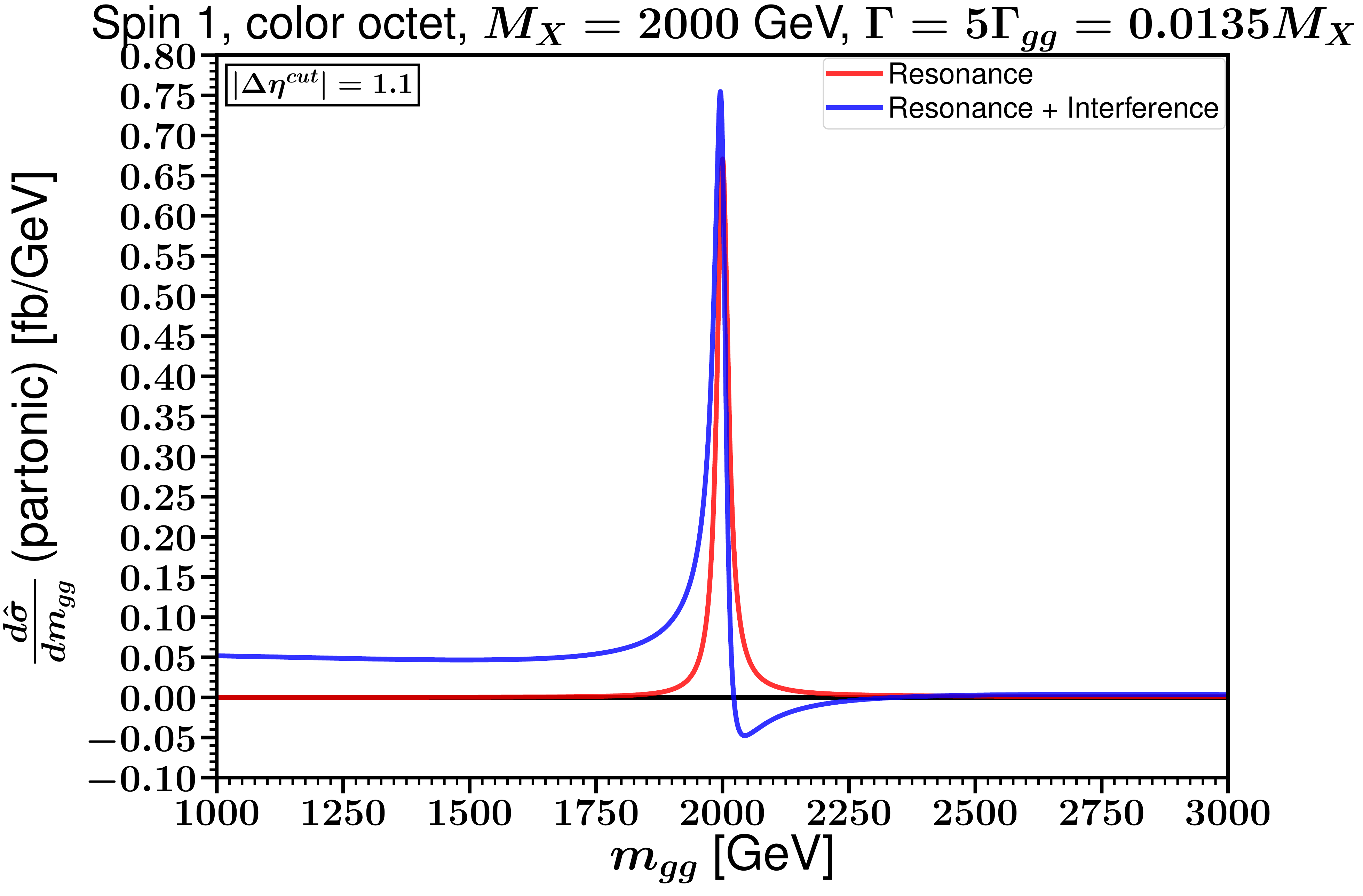}
  \end{minipage}
    \begin{minipage}[]{0.495\linewidth}
    \includegraphics[width=8.0cm,angle=0]{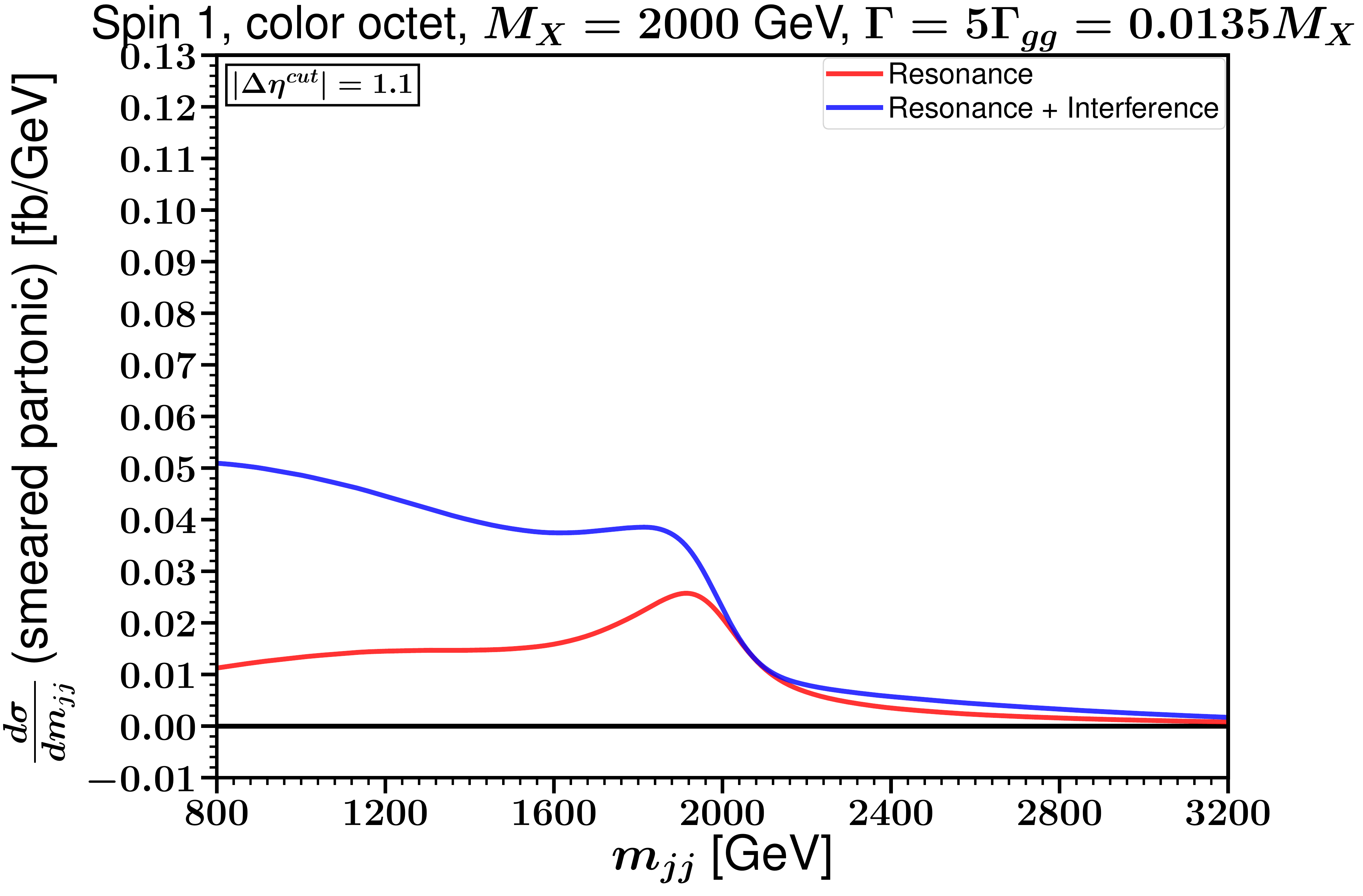}
  \end{minipage}

  \vspace{0.15cm}
  
  \begin{minipage}[]{0.495\linewidth}
    \includegraphics[width=8.0cm,angle=0]{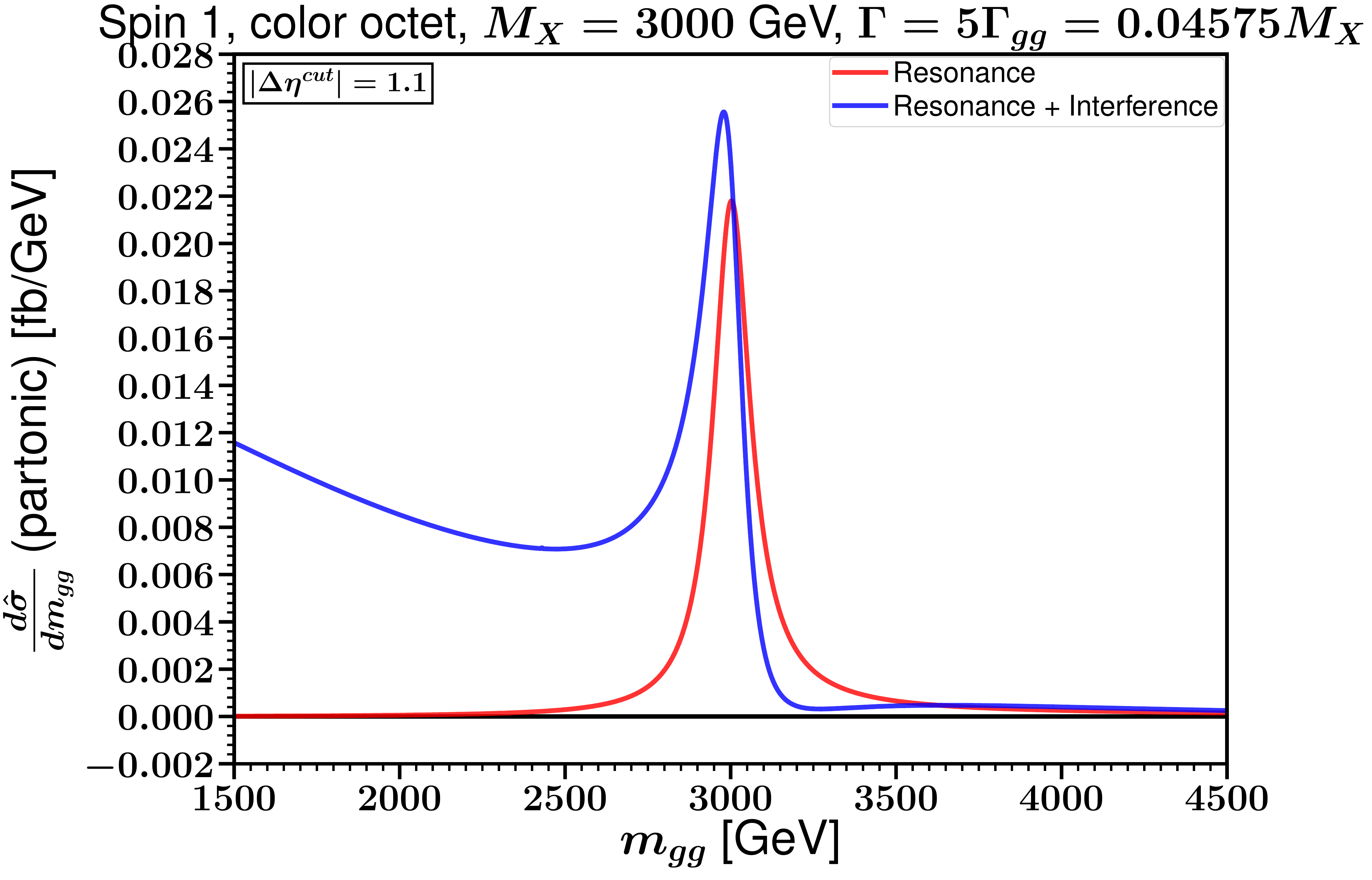}
  \end{minipage}
    \begin{minipage}[]{0.495\linewidth}
    \includegraphics[width=8.0cm,angle=0]{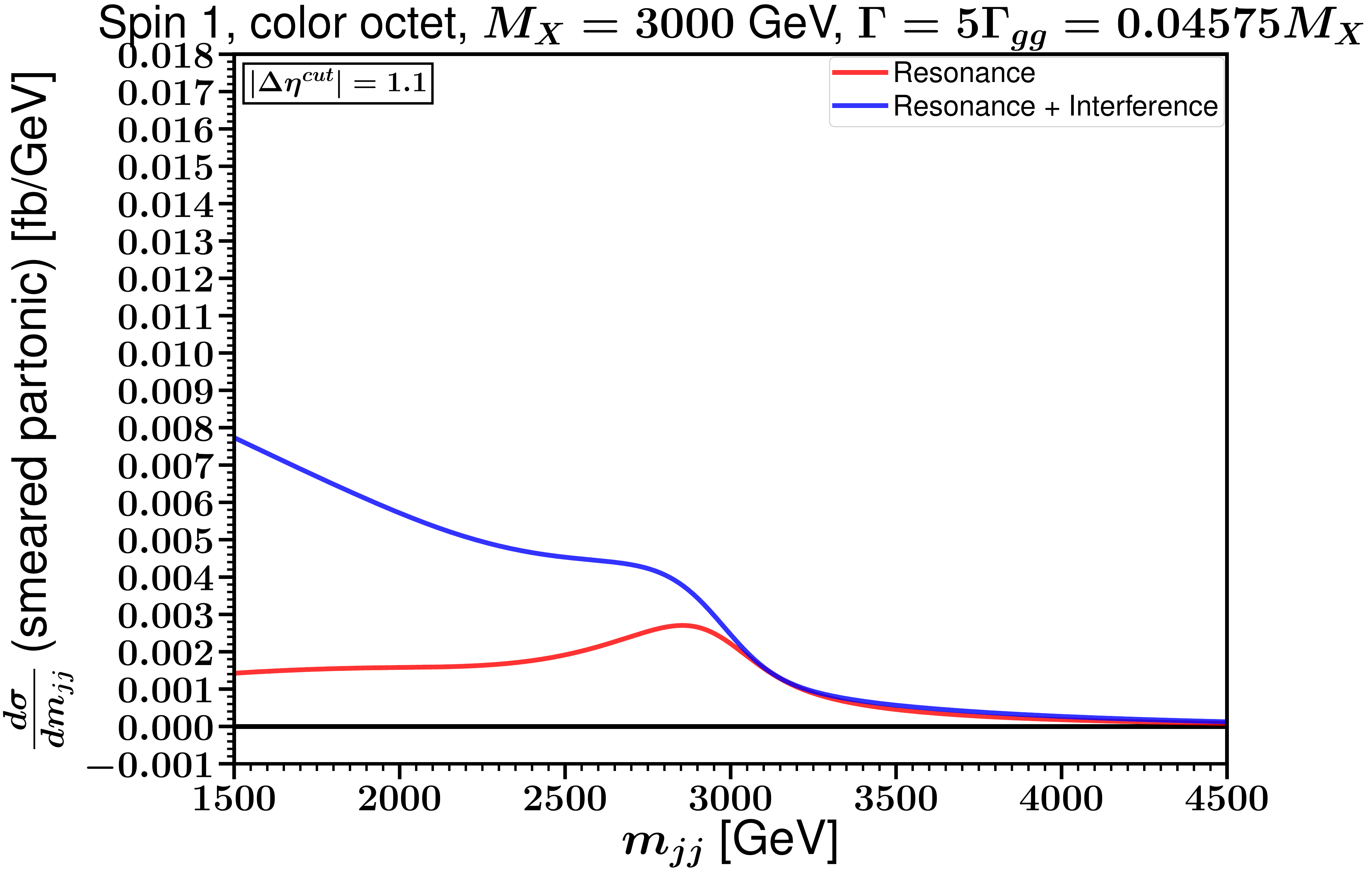}
  \end{minipage}
\begin{center}\begin{minipage}[]{0.95\linewidth}
\caption{\label{fig:gggg_s1c8_analytic_otherdecays}
Digluon invariant mass distributions, at the 13 TeV LHC, for benchmark spin-1, color-octet resonances 
from Table \ref{tab:benchmarks}, with $\Gamma_{gg} = \Gamma/5$, and 
$M_X = 1000$ GeV (top row), and 
2000 GeV (middle row) and 
3000 GeV (bottom row).
The parton-level distributions are shown in the left column panels. These are are then smeared by convolution with the estimated detector responses shown in Figure~\ref{fig:yield} to obtain the dijet invariant mass distributions in the right column panels. In all six panels, the red lines show the naive results for the resonant signal $g g \rightarrow X \rightarrow g g$, while the blue lines show the full results including the interferences with the  QCD background $g g \rightarrow g g$.}
\end{minipage}\end{center}
\end{figure}

\begin{figure}[!tb]
  \begin{minipage}[]{0.495\linewidth}
    \includegraphics[width=8.0cm,angle=0]{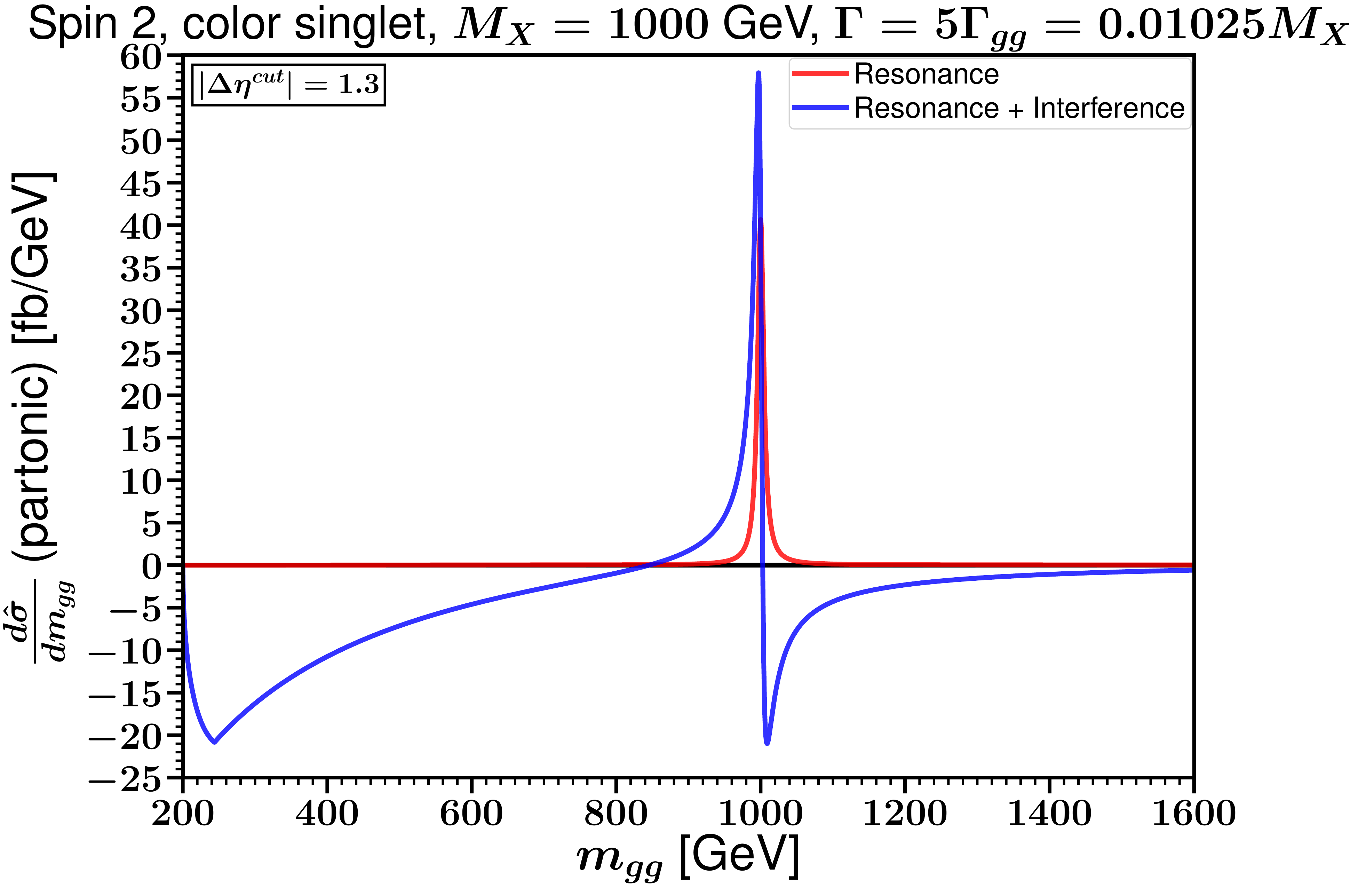}
  \end{minipage}
    \begin{minipage}[]{0.495\linewidth}
    \includegraphics[width=8.0cm,angle=0]{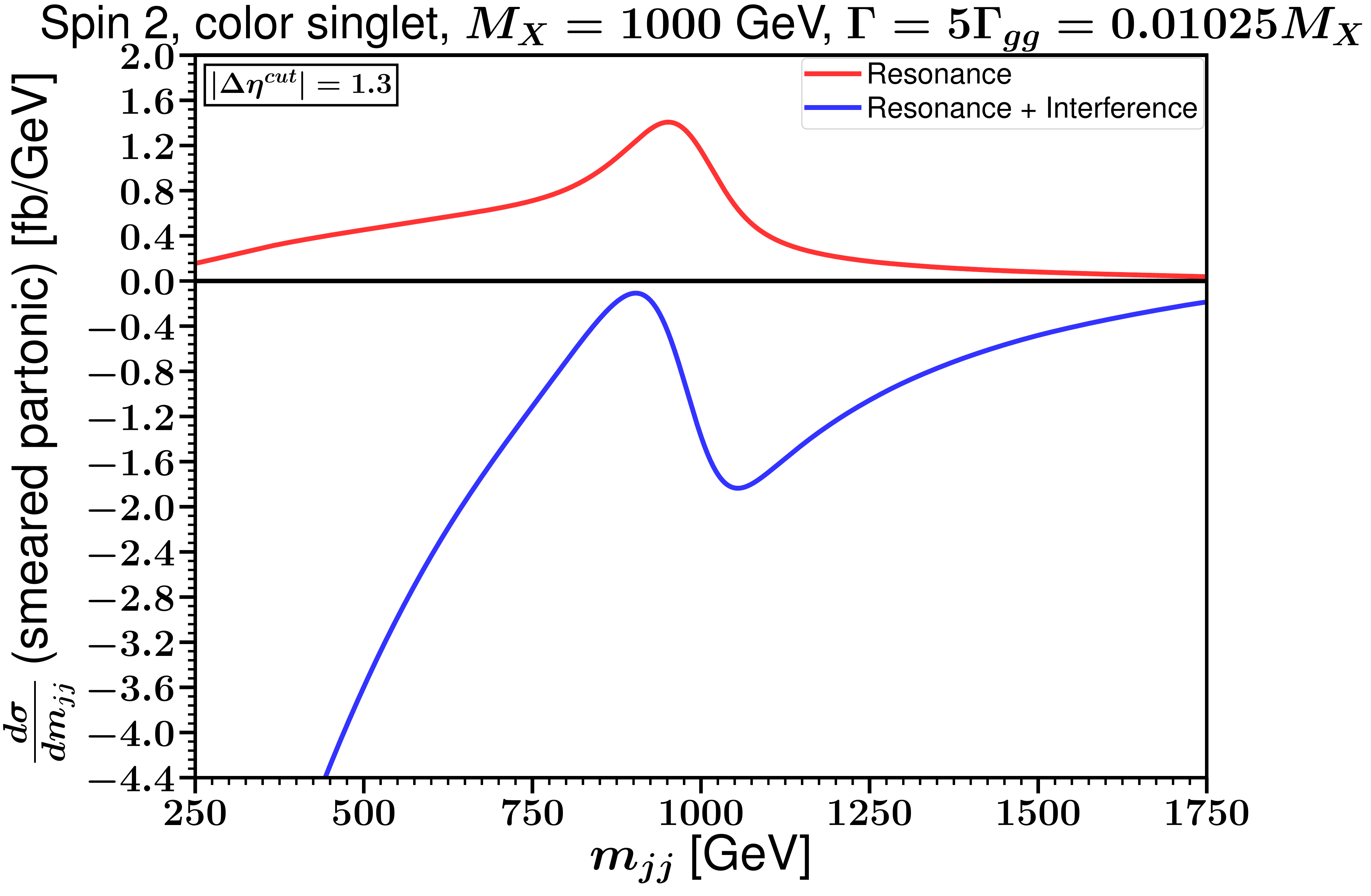}
  \end{minipage}

  \vspace{0.15cm}
  
  \begin{minipage}[]{0.495\linewidth}
    \includegraphics[width=8.0cm,angle=0]{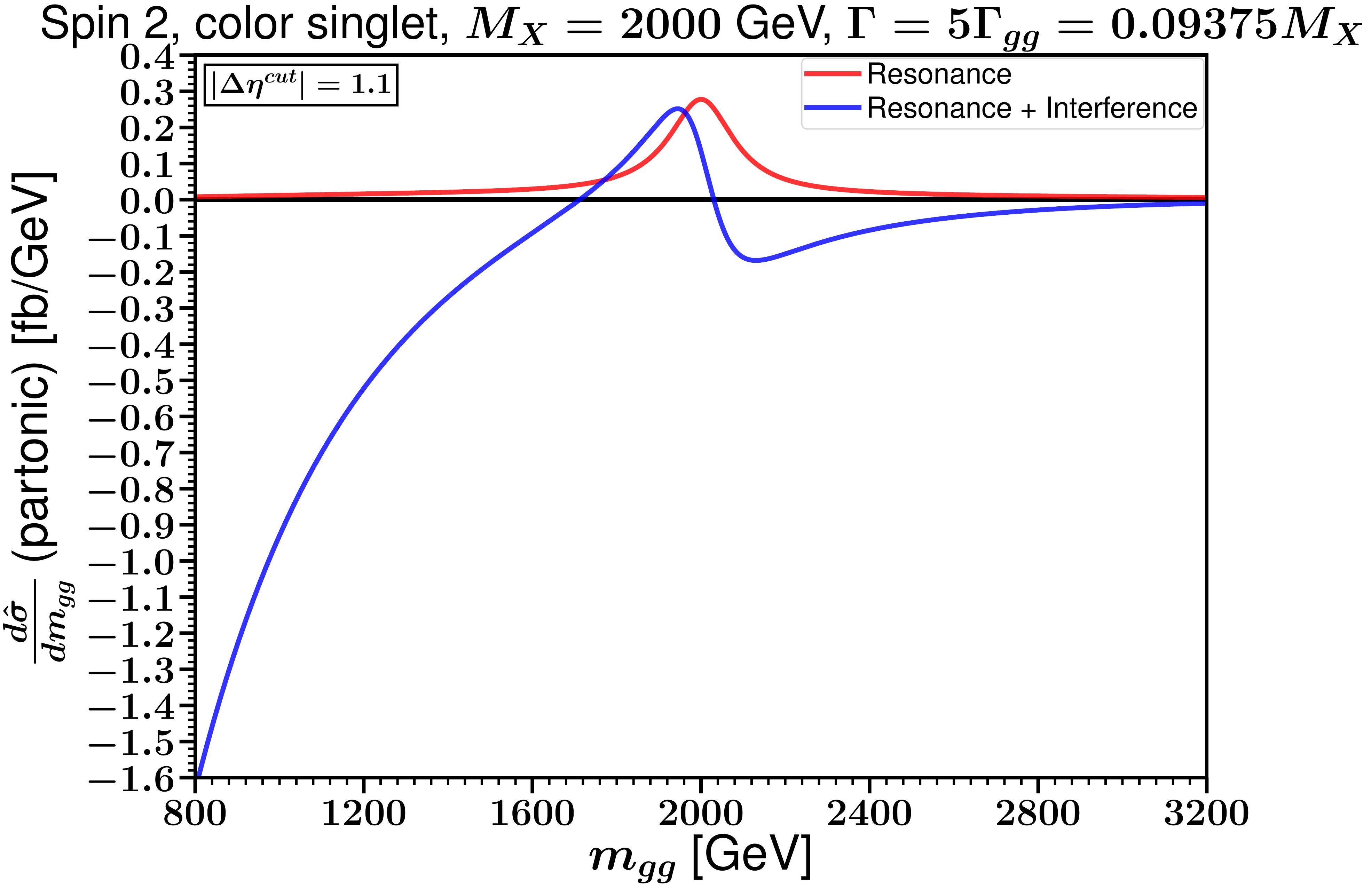}
  \end{minipage}
    \begin{minipage}[]{0.495\linewidth}
    \includegraphics[width=8.0cm,angle=0]{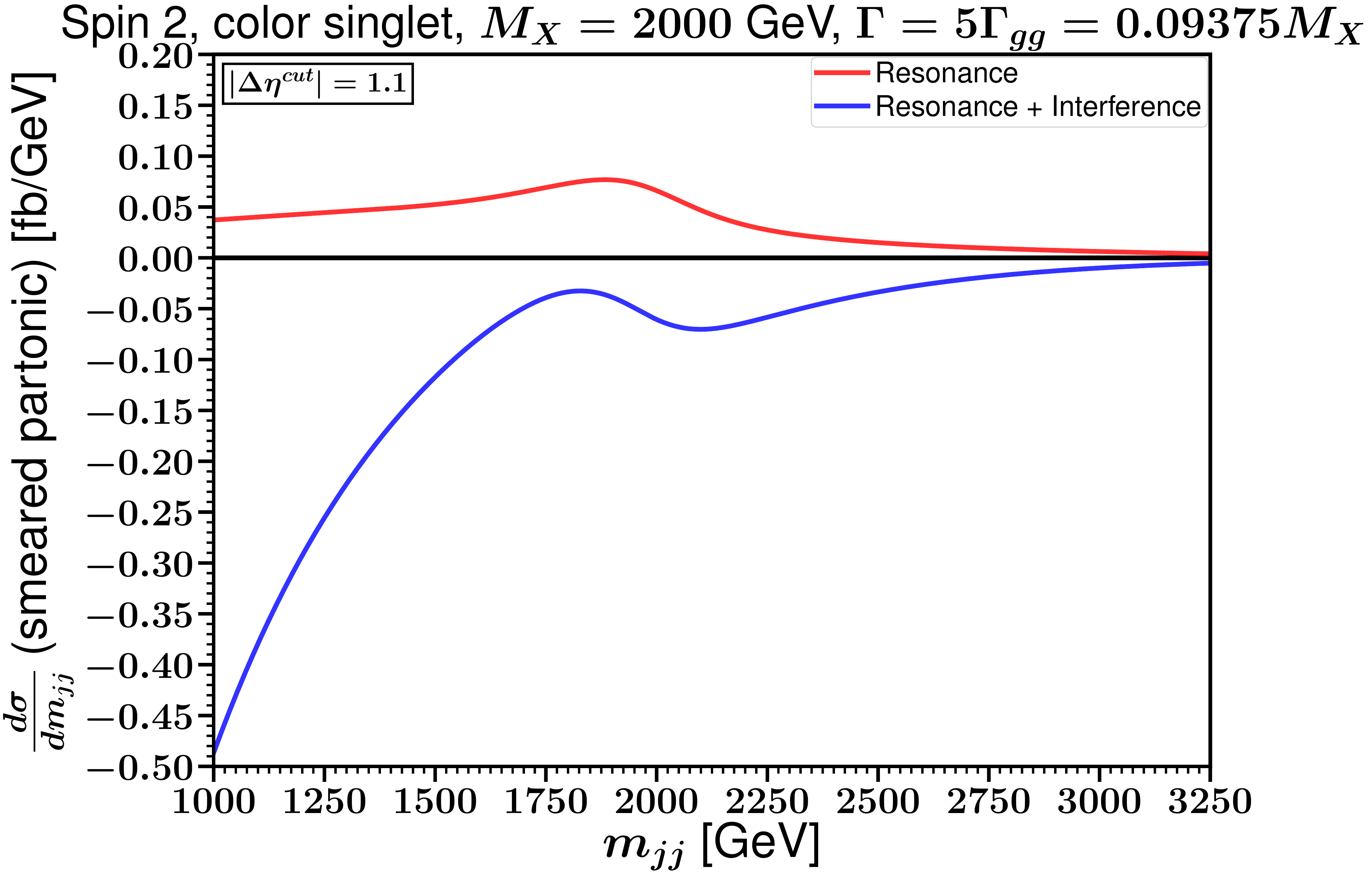}
  \end{minipage}

  \vspace{0.15cm}
  
  \begin{minipage}[]{0.495\linewidth}
    \includegraphics[width=8.0cm,angle=0]{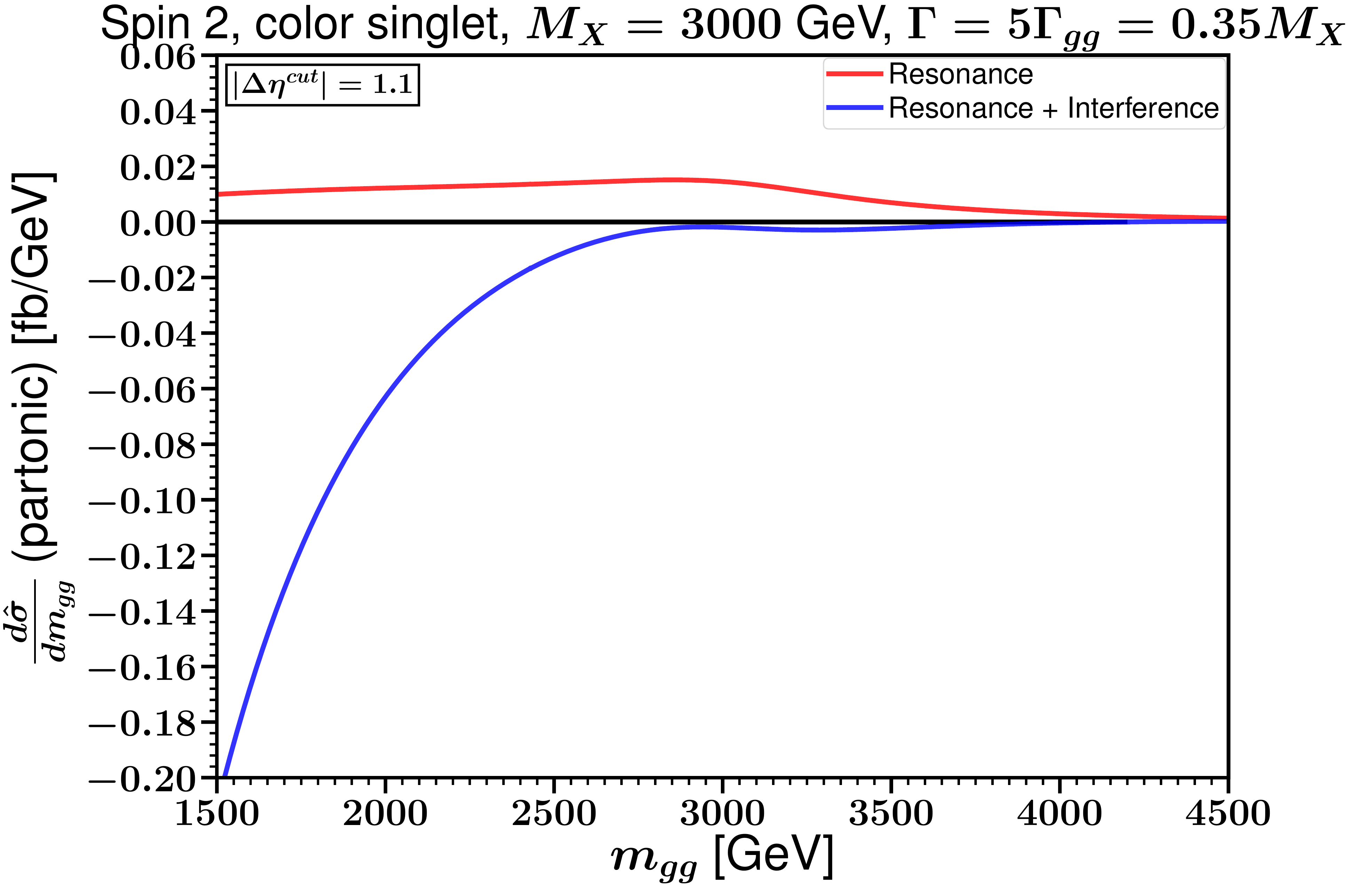}
  \end{minipage}
    \begin{minipage}[]{0.495\linewidth}
    \includegraphics[width=8.0cm,angle=0]{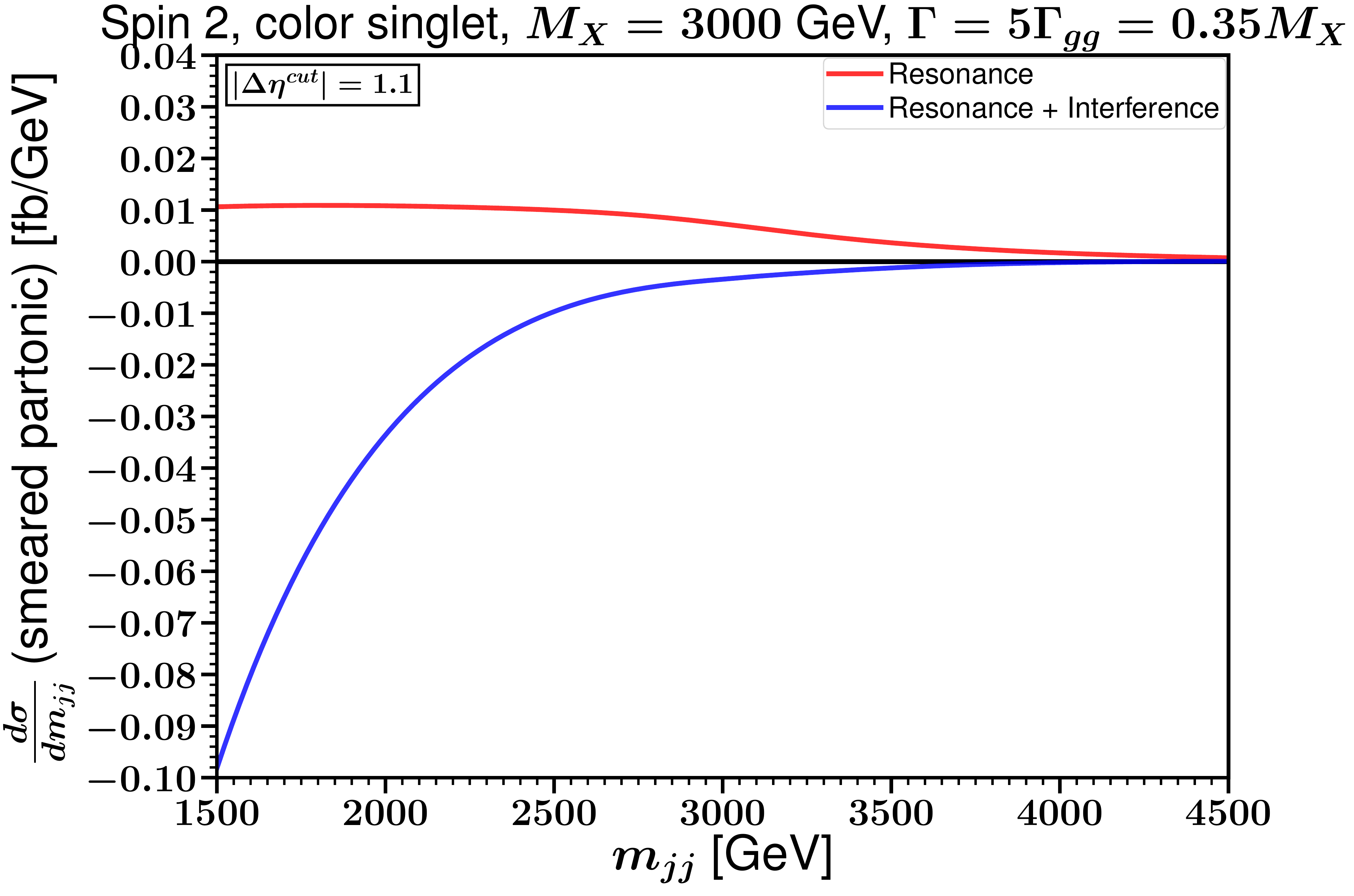}
  \end{minipage}
\begin{center}\begin{minipage}[]{0.95\linewidth}
\caption{\label{fig:gggg_s2c1_analytic_otherdecays}
Digluon invariant mass distributions, at the 13 TeV LHC, for benchmark spin-2, color-singlet resonances 
from Table \ref{tab:benchmarks}, with $\Gamma_{gg} = \Gamma/5$, and 
$M_X = 1000$ GeV (top row), and 
2000 GeV (middle row) and 
3000 GeV (bottom row).
The parton-level distributions are shown in the left column panels. These are are then smeared by convolution with the estimated detector responses shown in Figure~\ref{fig:yield} to obtain the dijet invariant mass distributions in the right column panels. In all six panels, the red lines show the naive results for the resonant signal $g g \rightarrow X \rightarrow g g$, while the blue lines show the full results including the interferences with the  QCD background $g g \rightarrow g g$.
The negative tails at small invariant mass come from the interference between the QCD amplitudes and the $t$- and $u$-channel $X$ exchange diagrams.}
\end{minipage}\end{center}
\end{figure}

The organization of results and structure of the figures below 
is exactly the same as in the previous section.
Thus in Figures \ref{fig:gggg_s0c1_analytic_otherdecays}, \ref{fig:gggg_s0c8_analytic_otherdecays},
\ref{fig:gggg_s1c8_analytic_otherdecays}, and \ref{fig:gggg_s2c1_analytic_otherdecays}, we show the 
parton-level distributions before (left columns) and after (right columns) smearing by convolution with the
detector response functions illustrated in Figure \ref{fig:yield}, for benchmark masses 1000 GeV (top rows), 2000 GeV (middle rows), and 3000 GeV (bottom rows). 
For $\Gamma_{gg} = \Gamma/5$, the benchmark width-to-mass ratios turn out to be very large for a few cases considered here. For example, $\Gamma/M_X = (1.75, 0.475)$ for a spin-0 color-singlet with $M_X = (3000, 2000)$ GeV, and $\Gamma/M_X = 0.35$ for a spin-2 color-singlet with $M_X = 3000$ GeV.
The case with $\Gamma/M_X = 1.75$ for a 3000 GeV massive color-singlet (pseudo-)scalar has unrealistically large width-to-mass ratio for a resonance, and is therefore omitted. On the other hand, cases with a large width-to-mass ratio but with $\Gamma/M_X < 0.55$ are not omitted as the CMS experiment has considered broad resonances with widths up to 55\% of the resonance mass in the reference~{\cite{CMSdijet2019}}.

\begin{figure}[!tb]
  \begin{minipage}[]{0.495\linewidth}
    \includegraphics[width=8.0cm,angle=0]{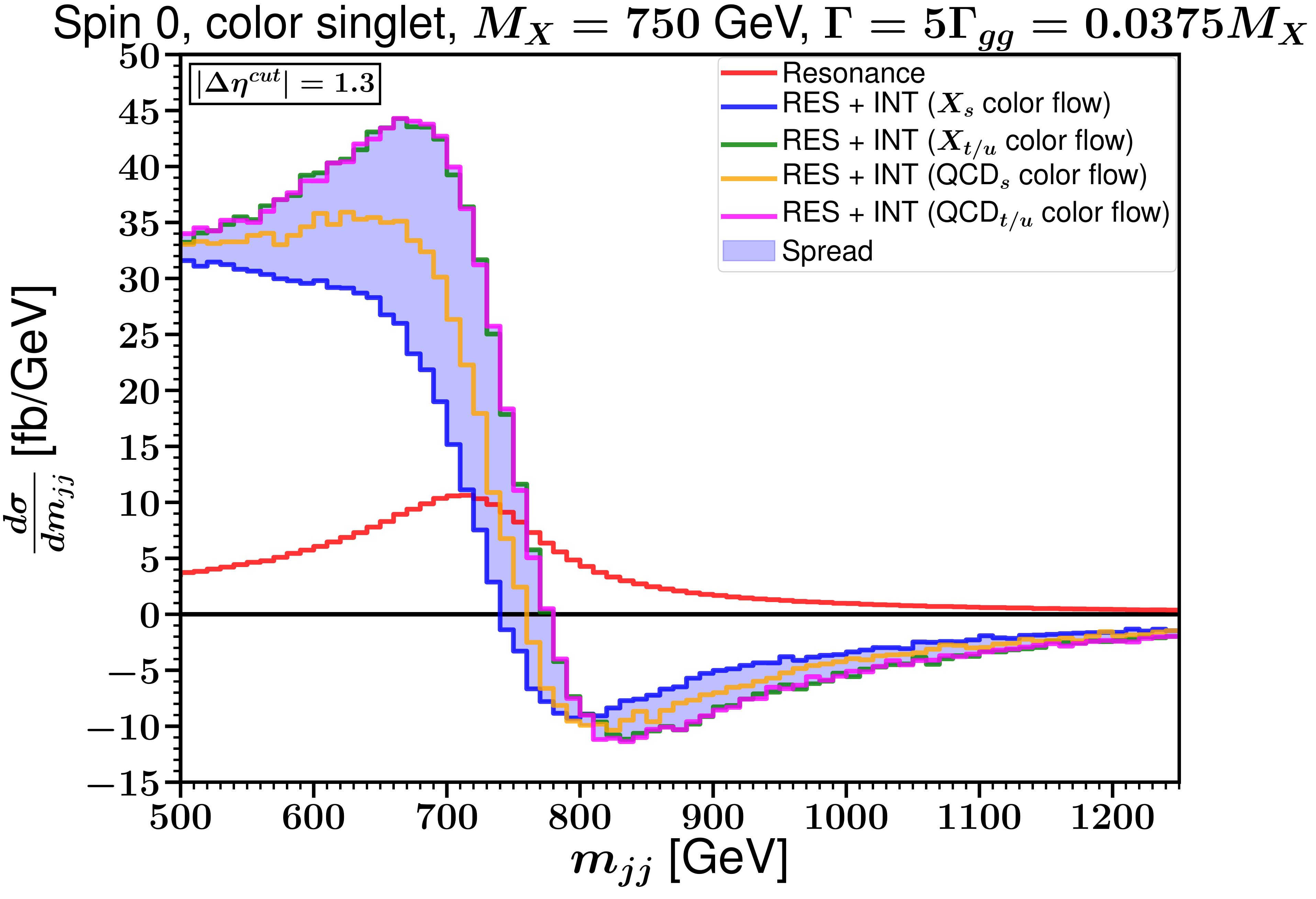}
  \end{minipage}
  \begin{minipage}[]{0.495\linewidth}
    \includegraphics[width=8.0cm,angle=0]{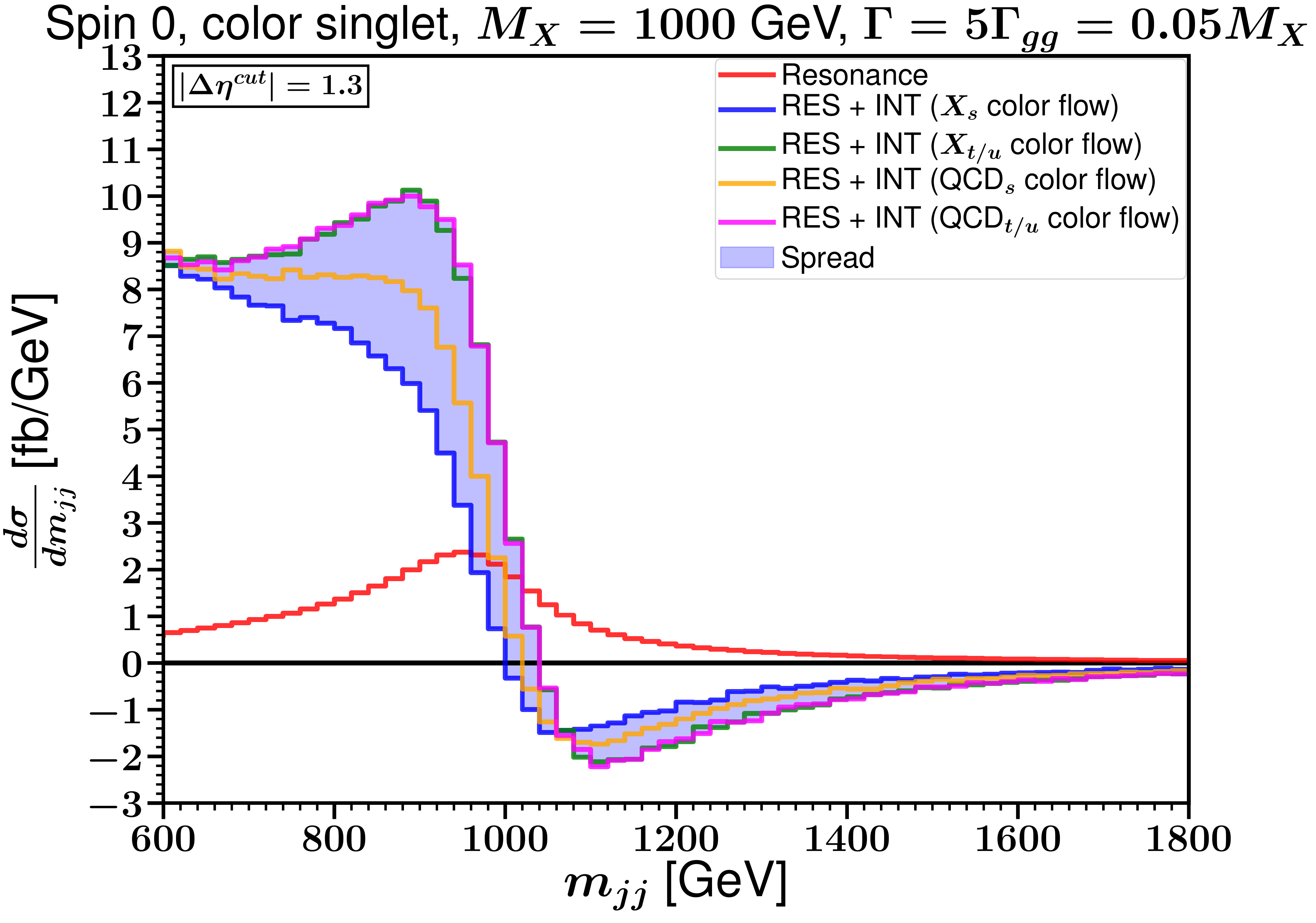}
  \end{minipage}

  \vspace{0.15cm}
    
  \begin{minipage}[]{0.495\linewidth}
    \includegraphics[width=8.0cm,angle=0]{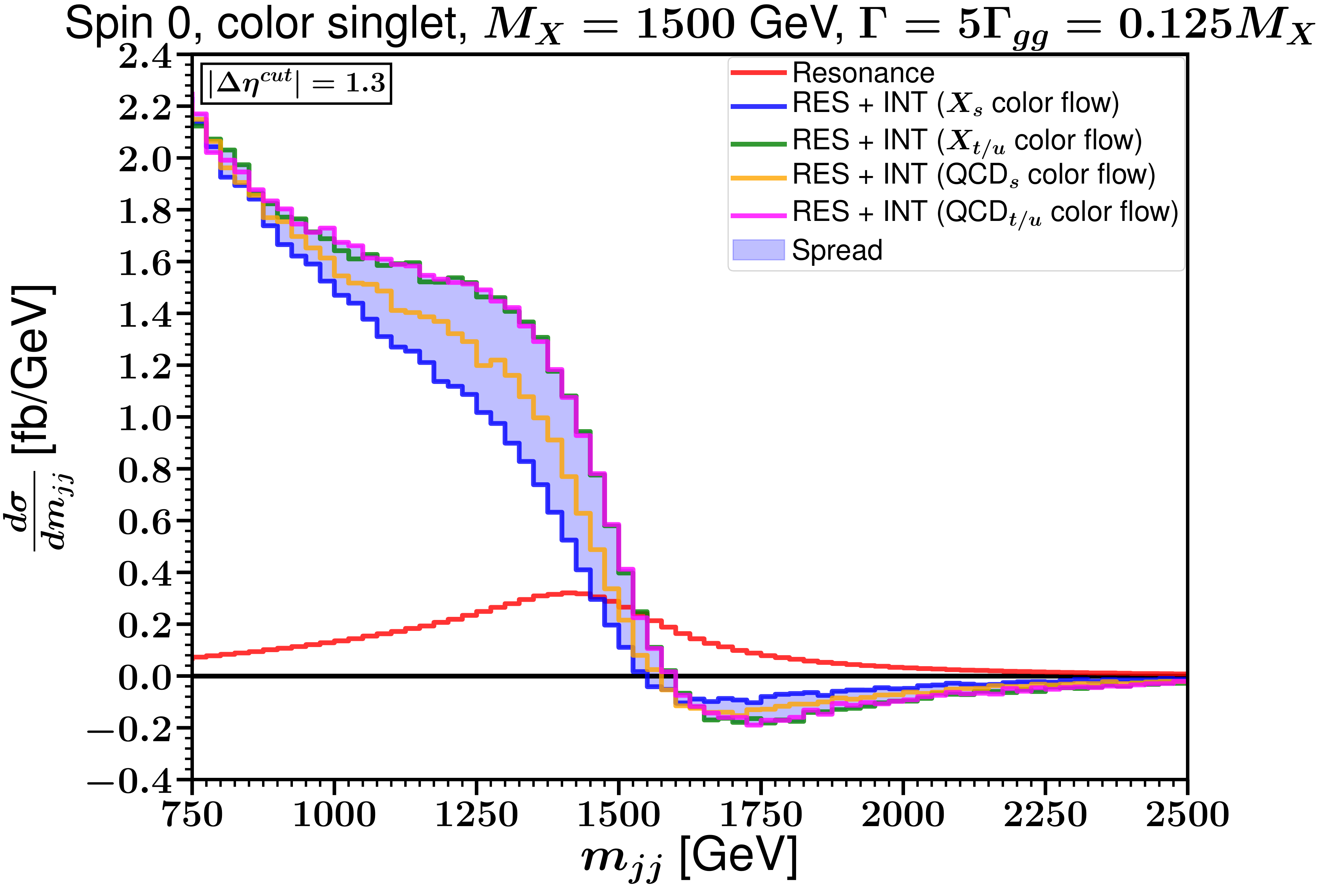}
  \end{minipage}
   \begin{minipage}[]{0.495\linewidth}
    \includegraphics[width=8.0cm,angle=0]{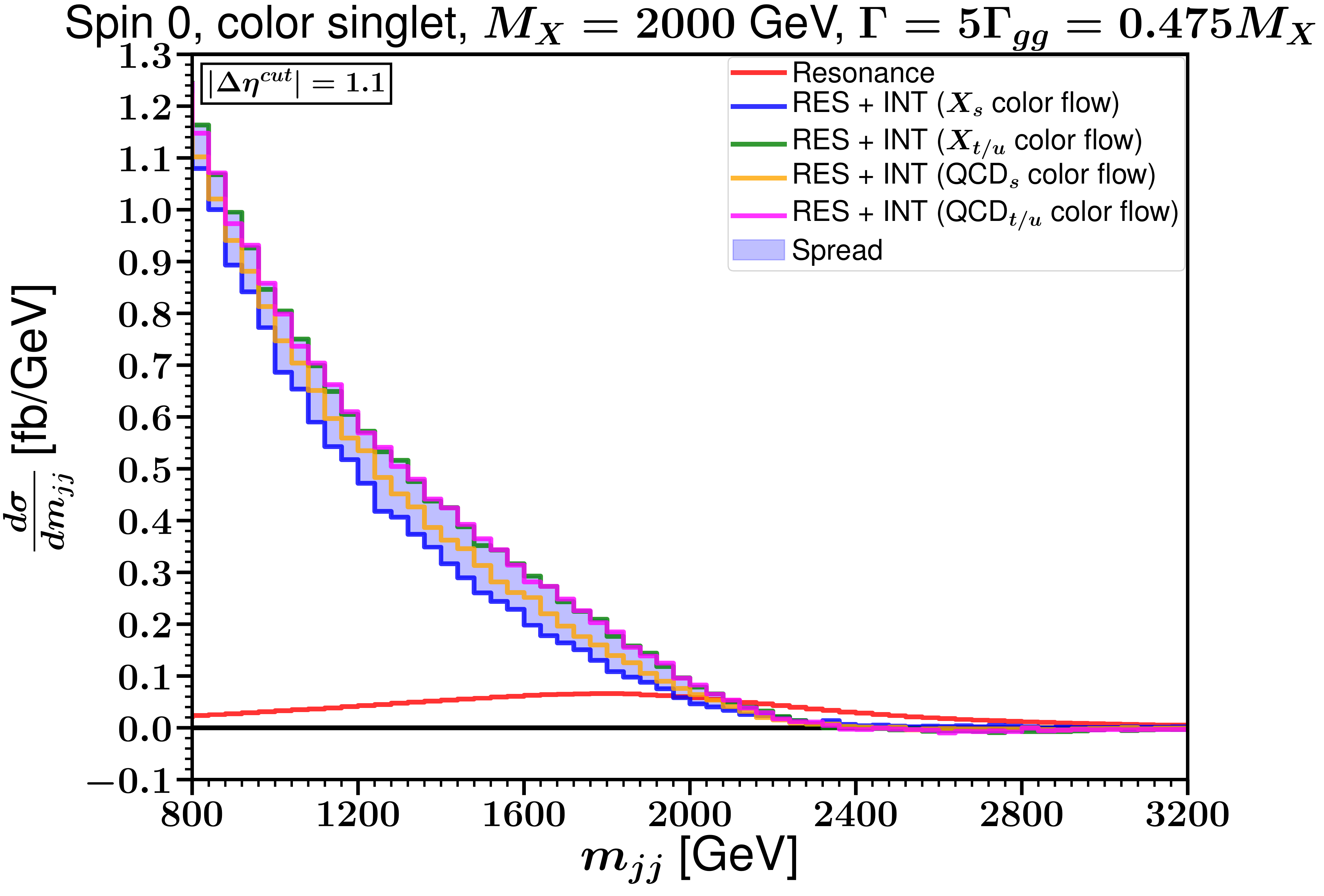}
  \end{minipage}
  
  \vspace{0.15cm}
    
  \begin{minipage}[]{0.495\linewidth}
    \includegraphics[width=8.0cm,angle=0]{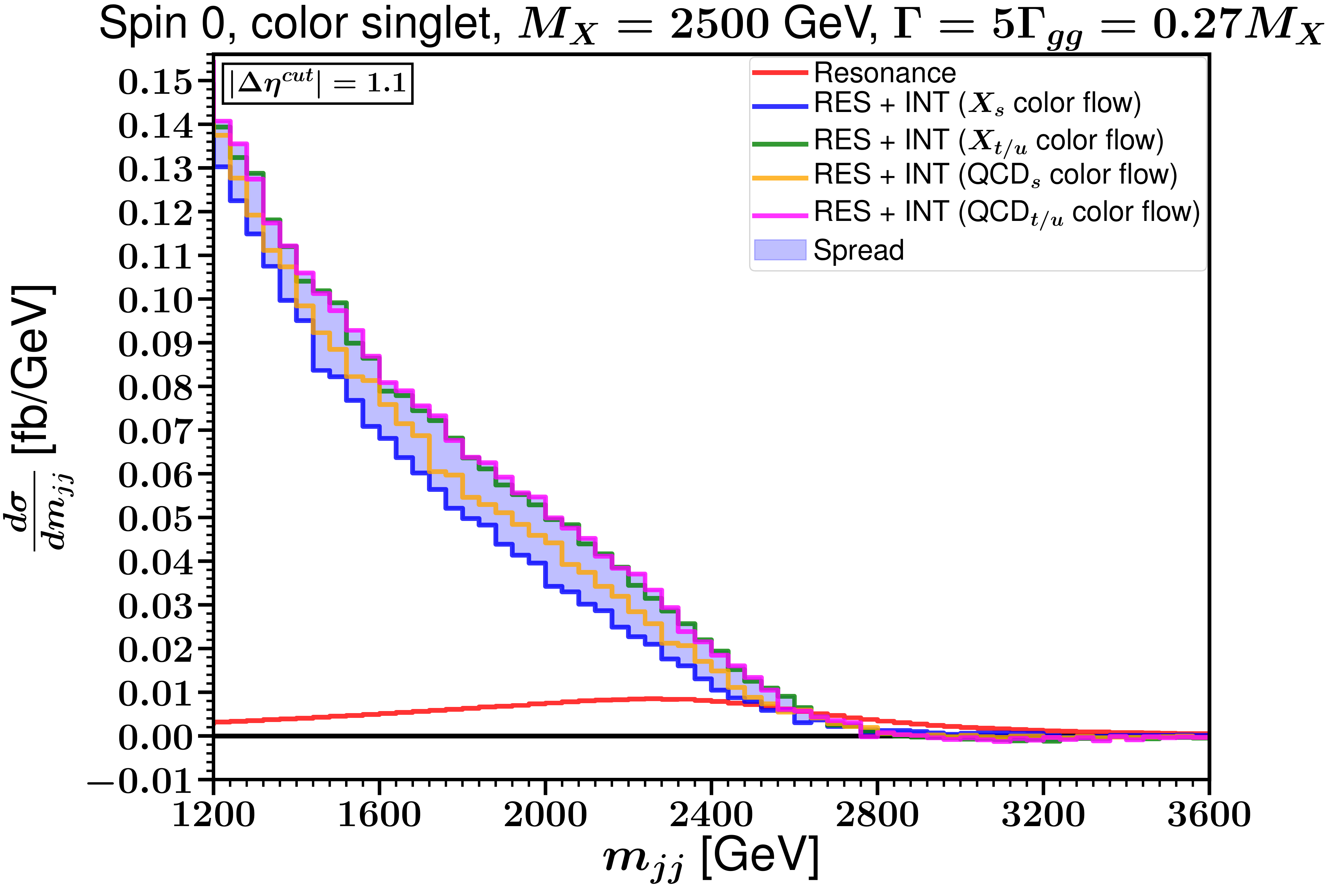}
  \end{minipage}
\begin{center}\begin{minipage}[]{0.95\linewidth}
\caption{\label{fig:gggg_s0c1_otherdecays} 
Dijet invariant mass distributions for the spin-0, color-singlet benchmarks of Table~\ref{tab:benchmarks} with $\Gamma_{gg}=\Gamma/5$, at the 13 TeV LHC, obtained with showering, hadronization and detector simulation. The red lines show the naive results with only the resonance diagrams of $g g \rightarrow X \rightarrow g g$ process (RES), which include the $s$-, $t$-, and $u$-channel exchanges of $X$, while the other four lines show the full results including interferences with the continuum QCD $g g \rightarrow g g$ amplitudes (INT) for all four color flows shown in Figure \ref{fig:colorflow}, as labeled. The shaded region shows the spread in the full result in each invariant mass bin for the different color flow choices. The cases shown in the right column of the top and the middle rows can be compared directly
to those in the right column of the previous Figure \ref{fig:gggg_s0c1_analytic_otherdecays} based on the more simplistic method of parton level with smearing.}
\end{minipage}\end{center}
\end{figure}

\begin{figure}[!tb]
  \begin{minipage}[]{0.495\linewidth}
    \includegraphics[width=8.0cm,angle=0]{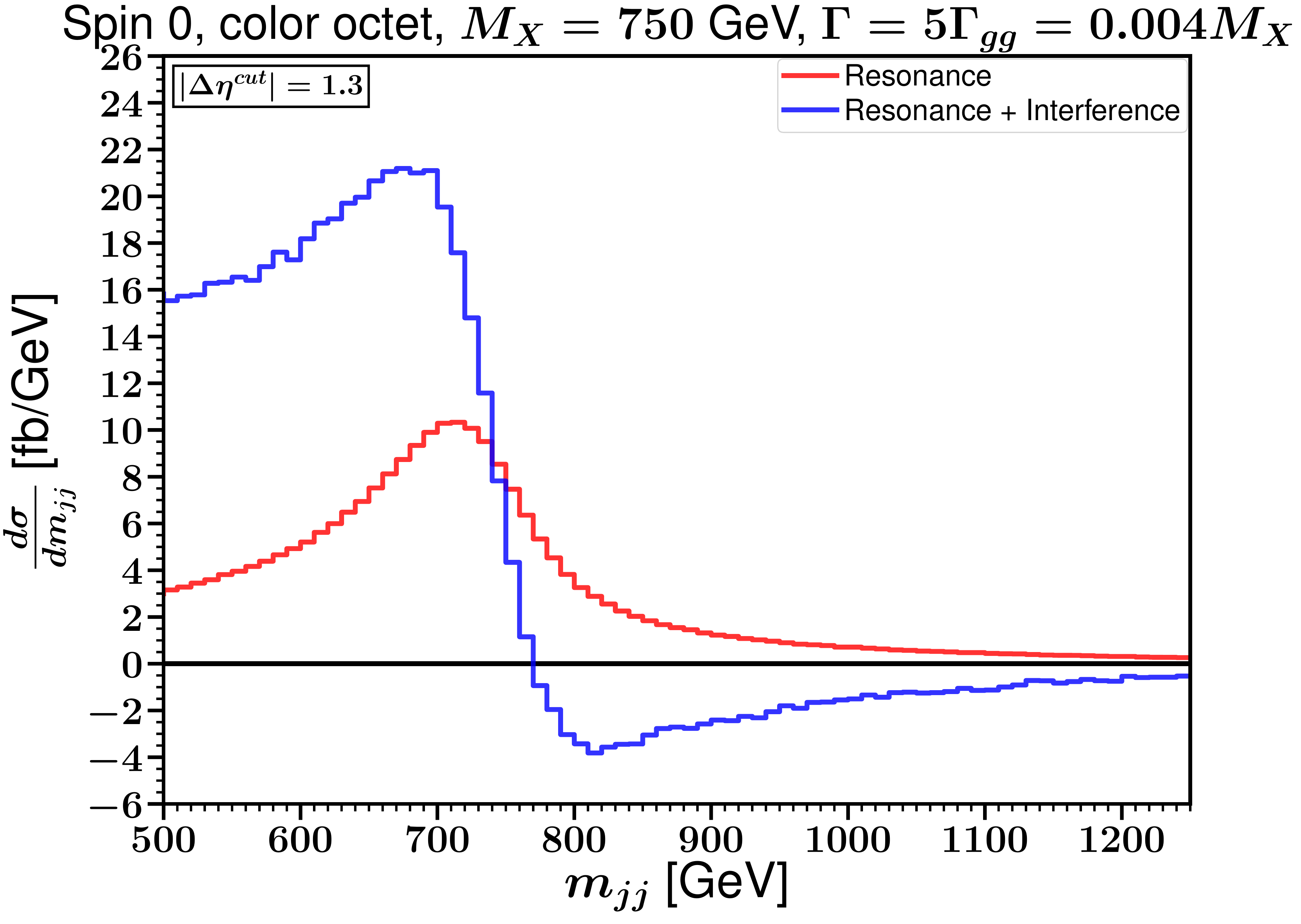}
  \end{minipage}
  \begin{minipage}[]{0.495\linewidth}
    \includegraphics[width=8.0cm,angle=0]{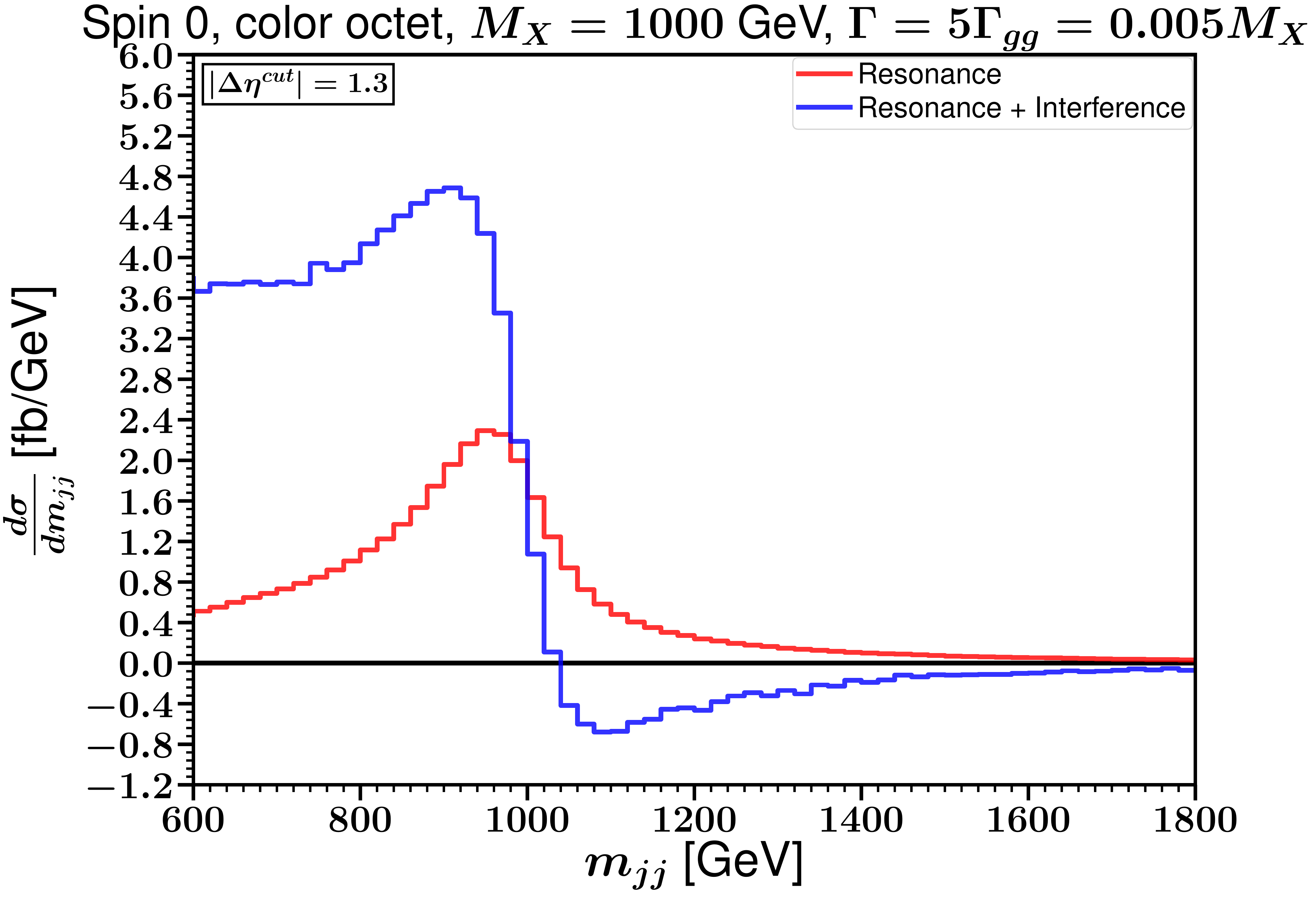}
  \end{minipage}

  \vspace{0.15cm}
    
  \begin{minipage}[]{0.495\linewidth}
    \includegraphics[width=8.0cm,angle=0]{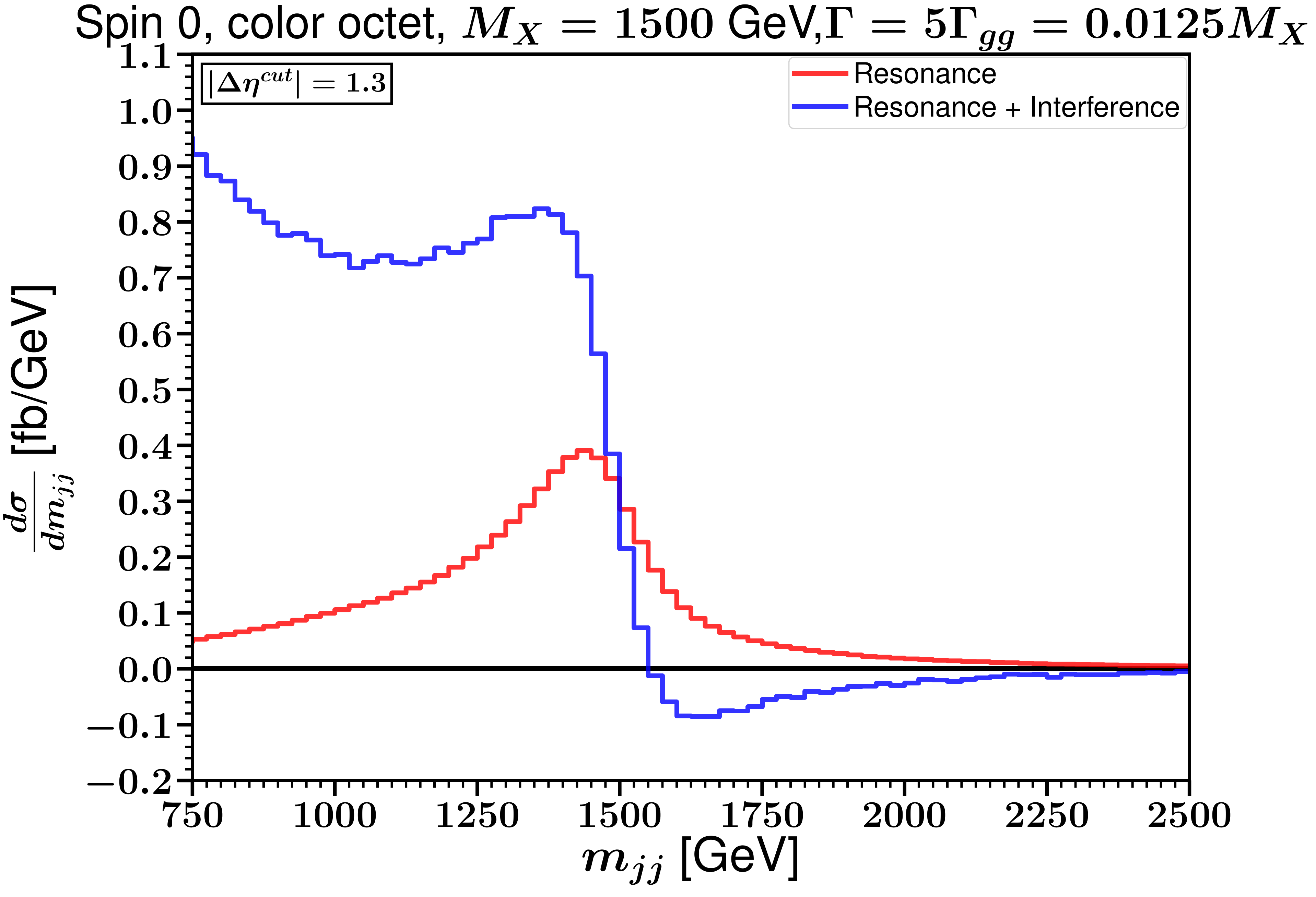}
  \end{minipage}
    \begin{minipage}[]{0.495\linewidth}
    \includegraphics[width=8.0cm,angle=0]{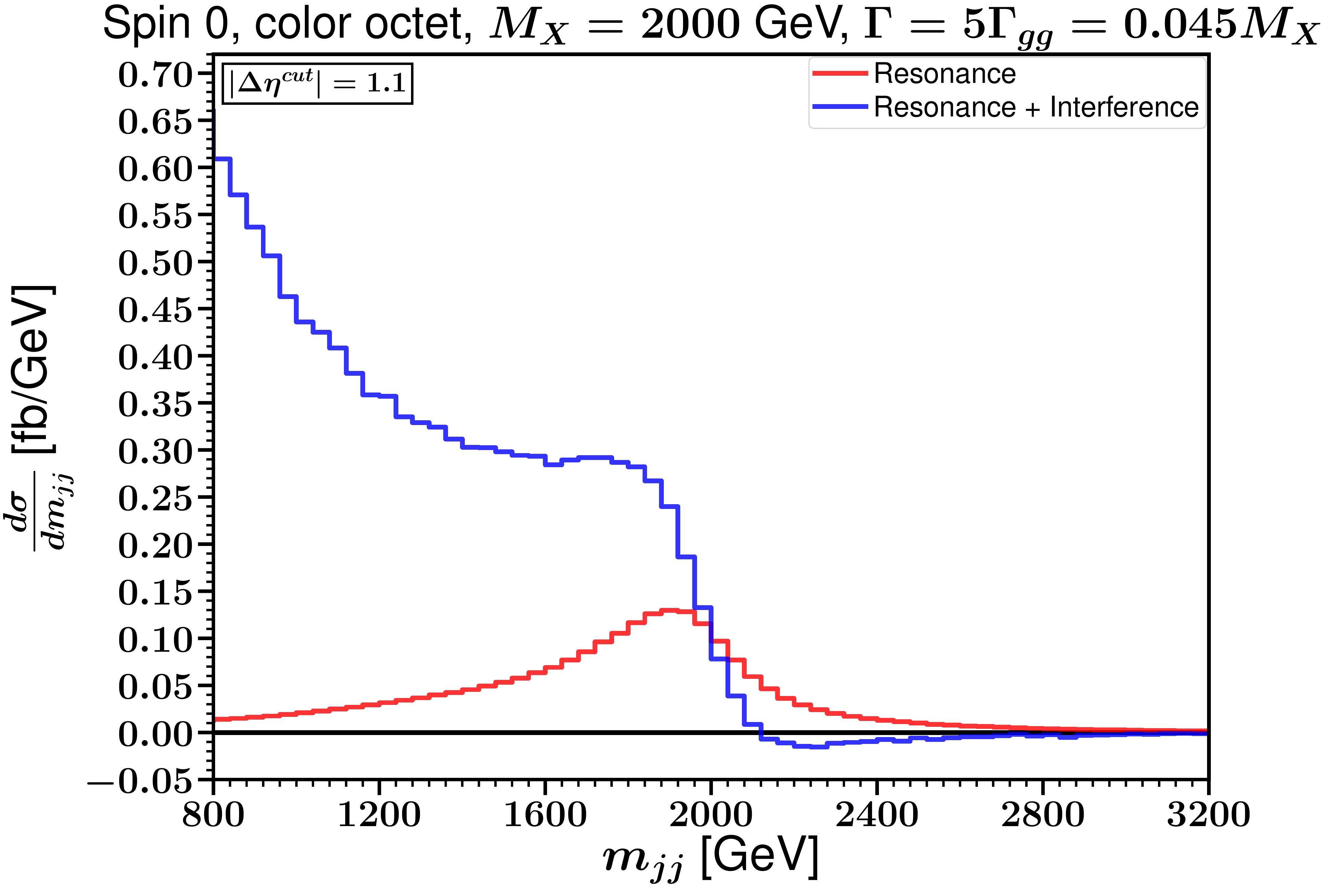}
  \end{minipage}

  \vspace{0.15cm}
    
  \begin{minipage}[]{0.495\linewidth}
    \includegraphics[width=8.0cm,angle=0]{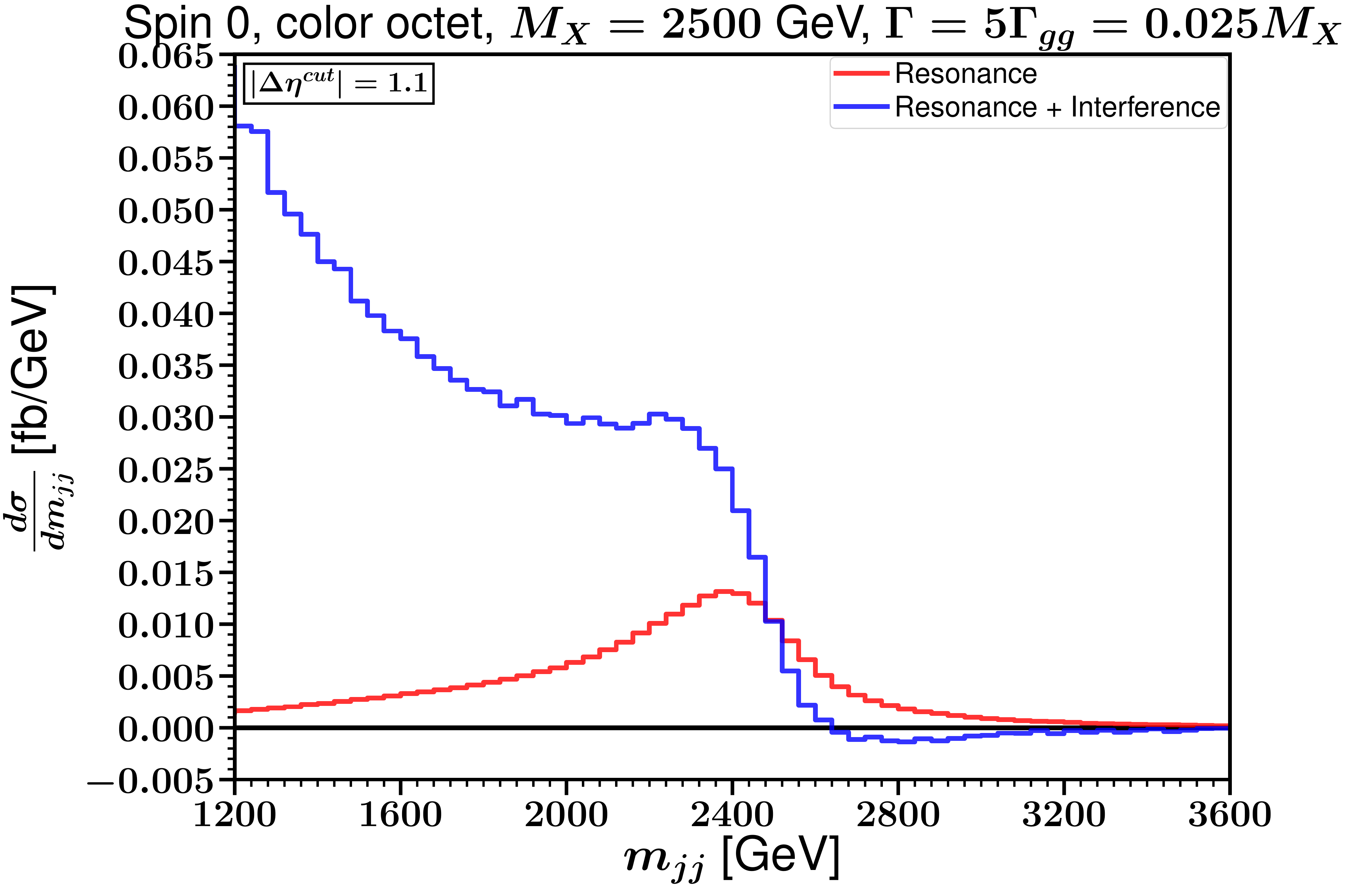}
  \end{minipage}
    \begin{minipage}[]{0.495\linewidth}
    \includegraphics[width=8.0cm,angle=0]{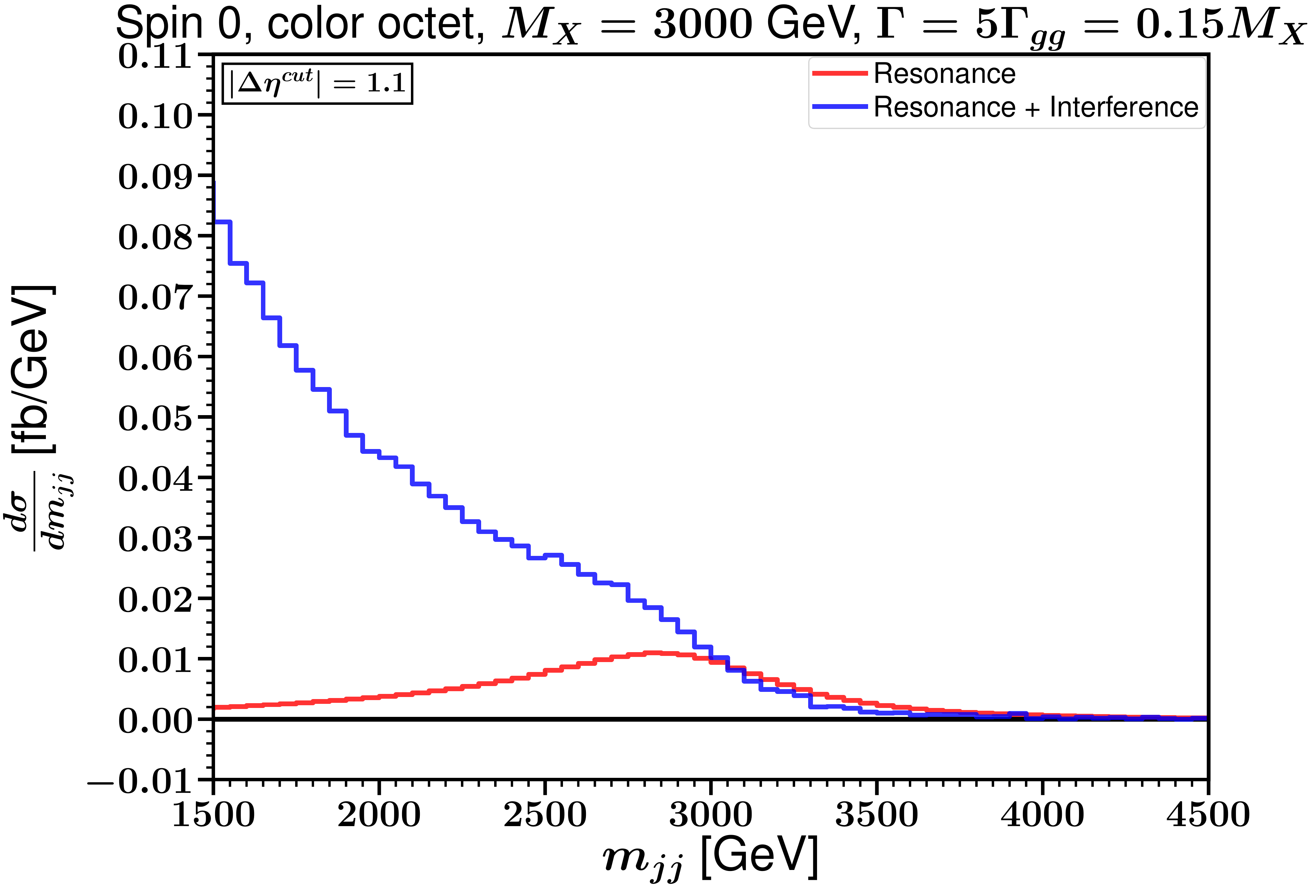}
  \end{minipage}
\begin{center}\begin{minipage}[]{0.95\linewidth}
\caption{\label{fig:gggg_s0c8_otherdecays} 
Dijet invariant mass distributions for the spin-0, color-octet benchmarks of Table~\ref{tab:benchmarks} with $\Gamma_{gg}=\Gamma/5$, at the 13 TeV LHC, obtained with showering, hadronization and detector simulation. The red lines show the naive results with only the resonance diagrams of $g g \rightarrow X \rightarrow g g$ process (RES), which include the $s$-, $t$-, and $u$-channel exchanges of $X$, while the blue lines show the full results including interferences with the continuum QCD $g g \rightarrow g g$ amplitudes (INT). The cases shown in the right column can be compared directly
to those in the right column of the previous Figure \ref{fig:gggg_s0c8_analytic_otherdecays} based on the more simplistic method of parton level with smearing.}
\end{minipage}\end{center}
\end{figure}

\begin{figure}[!tb]
  \begin{minipage}[]{0.495\linewidth}
    \includegraphics[width=8.0cm,angle=0]{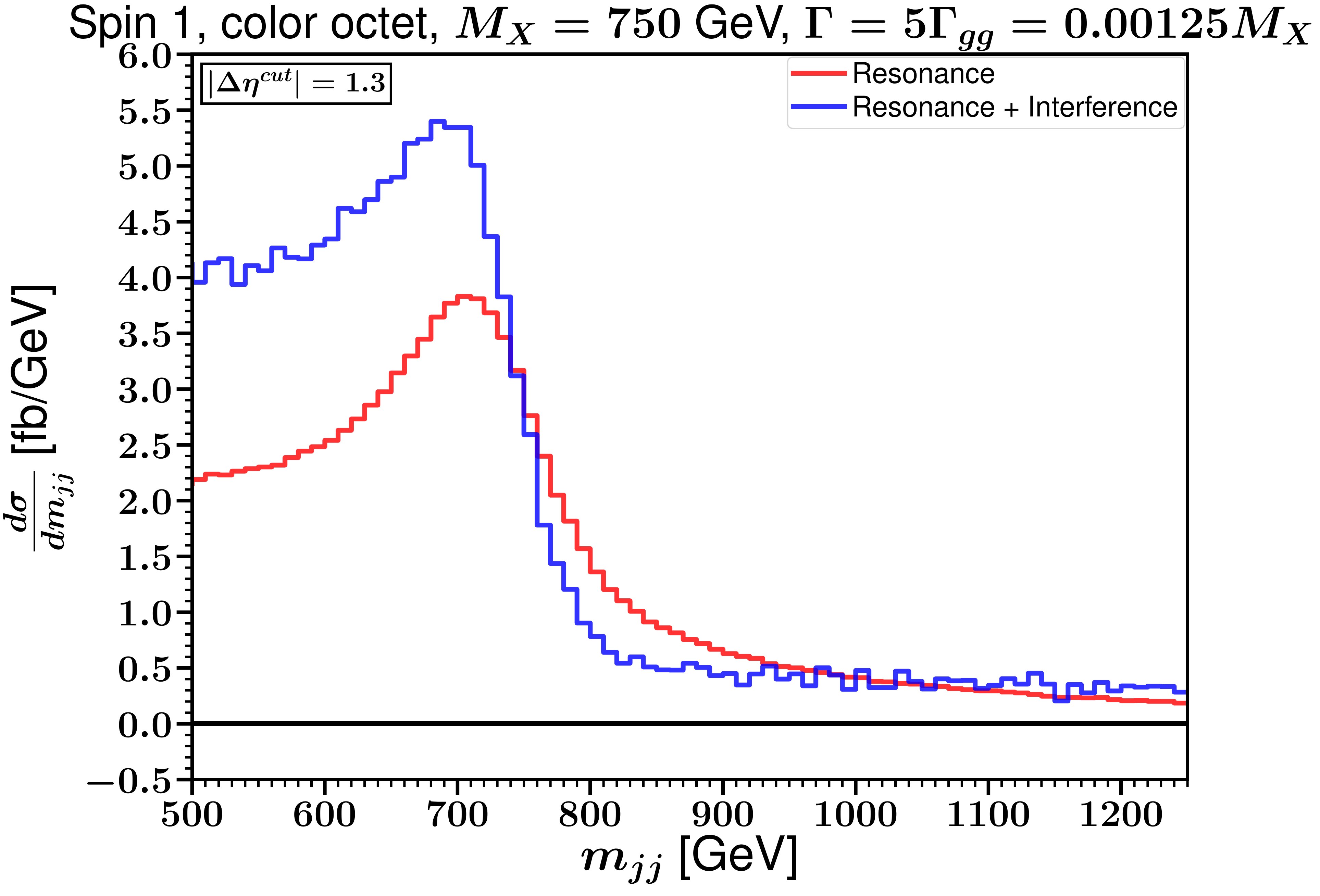}
  \end{minipage}
  \begin{minipage}[]{0.495\linewidth}
    \includegraphics[width=8.0cm,angle=0]{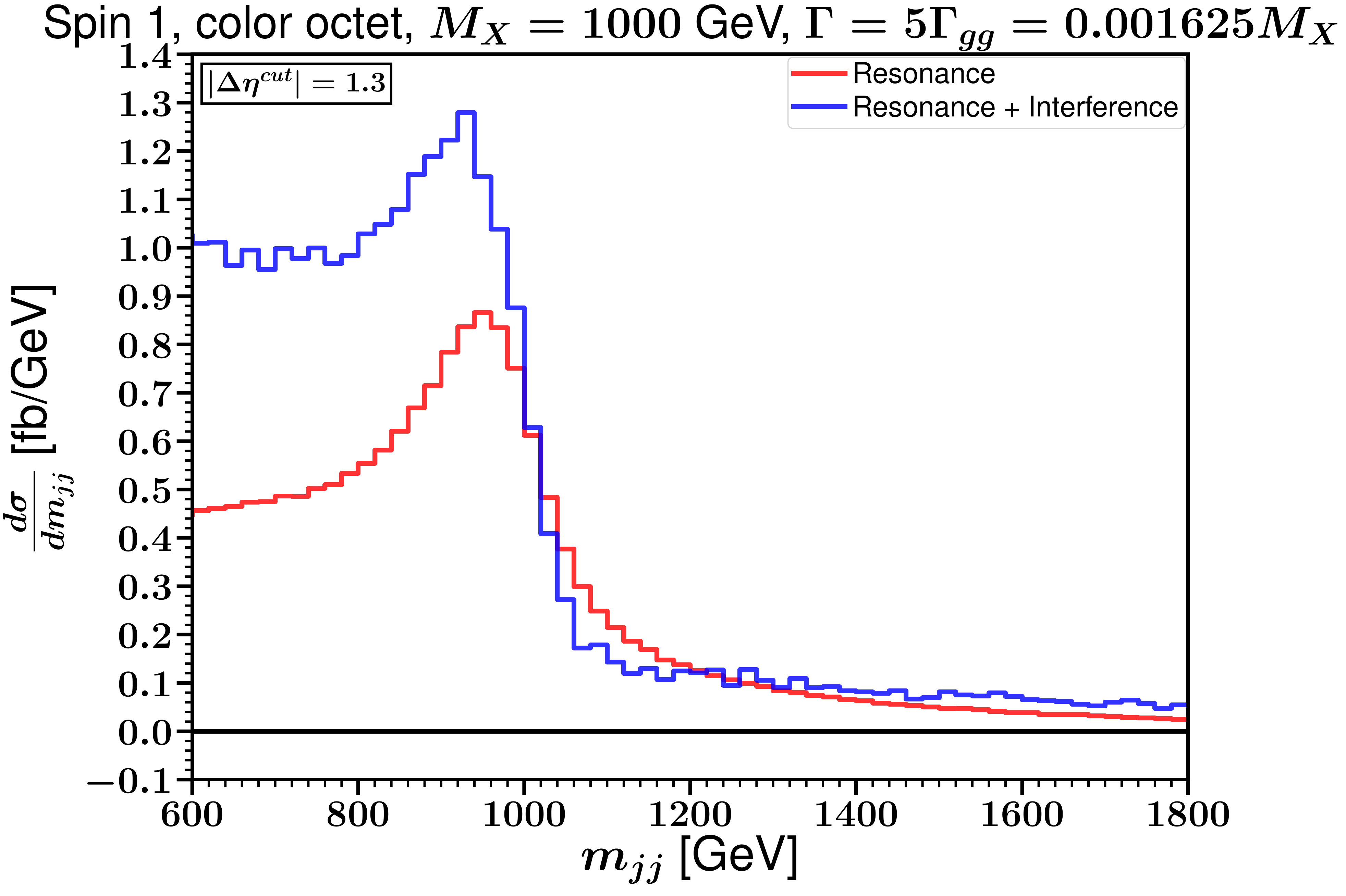}
  \end{minipage}

  \vspace{0.15cm}
    
  \begin{minipage}[]{0.495\linewidth}
    \includegraphics[width=8.0cm,angle=0]{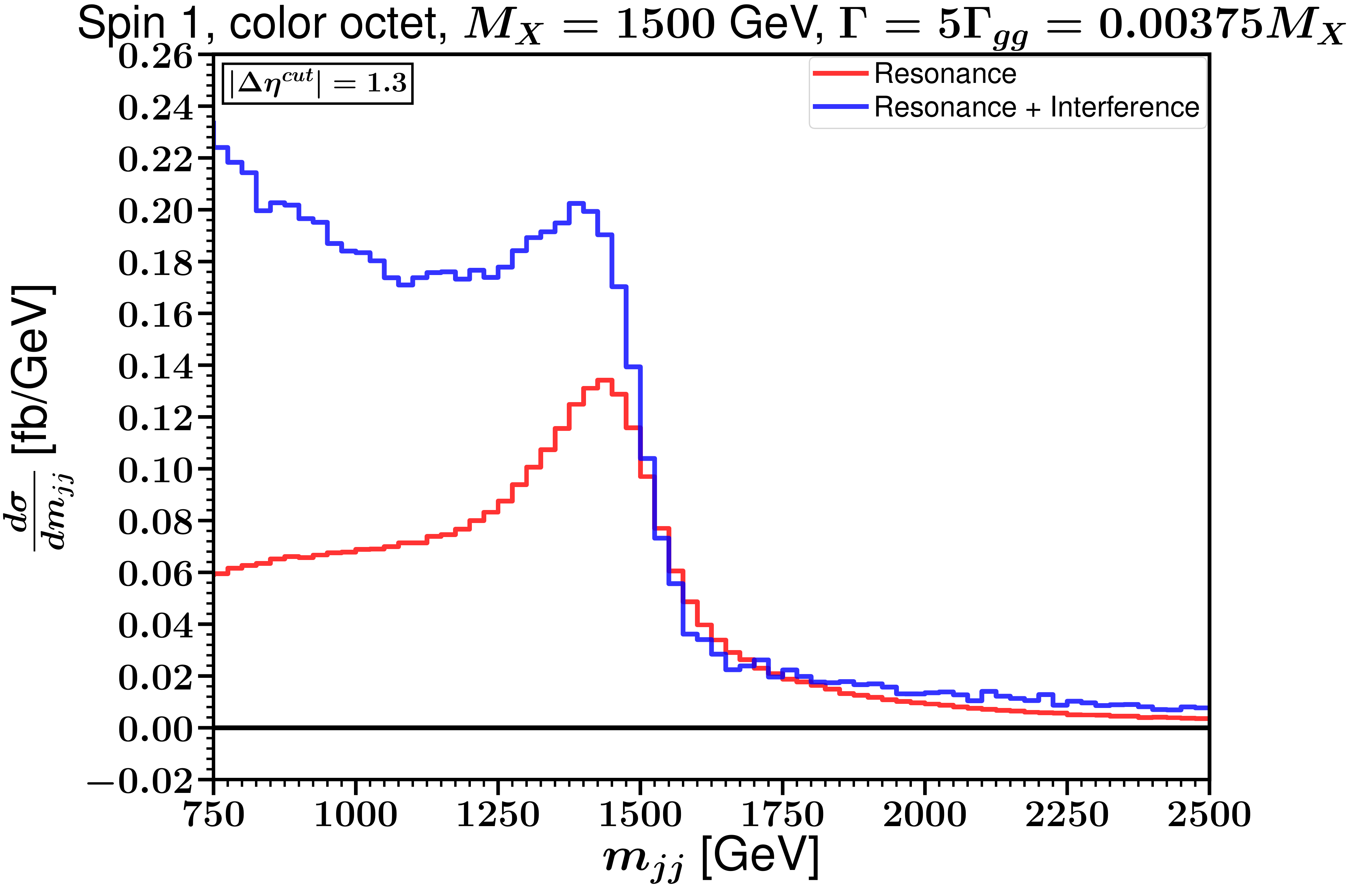}
  \end{minipage}
    \begin{minipage}[]{0.495\linewidth}
    \includegraphics[width=8.0cm,angle=0]{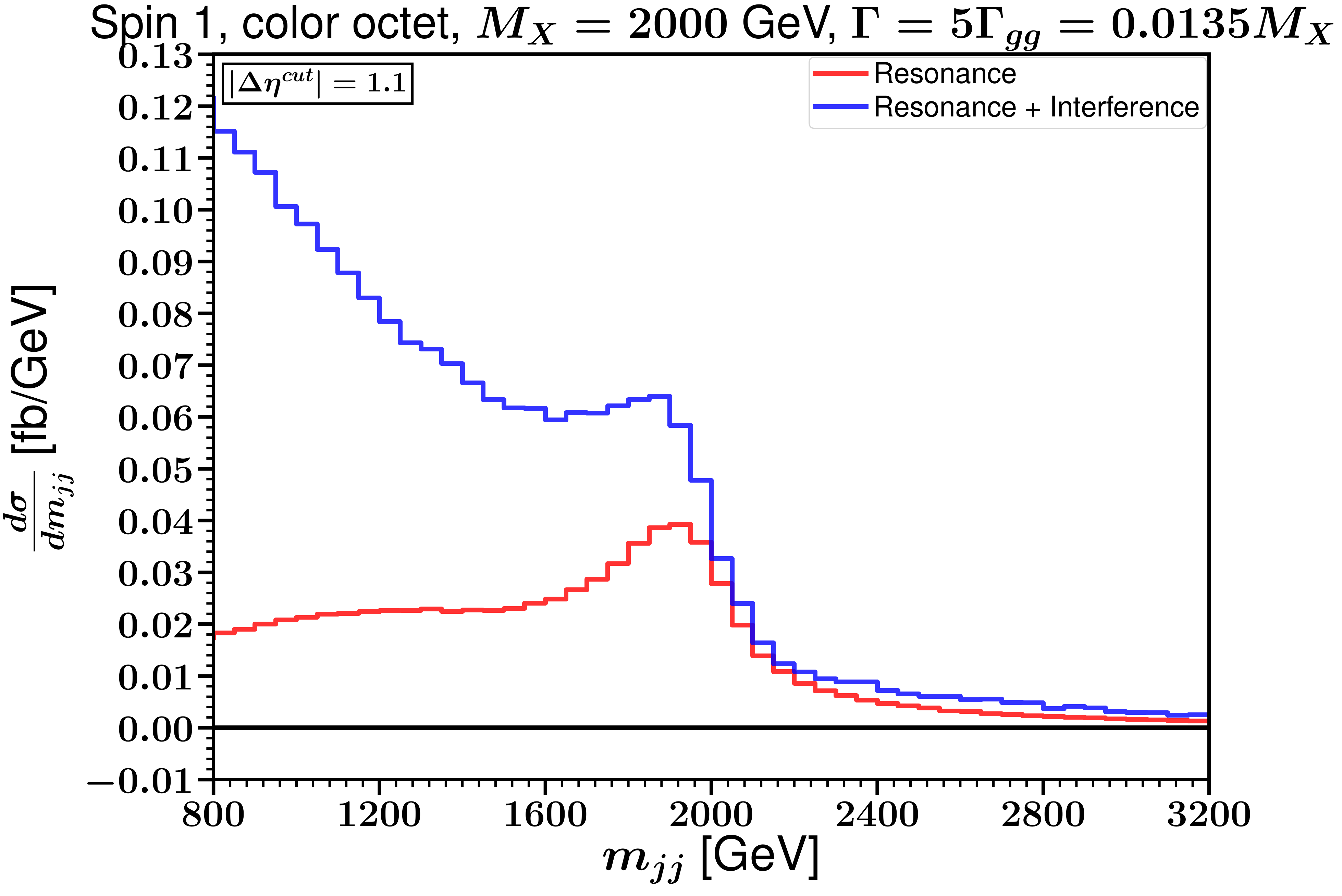}
  \end{minipage}

  \vspace{0.15cm}
    
  \begin{minipage}[]{0.495\linewidth}
    \includegraphics[width=8.0cm,angle=0]{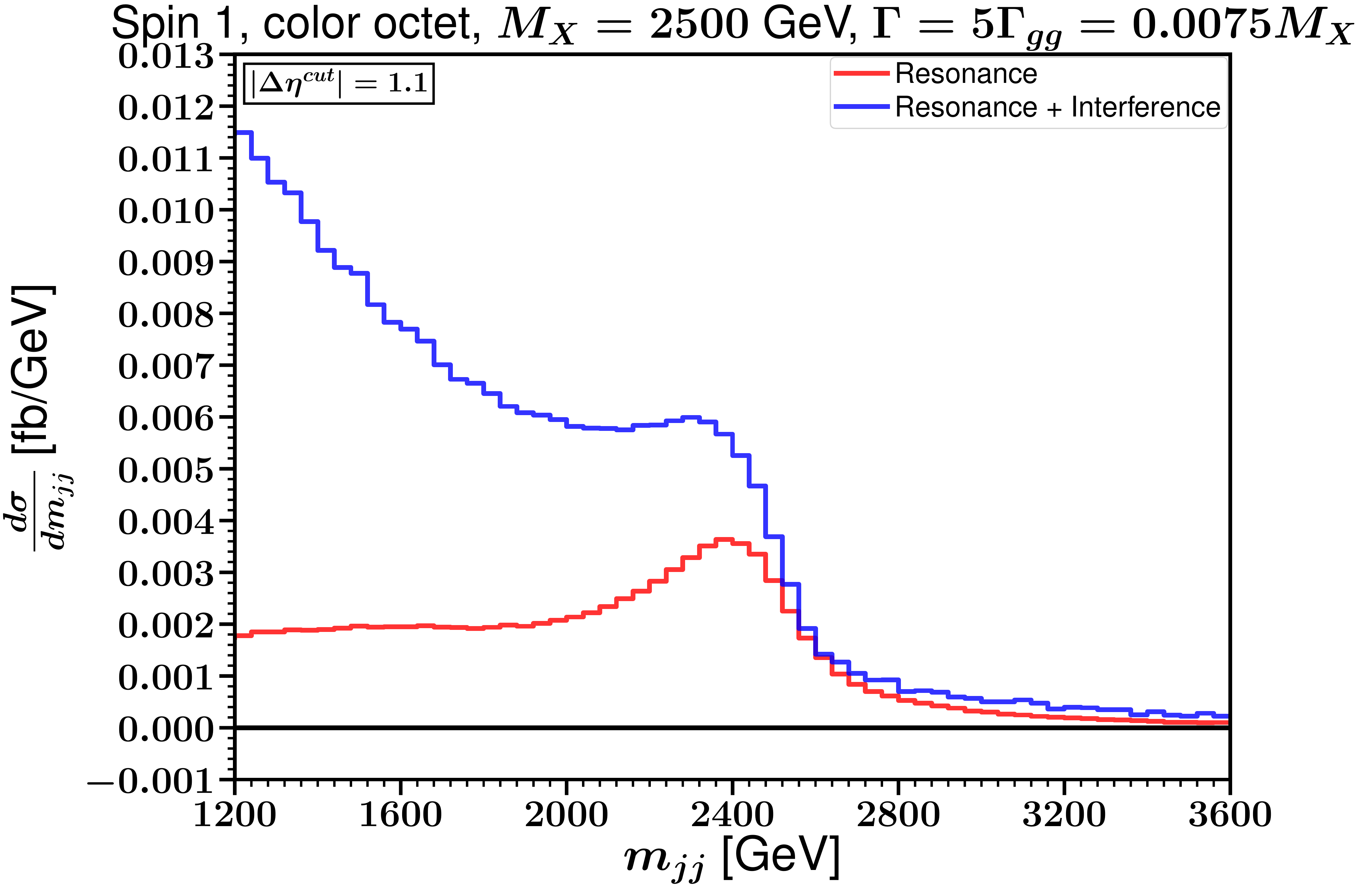}
  \end{minipage}
    \begin{minipage}[]{0.495\linewidth}
    \includegraphics[width=8.0cm,angle=0]{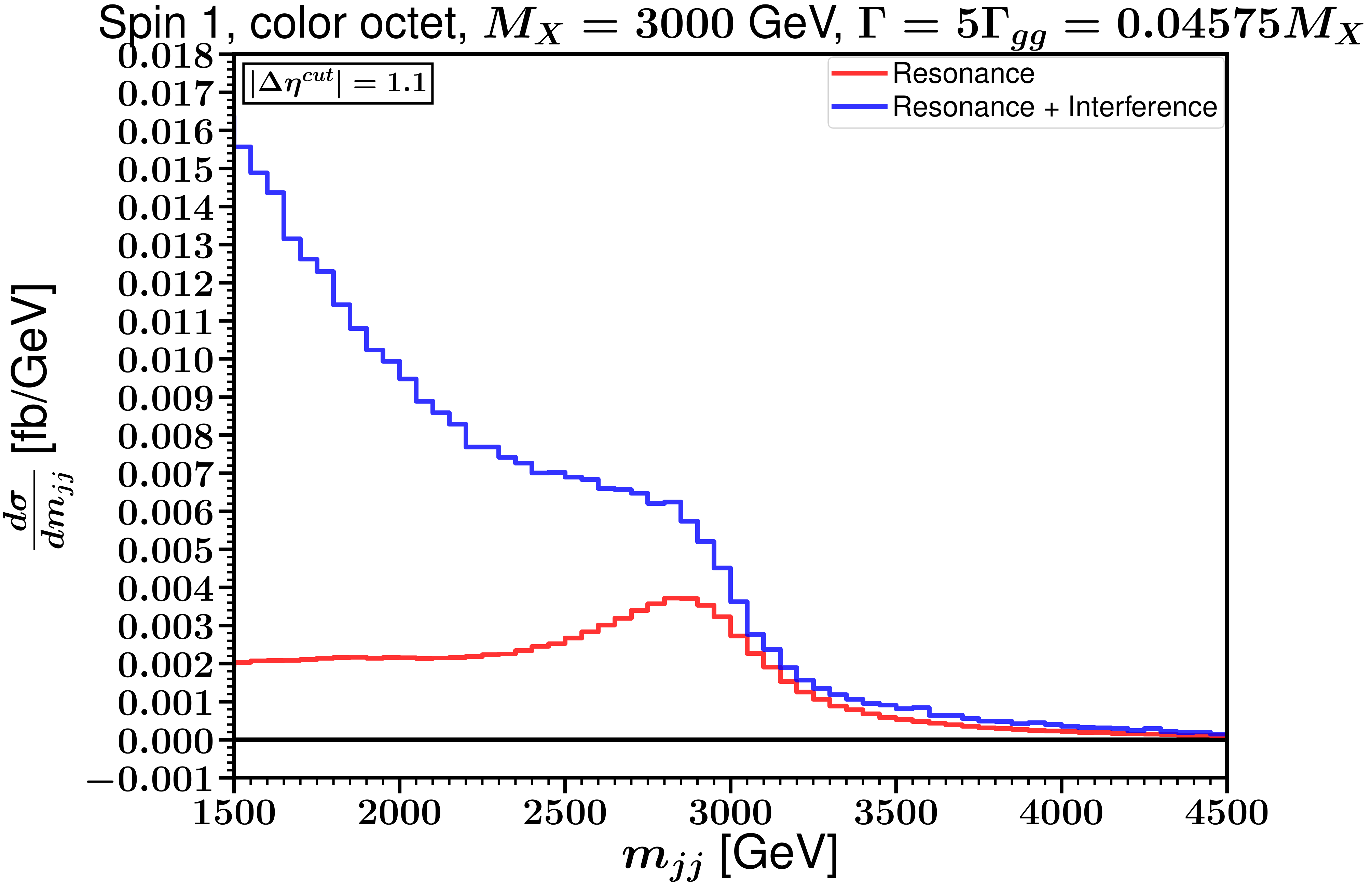}
  \end{minipage}
\begin{center}\begin{minipage}[]{0.95\linewidth}
\caption{\label{fig:gggg_s1c8_otherdecays} 
Dijet invariant mass distributions for the spin-1, color-octet benchmarks of Table~\ref{tab:benchmarks} with $\Gamma_{gg}=\Gamma/5$, at the 13 TeV LHC, obtained with showering, hadronization and detector simulation. The red lines show the naive results with only the resonance diagrams of $g g \rightarrow X \rightarrow g g$ process (RES), which include the $s$-, $t$-, and $u$-channel exchanges of $X$, while the blue lines show the full results including interferences with the continuum QCD $g g \rightarrow g g$ amplitudes (INT). 
The cases shown in the right column can be compared directly
to those in the right column of the previous Figure \ref{fig:gggg_s1c8_analytic_otherdecays} based on the more simplistic method of parton level with smearing.}
\end{minipage}\end{center}
\end{figure}

\begin{figure}[!tb]
  \begin{minipage}[]{0.495\linewidth}
    \includegraphics[width=8.0cm,angle=0]{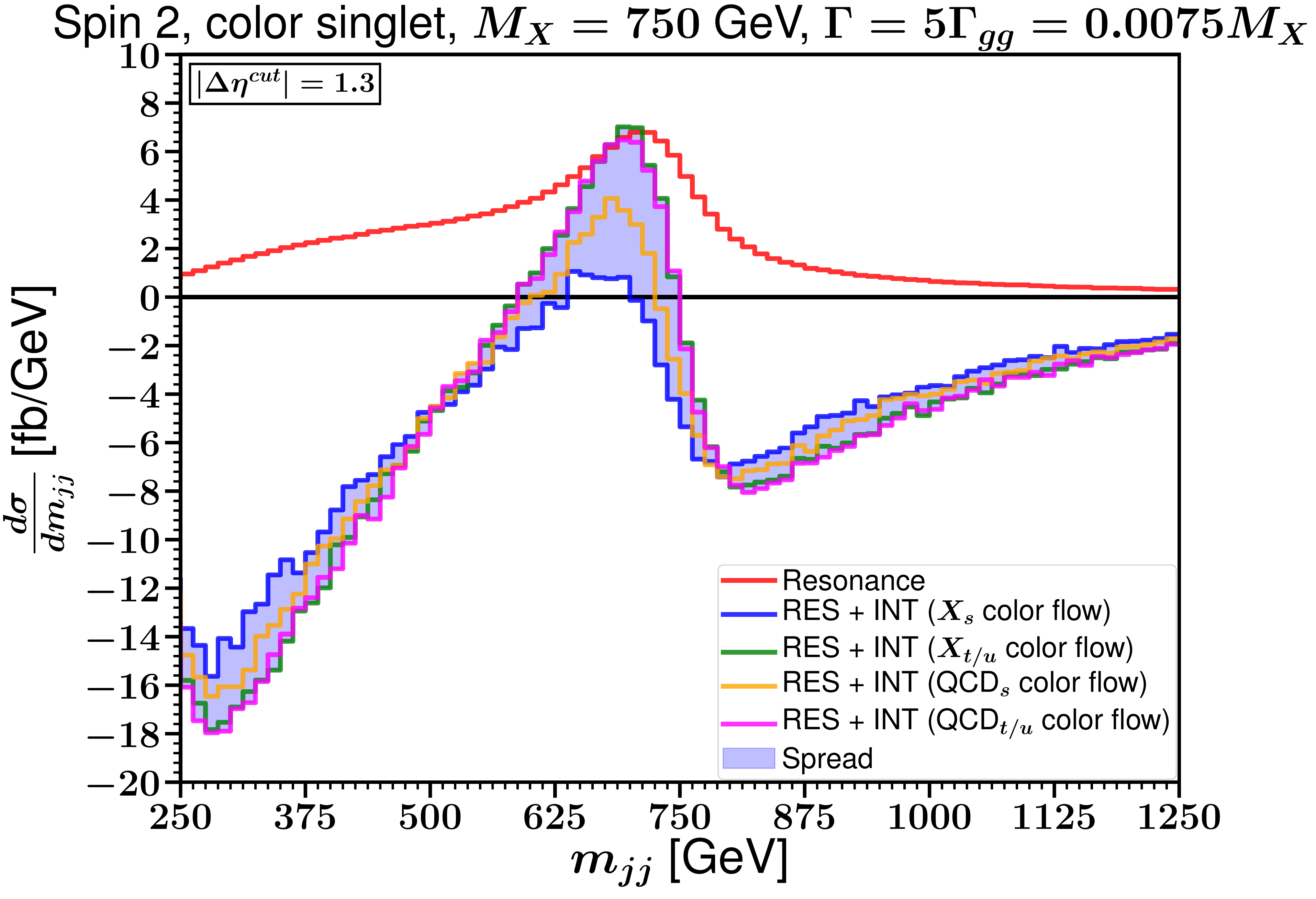}
  \end{minipage}
  \begin{minipage}[]{0.495\linewidth}
    \includegraphics[width=8.0cm,angle=0]{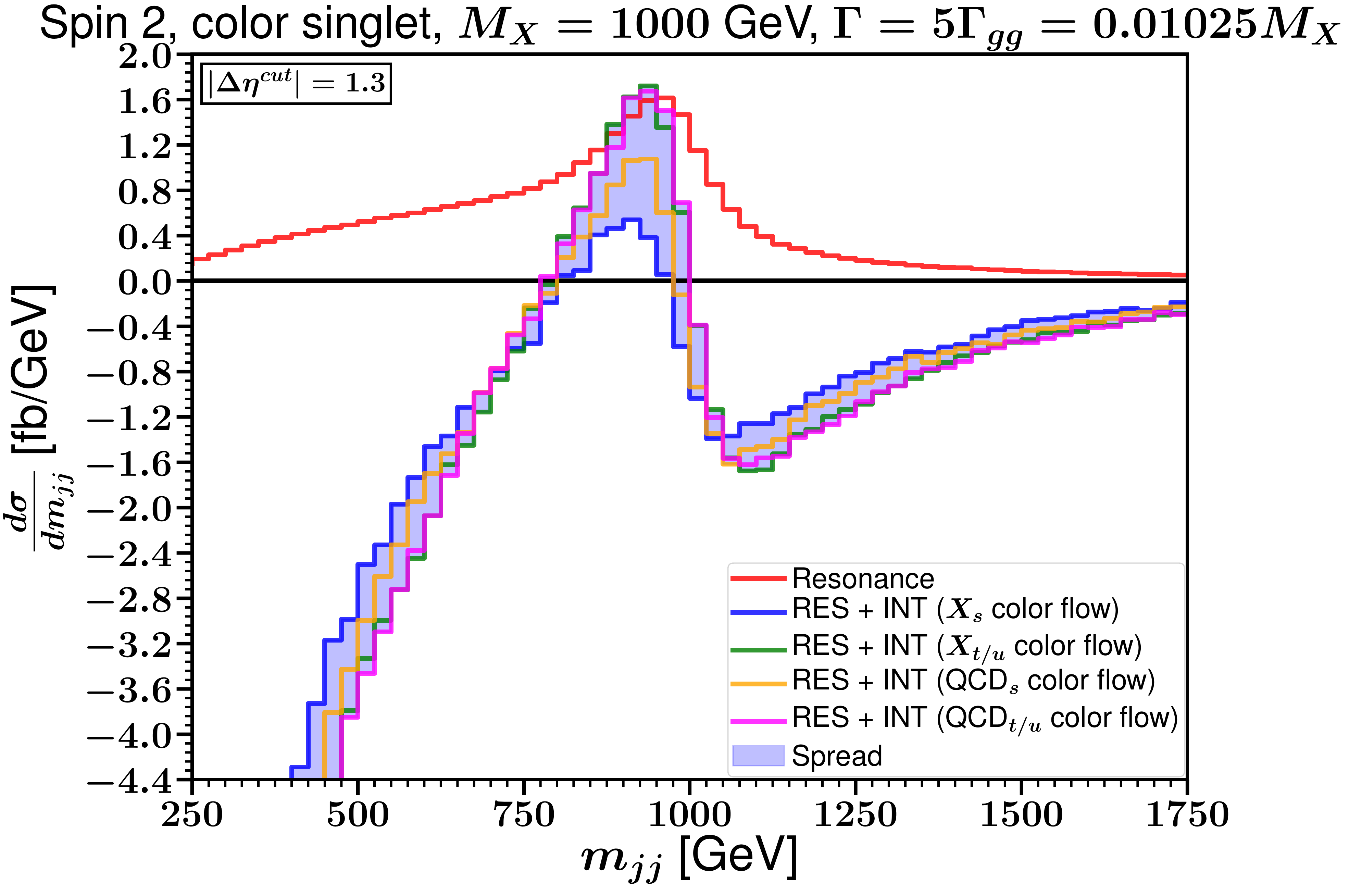}
  \end{minipage}

  \vspace{0.15cm}
    
  \begin{minipage}[]{0.495\linewidth}
    \includegraphics[width=8.0cm,angle=0]{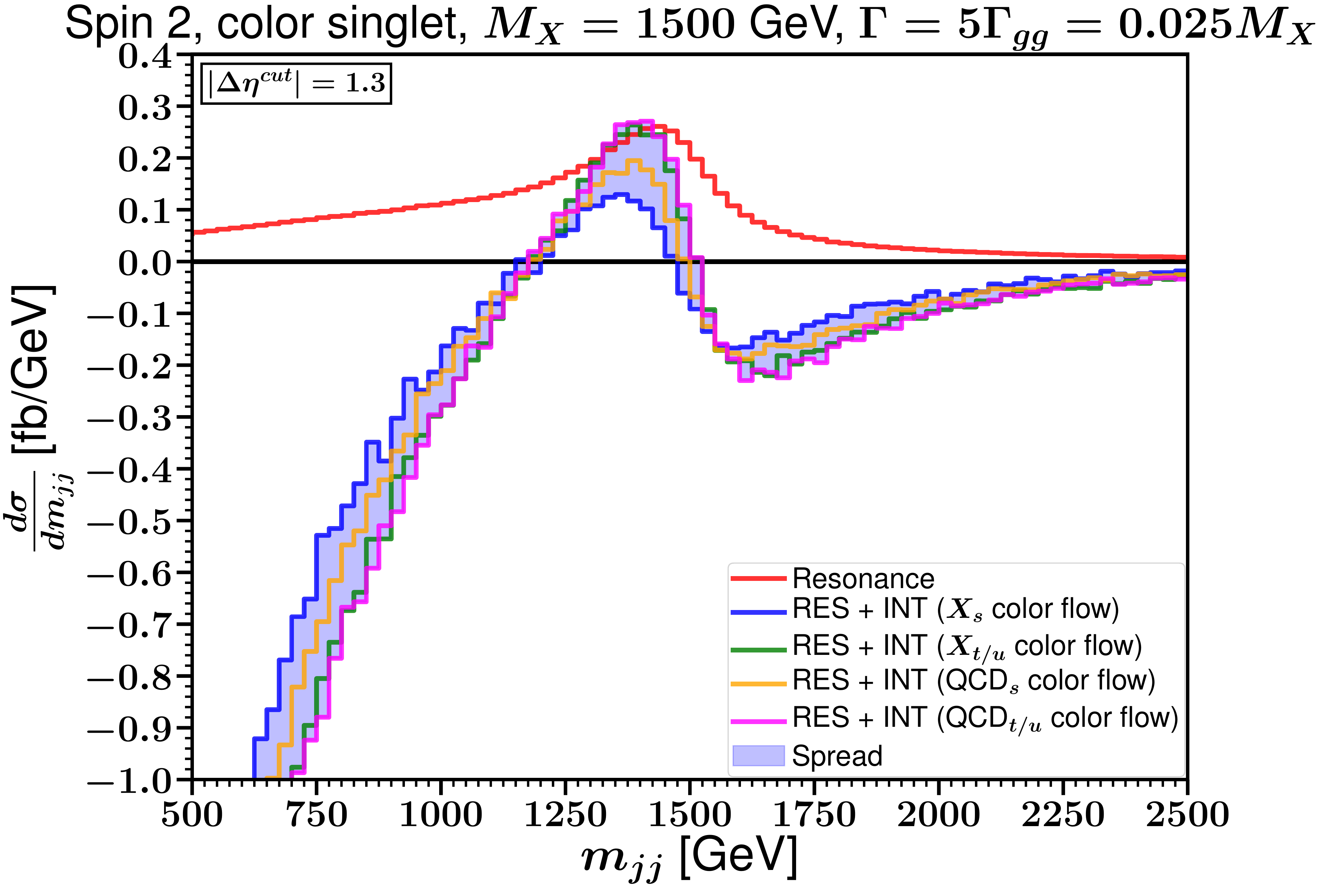}
  \end{minipage}
    \begin{minipage}[]{0.495\linewidth}
    \includegraphics[width=8.0cm,angle=0]{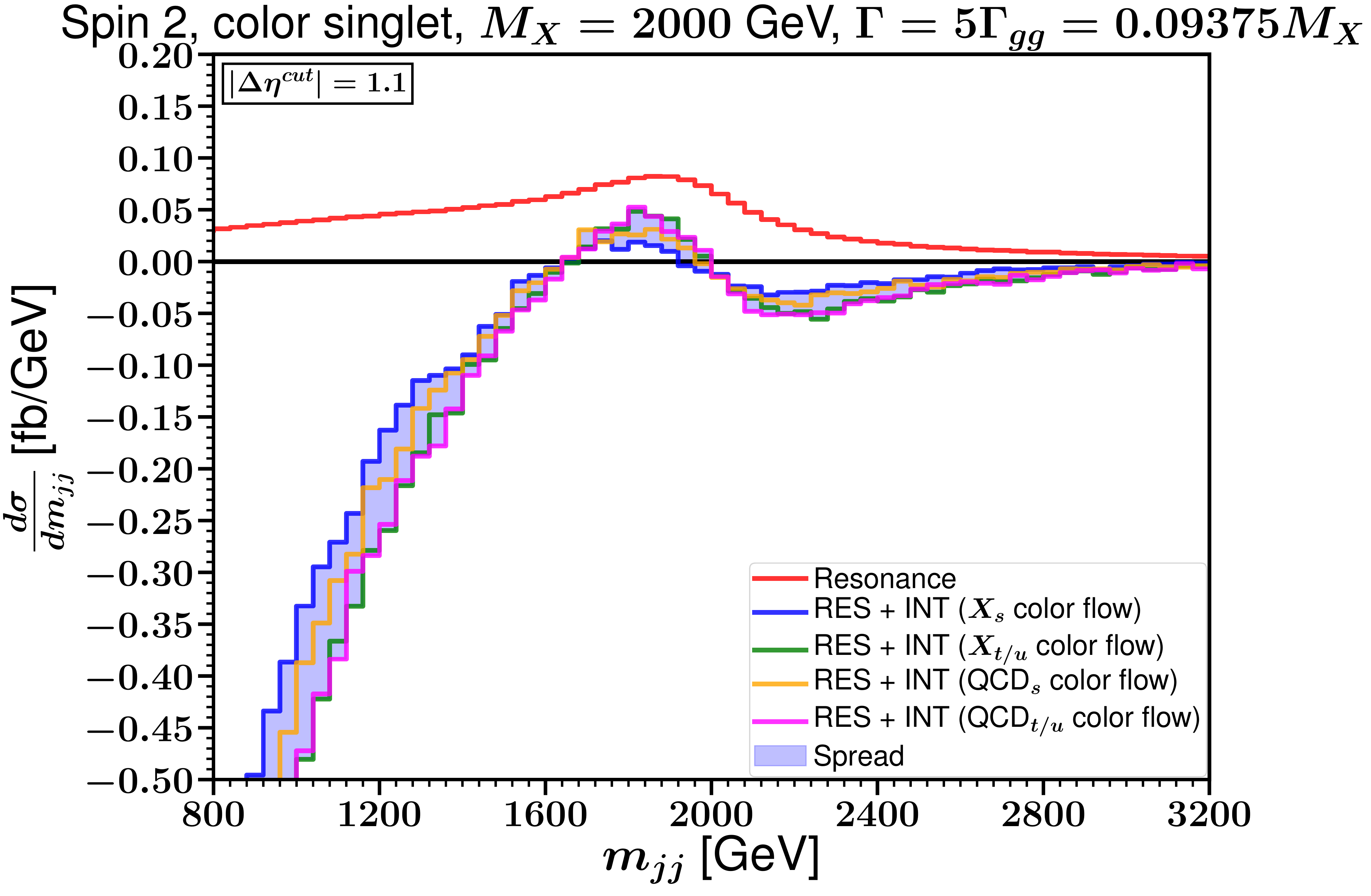}
  \end{minipage}

  \vspace{0.15cm}
    
  \begin{minipage}[]{0.495\linewidth}
    \includegraphics[width=8.0cm,angle=0]{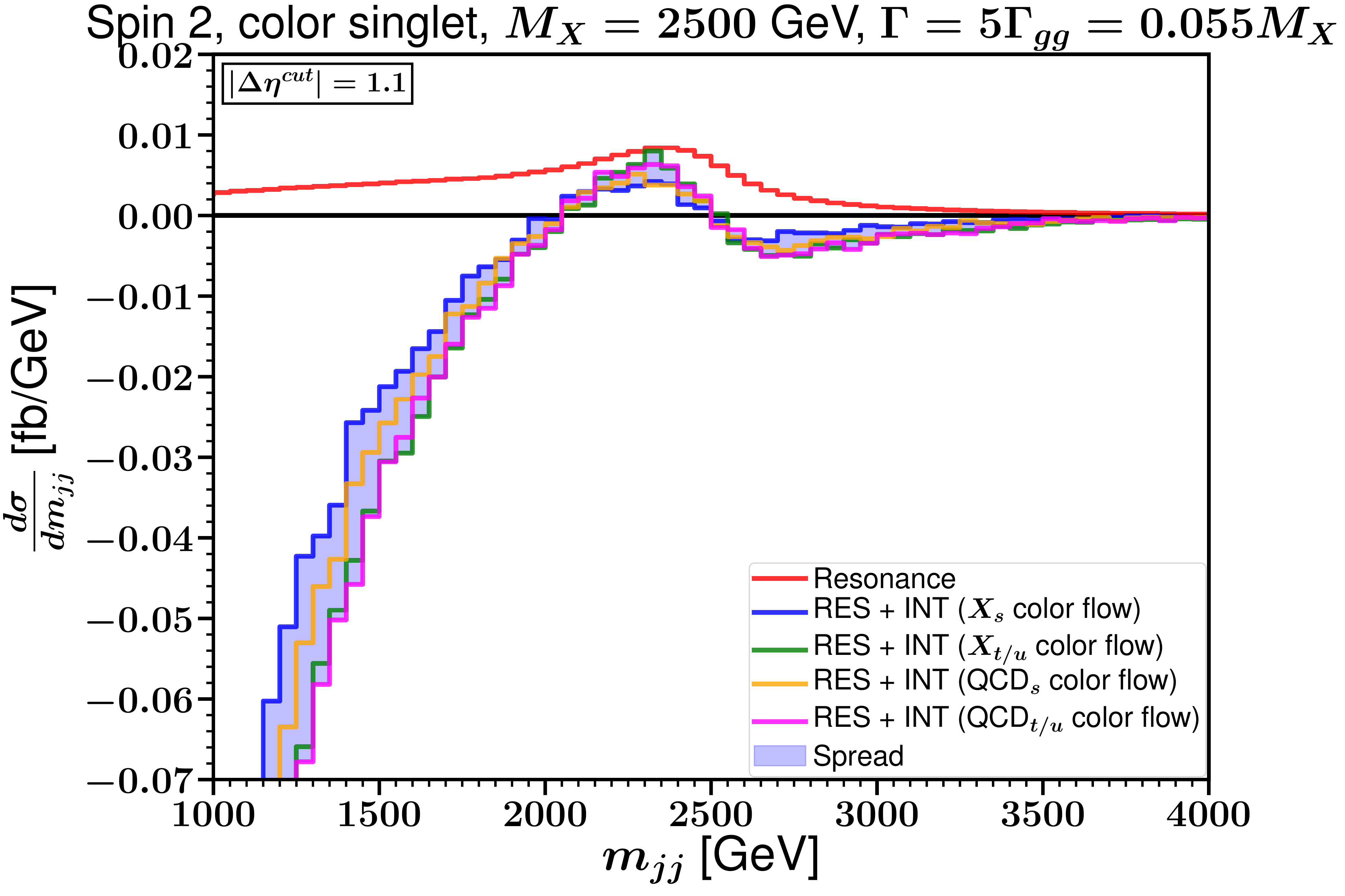}
  \end{minipage}
    \begin{minipage}[]{0.495\linewidth}
    \includegraphics[width=8.0cm,angle=0]{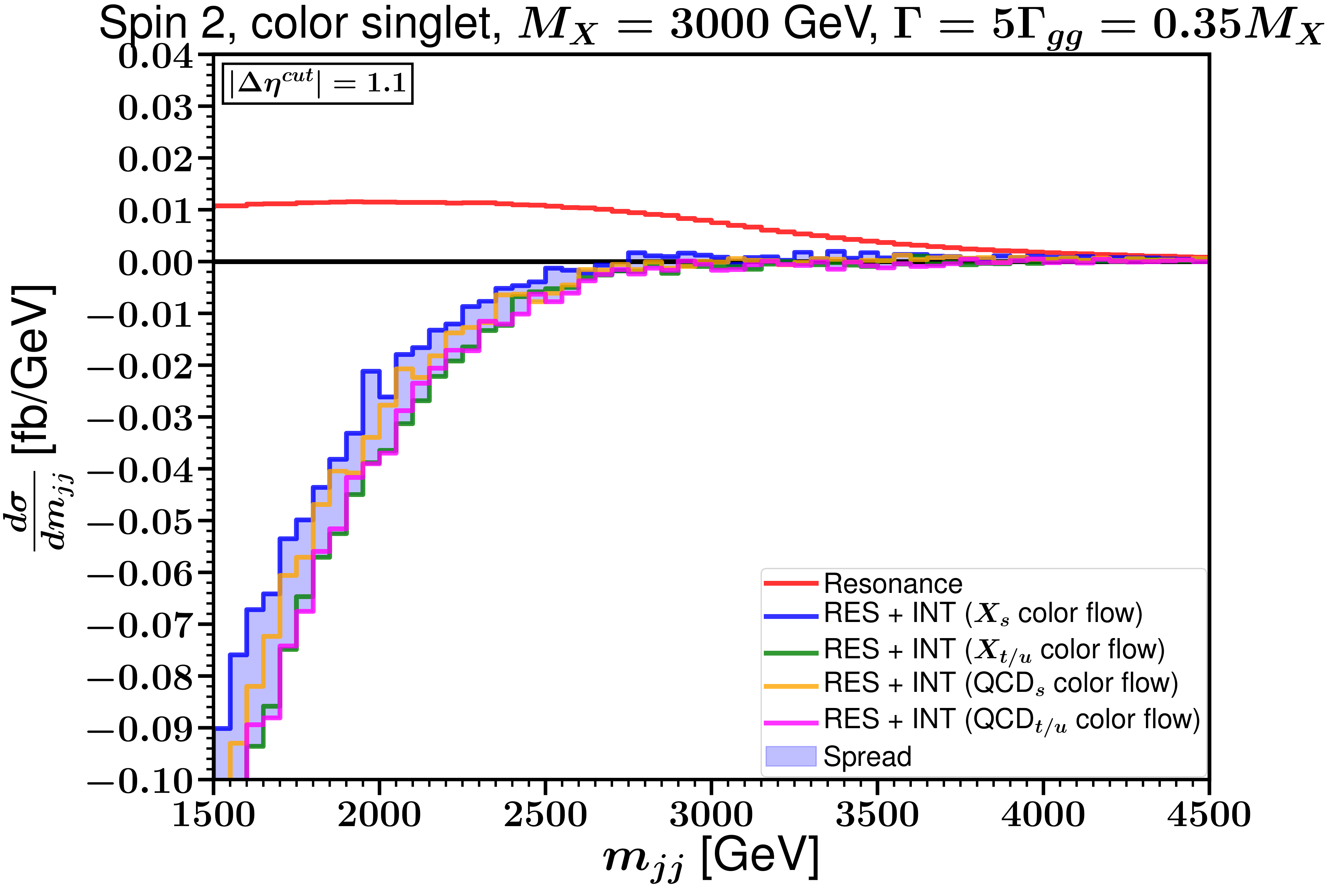}
  \end{minipage}
\begin{center}\begin{minipage}[]{0.95\linewidth}
\caption{\label{fig:gggg_s2c1_otherdecays}
Dijet invariant mass distributions for the spin-2, color-singlet benchmarks of Table~\ref{tab:benchmarks} with $\Gamma_{gg}=\Gamma/5$, at the 13 TeV LHC, obtained with showering, hadronization and detector simulation. The red lines show the naive results with only the resonance diagrams of $g g \rightarrow X \rightarrow g g$ process (RES), which include the $s$-, $t$-, and $u$-channel exchanges of $X$, while the other four colored lines show the full results including interferences with the continuum QCD $g g \rightarrow g g$ amplitudes (INT) for all four color flows shown in Figure \ref{fig:colorflow}, as labeled. The shaded region shows the spread in the full result in each invariant mass bin for the different color flow choices. 
The negative tails at small invariant mass come from the interference between the QCD amplitudes and the $t$- and $u$-channel $X$ exchange diagrams.
The cases shown in the right column can be compared directly
to those in the right column of the previous Figure \ref{fig:gggg_s2c1_analytic_otherdecays} based on the more simplistic method of parton level with smearing.}
\end{minipage}\end{center}
\end{figure}

In Figures \ref{fig:gggg_s0c1_otherdecays}, 
\ref{fig:gggg_s0c8_otherdecays}, 
\ref{fig:gggg_s1c8_otherdecays}, and 
\ref{fig:gggg_s2c1_otherdecays}, we show the results obtained by our full event simulation including showering,
hadronization, and detector simulation. The right columns of pairs of figures (\ref{fig:gggg_s0c1_analytic_otherdecays} and \ref{fig:gggg_s0c1_otherdecays} for spin-0 color-singlet,
\ref{fig:gggg_s0c8_analytic_otherdecays} and \ref{fig:gggg_s0c8_otherdecays} for spin-0 color-octet,
\ref{fig:gggg_s1c8_analytic_otherdecays} and \ref{fig:gggg_s1c8_otherdecays} for spin-1 color-octet,
and
\ref{fig:gggg_s2c1_analytic_otherdecays} and \ref{fig:gggg_s2c1_otherdecays} for spin-2 color-singlet)
can be directly compared, as they feature the same masses and widths. The agreement
between these sets of figures appears to be good at the qualitative level, but with differing yields at up to the level of tens of percent, and somewhat different shapes in some cases. In particular, for spin 1,
the full event simulation produces larger low-mass tails than the smeared parton-level results.
In addition, the color-flow choice in the color-singlet interference cases is seen to be a non-trivial effect.

A general feature that can be seen in all of these figures is that when $\Gamma_{gg} = \Gamma/5$, the  distributions including interference can bear little resemblance to the naive resonance-only results.
In each of the spin-0 and spin-1 cases, there is a very large low-mass tail from the interference. For lower $M_X$ and spin-0, we find a pronounced dip for $m_{jj}$ above $M_X$, but this tends to be washed out
for larger $M_X$ and higher spin.
In practice, the falling distribution well below $M_X$
will be partly absorbed into the QCD background fit, 
and, if present, can therefore have a significant effect on that fit. The result could be a peak/dip
or dip shape in the residual fit.
We note that
the magnitude of the effective dip after background fitting 
could easily be as large or larger than the peak that would be naively expected from the resonance if one
ignored the interference effect. We also emphasize that the low-mass tail from interference 
is much larger than the width effect in the pure resonant contribution (visible as the broad distribution
of the red curves). The latter effect has been considered in experimental searches \cite{CMSdijet2018,CMSdijet2019} with $\Gamma > \Gamma_{gg}$, but the much larger former effect has not.

In the spin-2 color-singlet case (see Figure \ref{fig:gggg_s2c1_otherdecays}), the distribution
shape tends to feature a
large deficit at low masses, a peak just below $M_X$, and then another deficit above $M_X$. After fitting
the QCD background, this should lead to an enhancement of the peak compared to the naive resonance-only
distribution, so we expect that the actual limits attainable would likely be stronger than those inferred
without considering interference.




\clearpage


\clearpage

\section{Outlook\label{sec:outlook}}
\setcounter{equation}{0}
\setcounter{figure}{0}
\setcounter{table}{0}
\setcounter{footnote}{1}

In this paper, we studied the importance of the interference between the digluon resonant signal and the QCD background amplitude in LHC searches.
We showed that the interference terms change the naive Breit-Wigner resonance peak to more like a peak-dip structure around the resonance mass.
However, the particular characteristic shape depends on the spin and the color of the digluon resonance.
The interference effects were studied for scalar and pseudo-scalar resonances in both singlet and octet color representations, spin-1 color-octets, and color-singlet massive gravitons.
To show the importance of the interference effects, we considered a few benchmark examples for various resonance masses, such that their production cross sections are close to the claimed exclusions of the CMS experiment in refs.~\cite{CMSdijet2019,CMSdijet2018}.

We found that the effects of interference were larger for the spin-0 color singlet case than for the
spin-0 color-octet and spin-1 color-octet cases, as was expected from consideration of the parton-level
differential cross-sections. We also note that the relative impact of the interference stays nearly constant for a fixed $M_X$ as the resonant cross-section decreases, but tends to increase
for larger $M_X$. It can also increase dramatically if the resonance has other decays contributing to its
width that are not detectable for some reason. 
Our results still contain significant uncertainties, in particular from the color-flow ambiguities in the color-singlet cases, and from the fact that in this paper we have not included any NLO effects. It would be interesting to go beyond the approximations used in this paper in order to reduce
these sources of uncertainty.

After performing a fit to the QCD background, the residual signal for a digluon resonance could have
a shape in the invariant mass distribution that appears to resemble a peak/dip, a shelf, an enhanced peak,
or even a pure dip. For all cases except spin-2, there is a large positive tail at invariant masses
below $M_X$. Because the magnitude of the QCD background amplitude falls with $\sqrt{\hat s}$, the low-mass tail tends to be more significant than the high-mass deficit from the interference. In the spin-2 case,
we found that the low mass tail is larger in magnitude but actually switches signs, and is negative far below $M_X$, due to interference between the QCD background and the $t$- and $u$-channel $X$ exchange. After QCD fitting and subtraction, this could lead to an enhanced peak in the spin-2 case, compared to the naive pure-resonance distribution expected if one neglects the interference.

More generally, the results found here point to the appropriateness of a flexible approach to searching for
dijet resonances. Although one is searching for a resonance, we have seen that there is a considerable diversity of possible invariant mass distribution signals even for resonances with rather narrow widths, depending on the resonance quantum numbers, width, and branching ratios, as can be seen from the figures above. Perhaps advanced data analysis and 
machine learning techniques can be brought to bear on the problem of identifying or 
setting limits on new-physics anomalies in mass distributions in a general and efficient way (see, for example, refs.~\cite{Collins:2019jip,Lillard:2019exp,Beauchesne:2019tpx,Nachman:2020lpy,Andreassen:2020nkr} for recent developments).
In any case, it seems necessary to consider a variety of possible anomalies in the dijet mass distribution, without undue prejudice towards a simple resonance peak.

{\it Acknowledgments:} P.N.B. thanks Olivier Mattelaer, Richard Efrain Ruiz, and Leif Gellersen
for helpful discussions at {\sc Madgraph} School 2019 (India).
P.N.B. thanks Sergey A.~Uzunyan
for his help and support in using the NICADD compute cluster
at Northern Illinois University.
P.N.B. also thanks Jeremiah Mitchell, Manoj Kumar Mandal, and  Chien-Yi Chen for technical advice.
S.P.M. thanks Zhen Liu for helpful discussions.
This work was supported in part by the National
Science Foundation grant number PHY-1719273.


\end{document}